\documentclass[submission, Phys]{SciPost}
\hypersetup{
    colorlinks,
    linkcolor={red!50!black},
    citecolor={blue!50!black},
    urlcolor={blue!80!black}
}

\usepackage[bitstream-charter]{mathdesign}
\DeclareSymbolFont{usualmathcal}{OMS}{cmsy}{m}{n}
\DeclareSymbolFontAlphabet{\mathcal}{usualmathcal}

\usepackage[T1]{fontenc} % if needed
\usepackage{amsmath,amsthm}
\usepackage{bbm}
\usepackage{graphicx}
\usepackage{dsfont}
\usepackage{relsize}
\usepackage{stmaryrd}
\usepackage{lmodern}
\usepackage{slantsc}
\usepackage{scalefnt}
\usepackage{xspace}
\usepackage{boxedminipage}
\usepackage{ifpdf}
\usepackage{multirow, hhline}
\usepackage{xifthen}
\usepackage{tensor}
\usepackage{array, nicematrix}
\usepackage{tabu}
\usepackage{pbox}
\usepackage{multirow}
\usepackage{tikz}
\usepackage{overpic}
\usepackage{color}
\usepackage{diagbox}
\usepackage{makecell}
\usetikzlibrary{arrows, matrix, calc, scopes, decorations.markings}
\usepackage{booktabs}
\usepackage{autobreak}
\usepackage{subfig}
\usepackage{diagbox}

\newtheorem*{ansatz*}{Ansatz}

\newcommand{\be}{\begin{equation}}
\newcommand{\ee}{\end{equation}}

\newcommand{\bse}{\begin{subequations}}
\newcommand{\ese}{\end{subequations}}
\newcommand{\ket}[1]{\left\vert{#1}\right\rangle}

\newcommand{\Z}{\mathbb{Z}}
\newcommand{\R}{\mathbb{R}}

\newcommand{\Hil}{\mathcal{H}}

\newcommand{\hs}{{\ }}

\newcommand{\bpm}{\begin{pmatrix}}
\newcommand{\epm}{\end{pmatrix}}
\newcommand{\bmm}{\begin{matrix}}
\newcommand{\emm}{\end{matrix}}

\newcommand{\dash}{{\hspace{1.5pt}\raisebox{0.3pt}{\text{-}}\hspace{1.5pt}}}

\usetikzlibrary{shapes.geometric, snakes}
\allowdisplaybreaks

\tikzset{>=latex}
\usetikzlibrary{arrows, matrix, calc, scopes, decorations.markings}
\usetikzlibrary{decorations.pathmorphing, positioning}
\tikzset{snake it/.style={decorate, decoration={snake,amplitude=0.2mm,segment length=1mm}}}
\tikzset{->-/.style={decoration={
			 markings,
			 mark=at position .5*\pgfdecoratedpathlength+2pt with {\arrow{>}}},postaction={decorate}}}

\tikzset{-<-/.style={decoration={
			 markings,
			 mark=at position .5*\pgfdecoratedpathlength+2pt with {\arrow{<}}},postaction={decorate}}}

\newtabulinestyle{dash=off 2pt}

{\null\,\vcenter\bgroup\scriptsize
\arraycolsep=.13885em
\hbox\bgroup$\array{@{}#1@{}}}
{\endarray$\egroup\egroup\,\null}

\newcolumntype{C}{>{\centering\arraybackslash} m{1.5em} }

\makeatletter
\newcommand*{\Relbarfill@}{\arrowfill@\Relbar\Relbar\Relbar}
\newcommand*{\xeq}[2][]{\ext@arrow 0055\Relbarfill@{#1}{#2}}
\makeatother
\allowdisplaybreaks[4]
\newcommand{\FigureModelLeft}{
\begin{tikzpicture}[baseline={([yshift=-.8ex]current bounding box.center)},scale=0.95]
    \draw [red, thick] (2.4,7.8) -- (2.8,7.8);
    \draw [red, thick] (2.4,7.4) -- (2.8,7.4);
    \draw [red, thick] (2.4,7.2) -- (2.4,5.6);
    \draw [red, thick] (2.4,6.4) -- (3.2,6.4);
    \draw [red, thick] (2.4,6.2) -- (2.8,6.2);
    \draw [red, thick] (2.4,5.8) -- (2.8,5.8);
    \draw [red, thick] (1.6,7.2) -- (2.4,7.2);
    \draw [red, thick] (1.6,7.) -- (2.,7.);
    \draw [red, thick] (1.6,6.6) -- (2.,6.6);
    \draw [red, thick] (1.6,6.4) -- (1.6,4.8);
    \draw [red, thick] (1.6,5.6) -- (2.4,5.6);
    \draw [red, thick] (1.6,5.4) -- (2.,5.4);
    \draw [red, thick] (1.6,5.) -- (2.,5.);
    \draw [red, thick] (1.6,4.) -- (2.4,4.);
    \draw [red, thick] (1.6,3.8) -- (2.,3.8);
    \draw [red, thick] (1.6,3.4) -- (2.,3.4);
    \draw [red, thick] (2.4,5.6) -- (2.4,4.);
    \draw [red, thick] (2.4,4.8) -- (3.2,4.8);
    \draw [red, thick] (2.4,4.6) -- (2.8,4.6);
    \draw [red, thick] (2.4,4.2) -- (2.8,4.2);
    \draw [red, thick] (2.4,3.2) -- (3.2,3.2);
    \draw [red, thick] (2.4,3.) -- (2.8,3.);
    \draw [red, thick] (3.2,7.2) -- (4.,7.2);
    \draw [red, thick] (3.2,7.) -- (3.6,7.);
    \draw [red, thick] (3.2,6.6) -- (3.6,6.6);
    \draw [red, thick] (3.2,6.4) -- (3.2,4.8);
    \draw [red, thick] (3.2,5.6) -- (4.,5.6);
    \draw [red, thick] (3.2,5.4) -- (3.6,5.4);
    \draw [red, thick] (3.2,5.) -- (3.6,5.);
    \draw [red, thick] (3.2,4.8) -- (3.2,3.2);
    \draw [red, thick] (3.2,4.) -- (4.,4.);
    \draw [red, thick] (3.2,3.8) -- (3.6,3.8);
    \draw [red, thick] (3.2,3.4) -- (3.6,3.4);
    \draw [red, thick] (4.,7.2) -- (4.,5.6);
    \draw [red, thick] (4.,6.4) -- (4.8,6.4);
    \draw [red, thick] (4.,6.2) -- (4.4,6.2);
    \draw [red, thick] (4.,5.8) -- (4.4,5.8);
    \draw [red, thick] (4.,7.8) -- (4.4,7.8);
    \draw [red, thick] (4.,7.4) -- (4.4,7.4);
    \draw [red, thick] (4.,5.6) -- (4.,4.);
    \draw [red, thick] (4.,4.8) -- (4.8,4.8);
    \draw [red, thick] (4.,4.6) -- (4.4,4.6);
    \draw [red, thick] (4.,4.2) -- (4.4,4.2);
    \draw [red, thick] (4.,3.2) -- (4.8,3.2);
    \draw [red, thick] (4.,3.) -- (4.4,3.);
    \draw [red, thick] (4.,8.) -- (4.8,8.);
    \draw [red, thick] (4.8,7.2) -- (5.6,7.2);
    \draw [red, thick] (4.8,7.) -- (5.2,7.);
    \draw [red, thick] (4.8,6.6) -- (5.2,6.6);
    \draw [red, thick] (4.8,6.4) -- (4.8,4.8);
    \draw [red, thick] (4.8,5.6) -- (5.6,5.6);
    \draw [red, thick] (4.8,5.4) -- (5.2,5.4);
    \draw [red, thick] (4.8,8.) -- (4.8,6.4);
    \draw [red, thick] (4.8,5.) -- (5.2,5.);
    \draw [red, thick] (4.8,4.8) -- (4.8,3.2);
    \draw [red, thick] (4.8,4.) -- (5.6,4.);
    \draw [red, thick] (4.8,3.8) -- (5.2,3.8);
    \draw [red, thick] (4.8,3.4) -- (5.2,3.4);
    \draw [red, thick] (5.6,7.2) -- (5.6,5.6);
    \draw [red, thick] (5.6,6.4) -- (6.4,6.4);
    \draw [red, thick] (5.6,6.2) -- (6.,6.2);
    \draw [red, thick] (5.6,5.8) -- (6.,5.8);
    \draw [red, thick] (5.6,8.) -- (6.4,8.);
    \draw [red, thick] (5.6,7.8) -- (6.,7.8);
    \draw [red, thick] (5.6,7.4) -- (6.,7.4);
    \draw [red, thick] (5.6,5.6) -- (5.6,4.);
    \draw [red, thick] (5.6,4.8) -- (6.4,4.8);
    \draw [red, thick] (5.6,4.6) -- (6.,4.6);
    \draw [red, thick] (5.6,4.2) -- (6.,4.2);
    \draw [red, thick] (5.6,3.2) -- (6.4,3.2);
    \draw [red, thick] (5.6,3.) -- (6.,3.);
    \draw [red, thick] (6.4,8.) -- (6.4,6.4);
    \draw [red, thick] (6.4,7.2) -- (7.2,7.2);
    \draw [red, thick] (6.4,7.) -- (6.8,7.);
    \draw [red, thick] (6.4,6.6) -- (6.8,6.6);
    \draw [red, thick] (6.4,6.4) -- (6.4,4.8);
    \draw [red, thick] (6.4,5.6) -- (7.2,5.6);
    \draw [red, thick] (6.4,5.4) -- (6.8,5.4);
    \draw [red, thick] (6.4,5.) -- (6.8,5.);
    \draw [red, thick] (6.4,4.8) -- (6.4,3.2);
    \draw [red, thick] (6.4,4.) -- (7.2,4.);
    \draw [red, thick] (6.4,3.8) -- (6.8,3.8);
    \draw [red, thick] (6.4,3.4) -- (6.8,3.4);
    \draw [red, thick] (2.4,8.) -- (3.2,8.);
    \draw [red, thick] (1.6,8.2) -- (2.,8.2);
    \draw [red, thick] (3.2,8.2) -- (3.6,8.2);
    \draw [red, thick] (4.8,8.2) -- (5.2,8.2);
    \draw [red, thick] (6.4,8.2) -- (6.9,8.2);
    \draw [red, thick] (7.2,7.2) -- (7.2,5.6);
    \draw [red, thick] (7.2,6.2) -- (7.6,6.2);
    \draw [red, thick] (7.2,5.8) -- (7.6,5.8);
    \draw [red, thick] (7.2,5.6) -- (7.2,4.);
    \draw [red, thick] (7.2,4.6) -- (7.6,4.6);
    \draw [red, thick] (7.2,4.2) -- (7.6,4.2);
    \draw [red, thick] (7.2,3.) -- (7.6,3.);
    \draw [red, thick] (7.2,7.8) -- (7.6,7.8);
    \draw [red, thick] (7.2,7.4) -- (7.6,7.4);
    \draw [red, thick] (7.2,8.) -- (7.8,8.);
    \draw [red, thick] (7.2,6.4) -- (7.8,6.4);
    \draw [red, thick] (7.2,4.8) -- (7.8,4.8);
    \draw [red, thick] (7.2,3.2) -- (7.8,3.2);
    \draw [red, thick] (7.2,4.) -- (7.2,2.6);
    \draw [red, thick] (6.4,3.2) -- (6.4,2.6);
    \draw [red, thick] (5.6,4.) -- (5.6,2.6);
    \draw [red, thick] (4.8,3.2) -- (4.8,2.6);
    \draw [red, thick] (1.6,6.4) -- (1.6,8.6);
    \draw [red, thick] (2.4,7.2) -- (2.4,8.6);
    \draw [red, thick] (3.2,6.4) -- (3.2,8.6);
    \draw [red, thick] (4.,7.2) -- (4.,8.6);
    \draw [red, thick] (4.8,8.) -- (4.8,8.6);
    \draw [red, thick] (5.6,7.2) -- (5.6,8.6);
    \draw [red, thick] (6.4,8.) -- (6.4,8.6);
    \draw [red, thick] (7.2,7.2) -- (7.2,8.6);
    \draw [red, thick] (4.,4.) -- (4.,2.6);
    \draw [red, thick] (3.2,3.2) -- (3.2,2.6);
    \draw [red, thick] (2.4,4.) -- (2.4,2.6);
    \draw [red, thick] (1.6,4.8) -- (1.6,2.6);
    \draw [red, thick] (1.,8.) -- (1.6,8.);
    \draw [red, thick] (1.,6.4) -- (1.6,6.4);
    \draw [red, thick] (1.,4.8) -- (1.6,4.8);
    \draw [red, thick] (1.,3.2) -- (1.6,3.2);
    \filldraw [red, opacity=0.2] (1.,2.6) -- (1.,8.6) -- (7.8,8.6)--(7.8,2.6)--cycle;
    \node [centered, rotate=0.] at (4.4,1.6) {\scriptsize {(a)}};
    \draw [white] (1.,9.3) -- (1.6,9.3);
\end{tikzpicture}
}

\newcommand{\FigureModelRight}{
\begin{tikzpicture}[baseline={([yshift=-.8ex]current bounding box.center)},scale=0.95]
    \draw [red, thick] (2.4,7.8) -- (2.8,7.8);
    \draw [red, thick] (2.4,7.4) -- (2.8,7.4);
    \draw [red, thick] (2.4,7.2) -- (2.4,5.6);
    \draw [red, thick] (2.4,6.4) -- (3.2,6.4);
    \draw [red, thick] (2.4,6.2) -- (2.8,6.2);
    \draw [red, thick] (2.4,5.8) -- (2.8,5.8);
    \draw [red, thick] (1.6,7.2) -- (2.4,7.2);
    \draw [red, thick] (1.6,7.) -- (2.,7.);
    \draw [red, thick] (1.6,6.6) -- (2.,6.6);
    \draw [red, thick] (1.6,6.4) -- (1.6,4.8);
    \draw [red, thick] (1.6,5.6) -- (2.4,5.6);
    \draw [red, thick] (1.6,5.4) -- (2.,5.4);
    \draw [red, thick] (1.6,5.) -- (2.,5.);
    \draw [red, thick] (1.6,4.) -- (2.4,4.);
    \draw [red, thick] (1.6,3.8) -- (2.,3.8);
    \draw [red, thick] (1.6,3.4) -- (2.,3.4);
    \draw [red, thick] (2.4,5.6) -- (2.4,4.);
    \draw [red, thick] (2.4,4.8) -- (3.2,4.8);
    \draw [red, thick] (2.4,4.6) -- (2.8,4.6);
    \draw [red, thick] (2.4,4.2) -- (2.8,4.2);
    \draw [red, thick] (2.4,3.2) -- (3.2,3.2);
    \draw [red, thick] (2.4,3.) -- (2.8,3.);
    \draw [red, thick] (3.2,7.2) -- (4.,7.2);
    \draw [red, thick] (3.2,7.) -- (3.6,7.);
    \draw [red, thick] (3.2,6.6) -- (3.6,6.6);
    \draw [red, thick] (3.2,6.4) -- (3.2,4.8);
    \draw [red, thick] (3.2,5.6) -- (4.,5.6);
    \draw [red, thick] (3.2,5.4) -- (3.6,5.4);
    \draw [red, thick] (3.2,5.) -- (3.6,5.);
    \draw [red, thick] (3.2,4.8) -- (3.2,3.2);
    \draw [red, thick] (3.2,4.) -- (4.,4.);
    \draw [red, thick] (3.2,3.8) -- (3.6,3.8);
    \draw [red, thick] (3.2,3.4) -- (3.6,3.4);
    \draw [red, thick] (4.,7.2) -- (4.,5.6);
    \draw [red, thick] (4.,6.4) -- (4.8,6.4);
    \draw [red, thick] (4.,6.2) -- (4.4,6.2);
    \draw [red, thick] (4.,5.8) -- (4.4,5.8);
    \draw [red, thick] (4.,7.8) -- (4.4,7.8);
    \draw [red, thick] (4.,7.4) -- (4.4,7.4);
    \draw [red, thick] (4.,5.6) -- (4.,4.);
    \draw [red, thick] (4.,4.8) -- (4.8,4.8);
    \draw [red, thick] (4.,4.6) -- (4.4,4.6);
    \draw [red, thick] (4.,4.2) -- (4.4,4.2);
    \draw [red, thick] (4.,3.2) -- (4.8,3.2);
    \draw [red, thick] (4.,3.) -- (4.4,3.);
    \draw [red, thick] (4.,8.) -- (4.8,8.);
    \draw [blue] (4.8,8.) -- (4.8,6.4);
    \draw [blue] (4.8,7.2) -- (5.6,7.2);
    \draw [blue] (4.8,7.) -- (5.2,7.);
    \draw [blue] (4.8,6.6) -- (5.2,6.6);
    \draw [blue] (4.8,6.4) -- (4.8,4.8);
    \draw [blue] (4.8,5.6) -- (5.6,5.6);
    \draw [blue] (4.8,5.4) -- (5.2,5.4);
    \draw [blue] (4.8,5.) -- (5.2,5.);
    \draw [blue] (4.8,4.8) -- (4.8,3.2);
    \draw [blue] (4.8,4.) -- (5.6,4.);
    \draw [blue] (4.8,3.8) -- (5.2,3.8);
    \draw [blue] (4.8,3.4) -- (5.2,3.4);
    \draw [blue] (5.6,7.2) -- (5.6,5.6);
    \draw [blue] (5.6,6.4) -- (6.4,6.4);
    \draw [blue] (5.6,6.2) -- (6.,6.2);
    \draw [blue] (5.6,5.8) -- (6.,5.8);
    \draw [blue] (5.6,8.) -- (6.4,8.);
    \draw [blue] (5.6,7.8) -- (6.,7.8);
    \draw [blue] (5.6,7.4) -- (6.,7.4);
    \draw [blue] (5.6,5.6) -- (5.6,4.);
    \draw [blue] (5.6,4.8) -- (6.4,4.8);
    \draw [blue] (5.6,4.6) -- (6.,4.6);
    \draw [blue] (5.6,4.2) -- (6.,4.2);
    \draw [blue] (5.6,3.2) -- (6.4,3.2);
    \draw [blue] (5.6,3.) -- (6.,3.);
    \draw [blue] (6.4,8.) -- (6.4,6.4);
    \draw [blue] (6.4,7.2) -- (7.2,7.2);
    \draw [blue] (6.4,7.) -- (6.8,7.);
    \draw [blue] (6.4,6.6) -- (6.8,6.6);
    \draw [blue] (6.4,6.4) -- (6.4,4.8);
    \draw [blue] (6.4,5.6) -- (7.2,5.6);
    \draw [blue] (6.4,5.4) -- (6.8,5.4);
    \draw [blue] (6.4,5.) -- (6.8,5.);
    \draw [blue] (6.4,4.8) -- (6.4,3.2);
    \draw [blue] (6.4,4.) -- (7.2,4.);
    \draw [blue] (6.4,3.8) -- (6.8,3.8);
    \draw [blue] (6.4,3.4) -- (6.8,3.4);
    \draw [red, thick] (2.4,8.) -- (3.2,8.);
    \draw [red, thick] (1.6,8.2) -- (2.,8.2);
    \draw [red, thick] (3.2,8.2) -- (3.6,8.2);
    \draw [blue] (4.8,8.2) -- (5.2,8.2);
    \draw [blue] (6.4,8.2) -- (6.9,8.2);
    \draw [blue] (7.2,7.2) -- (7.2,5.6);
    \draw [blue] (7.2,6.2) -- (7.6,6.2);
    \draw [blue] (7.2,5.8) -- (7.6,5.8);
    \draw [blue] (7.2,5.6) -- (7.2,4.);
    \draw [blue] (7.2,4.6) -- (7.6,4.6);
    \draw [blue] (7.2,4.2) -- (7.6,4.2);
    \draw [blue] (7.2,3.) -- (7.6,3.);
    \draw [blue] (7.2,7.8) -- (7.6,7.8);
    \draw [blue] (7.2,7.4) -- (7.6,7.4);
    \draw [blue] (7.2,8.) -- (7.8,8.);
    \draw [blue] (7.2,6.4) -- (7.8,6.4);
    \draw [blue] (7.2,4.8) -- (7.8,4.8);
    \draw [blue] (7.2,3.2) -- (7.8,3.2);
    \draw [blue] (7.2,4.) -- (7.2,2.6);
    \draw [blue] (6.4,3.2) -- (6.4,2.6);
    \draw [blue] (5.6,4.) -- (5.6,2.6);
    \draw [blue] (4.8,3.2) -- (4.8,2.6);
    \draw [red, thick] (1.6,6.4) -- (1.6,8.6);
    \draw [red, thick] (2.4,7.2) -- (2.4,8.6);
    \draw [red, thick] (3.2,6.4) -- (3.2,8.6);
    \draw [red, thick] (4.,7.2) -- (4.,8.6);
    \draw [blue] (4.8,8.) -- (4.8,8.6);
    \draw [blue] (5.6,7.2) -- (5.6,8.6);
    \draw [blue] (6.4,8.) -- (6.4,8.6);
    \draw [blue] (7.2,7.2) -- (7.2,8.6);
    \draw [red, thick] (4.,4.) -- (4.,2.6);
    \draw [red, thick] (3.2,3.2) -- (3.2,2.6);
    \draw [red, thick] (2.4,4.) -- (2.4,2.6);
    \draw [red, thick] (1.6,4.8) -- (1.6,2.6);
    \draw [red, thick] (1.,8.) -- (1.6,8.);
    \draw [red, thick] (1.,6.4) -- (1.6,6.4);
    \draw [red, thick] (1.,4.8) -- (1.6,4.8);
    \draw [red, thick] (1.,3.2) -- (1.6,3.2);
    \filldraw [red, opacity=0.2] (1.,2.6) -- (1.,8.6) -- (4.,8.6)--(4.,2.6)--cycle;
    \filldraw [blue, opacity=0.2] (4.8,2.6) -- (4.8,8.6) -- (7.8,8.6)--(7.8,2.6)--cycle;
    \filldraw [black, opacity=0.2] (4.,2.6) -- (4.,8.6) -- (4.8,8.6)--(4.8,2.6)--cycle;
    \node [centered, rotate=0.] at (2.45,2.3) {\scriptsize {Doubled Ising domain}};
    \node [centered, rotate=0.] at (6.4,2.3) {\scriptsize {Toric code domain}};
    \node [centered, rotate=0.] at (4.4,8.9) {\scriptsize {Domain wall}};
    \node [centered, rotate=0.] at (4.4,1.6) {\scriptsize {(b)}};
    \draw [white] (1.,9.3) -- (1.6,9.3);
\end{tikzpicture}
}

\newcommand{\EqQv}{
\hs\begin{tikzpicture}[baseline={([yshift=-.8ex]current bounding box.center)},scale=.4]
    \draw [] (3.2,9.) -- (3.2,5.8);
    \draw [] (6.3,9.) -- (7.1,7.4);
    \draw [] (6.3,5.8) -- (7.1,7.4);
    \draw [] (3.4,8.8) -- (4.8,8.);
    \draw [] (6.2,8.8) -- (4.8,8.);
    \draw [] (4.8,7.) -- (4.8,8.);
    \draw [] (4.8,6.) -- (4.8,7.);
    \draw [] (4.8,7.) -- (6.,7.);
    \node [centered, rotate=0.] at (6.,7.4) {\scriptsize $p$};
    \node [centered, rotate=0.] at (4.9,8.6) {\scriptsize $V$};
\end{tikzpicture}\hs
}

\newcommand{\EqBpsLeft}{
\hs\begin{tikzpicture}[baseline={([yshift=-.8ex]current bounding box.center)},scale=.4]
    \draw [] (2.,6.4) -- (2.,5.6);
    \draw [] (2.,5.6) -- (2.,4.8);
    \draw [] (2.,4.8) -- (2.,4.);
    \draw [] (2.,4.) -- (4.,2.8);
    \draw [] (4.,2.8) -- (6.,4.);
    \draw [] (6.,4.) -- (6.,4.8);
    \draw [] (6.,4.8) -- (6.,5.6);
    \draw [] (6.,5.6) -- (6.,6.4);
    \draw [] (6.,6.4) -- (4.,7.6);
    \draw [] (4.,7.6) -- (2.,6.4);
    \draw [] (0.8,7.2) -- (2.,6.4);
    \draw [] (0.8,3.2) -- (2.,4.);
    \draw [] (7.2,3.2) -- (6.,4.);
    \draw [] (7.2,7.2) -- (6.,6.4);
    \draw [] (4.,8.8) -- (4.,7.6);
    \draw [] (4.,1.6) -- (4.,2.8);
    \draw [] (2.,5.6) -- (3.2,5.6);
    \draw [] (2.,4.8) -- (3.2,4.8);
    \draw [] (6.,5.6) -- (7.2,5.6);
    \draw [] (6.,4.8) -- (7.2,4.8);
    \draw [] (7.6,8.8) -- (9.6,5.2);
    \draw [] (9.6,5.2) -- (7.6,1.6);
    \draw [] (0.,8.8) -- (0.,1.6);
    \node [centered, rotate=0.] at (1.4,6.) {\scriptsize $i_0$};
    \node [centered, rotate=0.] at (1.4,5.2) {\scriptsize $k_0$};
    \node [centered, rotate=0.] at (1.4,4.4) {\scriptsize $l_0$};
    \node [centered, rotate=0.] at (2.6,3.) {\scriptsize $i_1$};
    \node [centered, rotate=0.] at (5.4,3.) {\scriptsize $i_2$};
    \node [centered, rotate=0.] at (5.4,4.4) {\scriptsize $i_3$};
    \node [centered, rotate=0.] at (5.4,5.2) {\scriptsize $i_4$};
    \node [centered, rotate=0.] at (5.4,6.) {\scriptsize $i_5$};
    \node [centered, rotate=0.] at (5.4,7.4) {\scriptsize $i_6$};
    \node [centered, rotate=0.] at (2.6,7.4) {\scriptsize $i_7$};
    \node [centered, rotate=0.] at (0.8,7.7) {\scriptsize $e_0$};
    \node [centered, rotate=0.] at (0.8,2.7) {\scriptsize $e_1$};
    \node [centered, rotate=0.] at (4.6,2.) {\scriptsize $e_2$};
    \node [centered, rotate=0.] at (7.2,2.7) {\scriptsize $e_3$};
    \node [centered, rotate=0.] at (7.2,7.7) {\scriptsize $e_6$};
    \node [centered, rotate=0.] at (3.3,8.4) {\scriptsize $e_7$};
    \node [centered, rotate=0.] at (3.8,5.8) {\scriptsize $p$};
    \node [centered, rotate=0.] at (3.8,4.7) {\scriptsize $q$};
    \node [centered, rotate=0.] at (7.8,4.7) {\scriptsize $e_4$};
    \node [centered, rotate=0.] at (7.8,5.8) {\scriptsize $e_5$};
\end{tikzpicture}\hs
}

\newcommand{\EqBpsRight}{ 
\hs\begin{tikzpicture}[baseline={([yshift=-.8ex]current bounding box.center)},scale=.4]
    \draw [] (2.,6.4) -- (2.,5.6);
    \draw [] (2.,5.6) -- (2.,4.8);
    \draw [] (2.,4.8) -- (2.,4.);
    \draw [] (2.,4.) -- (4.,2.8);
    \draw [] (4.,2.8) -- (6.,4.);
    \draw [] (6.,4.) -- (6.,4.8);
    \draw [] (6.,4.8) -- (6.,5.6);
    \draw [] (6.,5.6) -- (6.,6.4);
    \draw [] (6.,6.4) -- (4.,7.6);
    \draw [] (4.,7.6) -- (2.,6.4);
    \draw [] (0.8,7.2) -- (2.,6.4);
    \draw [] (0.8,3.2) -- (2.,4.);
    \draw [] (7.2,3.2) -- (6.,4.);
    \draw [] (7.2,7.2) -- (6.,6.4);
    \draw [] (4.,8.8) -- (4.,7.6);
    \draw [] (4.,1.6) -- (4.,2.8);
    \draw [] (2.,5.6) -- (3.2,5.6);
    \draw [] (2.,4.8) -- (3.2,4.8);
    \draw [] (6.,5.6) -- (7.2,5.6);
    \draw [] (6.,4.8) -- (7.2,4.8);
    \draw [] (7.6,8.8) -- (9.6,5.2);
    \draw [] (9.6,5.2) -- (7.6,1.6);
    \draw [] (0.,8.8) -- (0.,1.6);
    \node [centered, rotate=0.] at (1.4,6.) {\scriptsize $j_0$};
    \node [centered, rotate=0.] at (1.4,5.2) {\scriptsize $j_0$};
    \node [centered, rotate=0.] at (1.4,4.4) {\scriptsize $j_0$};
    \node [centered, rotate=0.] at (2.6,3.) {\scriptsize $j_1$};
    \node [centered, rotate=0.] at (5.4,3.) {\scriptsize $j_2$};
    \node [centered, rotate=0.] at (5.4,4.4) {\scriptsize $j_3$};
    \node [centered, rotate=0.] at (5.4,5.2) {\scriptsize $j_4$};
    \node [centered, rotate=0.] at (5.4,6.) {\scriptsize $j_5$};
    \node [centered, rotate=0.] at (5.4,7.4) {\scriptsize $j_6$};
    \node [centered, rotate=0.] at (2.6,7.4) {\scriptsize $j_7$};
    \node [centered, rotate=0.] at (0.8,7.7) {\scriptsize $e_0$};
    \node [centered, rotate=0.] at (0.8,2.7) {\scriptsize $e_1$};
    \node [centered, rotate=0.] at (4.6,2.) {\scriptsize $e_2$};
    \node [centered, rotate=0.] at (7.2,2.7) {\scriptsize $e_3$};
    \node [centered, rotate=0.] at (7.2,7.7) {\scriptsize $e_4$};
    \node [centered, rotate=0.] at (3.3,8.4) {\scriptsize $e_5$};
    \node [centered, rotate=0.] at (3.8,5.8) {\scriptsize $1$};
    \node [centered, rotate=0.] at (3.8,4.7) {\scriptsize $1$};
    \node [centered, rotate=0.] at (7.8,4.7) {\scriptsize $e_4$};
    \node [centered, rotate=0.] at (7.8,5.8) {\scriptsize $e_5$};
\end{tikzpicture}\hs
}

\newcommand{\EqConstraint}{
\hs\begin{tikzpicture}[baseline={([yshift=-.8ex]current bounding box.center)},scale=.7]
    \draw [thick] (4.8,8.) -- (4.8,6.4);
    \draw [thick] (4.8,8.) -- (4.2,8.4);
    \draw [thick] (4.8,8.) -- (5.4,8.4);
    \draw [thick] (4.8,6.4) -- (4.2,6.);
    \draw [thick] (4.8,6.4) -- (5.4,6.);
    \draw [thick] (4.8,6.8) -- (5.8,6.8);
    \draw [thick] (4.8,7.6) -- (5.8,7.6);
    \draw [] (3.8,8.6) -- (3.8,5.8);
    \draw [] (5.6,8.6) -- (6.4,7.2);
    \draw [] (6.4,7.2) -- (5.6,5.8);
    \node [centered, rotate=0.] at (4.5,7.2) {\scriptsize $j_E$};
    \node [centered, rotate=0.] at (4.4,7.8) {\scriptsize $i$};
    \node [centered, rotate=0.] at (4.5,6.6) {\scriptsize $k$};
\end{tikzpicture}\hs
}

\newcommand{\FigureRibbonLeft}{
\begin{tikzpicture}[baseline={([yshift=-.8ex]current bounding box.center)},scale=1.]
    \draw [red, thick] (1.6,8.6) -- (1.6,4.2);
    \draw [red, thick] (1.6,8.) -- (2.4,8.);
    \draw [red, thick] (1.6,6.4) -- (2.4,6.4);
    \draw [red, thick] (1.6,4.8) -- (2.4,4.8);
    \draw [red, thick] (2.4,8.6) -- (2.4,4.2);
    \draw [red, thick] (2.4,7.2) -- (3.2,7.2);
    \draw [red, thick] (2.4,5.6) -- (3.2,5.6);
    \draw [red, thick] (3.2,8.6) -- (3.2,4.2);
    \draw [red, thick] (3.2,8.) -- (4.,8.);
    \draw [red, thick] (3.2,6.4) -- (4.,6.4);
    \draw [red, thick] (3.2,4.8) -- (4.,4.8);
    \draw [red, thick] (1.6,7.2) -- (1.,7.2);
    \draw [red, thick] (1.6,5.6) -- (1.,5.6);
    \draw [red, thick] (4.,8.6) -- (4.,4.2);
    \draw [red, thick] (4.,7.2) -- (4.8,7.2);
    \draw [red, thick] (4.,5.6) -- (4.8,5.6);
    \draw [red, thick] (4.8,8.6) -- (4.8,4.2);
    \draw [red, thick] (4.8,8.) -- (5.6,8.);
    \draw [red, thick] (4.8,6.4) -- (5.6,6.4);
    \draw [red, thick] (4.8,4.8) -- (5.6,4.8);
    \draw [red, thick] (5.6,8.6) -- (5.6,4.2);
    \draw [red, thick] (5.6,7.2) -- (6.2,7.2);
    \draw [red, thick] (5.6,5.6) -- (6.2,5.6);
    \draw [ultra thick] (2.,7.2) -- (2.,6.);
    \draw [ultra thick] (2.,6.) -- (4.4,6.);
    \node [centered, rotate=0.] at (2.,7.4) {\scriptsize $(J_\text{DI},p)$};
    \node [centered, rotate=0.] at (5.,6.) {\scriptsize $(J_\text{DI},q)$};
    \node [centered, rotate=0.] at (3.6,3.75) {\scriptsize {(a)}};
    \filldraw[red,opacity=0.2] (1.,8.6) -- (1.,4.2) -- (3.2,4.2) -- (3.2, 8.6);
    \filldraw[red,opacity=0.2] (4.,8.6) -- (4.,4.2) -- (3.2,4.2) -- (3.2, 8.6);
    \filldraw[red,opacity=0.2] (4.,8.6) -- (4.,4.2) -- (6.2,4.2) -- (6.2, 8.6);
\end{tikzpicture}
}

\newcommand{\FigureRibbonRight}{
\begin{tikzpicture}[baseline={([yshift=-.8ex]current bounding box.center)},scale=1.]
    \draw [red, thick] (1.6,8.6) -- (1.6,4.2);
    \draw [red, thick] (1.6,8.) -- (2.4,8.);
    \draw [red, thick] (1.6,6.4) -- (2.4,6.4);
    \draw [red, thick] (1.6,4.8) -- (2.4,4.8);
    \draw [red, thick] (2.4,8.6) -- (2.4,4.2);
    \draw [red, thick] (2.4,7.2) -- (3.2,7.2);
    \draw [red, thick] (2.4,5.6) -- (3.2,5.6);
    \draw [red, thick] (3.2,8.6) -- (3.2,4.2);
    \draw [red, thick] (3.2,8.) -- (4.,8.);
    \draw [red, thick] (3.2,6.4) -- (4.,6.4);
    \draw [red, thick] (3.2,4.8) -- (4.,4.8);
    \draw [red, thick] (1.6,7.2) -- (1.,7.2);
    \draw [red, thick] (1.6,5.6) -- (1.,5.6);
    \draw [red, thick] (4.,8.6) -- (4.,4.2);
    \draw [red, thick] (4.,7.2) -- (4.8,7.2);
    \draw [red, thick] (4.,5.6) -- (4.8,5.6);
    \draw [red, thick] (4.8,8.6) -- (4.8,4.2);
    \draw [red, thick] (4.8,8.) -- (5.6,8.);
    \draw [red, thick] (4.8,6.4) -- (5.6,6.4);
    \draw [red, thick] (4.8,4.8) -- (5.6,4.8);
    \draw [red, thick] (5.6,8.6) -- (5.6,4.2);
    \draw [red, thick] (5.6,7.2) -- (6.2,7.2);
    \draw [red, thick] (5.6,5.6) -- (6.2,5.6);
    \draw [ultra thick] (2.,7.2) -- (2.,5.2);
    \draw [ultra thick] (2.,5.2) -- (2.8,5.2);
    \draw [ultra thick] (2.8,5.2) -- (2.8,8.4);
    \draw [ultra thick] (2.8,8.4) -- (5.2,8.4);
    \draw [ultra thick] (5.2,8.4) -- (5.2,6.8);
    \draw [ultra thick] (5.2,6.8) -- (3.6,6.8);
    \draw [ultra thick] (3.6,6.8) -- (3.6,6.);
    \draw [ultra thick] (3.6,6.) -- (4.4,6.);
    \node [centered, rotate=0.] at (2.,7.4) {\scriptsize $(J_\text{DI},p)$};
    \node [centered, rotate=0.] at (5.,6.) {\scriptsize $(J_\text{DI},q)$};
    \node [centered, rotate=0.] at (3.6,3.75) {\scriptsize {(b)}};
    \filldraw[red,opacity=0.2] (1.,8.6) -- (1.,4.2) -- (3.2,4.2) -- (3.2, 8.6);
    \filldraw[red,opacity=0.2] (4.,8.6) -- (4.,4.2) -- (3.2,4.2) -- (3.2, 8.6);
    \filldraw[red,opacity=0.2] (4.,8.6) -- (4.,4.2) -- (6.2,4.2) -- (6.2, 8.6);
\end{tikzpicture}
}

\newcommand{\EqDoubledIsingState}{
\hs\begin{tikzpicture}[baseline={([yshift=-.8ex]current bounding box.center)},scale=.43]
    \draw [] (1.,3.8) -- (1.,9.);
    \draw [] (8.4,3.8) -- (9.7,6.4);
    \draw [] (9.7,6.4) -- (8.4,9.);
    \draw [red,thick] (2.,7.2) -- (2.,5.6);
    \draw [red,thick] (2.,5.6) -- (3.4,4.8);
    \draw [red,thick] (3.4,4.8) -- (4.8,5.6);
    \draw [red,thick] (4.8,5.6) -- (4.8,7.2);
    \draw [red,thick] (4.8,7.2) -- (3.4,8.);
    \draw [red,thick] (3.4,8.) -- (2.,7.2);
    \draw [red,thick] (3.4,8.6) -- (3.4,8.);
    \draw [red,thick] (1.3,7.6) -- (2.,7.2);
    \draw [red,thick] (1.3,5.2) -- (2.,5.6);
    \draw [red,thick] (3.4,4.2) -- (3.4,4.8);
    \draw [red,thick] (6.2,4.8) -- (4.8,5.6);
    \draw [red,thick] (6.2,4.2) -- (6.2,4.8);
    \draw [red,thick] (4.8,7.2) -- (6.2,8.);
    \draw [red,thick] (7.6,5.6) -- (6.2,4.8);
    \draw [ultra thick] (3.4,6.4) -- (6.2,6.4);
    \draw [red,thick] (6.2,8.) -- (7.6,7.2);
    \draw [red,thick] (7.6,7.2) -- (7.6,5.6);
    \draw [red,thick] (8.3,5.2) -- (7.6,5.6);
    \draw [red,thick] (8.3,7.6) -- (7.6,7.2);
    \draw [red,thick] (6.2,8.6) -- (6.2,8.);
    \filldraw [red, opacity=0.2] (1.3,5.2) -- (1.3, 8.6) -- (3.4,8.6) -- (3.4, 8.) -- (4.8,7.2) -- (4.8,5.6) -- (6.2,4.8) -- (6.2,4.2) -- (1.3, 4.2);
    \filldraw [red, opacity=0.2] (3.4,8.6) -- (3.4, 8.) -- (4.8,7.2) -- (4.8,5.6) -- (6.2,4.8) -- (6.2,4.2) -- (8.3,4.2) -- (8.3,5.2) -- (7.6,5.6) -- (7.6,7.2) -- (6.2,8.) -- (6.2,8.6);
    \filldraw [red, opacity=0.2] (8.3,5.2) -- (7.6,5.6) -- (7.6,7.2) -- (6.2,8.) -- (6.2,8.6) -- (8.3, 8.6);
    \node [centered, rotate=0.] at (2.15,6.4) {\scriptsize $(J_\text{DI},p)$};
    \node [centered, rotate=0.] at (7.4,6.4) {\scriptsize $(J_\text{DI},q)$};
    \node [centered, rotate=0.] at (4.4,6.) {\scriptsize $E$};
\end{tikzpicture}\hs
}

\newcommand{\EqDoubledIsingRibbonLeft}{
\hs\begin{tikzpicture}[baseline={([yshift=-.8ex]current bounding box.center)},scale=.43]
    \draw [] (1.,3.8) -- (1.,9.);
    \draw [] (8.4,3.8) -- (9.7,6.4);
    \draw [] (9.7,6.4) -- (8.4,9.);
    \draw [red,thick] (3.4,4.8) -- (4.8,5.6);
    \draw [red,thick] (1.3,5.2) -- (2.,5.6);
    \draw [red,thick] (3.4,4.2) -- (3.4,4.8);
    \draw [red,thick] (6.2,4.2) -- (6.2,4.8);
    \draw [red,thick] (8.3,5.2) -- (7.6,5.6);
    \draw [red,thick] (6.2,4.8) -- (4.8,5.6);
    \draw [red,thick] (7.6,5.6) -- (6.2,4.8);
    \draw [red,thick] (2.,5.6) -- (3.4,4.8);
    \draw [red,thick] (1.3,7.6) -- (2.,7.2);
    \draw [red,thick] (3.4,8.) -- (2.,7.2);
    \draw [red,thick] (3.4,8.6) -- (3.4,8.);
    \draw [red,thick] (4.8,7.2) -- (3.4,8.);
    \draw [red,thick] (8.3,7.6) -- (7.6,7.2);
    \draw [red,thick] (4.8,7.2) -- (6.2,8.);
    \draw [red,thick] (6.2,8.) -- (7.6,7.2);
    \draw [red,thick] (6.2,8.6) -- (6.2,8.);
    \draw [red,thick] (7.6,7.2) -- (7.6,5.6);
    \draw [red,thick] (4.8,5.6) -- (4.8,7.2);
    \draw [red,thick] (2.,7.2) -- (2.,5.6);
    \filldraw [red, opacity=0.2] (1.3,5.2) -- (2.,5.6) -- (3.4,4.8) -- (4.8,5.6) -- (6.2,4.8) -- (7.6,5.6) --(8.3,5.2) -- (8.3,4.2) -- (1.3,4.2);
    \filldraw [red, opacity=0.2] (1.3,7.6) -- (2.,7.2) -- (3.4,8.) -- (4.8,7.2) -- (6.2,8.) -- (7.6,7.2)--(8.3,7.6) --(8.3,5.2) -- (7.6,5.6) -- (6.2,4.8) -- (4.8,5.6) -- (3.4,4.8) -- (2.,5.6) -- (1.3,5.2);
    \filldraw [red, opacity=0.2] (1.3,7.6) -- (2.,7.2) -- (3.4,8.) -- (4.8,7.2) -- (6.2,8.) -- (7.6,7.2)--(8.3,7.6) -- (8.3,8.6) -- (1.3,8.6);
    \node [centered, rotate=0.] at (4.2,6.4) {\scriptsize $j_E$};
    \end{tikzpicture}\hs
}

\newcommand{\EqDoubledIsingRibbonRight}{
\hs\begin{tikzpicture}[baseline={([yshift=-.8ex]current bounding box.center)},scale=.43]
    \draw [] (1.,3.8) -- (1.,9.);
    \draw [] (8.4,3.8) -- (9.7,6.4);
    \draw [] (9.7,6.4) -- (8.4,9.);
    \draw [red,thick] (3.4,4.8) -- (4.8,5.6);
    \draw [red,thick] (1.3,5.2) -- (2.,5.6);
    \draw [red,thick] (3.4,4.2) -- (3.4,4.8);
    \draw [red,thick] (6.2,4.2) -- (6.2,4.8);
    \draw [red,thick] (8.3,5.2) -- (7.6,5.6);
    \draw [red,thick] (6.2,4.8) -- (4.8,5.6);
    \draw [red,thick] (7.6,5.6) -- (6.2,4.8);
    \draw [red,thick] (2.,5.6) -- (3.4,4.8);
    \draw [red,thick] (1.3,7.6) -- (2.,7.2);
    \draw [red,thick] (3.4,8.) -- (2.,7.2);
    \draw [red,thick] (3.4,8.6) -- (3.4,8.);
    \draw [red,thick] (4.8,7.2) -- (3.4,8.);
    \draw [red,thick] (8.3,7.6) -- (7.6,7.2);
    \draw [red,thick] (4.8,7.2) -- (6.2,8.);
    \draw [red,thick] (6.2,8.) -- (7.6,7.2);
    \draw [red,thick] (6.2,8.6) -- (6.2,8.);
    \draw [red,thick] (4.8,5.6) -- (4.8,7.2);
    \draw [red,thick] (7.6,7.2) -- (7.6,5.6);
    \draw [red,thick] (2.,7.2) -- (2.,5.6);
    \draw [red,thick] (4.8,6.1) -- (3.6,6.1);
    \draw [red,thick] (4.8,6.7) -- (6.,6.7);
    \draw [red,thick] (6.2,4.8) -- (4.8,5.6);
    \filldraw [red, opacity=0.2] (1.3,5.2) -- (2.,5.6) -- (3.4,4.8) -- (4.8,5.6) -- (6.2,4.8) -- (7.6,5.6) --(8.3,5.2) -- (8.3,4.2) -- (1.3,4.2);
    \filldraw [red, opacity=0.2] (1.3,7.6) -- (2.,7.2) -- (3.4,8.) -- (4.8,7.2) -- (6.2,8.) -- (7.6,7.2)--(8.3,7.6) --(8.3,5.2) -- (7.6,5.6) -- (6.2,4.8) -- (4.8,5.6) -- (3.4,4.8) -- (2.,5.6) -- (1.3,5.2);
    \filldraw [red, opacity=0.2] (1.3,7.6) -- (2.,7.2) -- (3.4,8.) -- (4.8,7.2) -- (6.2,8.) -- (7.6,7.2)--(8.3,7.6) -- (8.3,8.6) -- (1.3,8.6);
    \node [centered, rotate=0.] at (4.4,7) {\scriptsize $j_E$};
    \node [centered, rotate=0.] at (5.3,5.8) {\scriptsize $j_E$};
    \node [centered, rotate=0.] at (4.4,6.4) {\scriptsize $k$};
    \node [centered, rotate=0.] at (3.3,6.1) {\scriptsize $p$};
    \node [centered, rotate=0.] at (6.3,6.7) {\scriptsize $q$};
\end{tikzpicture}\hs
}

\newcommand{\TableQuasiparticleDI}{
\begin{tabular}{|c|c|c|}
    \hline Quasiparticle & Anyon species & Charge \\ \hline
    $(1\overline{1}, 1)$ & $1\overline{1}$ & $1$ \\ \hline
    $(1\overline{\sigma}, \sigma)$ & $1\overline{\sigma}$ & $\sigma$ \\ \hline
    $(1\overline{\psi}, \psi)$ & $1\overline{\psi}$ & $\psi$ \\ \hline
    $(\sigma\overline{1}, \sigma)$ & $\sigma\overline{1}$ & $\sigma$ \\ \hline
    $(\sigma\overline{\sigma}, 1)$ & \multirow{2}{*}{$\sigma\overline{\sigma}$} & $1$ \\ \cline{1-1} \cline{3-3}
    $(\sigma\overline{\sigma}, \psi)$ & & $\psi$ \\ \hline
    $(\sigma\overline{\psi}, \sigma)$ & $\sigma\overline{\psi}$ & $\sigma$ \\ \hline
    $(\psi\overline{1}, \psi)$ & $\psi\overline{1}$ & $\psi$ \\ \hline
    $(\psi\overline{\sigma}, \sigma)$ & $\psi\overline{\sigma}$ & $\sigma$ \\ \hline
    $(\psi\overline{\psi}, 1)$ & $\psi\overline{\psi}$ & $1$ \\\hline
\end{tabular}
}

\newcommand{\FigureSigmaSigmaBarA}{
\begin{tikzpicture}[baseline={([yshift=-.8ex]current bounding box.center)},scale=.49]
    \draw [red,thick] (3.4,4.8) -- (4.8,5.6);
    \draw [red,thick] (1.3,5.2) -- (2.,5.6);
    \draw [red,thick] (3.4,4.2) -- (3.4,4.8);
    \draw [red,thick] (6.2,4.2) -- (6.2,4.8);
    \draw [red,thick] (8.3,5.2) -- (7.6,5.6);
    \draw [red,thick] (6.2,4.8) -- (4.8,5.6);
    \draw [red,thick] (7.6,5.6) -- (6.2,4.8);
    \draw [red,thick] (2.,5.6) -- (3.4,4.8);
    \draw [red,thick] (1.3,7.6) -- (2.,7.2);
    \draw [red,thick] (3.4,8.) -- (2.,7.2);
    \draw [red,thick] (3.4,8.6) -- (3.4,8.);
    \draw [red,thick] (4.8,7.2) -- (3.4,8.);
    \draw [red,thick] (8.3,7.6) -- (7.6,7.2);
    \draw [red,thick] (4.8,7.2) -- (6.2,8.);
    \draw [red,thick] (6.2,8.) -- (7.6,7.2);
    \draw [red,thick] (6.2,8.6) -- (6.2,8.);
    \draw [red,thick] (4.8,5.6) -- (4.8,7.2);
    \draw [red,thick] (7.6,7.2) -- (7.6,5.6);
    \draw [red,thick] (2.,7.2) -- (2.,5.6);
    \draw [red,thick] (6.2,4.8) -- (4.8,5.6);
    \draw [ultra thick] (3.6,6.4) -- (6.,6.4);
    \filldraw [red, opacity=0.2] (1.3,5.2) -- (2.,5.6) -- (3.4,4.8) -- (4.8,5.6) -- (6.2,4.8) -- (7.6,5.6) --(8.3,5.2) -- (8.3,4.2) -- (1.3,4.2);
    \filldraw [red, opacity=0.2] (1.3,7.6) -- (2.,7.2) -- (3.4,8.) -- (4.8,7.2) -- (6.2,8.) -- (7.6,7.2)--(8.3,7.6) --(8.3,5.2) -- (7.6,5.6) -- (6.2,4.8) -- (4.8,5.6) -- (3.4,4.8) -- (2.,5.6) -- (1.3,5.2);
    \filldraw [red, opacity=0.2] (1.3,7.6) -- (2.,7.2) -- (3.4,8.) -- (4.8,7.2) -- (6.2,8.) -- (7.6,7.2)--(8.3,7.6) -- (8.3,8.6) -- (1.3,8.6);
    \node [centered, rotate=0.] at (2.6,6.4) {\scriptsize $(\sigma\overline{\sigma},1)$};
    \node [centered, rotate=0.] at (7.,6.4) {\scriptsize $(\sigma\overline{\sigma},1)$};
    \node [centered, rotate=0.] at (4.8,3.6) {\scriptsize {(a)}};
\end{tikzpicture}
}

\newcommand{\FigureSigmaSigmaBarB}{
\begin{tikzpicture}[baseline={([yshift=-.8ex]current bounding box.center)},scale=.49]
    \draw [red,thick] (3.4,4.8) -- (4.8,5.6);
    \draw [red,thick] (1.3,5.2) -- (2.,5.6);
    \draw [red,thick] (3.4,4.2) -- (3.4,4.8);
    \draw [red,thick] (6.2,4.2) -- (6.2,4.8);
    \draw [red,thick] (8.3,5.2) -- (7.6,5.6);
    \draw [red,thick] (6.2,4.8) -- (4.8,5.6);
    \draw [red,thick] (7.6,5.6) -- (6.2,4.8);
    \draw [red,thick] (2.,5.6) -- (3.4,4.8);
    \draw [red,thick] (1.3,7.6) -- (2.,7.2);
    \draw [red,thick] (3.4,8.) -- (2.,7.2);
    \draw [red,thick] (3.4,8.6) -- (3.4,8.);
    \draw [red,thick] (4.8,7.2) -- (3.4,8.);
    \draw [red,thick] (8.3,7.6) -- (7.6,7.2);
    \draw [red,thick] (4.8,7.2) -- (6.2,8.);
    \draw [red,thick] (6.2,8.) -- (7.6,7.2);
    \draw [red,thick] (6.2,8.6) -- (6.2,8.);
    \draw [red,thick] (4.8,5.6) -- (4.8,7.2);
    \draw [red,thick] (7.6,7.2) -- (7.6,5.6);
    \draw [red,thick] (2.,7.2) -- (2.,5.6);
    \draw [red,thick] (6.2,4.8) -- (4.8,5.6);
    \draw [ultra thick] (3.6,6.4) -- (6.,6.4);
    \filldraw [red, opacity=0.2] (1.3,5.2) -- (2.,5.6) -- (3.4,4.8) -- (4.8,5.6) -- (6.2,4.8) -- (7.6,5.6) --(8.3,5.2) -- (8.3,4.2) -- (1.3,4.2);
    \filldraw [red, opacity=0.2] (1.3,7.6) -- (2.,7.2) -- (3.4,8.) -- (4.8,7.2) -- (6.2,8.) -- (7.6,7.2)--(8.3,7.6) --(8.3,5.2) -- (7.6,5.6) -- (6.2,4.8) -- (4.8,5.6) -- (3.4,4.8) -- (2.,5.6) -- (1.3,5.2);
    \filldraw [red, opacity=0.2] (1.3,7.6) -- (2.,7.2) -- (3.4,8.) -- (4.8,7.2) -- (6.2,8.) -- (7.6,7.2)--(8.3,7.6) -- (8.3,8.6) -- (1.3,8.6);
    \node [centered, rotate=0.] at (2.6,6.4) {\scriptsize $(\sigma\overline{\sigma},1)$};
    \node [centered, rotate=0.] at (7.,6.4) {\scriptsize $(\sigma\overline{\sigma},\psi)$};
    \node [centered, rotate=0.] at (4.8,3.6) {\scriptsize {(b)}};
\end{tikzpicture}
}

\newcommand{\FigureSigmaSigmaBarC}{
\begin{tikzpicture}[baseline={([yshift=-.8ex]current bounding box.center)},scale=.49]
    \draw [red,thick] (3.4,4.8) -- (4.8,5.6);
    \draw [red,thick] (1.3,5.2) -- (2.,5.6);
    \draw [red,thick] (3.4,4.2) -- (3.4,4.8);
    \draw [red,thick] (6.2,4.2) -- (6.2,4.8);
    \draw [red,thick] (8.3,5.2) -- (7.6,5.6);
    \draw [red,thick] (6.2,4.8) -- (4.8,5.6);
    \draw [red,thick] (7.6,5.6) -- (6.2,4.8);
    \draw [red,thick] (2.,5.6) -- (3.4,4.8);
    \draw [red,thick] (1.3,7.6) -- (2.,7.2);
    \draw [red,thick] (3.4,8.) -- (2.,7.2);
    \draw [red,thick] (3.4,8.6) -- (3.4,8.);
    \draw [red,thick] (4.8,7.2) -- (3.4,8.);
    \draw [red,thick] (8.3,7.6) -- (7.6,7.2);
    \draw [red,thick] (4.8,7.2) -- (6.2,8.);
    \draw [red,thick] (6.2,8.) -- (7.6,7.2);
    \draw [red,thick] (6.2,8.6) -- (6.2,8.);
    \draw [red,thick] (4.8,5.6) -- (4.8,7.2);
    \draw [red,thick] (7.6,7.2) -- (7.6,5.6);
    \draw [red,thick] (2.,7.2) -- (2.,5.6);
    \draw [red,thick] (6.2,4.8) -- (4.8,5.6);
    \draw [ultra thick] (3.6,6.4) -- (6.,6.4);
    \filldraw [red, opacity=0.2] (1.3,5.2) -- (2.,5.6) -- (3.4,4.8) -- (4.8,5.6) -- (6.2,4.8) -- (7.6,5.6) --(8.3,5.2) -- (8.3,4.2) -- (1.3,4.2);
    \filldraw [red, opacity=0.2] (1.3,7.6) -- (2.,7.2) -- (3.4,8.) -- (4.8,7.2) -- (6.2,8.) -- (7.6,7.2)--(8.3,7.6) --(8.3,5.2) -- (7.6,5.6) -- (6.2,4.8) -- (4.8,5.6) -- (3.4,4.8) -- (2.,5.6) -- (1.3,5.2);
    \filldraw [red, opacity=0.2] (1.3,7.6) -- (2.,7.2) -- (3.4,8.) -- (4.8,7.2) -- (6.2,8.) -- (7.6,7.2)--(8.3,7.6) -- (8.3,8.6) -- (1.3,8.6);
    \node [centered, rotate=0.] at (2.6,6.4) {\scriptsize $(\sigma\overline{\sigma},\psi)$};
    \node [centered, rotate=0.] at (7.,6.4) {\scriptsize $(\sigma\overline{\sigma},1)$};
    \node [centered, rotate=0.] at (4.8,3.6) {\scriptsize {(c)}};
\end{tikzpicture}
}

\newcommand{\FigureSigmaSigmaBarD}{
\begin{tikzpicture}[baseline={([yshift=-.8ex]current bounding box.center)},scale=.49]
    \draw [red,thick] (3.4,4.8) -- (4.8,5.6);
    \draw [red,thick] (1.3,5.2) -- (2.,5.6);
    \draw [red,thick] (3.4,4.2) -- (3.4,4.8);
    \draw [red,thick] (6.2,4.2) -- (6.2,4.8);
    \draw [red,thick] (8.3,5.2) -- (7.6,5.6);
    \draw [red,thick] (6.2,4.8) -- (4.8,5.6);
    \draw [red,thick] (7.6,5.6) -- (6.2,4.8);
    \draw [red,thick] (2.,5.6) -- (3.4,4.8);
    \draw [red,thick] (1.3,7.6) -- (2.,7.2);
    \draw [red,thick] (3.4,8.) -- (2.,7.2);
    \draw [red,thick] (3.4,8.6) -- (3.4,8.);
    \draw [red,thick] (4.8,7.2) -- (3.4,8.);
    \draw [red,thick] (8.3,7.6) -- (7.6,7.2);
    \draw [red,thick] (4.8,7.2) -- (6.2,8.);
    \draw [red,thick] (6.2,8.) -- (7.6,7.2);
    \draw [red,thick] (6.2,8.6) -- (6.2,8.);
    \draw [red,thick] (4.8,5.6) -- (4.8,7.2);
    \draw [red,thick] (7.6,7.2) -- (7.6,5.6);
    \draw [red,thick] (2.,7.2) -- (2.,5.6);
    \draw [red,thick] (6.2,4.8) -- (4.8,5.6);
    \draw [ultra thick] (3.6,6.4) -- (6.,6.4);
    \filldraw [red, opacity=0.2] (1.3,5.2) -- (2.,5.6) -- (3.4,4.8) -- (4.8,5.6) -- (6.2,4.8) -- (7.6,5.6) --(8.3,5.2) -- (8.3,4.2) -- (1.3,4.2);
    \filldraw [red, opacity=0.2] (1.3,7.6) -- (2.,7.2) -- (3.4,8.) -- (4.8,7.2) -- (6.2,8.) -- (7.6,7.2)--(8.3,7.6) --(8.3,5.2) -- (7.6,5.6) -- (6.2,4.8) -- (4.8,5.6) -- (3.4,4.8) -- (2.,5.6) -- (1.3,5.2);
    \filldraw [red, opacity=0.2] (1.3,7.6) -- (2.,7.2) -- (3.4,8.) -- (4.8,7.2) -- (6.2,8.) -- (7.6,7.2)--(8.3,7.6) -- (8.3,8.6) -- (1.3,8.6);
    \node [centered, rotate=0.] at (2.6,6.4) {\scriptsize $(\sigma\overline{\sigma},\psi)$};
    \node [centered, rotate=0.] at (7.,6.4) {\scriptsize $(\sigma\overline{\sigma},\psi)$};
    \node [centered, rotate=0.] at (4.8,3.6) {\scriptsize {(d)}};
\end{tikzpicture}
}

\newcommand{\FigureConcatenationA}{
\begin{tikzpicture}[baseline={([yshift=-.8ex]current bounding box.center)},scale=.45]
    \draw [red,thick] (2.,7.2) -- (2.,5.6);
    \draw [red,thick] (2.,5.6) -- (3.4,4.8);
    \draw [red,thick] (3.4,4.8) -- (4.8,5.6);
    \draw [red,thick] (4.8,5.6) -- (4.8,7.2);
    \draw [red,thick] (4.8,7.2) -- (3.4,8.);
    \draw [red,thick] (3.4,8.) -- (2.,7.2);
    \draw [red,thick] (3.4,8.6) -- (3.4,8.);
    \draw [red,thick] (1.3,7.6) -- (2.,7.2);
    \draw [red,thick] (1.3,5.2) -- (2.,5.6);
    \draw [red,thick] (3.4,4.2) -- (3.4,4.8);
    \draw [red,thick] (6.2,4.8) -- (4.8,5.6);
    \draw [red,thick] (6.2,4.2) -- (6.2,4.8);
    \draw [red,thick] (7.6,5.6) -- (6.2,4.8);
    \draw [red,thick] (4.8,7.2) -- (6.2,8.);
    \draw [red,thick] (7.6,7.2) -- (7.6,5.6);
    \draw [red,thick] (9.,4.8) -- (7.6,5.6);
    \draw [red,thick] (9.,4.8) -- (9.,4.2);
    \draw [red,thick] (9.,8.) -- (7.6,7.2);
    \draw [red,thick] (9.,8.) -- (9.,8.6);
    \draw [red,thick] (9.,4.8) -- (10.4,5.6);
    \draw [red,thick] (9.,8.) -- (10.4,7.2);
    \draw [red,thick] (10.4,5.6) -- (10.4,7.2);
    \draw [red,thick] (10.4,5.6) -- (11.,5.2);
    \draw [red,thick] (11.,7.6) -- (10.4,7.2);
    \draw [red,thick] (6.2,8.) -- (7.6,7.2);
    \draw [red,thick] (6.2,8.6) -- (6.2,8.);
    \filldraw [red,opacity=0.2] (1.3,8.6) -- (1.3,4.2) -- (11.,4.2) -- (11.,8.6);
    \draw [ultra thick] (3.8,6.4) -- (5.8,6.4);
    \draw [ultra thick] (6.6,6.4) -- (8.6,6.4);
    \node [centered, rotate=0.] at (2.6,6.4) {\scriptsize $(J_\text{DI},p)$};
    \node [centered, rotate=0.] at (5.8,6.8) {\scriptsize $(J_\text{DI},q)$};
    \node [centered, rotate=0.] at (6.6,6.) {\scriptsize $(J_\text{DI},q)$};
    \node [centered, rotate=0.] at (9.8,6.4) {\scriptsize $(J_\text{DI},r)$};
    \node [centered, rotate=0.] at (6.2,7.3) {\scriptsize $P$};
    \node [centered, rotate=0.] at (6.2,3.6) {\scriptsize {(a)}};
\end{tikzpicture}
}

\newcommand{\FigureConcatenationB}{
\begin{tikzpicture}[baseline={([yshift=-.8ex]current bounding box.center)},scale=.45]
    \draw [red,thick] (2.,7.2) -- (2.,5.6);
    \draw [red,thick] (2.,5.6) -- (3.4,4.8);
    \draw [red,thick] (3.4,4.8) -- (4.8,5.6);
    \draw [red,thick] (4.8,5.6) -- (4.8,7.2);
    \draw [red,thick] (4.8,7.2) -- (3.4,8.);
    \draw [red,thick] (3.4,8.) -- (2.,7.2);
    \draw [red,thick] (3.4,8.6) -- (3.4,8.);
    \draw [red,thick] (1.3,7.6) -- (2.,7.2);
    \draw [red,thick] (1.3,5.2) -- (2.,5.6);
    \draw [red,thick] (3.4,4.2) -- (3.4,4.8);
    \draw [red,thick] (6.2,4.8) -- (4.8,5.6);
    \draw [red,thick] (6.2,4.2) -- (6.2,4.8);
    \draw [red,thick] (7.6,5.6) -- (6.2,4.8);
    \draw [red,thick] (4.8,7.2) -- (6.2,8.);
    \draw [red,thick] (7.6,7.2) -- (7.6,5.6);
    \draw [red,thick] (9.,4.8) -- (7.6,5.6);
    \draw [red,thick] (9.,4.8) -- (9.,4.2);
    \draw [red,thick] (9.,8.) -- (7.6,7.2);
    \draw [red,thick] (9.,8.) -- (9.,8.6);
    \draw [red,thick] (9.,4.8) -- (10.4,5.6);
    \draw [red,thick] (9.,8.) -- (10.4,7.2);
    \draw [red,thick] (10.4,5.6) -- (10.4,7.2);
    \draw [red,thick] (10.4,5.6) -- (11.,5.2);
    \draw [red,thick] (11.,7.6) -- (10.4,7.2);
    \draw [red,thick] (6.2,8.) -- (7.6,7.2);
    \draw [red,thick] (6.2,8.6) -- (6.2,8.);
    \filldraw [red,opacity=0.2] (1.3,8.6) -- (1.3,4.2) -- (11.,4.2) -- (11.,8.6);
    \draw [ultra thick] (3.8,6.4) -- (5.8,6.4);
    \draw [ultra thick] (6.6,6.4) -- (8.6,6.4);
    \draw [thick,densely dashed] (6.2,6.4) ellipse [x radius=1, y radius=1.1];
    \node [centered, rotate=0.] at (2.6,6.4) {\scriptsize $(J_\text{DI},p)$};
    \node [centered, rotate=0.] at (9.8,6.4) {\scriptsize $(J_\text{DI},r)$};
    \node [centered, rotate=0.] at (6.2,7.2) {\scriptsize $P$};
    \node [centered, rotate=0.] at (6.2,3.6) {\scriptsize {(b)}};
\end{tikzpicture}
}

\newcommand{\FigureConcatenationC}{
\begin{tikzpicture}[baseline={([yshift=-.8ex]current bounding box.center)},scale=.45]
    \draw [red,thick] (2.,7.2) -- (2.,5.6);
    \draw [red,thick] (2.,5.6) -- (3.4,4.8);
    \draw [red,thick] (3.4,4.8) -- (4.8,5.6);
    \draw [red,thick] (4.8,5.6) -- (4.8,7.2);
    \draw [red,thick] (4.8,7.2) -- (3.4,8.);
    \draw [red,thick] (3.4,8.) -- (2.,7.2);
    \draw [red,thick] (3.4,8.6) -- (3.4,8.);
    \draw [red,thick] (1.3,7.6) -- (2.,7.2);
    \draw [red,thick] (1.3,5.2) -- (2.,5.6);
    \draw [red,thick] (3.4,4.2) -- (3.4,4.8);
    \draw [red,thick] (6.2,4.8) -- (4.8,5.6);
    \draw [red,thick] (6.2,4.2) -- (6.2,4.8);
    \draw [red,thick] (7.6,5.6) -- (6.2,4.8);
    \draw [red,thick] (4.8,7.2) -- (6.2,8.);
    \draw [red,thick] (7.6,7.2) -- (7.6,5.6);
    \draw [red,thick] (9.,4.8) -- (7.6,5.6);
    \draw [red,thick] (9.,4.8) -- (9.,4.2);
    \draw [red,thick] (9.,8.) -- (7.6,7.2);
    \draw [red,thick] (9.,8.) -- (9.,8.6);
    \draw [red,thick] (9.,4.8) -- (10.4,5.6);
    \draw [red,thick] (9.,8.) -- (10.4,7.2);
    \draw [red,thick] (10.4,5.6) -- (10.4,7.2);
    \draw [red,thick] (10.4,5.6) -- (11.,5.2);
    \draw [red,thick] (11.,7.6) -- (10.4,7.2);
    \draw [red,thick] (6.2,8.) -- (7.6,7.2);
    \draw [red,thick] (6.2,8.6) -- (6.2,8.);
    \filldraw [red,opacity=0.2] (1.3,8.6) -- (1.3,4.2) -- (11.,4.2) -- (11.,8.6);
    \draw [ultra thick] (3.8,6.4) -- (8.6,6.4);
    \node [centered, rotate=0.] at (2.6,6.4) {\scriptsize $(J_\text{DI},p)$};
    \node [centered, rotate=0.] at (9.8,6.4) {\scriptsize $(J_\text{DI},r)$};
    \node [centered, rotate=0.] at (6.2,6.) {\scriptsize $L$};
    \node [centered, rotate=0.] at (6.2,7.3) {\scriptsize $P$};
    \node [centered, rotate=0.] at (6.2,3.6) {\scriptsize {(c)}};
\end{tikzpicture}
}

\newcommand{\FigureProjectorDoubledIsing}{
\begin{tikzpicture}[baseline={([yshift=-.8ex]current bounding box.center)},scale=0.7]
\draw [red, thick] (0.,7.6) -- (0.4,7.6);
\draw [red, thick] (1.2,8.8) -- (1.2,4.8);
\draw [red, thick] (0.4,8.8) -- (0.4,4.8);
\draw [red, thick] (0.4,8.4) -- (1.2,8.4);
\draw [red, thick] (0.4,6.8) -- (1.2,6.8);
\draw [red, thick] (0.4,5.2) -- (1.2,5.2);
\draw [red, thick] (0.4,6.) -- (0.,6.);
\filldraw [red, opacity=0.2] (0.0,8.8) -- (1.2,8.8) -- (1.2,4.8)--(0.0,4.8);

\draw [red, thick] (1.2,7.6) -- (2.,7.6);
\draw [red, thick] (1.2,6.) -- (2.,6.);
\filldraw [red, opacity=0.2] (2.0,8.8) -- (1.2,8.8) -- (1.2,4.8)--(2.0,4.8);

\draw [red, thick] (3.2,7.6) -- (2.8,7.6);
\draw [red, thick] (3.2,6.) -- (2.8,6.);
\draw [red, thick] (2.8,8.8) -- (2.8,4.8);
\draw [red, thick] (2.,8.8) -- (2.,4.8);
\draw [red, thick] (2.,8.4) -- (2.8,8.4);
\draw [red, thick] (2.,6.8) -- (2.8,6.8);
\draw [red, thick] (2.,5.2) -- (2.8,5.2);
\filldraw [red, opacity=0.2] (2.0,8.8) -- (3.2,8.8) -- (3.2,4.8)--(2.0,4.8);

\draw [ultra thick, ->] (3.3,6.8) -- (4.7,6.8);
\node [centered, rotate=0.] at (3.9,7.1) {\scriptsize $P_\text{eff}$};

\draw [red, thick] (4.8,7.6) -- (5.2,7.6);
\draw [red, thick] (6.,8.8) -- (6.,4.8);
\draw [red, thick] (5.2,8.8) -- (5.2,4.8);
\draw [red, thick] (5.2,8.4) -- (6.,8.4);
\draw [red, thick] (5.2,6.8) -- (6.,6.8);
\draw [red, thick] (5.2,5.2) -- (6.,5.2);
\draw [red, thick] (5.2,6.) -- (4.8,6.);
\filldraw [red, opacity=0.2] (4.8,8.8) -- (6.,8.8) -- (6.,4.8)--(4.8,4.8);

\draw [red, thick] (6.,7.6) -- (6.8,7.6);
\draw [red, thick] (6.,6.) -- (6.8,6.);
\filldraw [black, opacity=0.2] (6.8,8.8) -- (6.,8.8) -- (6.,4.8)--(6.8,4.8);

\draw [blue] (8.,7.6) -- (7.6,7.6);
\draw [blue] (8.,6.) -- (7.6,6.);
\draw [blue] (7.6,8.8) -- (7.6,4.8);
\draw [blue] (6.8,8.8) -- (6.8,4.8);
\draw [blue] (6.8,8.4) -- (7.6,8.4);
\draw [blue] (6.8,6.8) -- (7.6,6.8);
\draw [blue] (6.8,5.2) -- (7.6,5.2);
\filldraw [blue, opacity=0.2] (6.8,8.8) -- (8.,8.8) -- (8.,4.8)--(6.8,4.8);

\draw [ultra thick] (0.8,7.6) -- (0.8,6.);
\node [centered, rotate=0.] at (0.8,8.) {\scriptsize $(J_\text{DI},p)$};
\node [centered, rotate=0.] at (0.8,5.8) {\scriptsize $(J_\text{DI},q)$};

\draw [ultra thick] (5.6,7.6) -- (5.6,6.);
\node [centered, rotate=0.] at (5.6,8.) {\scriptsize $(J_\text{DI},p)$};
\node [centered, rotate=0.] at (5.6,5.8) {\scriptsize $(J_\text{DI},q)$};
\node [centered, rotate=0.] at (4.,4.2) {\scriptsize {(a)}};
\end{tikzpicture}
}

\newcommand{\FigureProjectorToricCode}{
\begin{tikzpicture}[baseline={([yshift=-.8ex]current bounding box.center)},scale=0.7]
\draw [red, thick] (0.,7.6) -- (0.4,7.6);
\draw [red, thick] (1.2,8.8) -- (1.2,4.8);
\draw [red, thick] (0.4,8.8) -- (0.4,4.8);
\draw [red, thick] (0.4,8.4) -- (1.2,8.4);
\draw [red, thick] (0.4,6.8) -- (1.2,6.8);
\draw [red, thick] (0.4,5.2) -- (1.2,5.2);
\draw [red, thick] (0.4,6.) -- (0.,6.);
\filldraw [red, opacity=0.2] (0.0,8.8) -- (1.2,8.8) -- (1.2,4.8)--(0.0,4.8);

\draw [red, thick] (1.2,7.6) -- (2.,7.6);
\draw [red, thick] (1.2,6.) -- (2.,6.);
\filldraw [red, opacity=0.2] (2.0,8.8) -- (1.2,8.8) -- (1.2,4.8)--(2.0,4.8);

\draw [red, thick] (3.2,7.6) -- (2.8,7.6);
\draw [red, thick] (3.2,6.) -- (2.8,6.);
\draw [red, thick] (2.8,8.8) -- (2.8,4.8);
\draw [red, thick] (2.,8.8) -- (2.,4.8);
\draw [red, thick] (2.,8.4) -- (2.8,8.4);
\draw [red, thick] (2.,6.8) -- (2.8,6.8);
\draw [red, thick] (2.,5.2) -- (2.8,5.2);
\filldraw [red, opacity=0.2] (2.0,8.8) -- (3.2,8.8) -- (3.2,4.8)--(2.0,4.8);

\draw [ultra thick, ->] (3.3,6.8) -- (4.7,6.8);
\node [centered, rotate=0.] at (3.9,7.1) {\scriptsize $P_\text{eff}$};

\draw [red, thick] (4.8,7.6) -- (5.2,7.6);
\draw [red, thick] (6.,8.8) -- (6.,4.8);
\draw [red, thick] (5.2,8.8) -- (5.2,4.8);
\draw [red, thick] (5.2,8.4) -- (6.,8.4);
\draw [red, thick] (5.2,6.8) -- (6.,6.8);
\draw [red, thick] (5.2,5.2) -- (6.,5.2);
\draw [red, thick] (5.2,6.) -- (4.8,6.);
\filldraw [red, opacity=0.2] (4.8,8.8) -- (6.,8.8) -- (6.,4.8)--(4.8,4.8);

\draw [red, thick] (6.,7.6) -- (6.8,7.6);
\draw [red, thick] (6.,6.) -- (6.8,6.);
\filldraw [black, opacity=0.2] (6.8,8.8) -- (6.,8.8) -- (6.,4.8)--(6.8,4.8);

\draw [blue] (8.,7.6) -- (7.6,7.6);
\draw [blue] (8.,6.) -- (7.6,6.);
\draw [blue] (7.6,8.8) -- (7.6,4.8);
\draw [blue] (6.8,8.8) -- (6.8,4.8);
\draw [blue] (6.8,8.4) -- (7.6,8.4);
\draw [blue] (6.8,6.8) -- (7.6,6.8);
\draw [blue] (6.8,5.2) -- (7.6,5.2);
\filldraw [blue, opacity=0.2] (6.8,8.8) -- (8.,8.8) -- (8.,4.8)--(6.8,4.8);

\draw [ultra thick] (2.4,7.6) -- (2.4,6.);
\node [centered, rotate=0.] at (2.4,8.) {\scriptsize $(J_\text{DI},p)$};
\node [centered, rotate=0.] at (2.4,5.8) {\scriptsize $(J_\text{DI},p)$};

\draw [ultra thick] (7.2,7.6) -- (7.2,6.);
\node [centered, rotate=0.] at (7.2,8.) {\scriptsize $(J_\text{TC},p)$};
\node [centered, rotate=0.] at (7.2,5.8) {\scriptsize $(J_\text{TC},p)$};
\node [centered, rotate=0.] at (4.,4.2) {\scriptsize {(b)}};
\end{tikzpicture}
}

\newcommand{\FigureProjectorInterDomain}{
\begin{tikzpicture}[baseline={([yshift=-.8ex]current bounding box.center)},scale=0.7]
\draw [red, thick] (0.,7.6) -- (0.4,7.6);
\draw [red, thick] (1.2,8.8) -- (1.2,4.8);
\draw [red, thick] (0.4,8.8) -- (0.4,4.8);
\draw [red, thick] (0.4,8.4) -- (1.2,8.4);
\draw [red, thick] (0.4,6.8) -- (1.2,6.8);
\draw [red, thick] (0.4,5.2) -- (1.2,5.2);
\draw [red, thick] (0.4,6.) -- (0.,6.);
\filldraw [red, opacity=0.2] (0.0,8.8) -- (1.2,8.8) -- (1.2,4.8)--(0.0,4.8);

\draw [red, thick] (1.2,7.6) -- (2.,7.6);
\draw [red, thick] (1.2,6.) -- (2.,6.);
\filldraw [red, opacity=0.2] (2.0,8.8) -- (1.2,8.8) -- (1.2,4.8)--(2.0,4.8);

\draw [red, thick] (3.2,7.6) -- (2.8,7.6);
\draw [red, thick] (3.2,6.) -- (2.8,6.);
\draw [red, thick] (2.8,8.8) -- (2.8,4.8);
\draw [red, thick] (2.,8.8) -- (2.,4.8);
\draw [red, thick] (2.,8.4) -- (2.8,8.4);
\draw [red, thick] (2.,6.8) -- (2.8,6.8);
\draw [red, thick] (2.,5.2) -- (2.8,5.2);
\filldraw [red, opacity=0.2] (2.0,8.8) -- (3.2,8.8) -- (3.2,4.8)--(2.0,4.8);

\draw [ultra thick, ->] (3.3,6.8) -- (4.7,6.8);
\node [centered, rotate=0.] at (3.9,7.1) {\scriptsize $P_\text{eff}$};

\draw [red, thick] (4.8,7.6) -- (5.2,7.6);
\draw [red, thick] (6.,8.8) -- (6.,4.8);
\draw [red, thick] (5.2,8.8) -- (5.2,4.8);
\draw [red, thick] (5.2,8.4) -- (6.,8.4);
\draw [red, thick] (5.2,6.8) -- (6.,6.8);
\draw [red, thick] (5.2,5.2) -- (6.,5.2);
\draw [red, thick] (5.2,6.) -- (4.8,6.);
\filldraw [red, opacity=0.2] (4.8,8.8) -- (6.,8.8) -- (6.,4.8)--(4.8,4.8);

\draw [red, thick] (6.,7.6) -- (6.8,7.6);
\draw [red, thick] (6.,6.) -- (6.8,6.);
\filldraw [black, opacity=0.2] (6.8,8.8) -- (6.,8.8) -- (6.,4.8)--(6.8,4.8);

\draw [blue] (8.,7.6) -- (7.6,7.6);
\draw [blue] (8.,6.) -- (7.6,6.);
\draw [blue] (7.6,8.8) -- (7.6,4.8);
\draw [blue] (6.8,8.8) -- (6.8,4.8);
\draw [blue] (6.8,8.4) -- (7.6,8.4);
\draw [blue] (6.8,6.8) -- (7.6,6.8);
\draw [blue] (6.8,5.2) -- (7.6,5.2);
\filldraw [blue, opacity=0.2] (6.8,8.8) -- (8.,8.8) -- (8.,4.8)--(6.8,4.8);

\draw [ultra thick] (0.8,7.2) -- (2.4,7.2);
\node [centered, rotate=0.] at (0.7,7.5) {\scriptsize $(J_\text{DI},p)$};
\node [centered, rotate=0.] at (2.5,7.5) {\scriptsize $(J_\text{DI},q)$};

\draw [ultra thick] (5.6,7.2) -- (7.2,7.2);
\node [centered, rotate=0.] at (5.5,7.5) {\scriptsize $(J_\text{DI},p)$};
\node [centered, rotate=0.] at (7.3,7.5) {\scriptsize $(J_\text{TC},q)$};
\node [centered, rotate=0.] at (4.,4.2) {\scriptsize {(c)}};
\end{tikzpicture}
}

\newcommand{\FigureProjectorDomainWall}{
\begin{tikzpicture}[baseline={([yshift=-.8ex]current bounding box.center)},scale=0.7]
\draw [red, thick] (0.,8.4) -- (0.4,8.4);
\draw [red, thick] (1.2,8.8) -- (1.2,4.8);
\draw [red, thick] (0.4,8.8) -- (0.4,4.8);
\draw [red, thick] (0.4,7.6) -- (1.2,7.6);
\draw [red, thick] (0.4,6.) -- (1.2,6.);
\draw [red, thick] (1.2,8.4) -- (2.,8.4);
\draw [red, thick] (3.2,8.4) -- (2.8,8.4);
\draw [red, thick] (2.8,8.8) -- (2.8,4.8);
\draw [red, thick] (2.,8.8) -- (2.,4.8);
\draw [red, thick] (2.,7.6) -- (2.8,7.6);
\draw [red, thick] (2.,6.) -- (2.8,6.);
\draw [red, thick] (0.4,6.8) -- (0.,6.8);
\draw [red, thick] (1.2,6.8) -- (2.,6.8);
\draw [red, thick] (3.2,6.8) -- (2.8,6.8);
\draw [red, thick] (0.4,5.2) -- (0.,5.2);
\draw [red, thick] (1.2,5.2) -- (2.,5.2);
\draw [red, thick] (3.2,5.2) -- (2.8,5.2);
\draw [red, thick] (4.8,8.4) -- (5.2,8.4);
\draw [red, thick] (6.,8.8) -- (6.,4.8);
\draw [red, thick] (5.2,8.8) -- (5.2,4.8);
\draw [red, thick] (5.2,7.6) -- (6.,7.6);
\draw [red, thick] (5.2,6.) -- (6.,6.);
\draw [red, thick] (6.,8.4) -- (6.8,8.4);
\draw [blue] (8.,8.4) -- (7.6,8.4);
\draw [blue] (7.6,8.8) -- (7.6,4.8);
\draw [blue] (6.8,8.8) -- (6.8,4.8);
\draw [blue] (6.8,7.6) -- (7.6,7.6);
\draw [blue] (6.8,6.) -- (7.6,6.);
\draw [red, thick] (5.2,6.8) -- (4.8,6.8);
\draw [red, thick] (6.,6.8) -- (6.8,6.8);
\draw [blue] (8.,6.8) -- (7.6,6.8);
\draw [red, thick] (5.2,5.2) -- (4.8,5.2);
\draw [red, thick] (6.,5.2) -- (6.8,5.2);
\draw [blue] (8.,5.2) -- (7.6,5.2);
\filldraw [red, opacity=0.2] (0.0,8.8) -- (1.2,8.8) -- (1.2,4.8)--(0.0,4.8);
\filldraw [red, opacity=0.2] (2.0,8.8) -- (1.2,8.8) -- (1.2,4.8)--(2.0,4.8);
\filldraw [red, opacity=0.2] (2.0,8.8) -- (3.2,8.8) -- (3.2,4.8)--(2.0,4.8);
\draw [ultra thick, ->] (3.3,6.8) -- (4.7,6.8);
\node [centered, rotate=0.] at (3.9,7.1) {\scriptsize $P_\text{eff}$};
\filldraw [red, opacity=0.2] (4.8,8.8) -- (6.,8.8) -- (6.,4.8)--(4.8,4.8);
\filldraw [black, opacity=0.2] (6.8,8.8) -- (6.,8.8) -- (6.,4.8)--(6.8,4.8);
\filldraw [blue, opacity=0.2] (6.8,8.8) -- (8.,8.8) -- (8.,4.8)--(6.8,4.8);

\draw [ultra thick] (1.6,7.6) -- (1.6,6.);
\node [centered, rotate=0.] at (1.6,8.) {\scriptsize $(J_\text{DI},p)$};
\node [centered, rotate=0.] at (1.6,5.8) {\scriptsize $(J_\text{DI},p)$};

\draw [ultra thick] (6.4,7.6) -- (6.4,6.);
\node [centered, rotate=0.] at (6.4,8.) {\scriptsize $(J_\text{DW},p)$};
\node [centered, rotate=0.] at (6.4,5.8) {\scriptsize $(J_\text{DW},p)$};
\node [centered, rotate=0.] at (4.,4.2) {\scriptsize {(d)}};
\end{tikzpicture}
}

\newcommand{\FigureProjectorDiDw}{
\begin{tikzpicture}[baseline={([yshift=-.8ex]current bounding box.center)},scale=0.7]
\draw [red, thick] (0.,8.4) -- (0.4,8.4);
\draw [red, thick] (1.2,8.8) -- (1.2,4.8);
\draw [red, thick] (0.4,8.8) -- (0.4,4.8);
\draw [red, thick] (0.4,7.6) -- (1.2,7.6);
\draw [red, thick] (0.4,6.) -- (1.2,6.);
\draw [red, thick] (1.2,8.4) -- (2.,8.4);
\draw [red, thick] (3.2,8.4) -- (2.8,8.4);
\draw [red, thick] (2.8,8.8) -- (2.8,4.8);
\draw [red, thick] (2.,8.8) -- (2.,4.8);
\draw [red, thick] (2.,7.6) -- (2.8,7.6);
\draw [red, thick] (2.,6.) -- (2.8,6.);
\draw [red, thick] (0.4,6.8) -- (0.,6.8);
\draw [red, thick] (1.2,6.8) -- (2.,6.8);
\draw [red, thick] (3.2,6.8) -- (2.8,6.8);
\draw [red, thick] (0.4,5.2) -- (0.,5.2);
\draw [red, thick] (1.2,5.2) -- (2.,5.2);
\draw [red, thick] (3.2,5.2) -- (2.8,5.2);
\draw [red, thick] (4.8,8.4) -- (5.2,8.4);
\draw [red, thick] (6.,8.8) -- (6.,4.8);
\draw [red, thick] (5.2,8.8) -- (5.2,4.8);
\draw [red, thick] (5.2,7.6) -- (6.,7.6);
\draw [red, thick] (5.2,6.) -- (6.,6.);
\draw [red, thick] (6.,8.4) -- (6.8,8.4);
\draw [blue] (8.,8.4) -- (7.6,8.4);
\draw [blue] (7.6,8.8) -- (7.6,4.8);
\draw [blue] (6.8,8.8) -- (6.8,4.8);
\draw [blue] (6.8,7.6) -- (7.6,7.6);
\draw [blue] (6.8,6.) -- (7.6,6.);
\draw [red, thick] (5.2,6.8) -- (4.8,6.8);
\draw [red, thick] (6.,6.8) -- (6.8,6.8);
\draw [blue] (8.,6.8) -- (7.6,6.8);
\draw [red, thick] (5.2,5.2) -- (4.8,5.2);
\draw [red, thick] (6.,5.2) -- (6.8,5.2);
\draw [blue] (8.,5.2) -- (7.6,5.2);
\filldraw [red, opacity=0.2] (0.0,8.8) -- (1.2,8.8) -- (1.2,4.8)--(0.0,4.8);
\filldraw [red, opacity=0.2] (2.0,8.8) -- (1.2,8.8) -- (1.2,4.8)--(2.0,4.8);
\filldraw [red, opacity=0.2] (2.0,8.8) -- (3.2,8.8) -- (3.2,4.8)--(2.0,4.8);
\draw [ultra thick, ->] (3.3,6.8) -- (4.7,6.8);
\node [centered, rotate=0.] at (3.9,7.1) {\scriptsize $P_\text{eff}$};
\filldraw [red, opacity=0.2] (4.8,8.8) -- (6.,8.8) -- (6.,4.8)--(4.8,4.8);
\filldraw [black, opacity=0.2] (6.8,8.8) -- (6.,8.8) -- (6.,4.8)--(6.8,4.8);
\filldraw [blue, opacity=0.2] (6.8,8.8) -- (8.,8.8) -- (8.,4.8)--(6.8,4.8);

\draw [ultra thick] (0.8,7.2) -- (1.6,7.2);
\node [centered, rotate=0.] at (0.7,7.5) {\scriptsize $(J_\text{DI},p)$};
\node [centered, rotate=0.] at (2.4,7.2) {\scriptsize $(J_\text{DI},q)$};

\draw [ultra thick] (5.6,7.2) -- (6.4,7.2);
\node [centered, rotate=0.] at (5.5,7.5) {\scriptsize $(J_\text{DI},p)$};
\node [centered, rotate=0.] at (7.2,7.2) {\scriptsize $(J_\text{DW},q)$};
\node [centered, rotate=0.] at (4.,4.2) {\scriptsize {(e)}};
\end{tikzpicture}
}

\newcommand{\FigureProjectorDwTc}{
\begin{tikzpicture}[baseline={([yshift=-.8ex]current bounding box.center)},scale=0.7]
\draw [red, thick] (0.,8.4) -- (0.4,8.4);
\draw [red, thick] (1.2,8.8) -- (1.2,4.8);
\draw [red, thick] (0.4,8.8) -- (0.4,4.8);
\draw [red, thick] (0.4,7.6) -- (1.2,7.6);
\draw [red, thick] (0.4,6.) -- (1.2,6.);
\draw [red, thick] (1.2,8.4) -- (2.,8.4);
\draw [red, thick] (3.2,8.4) -- (2.8,8.4);
\draw [red, thick] (2.8,8.8) -- (2.8,4.8);
\draw [red, thick] (2.,8.8) -- (2.,4.8);
\draw [red, thick] (2.,7.6) -- (2.8,7.6);
\draw [red, thick] (2.,6.) -- (2.8,6.);
\draw [red, thick] (0.4,6.8) -- (0.,6.8);
\draw [red, thick] (1.2,6.8) -- (2.,6.8);
\draw [red, thick] (3.2,6.8) -- (2.8,6.8);
\draw [red, thick] (0.4,5.2) -- (0.,5.2);
\draw [red, thick] (1.2,5.2) -- (2.,5.2);
\draw [red, thick] (3.2,5.2) -- (2.8,5.2);
\draw [red, thick] (4.8,8.4) -- (5.2,8.4);
\draw [red, thick] (6.,8.8) -- (6.,4.8);
\draw [red, thick] (5.2,8.8) -- (5.2,4.8);
\draw [red, thick] (5.2,7.6) -- (6.,7.6);
\draw [red, thick] (5.2,6.) -- (6.,6.);
\draw [red, thick] (6.,8.4) -- (6.8,8.4);
\draw [blue] (8.,8.4) -- (7.6,8.4);
\draw [blue] (7.6,8.8) -- (7.6,4.8);
\draw [blue] (6.8,8.8) -- (6.8,4.8);
\draw [blue] (6.8,7.6) -- (7.6,7.6);
\draw [blue] (6.8,6.) -- (7.6,6.);
\draw [red, thick] (5.2,6.8) -- (4.8,6.8);
\draw [red, thick] (6.,6.8) -- (6.8,6.8);
\draw [blue] (8.,6.8) -- (7.6,6.8);
\draw [red, thick] (5.2,5.2) -- (4.8,5.2);
\draw [red, thick] (6.,5.2) -- (6.8,5.2);
\draw [blue] (8.,5.2) -- (7.6,5.2);
\filldraw [red, opacity=0.2] (0.0,8.8) -- (1.2,8.8) -- (1.2,4.8)--(0.0,4.8);
\filldraw [red, opacity=0.2] (2.0,8.8) -- (1.2,8.8) -- (1.2,4.8)--(2.0,4.8);
\filldraw [red, opacity=0.2] (2.0,8.8) -- (3.2,8.8) -- (3.2,4.8)--(2.0,4.8);
\draw [ultra thick, ->] (3.3,6.8) -- (4.7,6.8);
\node [centered, rotate=0.] at (3.9,7.1) {\scriptsize $P_\text{eff}$};
\filldraw [red, opacity=0.2] (4.8,8.8) -- (6.,8.8) -- (6.,4.8)--(4.8,4.8);
\filldraw [black, opacity=0.2] (6.8,8.8) -- (6.,8.8) -- (6.,4.8)--(6.8,4.8);
\filldraw [blue, opacity=0.2] (6.8,8.8) -- (8.,8.8) -- (8.,4.8)--(6.8,4.8);

\draw [ultra thick] (1.6,7.2) -- (2.4,7.2);
\node [centered, rotate=0.] at (0.8,7.2) {\scriptsize $(J_\text{DI},p)$};
\node [centered, rotate=0.] at (2.5,7.5) {\scriptsize $(J_\text{DI},p)$};

\draw [ultra thick] (6.4,7.2) -- (7.2,7.2);
\node [centered, rotate=0.] at (5.6,7.2) {\scriptsize $(J_\text{DW},p)$};
\node [centered, rotate=0.] at (7.3,7.5) {\scriptsize $(J_\text{TC},p)$};
\node [centered, rotate=0.] at (4.,4.2) {\scriptsize {(f)}};
\end{tikzpicture}
}

\newcommand{\TableQuasiparticleTC}{
\begin{tabular}{|c|c|c|}
    \hline Quasiparticle & Anyon species & Charge\\ \hline
    $(1,1)$ & $1$ & $1$\\ \hline
    $(e,\psi)$ & $e$ & $\psi$ \\ \hline
    $(m,1)$ & $m$ & $1$ \\ \hline
    $(\epsilon,\psi)$ & $\epsilon$ & $\psi$ \\ \hline
\end{tabular}
}

\newcommand{\EqToricCodeState}{
\hs\begin{tikzpicture}[baseline={([yshift=-.8ex]current bounding box.center)},scale=.43]
    \draw [] (1.,3.8) -- (1.,9.);
    \draw [] (8.4,3.8) -- (9.7,6.4);
    \draw [] (9.7,6.4) -- (8.4,9.);
    \draw [blue] (2.,7.2) -- (2.,5.6);
    \draw [blue] (1.3,7.6) -- (2.,7.2);
    \draw [blue] (1.3,5.2) -- (2.,5.6);
    \draw [blue] (3.4,4.2) -- (3.4,4.8);
    \draw [blue] (2.,5.6) -- (3.4,4.8);
    \draw [blue] (3.4,8.) -- (2.,7.2);
    \draw [blue] (3.4,4.8) -- (4.8,5.6);
    \draw [ultra thick] (3.4,6.4) -- (6.2,6.4);
    \draw [blue] (3.4,8.6) -- (3.4,8.);
    \draw [blue] (4.8,5.6) -- (4.8,7.2);
    \draw [blue] (4.8,7.2) -- (3.4,8.);
    \draw [blue] (6.2,4.8) -- (4.8,5.6);
    \draw [blue] (6.2,4.2) -- (6.2,4.8);
    \draw [blue] (7.6,7.2) -- (7.6,5.6);
    \draw [blue] (7.6,5.6) -- (6.2,4.8);
    \draw [blue] (8.3,5.2) -- (7.6,5.6);
    \draw [blue] (8.3,7.6) -- (7.6,7.2);
    \draw [blue] (4.8,7.2) -- (6.2,8.);
    \draw [blue] (6.2,8.) -- (7.6,7.2);
    \draw [blue] (6.2,8.6) -- (6.2,8.);
    \filldraw [blue, opacity=0.2] (1.3,4.2) -- (1.3, 7.6) -- (2.,7.2) -- (2.,5.6) -- (3.4,4.8) -- (3.4,4.2);
    \filldraw [blue, opacity=0.2] (1.3, 7.6) -- (1.3, 8.6) -- (3.4,8.6) -- (3.4, 8.) -- (4.8,7.2) -- (4.8,5.6) -- (6.2,4.8) -- (6.2,4.2) -- (3.4, 4.2) -- (3.4, 4.8) -- (2.,5.6) -- (2.,7.2);
    \filldraw [blue, opacity=0.2] (3.4,8.6) -- (3.4, 8.) -- (4.8,7.2) -- (4.8,5.6) -- (6.2,4.8) -- (6.2,4.2) --(8.3, 4.2) -- (8.3,5.2) -- (8.3, 8.6);
    \node [centered, rotate=0.] at (2.2,6.4) {\scriptsize $(J_\text{TC},p)$};
    \node [centered, rotate=0.] at (7.35,6.4) {\scriptsize $(J_\text{TC},p)$};
\end{tikzpicture}\hs
}

\newcommand{\EqToricCodeRibbonLeft}{
\hs\begin{tikzpicture}[baseline={([yshift=-.8ex]current bounding box.center)},scale=.43]
    \draw [] (1.,3.8) -- (1.,9.);
    \draw [] (8.4,3.8) -- (9.7,6.4);
    \draw [] (9.7,6.4) -- (8.4,9.);
    \draw [blue] (3.4,4.8) -- (4.8,5.6);
    \draw [blue] (1.3,5.2) -- (2.,5.6);
    \draw [blue] (3.4,4.2) -- (3.4,4.8);
    \draw [blue] (6.2,4.2) -- (6.2,4.8);
    \draw [blue] (8.3,5.2) -- (7.6,5.6);
    \draw [blue] (6.2,4.8) -- (4.8,5.6);
    \draw [blue] (7.6,5.6) -- (6.2,4.8);
    \draw [blue] (2.,5.6) -- (3.4,4.8);
    \draw [blue] (1.3,7.6) -- (2.,7.2);
    \draw [blue] (3.4,8.) -- (2.,7.2);
    \draw [blue] (3.4,8.6) -- (3.4,8.);
    \draw [blue] (4.8,7.2) -- (3.4,8.);
    \draw [blue] (8.3,7.6) -- (7.6,7.2);
    \draw [blue] (4.8,7.2) -- (6.2,8.);
    \draw [blue] (6.2,8.) -- (7.6,7.2);
    \draw [blue] (6.2,8.6) -- (6.2,8.);
    \draw [blue] (7.6,7.2) -- (7.6,5.6);
    \draw [blue] (4.8,5.6) -- (4.8,7.2);
    \draw [blue] (2.,7.2) -- (2.,5.6);
    \filldraw [blue, opacity=0.2] (1.3,5.2) -- (2.,5.6) -- (3.4,4.8) -- (4.8,5.6) -- (6.2,4.8) -- (7.6,5.6) --(8.3,5.2) -- (8.3,4.2) -- (1.3,4.2);
    \filldraw [blue, opacity=0.2] (1.3,7.6) -- (2.,7.2) -- (3.4,8.) -- (4.8,7.2) -- (6.2,8.) -- (7.6,7.2)--(8.3,7.6) --(8.3,5.2) -- (7.6,5.6) -- (6.2,4.8) -- (4.8,5.6) -- (3.4,4.8) -- (2.,5.6) -- (1.3,5.2);
    \filldraw [blue, opacity=0.2] (1.3,7.6) -- (2.,7.2) -- (3.4,8.) -- (4.8,7.2) -- (6.2,8.) -- (7.6,7.2)--(8.3,7.6) -- (8.3,8.6) -- (1.3,8.6);
    \node [centered, rotate=0.] at (4.2,6.4) {\scriptsize $j_E$};
\end{tikzpicture}\hs
}

\newcommand{\EqToricCodeRibbonRight}{
\hs\begin{tikzpicture}[baseline={([yshift=-.8ex]current bounding box.center)},scale=.43]
    \draw [] (1.,3.8) -- (1.,9.);
    \draw [] (8.4,3.8) -- (9.7,6.4);
    \draw [] (9.7,6.4) -- (8.4,9.);
    \draw [blue] (3.4,4.8) -- (4.8,5.6);
    \draw [blue] (1.3,5.2) -- (2.,5.6);
    \draw [blue] (3.4,4.2) -- (3.4,4.8);
    \draw [blue] (6.2,4.2) -- (6.2,4.8);
    \draw [blue] (8.3,5.2) -- (7.6,5.6);
    \draw [blue] (6.2,4.8) -- (4.8,5.6);
    \draw [blue] (7.6,5.6) -- (6.2,4.8);
    \draw [blue] (2.,5.6) -- (3.4,4.8);
    \draw [blue] (1.3,7.6) -- (2.,7.2);
    \draw [blue] (3.4,8.) -- (2.,7.2);
    \draw [blue] (3.4,8.6) -- (3.4,8.);
    \draw [blue] (4.8,7.2) -- (3.4,8.);
    \draw [blue] (8.3,7.6) -- (7.6,7.2);
    \draw [blue] (4.8,7.2) -- (6.2,8.);
    \draw [blue] (6.2,8.) -- (7.6,7.2);
    \draw [blue] (6.2,8.6) -- (6.2,8.);
    \draw [blue] (4.8,5.6) -- (4.8,7.2);
    \draw [blue] (7.6,7.2) -- (7.6,5.6);
    \draw [blue] (2.,7.2) -- (2.,5.6);
    \draw [blue] (4.8,6.1) -- (3.6,6.1);
    \draw [blue] (4.8,6.7) -- (6.,6.7);
    \draw [blue] (6.2,4.8) -- (4.8,5.6);
    \filldraw [blue, opacity=0.2] (1.3,5.2) -- (2.,5.6) -- (3.4,4.8) -- (4.8,5.6) -- (6.2,4.8) -- (7.6,5.6) --(8.3,5.2) -- (8.3,4.2) -- (1.3,4.2);
    \filldraw [blue, opacity=0.2] (1.3,7.6) -- (2.,7.2) -- (3.4,8.) -- (4.8,7.2) -- (6.2,8.) -- (7.6,7.2)--(8.3,7.6) --(8.3,5.2) -- (7.6,5.6) -- (6.2,4.8) -- (4.8,5.6) -- (3.4,4.8) -- (2.,5.6) -- (1.3,5.2);
    \filldraw [blue, opacity=0.2] (1.3,7.6) -- (2.,7.2) -- (3.4,8.) -- (4.8,7.2) -- (6.2,8.) -- (7.6,7.2)--(8.3,7.6) -- (8.3,8.6) -- (1.3,8.6);
    \node [centered, rotate=0.] at (4.4,7) {\scriptsize $j_E$};
    \node [centered, rotate=0.] at (5.3,5.8) {\scriptsize $j_E$};
    \node [centered, rotate=0.] at (4.4,6.4) {\scriptsize $k$};
    \node [centered, rotate=0.] at (3.3,6.1) {\scriptsize $p$};
    \node [centered, rotate=0.] at (6.3,6.7) {\scriptsize $p$};
    \end{tikzpicture}\hs
}

\newcommand{\TableQuasiparticleInter}{
\begin{tabular}{|c|c|c|}
\hline
Elementary excitation state & Doubled-Ising quasiparticle & Toric-code quasiparticle\\ \hline
$\ket{1\overline{1}\dash 1;1,1}$ & $(1\overline{1},1)$ & \multirow{2}{*}{$(1,1)$}\\
$\ket{\psi\overline{\psi}\dash 1;1,1}$ & $(\psi\overline{\psi},1)$ & \\ \hline
$\ket{\psi\overline{1}\dash \epsilon;\psi,\psi}$ & $(\psi\overline{1},\psi)$ & \multirow{2}{*}{$(\epsilon,\psi)$}\\
$\ket{1\overline{\psi}\dash \epsilon;\psi,\psi}$ & $(1\overline{\psi},\psi)$ & \\ \hline
$\ket{\sigma\overline{\sigma}\dash m;1,1}$ & $(\sigma\overline{\sigma},1)$ & \multirow{2}{*}{$(m,1)$}\\
$\ket{\sigma\overline{\sigma}\dash m;\psi,1}$ & $(\sigma\overline{\sigma},\psi)$ & \\ \cline{1-1}\cline{3-3}
$\ket{\sigma\overline{\sigma}\dash e;1,\psi}$ & $(\sigma\overline{\sigma},1)$ & \multirow{2}{*}{$(e,\psi)$}\\
$\ket{\sigma\overline{\sigma}\dash e;\psi,\psi}$ & $(\sigma\overline{\sigma},\psi)$ & \\ \hline
\end{tabular}
}

\newcommand{\EqInterStateLeft}{
\hs\begin{tikzpicture}[baseline={([yshift=-.8ex]current bounding box.center)},scale=.5]
    \draw [] (1.,8.8) -- (1.,4.);
    \draw [] (11.2,8.8) -- (12.4,6.4);
    \draw [] (12.4,6.4) -- (11.2,4.);
    \draw [red,thick] (2.,7.2) -- (2.,5.6);
    \draw [red,thick] (2.,5.6) -- (3.4,4.8);
    \draw [red,thick] (3.4,4.8) -- (4.8,5.6);
    \draw [red,thick] (4.8,5.6) -- (4.8,7.2);
    \draw [red,thick] (4.8,7.2) -- (3.4,8.);
    \draw [red,thick] (3.4,8.) -- (2.,7.2);
    \draw [red,thick] (3.4,8.6) -- (3.4,8.);
    \draw [red,thick] (1.3,7.6) -- (2.,7.2);
    \draw [red,thick] (1.3,5.2) -- (2.,5.6);
    \draw [red,thick] (3.4,4.2) -- (3.4,4.8);
    \draw [red,thick] (6.2,4.8) -- (4.8,5.6);
    \draw [red,thick] (6.2,4.2) -- (6.2,4.8);
    \draw [red,thick] (7.6,5.6) -- (6.2,4.8);
    \draw [red,thick] (4.8,7.2) -- (6.2,8.);
    \draw [blue] (7.6,7.2) -- (7.6,5.6);
    \draw [blue] (9.,4.8) -- (7.6,5.6);
    \draw [blue] (9.,4.8) -- (9.,4.2);
    \draw [blue] (9.,8.) -- (7.6,7.2);
    \draw [blue] (9.,8.) -- (9.,8.6);
    \draw [blue] (9.,4.8) -- (10.4,5.6);
    \draw [blue] (9.,8.) -- (10.4,7.2);
    \draw [blue] (10.4,5.6) -- (10.4,7.2);
    \draw [blue] (10.4,5.6) -- (11.,5.2);
    \draw [blue] (11.,7.6) -- (10.4,7.2);
    \draw [blue] (6.2,8.) -- (7.6,7.2);
    \draw [blue] (6.2,8.6) -- (6.2,8.);
    \filldraw [red, opacity=0.2] (1.3,5.2) -- (1.3, 8.6) -- (3.4,8.6) -- (3.4, 8.) -- (4.8,7.2) -- (4.8,5.6) -- (6.2,4.8) -- (6.2,4.2) -- (1.3, 4.2);
    \filldraw [black, opacity=0.2] (3.4,8.6) -- (3.4, 8.) -- (4.8,7.2) -- (4.8,5.6) -- (6.2,4.8) -- (6.2,4.2) -- (9.,4.2) -- (9.,4.8) -- (7.6,5.6) -- (7.6,7.2) -- (6.2,8.) -- (6.2,8.6);
    \filldraw [blue, opacity=0.2] (9.,4.2) -- (9.,4.8) -- (7.6,5.6) -- (7.6,7.2) -- (6.2,8.) -- (6.2,8.6) --(11.,8.6) -- (11.,4.2);
    \draw [ultra thick] (3.8,6.4) -- (8.6,6.4);
    \node [centered, rotate=0.] at (2.6,6.4) {\scriptsize $(J_\text{DI},p)$};
    \node [centered, rotate=0.] at (9.7,6.4) {\scriptsize $(J_\text{TC},q)$};
    \node [centered, rotate=0.] at (6.2,6.8) {\scriptsize $L$};
    \node [centered, rotate=0.] at (5.2,5.9) {\scriptsize $E_1$};
    \node [centered, rotate=0.] at (7.1,6.8) {\scriptsize $E_4$};
\end{tikzpicture}\hs
}

\newcommand{\EqInterStateRight}{
\hs\begin{tikzpicture}[baseline={([yshift=-.8ex]current bounding box.center)},scale=.5]
    \draw [] (1.,8.8) -- (1.,4.);
    \draw [] (11.2,8.8) -- (12.4,6.4);
    \draw [] (12.4,6.4) -- (11.2,4.);
    \draw [red,thick] (2.,7.2) -- (2.,5.6);
    \draw [red,thick] (2.,5.6) -- (3.4,4.8);
    \draw [red,thick] (3.4,4.8) -- (4.8,5.6);
    \draw [red,thick] (4.8,5.6) -- (4.8,7.2);
    \draw [red,thick] (4.8,7.2) -- (3.4,8.);
    \draw [red,thick] (3.4,8.) -- (2.,7.2);
    \draw [red,thick] (3.4,8.6) -- (3.4,8.);
    \draw [red,thick] (1.3,7.6) -- (2.,7.2);
    \draw [red,thick] (1.3,5.2) -- (2.,5.6);
    \draw [red,thick] (3.4,4.2) -- (3.4,4.8);
    \draw [red,thick] (6.2,4.8) -- (4.8,5.6);
    \draw [red,thick] (6.2,4.2) -- (6.2,4.8);
    \draw [red,thick] (7.6,5.6) -- (6.2,4.8);
    \draw [red,thick] (4.8,7.2) -- (6.2,8.);
    \draw [blue] (7.6,7.2) -- (7.6,5.6);
    \draw [blue] (9.,4.8) -- (7.6,5.6);
    \draw [blue] (9.,4.8) -- (9.,4.2);
    \draw [blue] (9.,8.) -- (7.6,7.2);
    \draw [blue] (9.,8.) -- (9.,8.6);
    \draw [blue] (9.,4.8) -- (10.4,5.6);
    \draw [blue] (9.,8.) -- (10.4,7.2);
    \draw [blue] (10.4,5.6) -- (10.4,7.2);
    \draw [blue] (10.4,5.6) -- (11.,5.2);
    \draw [blue] (11.,7.6) -- (10.4,7.2);
    \draw [blue] (6.2,8.) -- (7.6,7.2);
    \draw [blue] (6.2,8.6) -- (6.2,8.);
    \filldraw [red, opacity=0.2] (1.3,5.2) -- (1.3, 8.6) -- (3.4,8.6) -- (3.4, 8.) -- (4.8,7.2) -- (4.8,5.6) -- (6.2,4.8) -- (6.2,4.2) -- (1.3, 4.2);
    \filldraw [black, opacity=0.2] (3.4,8.6) -- (3.4, 8.) -- (4.8,7.2) -- (4.8,5.6) -- (6.2,4.8) -- (6.2,4.2) -- (9.,4.2) -- (9.,4.8) -- (7.6,5.6) -- (7.6,7.2) -- (6.2,8.) -- (6.2,8.6);
    \filldraw [blue, opacity=0.2] (9.,4.2) -- (9.,4.8) -- (7.6,5.6) -- (7.6,7.2) -- (6.2,8.) -- (6.2,8.6) --(11.,8.6) -- (11.,4.2);
    \node [centered, rotate=0.] at (4.2,6.4) {\scriptsize $E_1$};
    \node [centered, rotate=0.] at (8.3,6.4) {\scriptsize $E_4$};
\end{tikzpicture}\hs
}

\newcommand{\EqRibbonInterLeft}{
\hs\begin{tikzpicture}[baseline={([yshift=-.8ex]current bounding box.center)},scale=.5]
    \draw [] (1.,3.8) -- (1.,9.);
    \draw [] (11.2,3.8) -- (12.5,6.4);
    \draw [] (12.5,6.4) -- (11.2,9.);
    \draw [red,thick] (2.,7.2) -- (2.,5.6);
    \draw [red,thick] (2.,5.6) -- (3.4,4.8);
    \draw [red,thick] (3.4,4.8) -- (4.8,5.6);
    \draw [red,thick] (4.8,5.6) -- (4.8,7.2);
    \draw [red,thick] (4.8,7.2) -- (3.4,8.);
    \draw [red,thick] (3.4,8.) -- (2.,7.2);
    \draw [red,thick] (3.4,8.6) -- (3.4,8.);
    \draw [red,thick] (1.3,7.6) -- (2.,7.2);
    \draw [red,thick] (1.3,5.2) -- (2.,5.6);
    \draw [red,thick] (3.4,4.2) -- (3.4,4.8);
    \draw [red,thick] (6.2,4.8) -- (4.8,5.6);
    \draw [red,thick] (6.2,4.2) -- (6.2,4.8);
    \draw [red,thick] (7.6,5.6) -- (6.2,4.8);
    \draw [red,thick] (4.8,7.2) -- (6.2,8.);
    \draw [blue] (7.6,7.2) -- (7.6,5.6);
    \draw [blue] (9.,4.8) -- (7.6,5.6);
    \draw [blue] (9.,4.8) -- (9.,4.2);
    \draw [blue] (9.,8.) -- (7.6,7.2);
    \draw [blue] (9.,8.) -- (9.,8.6);
    \draw [blue] (9.,4.8) -- (10.4,5.6);
    \draw [blue] (9.,8.) -- (10.4,7.2);
    \draw [blue] (10.4,5.6) -- (10.4,7.2);
    \draw [blue] (10.4,5.6) -- (11.,5.2);
    \draw [blue] (11.,7.6) -- (10.4,7.2);
    \draw [blue] (6.2,8.) -- (7.6,7.2);
    \draw [blue] (6.2,8.6) -- (6.2,8.);
    \node [centered, rotate=0.] at (4.4,6.4) {\scriptsize $j_{1}$};
    \node [centered, rotate=0.] at (5.2,7.9) {\scriptsize $j_{2}$};
    \node [centered, rotate=0.] at (6.8,7.3) {\scriptsize $j_{3}$};
    \node [centered, rotate=0.] at (8.,6.4) {\scriptsize $j_{4}$};
    \node [centered, rotate=0.] at (3.8,7.4) {\scriptsize $j_{5}$};
    \node [centered, rotate=0.] at (6.6,8.3) {\scriptsize $j_{6}$};
    \node [centered, rotate=0.] at (8.7,7.4) {\scriptsize $j_{7}$};
    \filldraw [red, opacity=0.2] (1.3,5.2) -- (1.3, 8.6) -- (3.4,8.6) -- (3.4, 8.) -- (4.8,7.2) -- (4.8,5.6) -- (6.2,4.8) -- (6.2,4.2) -- (1.3, 4.2);
    \filldraw [black, opacity=0.2] (3.4,8.6) -- (3.4, 8.) -- (4.8,7.2) -- (4.8,5.6) -- (6.2,4.8) -- (6.2,4.2) -- (9.,4.2) -- (9.,4.8) -- (7.6,5.6) -- (7.6,7.2) -- (6.2,8.) -- (6.2,8.6);
    \filldraw [blue, opacity=0.2] (9.,4.2) -- (9.,4.8) -- (7.6,5.6) -- (7.6,7.2) -- (6.2,8.) -- (6.2,8.6) --(11.,8.6) -- (11.,4.2);
\end{tikzpicture}\hs
}

\newcommand{\EqRibbonInterRight}{ 
\hs\begin{tikzpicture}[baseline={([yshift=-.8ex]current bounding box.center)},scale=.5]
    \draw [] (1.,3.8) -- (1.,9.);
    \draw [] (11.2,3.8) -- (12.5,6.4);
    \draw [] (12.5,6.4) -- (11.2,9.);
    \draw [red,thick] (2.,7.2) -- (2.,5.6);
    \draw [red,thick] (2.,5.6) -- (3.4,4.8);
    \draw [red,thick] (3.4,4.8) -- (4.8,5.6);
    \draw [red,thick] (4.8,5.6) -- (4.8,7.2);
    \draw [red,thick] (4.8,7.2) -- (3.4,8.);
    \draw [red,thick] (3.4,8.) -- (2.,7.2);
    \draw [red,thick] (3.4,8.6) -- (3.4,8.);
    \draw [red,thick] (1.3,7.6) -- (2.,7.2);
    \draw [red,thick] (1.3,5.2) -- (2.,5.6);
    \draw [red,thick] (3.4,4.2) -- (3.4,4.8);
    \draw [red,thick] (6.2,4.8) -- (4.8,5.6);
    \draw [red,thick] (6.2,4.2) -- (6.2,4.8);
    \draw [red,thick] (7.6,5.6) -- (6.2,4.8);
    \draw [red,thick] (4.8,7.2) -- (6.2,8.);
    \draw [blue] (7.6,7.2) -- (7.6,5.6);
    \draw [blue] (9.,4.8) -- (7.6,5.6);
    \draw [blue] (9.,4.8) -- (9.,4.2);
    \draw [blue] (9.,8.) -- (7.6,7.2);
    \draw [blue] (9.,8.) -- (9.,8.6);
    \draw [blue] (9.,4.8) -- (10.4,5.6);
    \draw [blue] (9.,8.) -- (10.4,7.2);
    \draw [blue] (10.4,5.6) -- (10.4,7.2);
    \draw [blue] (10.4,5.6) -- (11.,5.2);
    \draw [blue] (11.,7.6) -- (10.4,7.2);
    \draw [blue] (6.2,8.) -- (7.6,7.2);
    \draw [blue] (6.2,8.6) -- (6.2,8.);
    \filldraw [red, opacity=0.2] (1.3,5.2) -- (1.3, 8.6) -- (3.4,8.6) -- (3.4, 8.) -- (4.8,7.2) -- (4.8,5.6) -- (6.2,4.8) -- (6.2,4.2) -- (1.3, 4.2);
    \filldraw [black, opacity=0.2] (3.4,8.6) -- (3.4, 8.) -- (4.8,7.2) -- (4.8,5.6) -- (6.2,4.8) -- (6.2,4.2) -- (9.,4.2) -- (9.,4.8) -- (7.6,5.6) -- (7.6,7.2) -- (6.2,8.) -- (6.2,8.6);
    \filldraw [blue, opacity=0.2] (9.,4.2) -- (9.,4.8) -- (7.6,5.6) -- (7.6,7.2) -- (6.2,8.) -- (6.2,8.6) --(11.,8.6) -- (11.,4.2);
    \draw [red,thick] (3.8,6.2) -- (4.8, 6.2);
    \draw [blue] (7.6,6.6) -- (8.6,6.6);
    \node [centered, rotate=0.] at (3.5,6.1) {\scriptsize $p$};
    \node [centered, rotate=0.] at (8.8,6.6) {\scriptsize $q$};
    \node [centered, rotate=0.] at (5.2,5.9) {\scriptsize $j_{1}$};
    \node [centered, rotate=0.] at (5.2,6.8) {\scriptsize $j_{1}'$};
    \node [centered, rotate=0.] at (5.2,8.) {\scriptsize $j_{2}'$};
    \node [centered, rotate=0.] at (7.4,7.9) {\scriptsize $j_{3}'$};
    \node [centered, rotate=0.] at (7.2,6.9) {\scriptsize $j_{4}''$};
    \node [centered, rotate=0.] at (7.2,6.) {\scriptsize $j_{4}$};
    \node [centered, rotate=0.] at (3.8,7.4) {\scriptsize $j_{5}$};
    \node [centered, rotate=0.] at (6.6,8.3) {\scriptsize $j_{6}$};
    \node [centered, rotate=0.] at (8.7,7.4) {\scriptsize $j_{7}$};
\end{tikzpicture}\hs
}

\newcommand{\TableQuasiparticleDW}{
\begin{tabular}{|c|c|c|}
    \hline Quasiparticle & Anyon species & Charge \\ \hline
    $(1,1)$ & $1$ & $1$ \\ \hline
    $({e},\psi)$ & ${e}$ & $\psi$ \\ \hline
    $({m},1)$ & ${m}$ & $1$ \\ \hline
    $({\epsilon},\psi)$ & ${\epsilon}$ & $\psi$ \\ \hline
    $(\chi,\sigma)$ & $\chi$ & $\sigma$ \\ \hline
    $(\overline{\chi},\sigma)$ & $\overline{\chi}$ & $\sigma$ \\ \hline
\end{tabular}
}

\newcommand{\EqStateDomainWall}{
\hs\begin{tikzpicture}[baseline={([yshift=-.8ex]current bounding box.center)},scale=.43]
    \draw [] (.9,3.8) -- (.9,9.);
    \draw [] (8.4,3.8) -- (9.7,6.4);
    \draw [] (9.7,6.4) -- (8.4,9.);
    \draw [red,thick] (2.,7.2) -- (2.,5.6);
    \draw [red,thick] (2.,5.6) -- (3.4,4.8);
    \draw [red,thick] (3.4,4.8) -- (4.8,5.6);
    \draw [red,thick] (4.8,5.6) -- (4.8,7.2);
    \draw [red,thick] (1.3,5.2) -- (2.,5.6);
    \draw [red,thick] (3.4,4.2) -- (3.4,4.8);
    \draw [red,thick] (6.2,4.8) -- (4.8,5.6);
    \draw [red,thick] (6.2,4.2) -- (6.2,4.8);
    \draw [red,thick] (7.6,5.6) -- (6.2,4.8);
    \draw [ultra thick] (3.4,6.4) -- (6.2,6.4);
    \draw [red,thick] (7.6,7.2) -- (7.6,5.6);
    \draw [red,thick] (8.3,5.2) -- (7.6,5.6);
    \draw [blue] (3.4,8.6) -- (3.4,8.);
    \draw [blue] (3.4,8.) -- (2.,7.2);
    \draw [blue] (4.8,7.2) -- (6.2,8.);
    \draw [blue] (1.3,7.6) -- (2.,7.2);
    \draw [blue] (4.8,7.2) -- (3.4,8.);
    \draw [blue] (6.2,8.) -- (7.6,7.2);
    \draw [blue] (8.3,7.6) -- (7.6,7.2);
    \draw [blue] (6.2,8.6) -- (6.2,8.);
    \filldraw [red, opacity=0.2] (1.3,5.2) -- (2.,5.6) -- (3.4,4.8) -- (4.8,5.6) -- (6.2,4.8) -- (7.6,5.6) --(8.3,5.2) -- (8.3,4.2) -- (1.3,4.2);
    \filldraw [black, opacity=0.2] (1.3,7.6) -- (2.,7.2) -- (3.4,8.) -- (4.8,7.2) -- (6.2,8.) -- (7.6,7.2)--(8.3,7.6) --(8.3,5.2) -- (7.6,5.6) -- (6.2,4.8) -- (4.8,5.6) -- (3.4,4.8) -- (2.,5.6) -- (1.3,5.2);
    \filldraw [blue, opacity=0.2] (1.3,7.6) -- (2.,7.2) -- (3.4,8.) -- (4.8,7.2) -- (6.2,8.) -- (7.6,7.2)--(8.3,7.6) -- (8.3,8.6) -- (1.3,8.6);
    \node [centered, rotate=0.] at (2.15,6.4) {\scriptsize $(J_\text{DW},p)$};
    \node [centered, rotate=0.] at (7.4,6.4) {\scriptsize $(J_\text{DW},p)$};
\end{tikzpicture}\hs
}

\newcommand{\EqRibbonDomainWallLeft}{
\hs\begin{tikzpicture}[baseline={([yshift=-.8ex]current bounding box.center)},scale=.43]
    \draw [] (1.,3.8) -- (1.,9.);
    \draw [] (8.4,3.8) -- (9.7,6.4);
    \draw [] (9.7,6.4) -- (8.4,9.);
    \draw [red,thick] (3.4,4.8) -- (4.8,5.6);
    \draw [red,thick] (1.3,5.2) -- (2.,5.6);
    \draw [red,thick] (3.4,4.2) -- (3.4,4.8);
    \draw [red,thick] (6.2,4.2) -- (6.2,4.8);
    \draw [red,thick] (8.3,5.2) -- (7.6,5.6);
    \draw [red,thick] (6.2,4.8) -- (4.8,5.6);
    \draw [red,thick] (7.6,5.6) -- (6.2,4.8);
    \draw [red,thick] (2.,5.6) -- (3.4,4.8);
    \draw [blue] (1.3,7.6) -- (2.,7.2);
    \draw [blue] (3.4,8.) -- (2.,7.2);
    \draw [blue] (3.4,8.6) -- (3.4,8.);
    \draw [blue] (4.8,7.2) -- (3.4,8.);
    \draw [blue] (8.3,7.6) -- (7.6,7.2);
    \draw [blue] (4.8,7.2) -- (6.2,8.);
    \draw [blue] (6.2,8.) -- (7.6,7.2);
    \draw [blue] (6.2,8.6) -- (6.2,8.);
    \draw [red,thick] (7.6,7.2) -- (7.6,5.6);
    \draw [red,thick] (4.8,5.6) -- (4.8,7.2);
    \draw [red,thick] (2.,7.2) -- (2.,5.6);
    \filldraw [red, opacity=0.2] (1.3,5.2) -- (2.,5.6) -- (3.4,4.8) -- (4.8,5.6) -- (6.2,4.8) -- (7.6,5.6) --(8.3,5.2) -- (8.3,4.2) -- (1.3,4.2);
    \filldraw [black, opacity=0.2] (1.3,7.6) -- (2.,7.2) -- (3.4,8.) -- (4.8,7.2) -- (6.2,8.) -- (7.6,7.2)--(8.3,7.6) --(8.3,5.2) -- (7.6,5.6) -- (6.2,4.8) -- (4.8,5.6) -- (3.4,4.8) -- (2.,5.6) -- (1.3,5.2);
    \filldraw [blue, opacity=0.2] (1.3,7.6) -- (2.,7.2) -- (3.4,8.) -- (4.8,7.2) -- (6.2,8.) -- (7.6,7.2)--(8.3,7.6) -- (8.3,8.6) -- (1.3,8.6);
    \node [centered, rotate=0.] at (4.2,6.4) {\scriptsize $j_E$};
\end{tikzpicture}\hs
}

\newcommand{\EqRibbonDomainWallRight}{
\hs\begin{tikzpicture}[baseline={([yshift=-.8ex]current bounding box.center)},scale=.43]
    \draw [] (1.,3.8) -- (1.,9.);
    \draw [] (8.4,3.8) -- (9.7,6.4);
    \draw [] (9.7,6.4) -- (8.4,9.);
    \draw [red,thick] (3.4,4.8) -- (4.8,5.6);
    \draw [red,thick] (1.3,5.2) -- (2.,5.6);
    \draw [red,thick] (3.4,4.2) -- (3.4,4.8);
    \draw [red,thick] (6.2,4.2) -- (6.2,4.8);
    \draw [red,thick] (8.3,5.2) -- (7.6,5.6);
    \draw [red,thick] (6.2,4.8) -- (4.8,5.6);
    \draw [red,thick] (7.6,5.6) -- (6.2,4.8);
    \draw [red,thick] (2.,5.6) -- (3.4,4.8);
    \draw [blue] (1.3,7.6) -- (2.,7.2);
    \draw [blue] (3.4,8.) -- (2.,7.2);
    \draw [blue] (3.4,8.6) -- (3.4,8.);
    \draw [blue] (4.8,7.2) -- (3.4,8.);
    \draw [blue] (8.3,7.6) -- (7.6,7.2);
    \draw [blue] (4.8,7.2) -- (6.2,8.);
    \draw [blue] (6.2,8.) -- (7.6,7.2);
    \draw [blue] (6.2,8.6) -- (6.2,8.);
    \draw [red,thick] (4.8,5.6) -- (4.8,7.2);
    \draw [red,thick] (7.6,7.2) -- (7.6,5.6);
    \draw [red,thick] (2.,7.2) -- (2.,5.6);
    \draw [red,thick] (4.8,6.1) -- (3.6,6.1);
    \draw [red,thick] (4.8,6.7) -- (6.,6.7);
    \draw [red,thick] (6.2,4.8) -- (4.8,5.6);
    \filldraw [red, opacity=0.2] (1.3,5.2) -- (2.,5.6) -- (3.4,4.8) -- (4.8,5.6) -- (6.2,4.8) -- (7.6,5.6) --(8.3,5.2) -- (8.3,4.2) -- (1.3,4.2);
    \filldraw [black, opacity=0.2] (1.3,7.6) -- (2.,7.2) -- (3.4,8.) -- (4.8,7.2) -- (6.2,8.) -- (7.6,7.2)--(8.3,7.6) --(8.3,5.2) -- (7.6,5.6) -- (6.2,4.8) -- (4.8,5.6) -- (3.4,4.8) -- (2.,5.6) -- (1.3,5.2);
    \filldraw [blue, opacity=0.2] (1.3,7.6) -- (2.,7.2) -- (3.4,8.) -- (4.8,7.2) -- (6.2,8.) -- (7.6,7.2)--(8.3,7.6) -- (8.3,8.6) -- (1.3,8.6);
    \node [centered, rotate=0.] at (4.4,7) {\scriptsize $j_E$};
    \node [centered, rotate=0.] at (5.3,5.8) {\scriptsize $j_E$};
    \node [centered, rotate=0.] at (4.4,6.4) {\scriptsize $k$};
    \node [centered, rotate=0.] at (3.3,6.1) {\scriptsize $p$};
    \node [centered, rotate=0.] at (6.3,6.7) {\scriptsize $p$};
    \end{tikzpicture}\hs
}

\newcommand{\TableQuasiparticleDiDw}{
\begin{tabular}{|c|c|c|}
    \hline Elementary excitation state & Doubled-Ising quasiparticle & Domainwall quasiparticle\\ \hline
    $\ket{1\overline{1}\dash 1;1,1}$ & $(1\overline{1},1)$ & \multirow{2}{*}{$(1,1)$}\\
    $\ket{\psi\overline{\psi}\dash 1;1,1}$ & $(\psi\overline{\psi},1)$ &\\ \hline
    $\ket{\psi\overline{1}\dash \epsilon; \psi,\psi}$ & $(\psi\overline{1},\psi)$ & \multirow{2}{*}{$(\epsilon,\psi)$}\\
    $\ket{1\overline{\psi}\dash \epsilon; \psi,\psi}$ & $(1\overline{\psi},\psi)$ &\\ \hline
    $\ket{\sigma\overline{1}\dash \chi; \sigma,\sigma}$ & $(\sigma\overline{1},\sigma)$ & \multirow{2}{*}{$(\chi,\sigma)$}\\
    $\ket{\sigma\overline{\psi}\dash \chi; \sigma,\sigma}$ & $(\sigma\overline{\psi},\sigma)$ &\\ \hline
    $\ket{1\overline{\sigma}\dash \bar\chi; \sigma,\sigma}$ & $(1\overline{\sigma},\sigma)$ & \multirow{2}{*}{$(\bar\chi,\sigma)$}\\
    $\ket{\psi\overline{\sigma}\dash \bar\chi; \sigma,\sigma}$ & $(\psi\overline{\sigma},\sigma)$ &\\ \hline
    $\ket{\sigma\overline{\sigma}\dash m; 1,1}$ & $(\sigma\overline{\sigma},1)$ & \multirow{2}{*}{$(m,1)$} \\
    $\ket{\sigma\overline{\sigma}\dash m; \psi,1}$ & $(\sigma\overline{\sigma},\psi)$&\\ \cline{1-1}\cline{3-3}
    $\ket{\sigma\overline{\sigma}\dash e; 1,\psi}$ & $(\sigma\overline{\sigma},1)$ & \multirow{2}{*}{$(e,\psi)$}\\
    $\ket{\sigma\overline{\sigma}\dash e; \psi,\psi}$ & $(\sigma\overline{\sigma},\psi)$ &\\ \hline
\end{tabular}
}

\newcommand{\EqRibbonDiDwLeft}{
\hs\begin{tikzpicture}[baseline={([yshift=-.8ex]current bounding box.center)},scale=.4]
    \draw [] (1.,3.8) -- (1.,9.);
    \draw [] (8.4,3.8) -- (9.7,6.4);
    \draw [] (9.7,6.4) -- (8.4,9.);
    \draw [red,thick] (2.,7.2) -- (2.,5.6);
    \draw [red,thick] (2.,5.6) -- (3.4,4.8);
    \draw [red,thick] (3.4,4.8) -- (4.8,5.6);
    \draw [red,thick] (4.8,5.6) -- (4.8,7.2);
    \draw [red,thick] (4.8,7.2) -- (3.4,8.);
    \draw [red,thick] (3.4,8.) -- (2.,7.2);
    \draw [red,thick] (3.4,8.6) -- (3.4,8.);
    \draw [red,thick] (1.3,7.6) -- (2.,7.2);
    \draw [red,thick] (1.3,5.2) -- (2.,5.6);
    \draw [red,thick] (3.4,4.2) -- (3.4,4.8);
    \draw [red,thick] (6.2,4.8) -- (4.8,5.6);
    \draw [red,thick] (6.2,4.2) -- (6.2,4.8);
    \draw [red,thick] (4.8,7.2) -- (6.2,8.);
    \draw [red,thick] (7.6,5.6) -- (6.2,4.8);
    \draw [blue] (6.2,8.) -- (7.6,7.2);
    \draw [blue] (7.6,7.2) -- (7.6,5.6);
    \draw [blue] (8.3,5.2) -- (7.6,5.6);
    \draw [blue] (8.3,7.6) -- (7.6,7.2);
    \draw [blue] (6.2,8.6) -- (6.2,8.);
    \filldraw [red, opacity=0.2] (1.3,5.2) -- (1.3, 8.6) -- (3.4,8.6) -- (3.4, 8.) -- (4.8,7.2) -- (4.8,5.6) -- (6.2,4.8) -- (6.2,4.2) -- (1.3, 4.2);
    \filldraw [black, opacity=0.2] (3.4,8.6) -- (3.4, 8.) -- (4.8,7.2) -- (4.8,5.6) -- (6.2,4.8) -- (6.2,4.2) -- (8.3,4.2) -- (8.3,5.2) -- (7.6,5.6) -- (7.6,7.2) -- (6.2,8.) -- (6.2,8.6);
    \filldraw [blue, opacity=0.2] (8.3,5.2) -- (7.6,5.6) -- (7.6,7.2) -- (6.2,8.) -- (6.2,8.6) -- (8.3, 8.6);
    \node [centered, rotate=0.] at (4.2,6.4) {\scriptsize $j_E$};
\end{tikzpicture}\hs
}

\newcommand{\EqRibbonDiDwRight}{
\hs\begin{tikzpicture}[baseline={([yshift=-.8ex]current bounding box.center)},scale=.4]
    \draw [] (1.,3.8) -- (1.,9.);
    \draw [] (8.4,3.8) -- (9.7,6.4);
    \draw [] (9.7,6.4) -- (8.4,9.);
    \draw [red,thick] (4.8,5.6) -- (4.8,7.2);
    \draw [red,thick] (4.8,7.2) -- (3.4,8.);
    \draw [red,thick] (3.4,8.) -- (2.,7.2);
    \draw [red,thick] (2.,7.2) -- (2.,5.6);
    \draw [red,thick] (2.,5.6) -- (3.4,4.8);
    \draw [red,thick] (3.4,4.8) -- (4.8,5.6);
    \draw [red,thick] (1.3,5.2) -- (2.,5.6);
    \draw [red,thick] (1.3,7.6) -- (2.,7.2);
    \draw [red,thick] (3.4,8.4) -- (3.4,8.);
    \draw [red,thick] (3.4,8.6) -- (3.4,8.4);
    \draw [red,thick] (6.2,4.2) -- (6.2,4.4);
    \draw [red,thick] (6.2,4.4) -- (6.2,4.8);
    \draw [red,thick] (3.4,4.2) -- (3.4,4.4);
    \draw [red,thick] (3.4,4.4) -- (3.4,4.8);
    \draw [red,thick] (4.8,6.8) -- (4.8,7.2);
    \draw [red,thick] (4.8,5.6) -- (4.8,6.);
    \draw [red,thick] (4.8,6.1) -- (3.6,6.1);
    \draw [red,thick] (4.8,6.7) -- (6.,6.7);
    \draw [red,thick] (6.2,4.8) -- (4.8,5.6);
    \draw [red,thick] (4.8,7.2) -- (6.2,8.);
    \draw [red,thick] (7.6,5.6) -- (6.2,4.8);
    \draw [blue] (7.6,7.2) -- (7.6,5.6);
    \draw [blue] (8.3,5.2) -- (7.6,5.6);
    \draw [blue] (8.3,7.6) -- (7.6,7.2);
    \draw [blue] (6.2,8.) -- (7.6,7.2);
    \draw [blue] (6.2,8.6) -- (6.2,8.);
    \filldraw [red, opacity=0.2] (1.3,5.2) -- (1.3, 8.6) -- (3.4,8.6) -- (3.4, 8.) -- (4.8,7.2) -- (4.8,5.6) -- (6.2,4.8) -- (6.2,4.2) -- (1.3, 4.2);
    \filldraw [black, opacity=0.2] (3.4,8.6) -- (3.4, 8.) -- (4.8,7.2) -- (4.8,5.6) -- (6.2,4.8) -- (6.2,4.2) -- (8.3,4.2) -- (8.3,5.2) -- (7.6,5.6) -- (7.6,7.2) -- (6.2,8.) -- (6.2,8.6);
    \filldraw [blue, opacity=0.2] (8.3,5.2) -- (7.6,5.6) -- (7.6,7.2) -- (6.2,8.) -- (6.2,8.6) -- (8.3, 8.6);
    \node [centered, rotate=0.] at (4.4,7) {\scriptsize $j_E$};
    \node [centered, rotate=0.] at (5.3,5.8) {\scriptsize $j_E$};
    \node [centered, rotate=0.] at (4.4,6.4) {\scriptsize $k$};
    \node [centered, rotate=0.] at (3.4,6.1) {\scriptsize $p$};
    \node [centered, rotate=0.] at (6.2,6.7) {\scriptsize $q$};
\end{tikzpicture}\hs
}

\newcommand{\TableQuasiparticleTcDw}{
\begin{tabular}{|c|c|c|}
    \hline Elementary excitation state & Domainwall quasiparticle & Toric-code quasiparticle\\ \hline
    $\ket{1\dash 1; 1,1}$ & $(1,1)$ & $(1,1)$\\ \hline
    $\ket{e\dash e; \psi,\psi}$ & $({e},\psi)$ & $(e,\psi)$ \\ \hline
    $\ket{m\dash m; 1,1}$ & $({m},1)$ & $(m,1)$\\ \hline
    $\ket{\epsilon\dash \epsilon; \psi,\psi}$ & $({\epsilon},\psi)$ & $(\epsilon,\psi)$ \\ \hline
\end{tabular}
}

\newcommand{\EqRibbonDwTcLeft}{
\hs\begin{tikzpicture}[baseline={([yshift=-.8ex]current bounding box.center)},scale=.4]
    \draw [] (1.,3.8) -- (1.,9.);
    \draw [] (8.4,3.8) -- (9.7,6.4);
    \draw [] (9.7,6.4) -- (8.4,9.);
    \draw [red,thick] (2.,7.2) -- (2.,5.6);
    \draw [red,thick] (1.3,7.6) -- (2.,7.2);
    \draw [red,thick] (1.3,5.2) -- (2.,5.6);
    \draw [red,thick] (3.4,4.2) -- (3.4,4.8);
    \draw [red,thick] (2.,5.6) -- (3.4,4.8);
    \draw [red,thick] (3.4,8.) -- (2.,7.2);
    \draw [red,thick] (3.4,4.8) -- (4.8,5.6);
    \draw [blue] (3.4,8.6) -- (3.4,8.);
    \draw [blue] (4.8,5.6) -- (4.8,7.2);
    \draw [blue] (4.8,7.2) -- (3.4,8.);
    \draw [blue] (6.2,4.8) -- (4.8,5.6);
    \draw [blue] (6.2,4.2) -- (6.2,4.8);
    \draw [blue] (7.6,7.2) -- (7.6,5.6);
    \draw [blue] (7.6,5.6) -- (6.2,4.8);
    \draw [blue] (8.3,5.2) -- (7.6,5.6);
    \draw [blue] (8.3,7.6) -- (7.6,7.2);
    \draw [blue] (4.8,7.2) -- (6.2,8.);
    \draw [blue] (6.2,8.) -- (7.6,7.2);
    \draw [blue] (6.2,8.6) -- (6.2,8.);
    \filldraw [red, opacity=0.2] (1.3,4.2) -- (1.3, 7.6) -- (2.,7.2) -- (2.,5.6) -- (3.4,4.8) -- (3.4,4.2);
    \filldraw [black, opacity=0.2] (1.3, 7.6) -- (1.3, 8.6) -- (3.4,8.6) -- (3.4, 8.) -- (4.8,7.2) -- (4.8,5.6) -- (6.2,4.8) -- (6.2,4.2) -- (3.4, 4.2) -- (3.4, 4.8) -- (2.,5.6) -- (2.,7.2);
    \filldraw [blue, opacity=0.2] (3.4,8.6) -- (3.4, 8.) -- (4.8,7.2) -- (4.8,5.6) -- (6.2,4.8) -- (6.2,4.2) --(8.3, 4.2) -- (8.3,5.2) -- (8.3, 8.6);
    \node [centered, rotate=0.] at (4.2,6.4) {\scriptsize $j_E$};
\end{tikzpicture}
}
\newcommand{\EqRibbonDwTcRight}{
\hs\begin{tikzpicture}[baseline={([yshift=-.8ex]current bounding box.center)},scale=.4]
    \draw [] (1.,3.8) -- (1.,9.);
    \draw [] (8.4,3.8) -- (9.7,6.4);
    \draw [] (9.7,6.4) -- (8.4,9.);
    \draw [red,thick] (2.,7.2) -- (2.,5.6);
    \draw [red,thick] (2.,5.6) -- (3.4,4.8);
    \draw [red,thick] (1.3,5.2) -- (2.,5.6);
    \draw [red,thick] (1.3,7.6) -- (2.,7.2);
    \draw [red,thick] (3.4,4.2) -- (3.4,4.4);
    \draw [red,thick] (3.4,4.4) -- (3.4,4.8);
    \draw [red,thick] (3.4,8.) -- (2.,7.2);
    \draw [red,thick] (3.4,4.8) -- (4.8,5.6);
    \draw [blue] (6.2,4.8) -- (4.8,5.6);
    \draw [blue] (4.8,6.8) -- (4.8,7.2);
    \draw [blue] (4.8,5.6) -- (4.8,6.);
    \draw [blue] (4.8,6.1) -- (3.6,6.1);
    \draw [blue] (4.8,6.7) -- (6.,6.7);
    \draw [blue] (3.4,8.4) -- (3.4,8.);
    \draw [blue] (3.4,8.6) -- (3.4,8.4);
    \draw [blue] (6.2,4.2) -- (6.2,4.4);
    \draw [blue] (6.2,4.4) -- (6.2,4.8);
    \draw [blue] (4.8,5.6) -- (4.8,7.2);
    \draw [blue] (4.8,7.2) -- (3.4,8.);
    \draw [blue] (7.6,7.2) -- (7.6,5.6);
    \draw [blue] (7.6,5.6) -- (6.2,4.8);
    \draw [blue] (8.3,5.2) -- (7.6,5.6);
    \draw [blue] (8.3,7.6) -- (7.6,7.2);
    \draw [blue] (4.8,7.2) -- (6.2,8.);
    \draw [blue] (6.2,8.) -- (7.6,7.2);
    \draw [blue] (6.2,8.6) -- (6.2,8.);
    \filldraw [red, opacity=0.2] (1.3,4.2) -- (1.3, 7.6) -- (2.,7.2) -- (2.,5.6) -- (3.4,4.8) -- (3.4,4.2);
    \filldraw [black, opacity=0.2] (1.3, 7.6) -- (1.3, 8.6) -- (3.4,8.6) -- (3.4, 8.) -- (4.8,7.2) -- (4.8,5.6) -- (6.2,4.8) -- (6.2,4.2) -- (3.4, 4.2) -- (3.4, 4.8) -- (2.,5.6) -- (2.,7.2);
    \filldraw [blue, opacity=0.2] (3.4,8.6) -- (3.4, 8.) -- (4.8,7.2) -- (4.8,5.6) -- (6.2,4.8) -- (6.2,4.2) --(8.3, 4.2) -- (8.3,5.2) -- (8.3, 8.6);
    \node [centered, rotate=0.] at (4.4,7) {\scriptsize $j_E$};
    \node [centered, rotate=0.] at (5.3,5.8) {\scriptsize $j_E$};
    \node [centered, rotate=0.] at (4.4,6.4) {\scriptsize $k$};
    \node [centered, rotate=0.] at (3.4,6.1) {\scriptsize $p$};
    \node [centered, rotate=0.] at (6.2,6.7) {\scriptsize $p$};
\end{tikzpicture}\hs
}

\newcommand{\FigureCorrespondence}{
\begin{tikzpicture}[baseline={([yshift=-.8ex]current bounding box.center)},scale=.95]

\filldraw [red, opacity=0.2] (-4.4,5.5) -- (-1.0,5.5) -- (0.6,7.1) -- (-2.8,7.1);
\node [centered, rotate=0.] at (-2.4,5.8) {\scriptsize {Doubled Ising Domain}};
\fill [red] (-1.4, 6.5) circle (0.1);
\node [centered, rotate=0.] at (-1.4,6.8) {\scriptsize $J_\text{DI}$};

\filldraw [black, opacity=0.2] (0.6,5.5) -- (-1.0,5.5) -- (0.6,7.1) -- (2.2,7.1);
\node [centered, rotate=45.] at (0.9,6.3) {\scriptsize {Domain Wall}};
\fill [black] (0.7, 6.5) circle (0.1);
\node [centered, rotate=0.] at (0.7,6.8) {\scriptsize $J_\text{DW}$};

\filldraw [blue, opacity=0.2] (0.6,5.5) -- (4.,5.5) -- (5.6,7.1) -- (2.2,7.1);
\node [centered, rotate=0.] at (2.5,5.8) {\scriptsize {Toric Code Domain}};
\fill [blue] (3.0, 6.5) circle (0.1);
\node [centered, rotate=0.] at (3.0,6.8) {\scriptsize $J_\text{TC}$};

\filldraw [red, opacity=0.2] (5.6,6.4) -- (9.0,6.4) -- (10.2,7.6) -- (6.8,7.6);
\node [centered, rotate=0.] at (7.5,6.7) {\scriptsize {Parent Phase}};
\fill [red] (9.0, 7.0) circle (0.1);
\node [centered, rotate=0.] at (9.3,7.3) {\scriptsize $J_\text{DI}$};

\filldraw [black, opacity=0.2] (5.6,5.6) -- (9.0,5.6) -- (10.2,6.8) -- (9.4,6.8) -- (9.0,6.4) -- (6.4,6.4);
\node [centered, rotate=0.] at (7.5,5.9) {\scriptsize {Intermediate Phase}};
\fill [black] (9.0, 6.) circle (0.1);
\node [centered, rotate=0.] at (9.3,6.3) {\scriptsize $J_\text{DW}$};

\filldraw [blue, opacity=0.2] (5.6,4.8) -- (9.0,4.8) -- (10.2,6.) -- (9.4,6.) -- (9.0,5.6) -- (6.4,5.6);
\node [centered, rotate=0.] at (7.5,5.1) {\scriptsize {Child Phase}};
\fill [blue] (9.0, 5.1) circle (0.1);
\node [centered, rotate=0.] at (9.2,5.4) {\scriptsize $J_\text{TC}$};

\draw [ultra thick, ->, orange] (-4.6,5.2) -- (4.8,5.2);
\node [centered, rotate=0.] at (-0.4,4.9) {\scriptsize $x$};
\node [centered, rotate=0.] at (-0.4,4.) {\scriptsize {(a)}};

\draw [ultra thick, ->, magenta] (9.9,8.) -- (9.9,4.8);
\node [centered, rotate=0.] at (10.1,6.3) {\scriptsize $t$};
\node [centered, rotate=0.] at (7.6,4.) {\scriptsize {(b)}};
\end{tikzpicture}
}

\newcommand{\TableAnyonCondensation}{
\begin{tikzpicture}[baseline={([yshift=-.8ex]current bounding box.center)},scale=1]
\draw [thick] (2.,7.6) -- (3.6,7.6);
\draw [thick] (2.,6.4) -- (2.8,6.);
\draw [thick] (2.,5.6) -- (2.8,6.);
\draw [thick] (2.,4.8) -- (2.8,4.4);
\draw [thick] (2.,4.) -- (2.8,4.4);
\draw [thick] (2.,3.2) -- (2.8,2.8);
\draw [thick] (2.,2.4) -- (2.8,2.8);
\draw [thick] (2.,1.6) -- (2.8,1.2);
\draw [thick] (2.,0.8) -- (2.8,1.2);
\draw [thick, ->] (3.6,7.6) -- (4.7,8.);
\draw [thick, ->] (3.6,7.6) -- (4.7,7.2);
\draw [thick, ->] (2.8,6.) -- (4.7,6.);
\draw [thick, ->] (2.8,4.4) -- (4.7,4.4);
\draw [thick, ->] (2.8,2.8) -- (4.7,2.8);
\draw [thick, ->] (2.8,1.2) -- (4.7,1.2);
\draw [thick, ->] (5.8,8.) -- (8.2,8.);
\draw [thick, ->] (5.8,7.2) -- (8.2,7.2);
\draw [thick, ->] (5.8,6.) -- (8.2,6.);
\draw [thick, ->] (5.8,4.4) -- (8.2,4.4);
\draw [thick, dashed, ->] (5.7,2.8) -- (7.7,2.8);
\draw [thick, dashed, ->] (5.7,1.2) -- (7.7,1.2);
\node [centered, rotate=0.] at (1.4,9.) {\scriptsize $J_\text{DI}$};
\node [centered, rotate=0.] at (5.2,9.) {\scriptsize $J_\text{DW}$};
\node [centered, rotate=0.] at (8.7,9.) {\scriptsize $J_\text{TC}$};
\node [centered, rotate=0.] at (1.4,7.6) {\scriptsize $\sigma\bar\sigma$};
\node [centered, rotate=0.] at (1.4,6.4) {\scriptsize $1\bar 1$};
\node [centered, rotate=0.] at (1.4,5.6) {\scriptsize $\psi\bar\psi$};
\node [centered, rotate=0.] at (1.4,4.8) {\scriptsize $\psi\bar 1$};
\node [centered, rotate=0.] at (1.4,4.) {\scriptsize $1\bar\psi$};
\node [centered, rotate=0.] at (1.4,3.2) {\scriptsize $\sigma\bar 1$};
\node [centered, rotate=0.] at (1.4,2.4) {\scriptsize $\sigma\bar\psi$};
\node [centered, rotate=0.] at (1.4,1.6) {\scriptsize $1\bar\sigma$};
\node [centered, rotate=0.] at (1.4,0.8) {\scriptsize $\psi\bar\sigma$};
\node [centered, rotate=0.] at (5.2,8.) {\scriptsize $m$};
\node [centered, rotate=0.] at (5.2,7.2) {\scriptsize $e$};
\node [centered, rotate=0.] at (8.7,8.) {\scriptsize $m$};
\node [centered, rotate=0.] at (8.7,7.2) {\scriptsize $e$};
\node [centered, rotate=0.] at (5.2,6.) {\scriptsize $1$};
\node [centered, rotate=0.] at (8.7,6.) {\scriptsize $1$};
\node [centered, rotate=0.] at (5.2,4.4) {\scriptsize $\epsilon$};
\node [centered, rotate=0.] at (8.7,4.4) {\scriptsize $\epsilon$};
\node [centered, rotate=0.] at (5.2,2.8) {\scriptsize $\chi$};
\node [centered, rotate=0.] at (5.2,1.2) {\scriptsize $\bar\chi$};
\node [centered, rotate=0.] at (2.8,7.9) {\scriptsize {Splitting}};
\node [centered, rotate=0.] at (3.6,6.3) {\scriptsize {Identification}};
\node [centered, rotate=0.] at (3.6,4.7) {\scriptsize {Identification}};
\node [centered, rotate=0.] at (3.6,3.1) {\scriptsize {Identification}};
\node [centered, rotate=0.] at (3.6,1.5) {\scriptsize {Identification}};
\node [centered, rotate=0.] at (8.7,2.8) {\scriptsize {(confined)}};
\node [centered, rotate=0.] at (8.7,1.2) {\scriptsize {(confined)}};
\node [centered, rotate=0.] at (6.6,3.1) {\scriptsize {Confinement}};
\node [centered, rotate=0.] at (6.6,1.5) {\scriptsize {Confinement}};

\draw [] (0.9,8.6) -- (9.7,8.6);
\draw [] (0.9,9.4) -- (9.7,9.4);
\draw [] (0.9,0.3) -- (9.7,0.3);
\draw [] (0.9,9.4) -- (0.9,0.3);
\draw [] (9.7,9.4) -- (9.7,0.3);
\end{tikzpicture}
}

\newcommand{\TableCorrespondenceTC}{
\begin{tabular}{|c|c|c|}
     \hline
     Phenomenon & Projection & Interdomain states\\ \hline
     \multirow{2}{*}{Splitting} & $P_\text{eff}\ket{\sigma\bar\sigma;1,1}_\text{DI} = \ket{m;1,1}$ & $\ket{\sigma\bar\sigma\dash m;1,1},\quad \ket{\sigma\bar\sigma\dash m;\psi,1}$\\ \cline{2-3}
     & $P_\text{eff}\ket{\sigma\bar\sigma;\psi,\psi}_\text{DI} = \ket{e;\psi,\psi}$ & $\ket{\sigma\bar\sigma\dash e;1,\psi},\quad \ket{\sigma\bar\sigma\dash e;\psi,\psi}$\\ \hline
     \multirow{4}{*}{Identification} & $P_\text{eff}\ket{1\bar 1;1,1}_\text{DI} = \ket{1;1,1}$ & $\ket{1\bar 1\dash 1;1,1}$\\
     & $P_\text{eff}\ket{\psi\bar\psi;1,1}_\text{DI} = \ket{1;1,1}$ & $\ket{\psi\bar\psi\dash 1;1,1}$\\ \cline{2-3}
     & $P_\text{eff}\ket{\psi\bar 1;\psi,\psi}_\text{DI} = \ket{\epsilon;\psi,\psi}$ & $\ket{\psi\bar 1\dash \epsilon;\psi,\psi}$ \\
     & $P_\text{eff}\ket{1\bar\psi;\psi,\psi}_\text{DI} = \ket{\epsilon;\psi,\psi}$ & $\ket{1\bar\psi\dash \epsilon;\psi,\psi}$\\ \hline
     \multirow{4}{*}{Confinement} & $P_\text{eff}\ket{\sigma\bar 1;\sigma,\sigma}_\text{DI} = 0$ & \\
     & $P_\text{eff}\ket{\sigma\bar\psi;1,1}_\text{DI} = 0$ & There is no interdomain state\\ 
     & $P_\text{eff}\ket{1\bar\sigma;\sigma,\sigma}_\text{DI} = 0$ &  with $J_\text{DI} = \sigma\bar 1,\ \sigma\bar\psi,\ 1\bar\sigma,\ \psi\bar\sigma$.\\
     & $P_\text{eff}\ket{\psi\bar\sigma;\sigma,\sigma}_\text{DI} = 0$ & \\ \hline
\end{tabular}
}

\newcommand{\FigureSplittingA}{
\begin{tikzpicture}[baseline={([yshift=-.8ex]current bounding box.center)},scale=.33]
    \draw [red,thick] (2.,7.2) -- (2.,5.6);
    \draw [red,thick] (2.,5.6) -- (3.4,4.8);
    \draw [red,thick] (3.4,4.8) -- (4.8,5.6);
    \draw [red,thick] (4.8,5.6) -- (4.8,7.2);
    \draw [red,thick] (4.8,7.2) -- (3.4,8.);
    \draw [red,thick] (3.4,8.) -- (2.,7.2);
    \draw [red,thick] (3.4,8.6) -- (3.4,8.);
    \draw [red,thick] (1.3,7.6) -- (2.,7.2);
    \draw [red,thick] (1.3,5.2) -- (2.,5.6);
    \draw [red,thick] (3.4,4.2) -- (3.4,4.8);
    \draw [red,thick] (6.2,4.8) -- (4.8,5.6);
    \draw [red,thick] (6.2,4.2) -- (6.2,4.8);
    \draw [red,thick] (7.6,5.6) -- (6.2,4.8);
    \draw [red,thick] (4.8,7.2) -- (6.2,8.);
    \draw [blue] (7.6,7.2) -- (7.6,5.6);
    \draw [blue] (9.,4.8) -- (7.6,5.6);
    \draw [blue] (9.,4.8) -- (9.,4.2);
    \draw [blue] (9.,8.) -- (7.6,7.2);
    \draw [blue] (9.,8.) -- (9.,8.6);
    \draw [blue] (9.,4.8) -- (10.4,5.6);
    \draw [blue] (9.,8.) -- (10.4,7.2);
    \draw [blue] (10.4,5.6) -- (10.4,7.2);
    \draw [blue] (10.4,5.6) -- (11.,5.2);
    \draw [blue] (11.,7.6) -- (10.4,7.2);
    \draw [blue] (6.2,8.) -- (7.6,7.2);
    \draw [blue] (6.2,8.6) -- (6.2,8.);
    \filldraw [red, opacity=0.2] (1.3,5.2) -- (1.3, 8.6) -- (3.4,8.6) -- (3.4, 8.) -- (4.8,7.2) -- (4.8,5.6) -- (6.2,4.8) -- (6.2,4.2) -- (1.3, 4.2);
    \filldraw [black, opacity=0.2] (3.4,8.6) -- (3.4, 8.) -- (4.8,7.2) -- (4.8,5.6) -- (6.2,4.8) -- (6.2,4.2) -- (9.,4.2) -- (9.,4.8) -- (7.6,5.6) -- (7.6,7.2) -- (6.2,8.) -- (6.2,8.6);
    \filldraw [blue, opacity=0.2] (9.,4.2) -- (9.,4.8) -- (7.6,5.6) -- (7.6,7.2) -- (6.2,8.) -- (6.2,8.6) --(11.,8.6) -- (11.,4.2);
    \draw [ultra thick] (3.8,6.4) -- (8.6,6.4);
    \node [centered, rotate=0.] at (2.5,6.4) {\scriptsize $(\sigma\bar\sigma,1)$};
    \node [centered, rotate=0.] at (9.75,6.4) {\scriptsize $(m,1)$};
    \node [centered, rotate=0.] at (6.2,3.5) {\scriptsize {(a)}};
\end{tikzpicture}
}

\newcommand{\FigureSplittingB}{
\begin{tikzpicture}[baseline={([yshift=-.8ex]current bounding box.center)},scale=.33]
    \draw [red,thick] (2.,7.2) -- (2.,5.6);
    \draw [red,thick] (2.,5.6) -- (3.4,4.8);
    \draw [red,thick] (3.4,4.8) -- (4.8,5.6);
    \draw [red,thick] (4.8,5.6) -- (4.8,7.2);
    \draw [red,thick] (4.8,7.2) -- (3.4,8.);
    \draw [red,thick] (3.4,8.) -- (2.,7.2);
    \draw [red,thick] (3.4,8.6) -- (3.4,8.);
    \draw [red,thick] (1.3,7.6) -- (2.,7.2);
    \draw [red,thick] (1.3,5.2) -- (2.,5.6);
    \draw [red,thick] (3.4,4.2) -- (3.4,4.8);
    \draw [red,thick] (6.2,4.8) -- (4.8,5.6);
    \draw [red,thick] (6.2,4.2) -- (6.2,4.8);
    \draw [red,thick] (7.6,5.6) -- (6.2,4.8);
    \draw [red,thick] (4.8,7.2) -- (6.2,8.);
    \draw [blue] (7.6,7.2) -- (7.6,5.6);
    \draw [blue] (9.,4.8) -- (7.6,5.6);
    \draw [blue] (9.,4.8) -- (9.,4.2);
    \draw [blue] (9.,8.) -- (7.6,7.2);
    \draw [blue] (9.,8.) -- (9.,8.6);
    \draw [blue] (9.,4.8) -- (10.4,5.6);
    \draw [blue] (9.,8.) -- (10.4,7.2);
    \draw [blue] (10.4,5.6) -- (10.4,7.2);
    \draw [blue] (10.4,5.6) -- (11.,5.2);
    \draw [blue] (11.,7.6) -- (10.4,7.2);
    \draw [blue] (6.2,8.) -- (7.6,7.2);
    \draw [blue] (6.2,8.6) -- (6.2,8.);
    \filldraw [red, opacity=0.2] (1.3,5.2) -- (1.3, 8.6) -- (3.4,8.6) -- (3.4, 8.) -- (4.8,7.2) -- (4.8,5.6) -- (6.2,4.8) -- (6.2,4.2) -- (1.3, 4.2);
    \filldraw [black, opacity=0.2] (3.4,8.6) -- (3.4, 8.) -- (4.8,7.2) -- (4.8,5.6) -- (6.2,4.8) -- (6.2,4.2) -- (9.,4.2) -- (9.,4.8) -- (7.6,5.6) -- (7.6,7.2) -- (6.2,8.) -- (6.2,8.6);
    \filldraw [blue, opacity=0.2] (9.,4.2) -- (9.,4.8) -- (7.6,5.6) -- (7.6,7.2) -- (6.2,8.) -- (6.2,8.6) --(11.,8.6) -- (11.,4.2);
    \draw [ultra thick] (3.8,6.4) -- (8.6,6.4);
    \node [centered, rotate=0.] at (2.5,6.4) {\scriptsize $(\sigma\bar\sigma,\psi)$};
    \node [centered, rotate=0.] at (9.75,6.4) {\scriptsize $(m,1)$};
    \node [centered, rotate=0.] at (6.2,3.5) {\scriptsize {(b)}};
\end{tikzpicture}
}

\newcommand{\FigureSplittingC}{
\begin{tikzpicture}[baseline={([yshift=-.8ex]current bounding box.center)},scale=.33]
    \draw [red,thick] (2.,7.2) -- (2.,5.6);
    \draw [red,thick] (2.,5.6) -- (3.4,4.8);
    \draw [red,thick] (3.4,4.8) -- (4.8,5.6);
    \draw [red,thick] (4.8,5.6) -- (4.8,7.2);
    \draw [red,thick] (4.8,7.2) -- (3.4,8.);
    \draw [red,thick] (3.4,8.) -- (2.,7.2);
    \draw [red,thick] (3.4,8.6) -- (3.4,8.);
    \draw [red,thick] (1.3,7.6) -- (2.,7.2);
    \draw [red,thick] (1.3,5.2) -- (2.,5.6);
    \draw [red,thick] (3.4,4.2) -- (3.4,4.8);
    \draw [red,thick] (6.2,4.8) -- (4.8,5.6);
    \draw [red,thick] (6.2,4.2) -- (6.2,4.8);
    \draw [red,thick] (7.6,5.6) -- (6.2,4.8);
    \draw [red,thick] (4.8,7.2) -- (6.2,8.);
    \draw [blue] (7.6,7.2) -- (7.6,5.6);
    \draw [blue] (9.,4.8) -- (7.6,5.6);
    \draw [blue] (9.,4.8) -- (9.,4.2);
    \draw [blue] (9.,8.) -- (7.6,7.2);
    \draw [blue] (9.,8.) -- (9.,8.6);
    \draw [blue] (9.,4.8) -- (10.4,5.6);
    \draw [blue] (9.,8.) -- (10.4,7.2);
    \draw [blue] (10.4,5.6) -- (10.4,7.2);
    \draw [blue] (10.4,5.6) -- (11.,5.2);
    \draw [blue] (11.,7.6) -- (10.4,7.2);
    \draw [blue] (6.2,8.) -- (7.6,7.2);
    \draw [blue] (6.2,8.6) -- (6.2,8.);
    \filldraw [red, opacity=0.2] (1.3,5.2) -- (1.3, 8.6) -- (3.4,8.6) -- (3.4, 8.) -- (4.8,7.2) -- (4.8,5.6) -- (6.2,4.8) -- (6.2,4.2) -- (1.3, 4.2);
    \filldraw [black, opacity=0.2] (3.4,8.6) -- (3.4, 8.) -- (4.8,7.2) -- (4.8,5.6) -- (6.2,4.8) -- (6.2,4.2) -- (9.,4.2) -- (9.,4.8) -- (7.6,5.6) -- (7.6,7.2) -- (6.2,8.) -- (6.2,8.6);
    \filldraw [blue, opacity=0.2] (9.,4.2) -- (9.,4.8) -- (7.6,5.6) -- (7.6,7.2) -- (6.2,8.) -- (6.2,8.6) --(11.,8.6) -- (11.,4.2);
    \draw [ultra thick] (3.8,6.4) -- (8.6,6.4);
    \node [centered, rotate=0.] at (2.5,6.4) {\scriptsize $(\sigma\bar\sigma,1)$};
    \node [centered, rotate=0.] at (9.7,6.4) {\scriptsize $(e,\psi)$};
    \node [centered, rotate=0.] at (6.2,3.5) {\scriptsize {(c)}};
\end{tikzpicture}
}

\newcommand{\FigureSplittingD}{
\begin{tikzpicture}[baseline={([yshift=-.8ex]current bounding box.center)},scale=.33]
    \draw [red,thick] (2.,7.2) -- (2.,5.6);
    \draw [red,thick] (2.,5.6) -- (3.4,4.8);
    \draw [red,thick] (3.4,4.8) -- (4.8,5.6);
    \draw [red,thick] (4.8,5.6) -- (4.8,7.2);
    \draw [red,thick] (4.8,7.2) -- (3.4,8.);
    \draw [red,thick] (3.4,8.) -- (2.,7.2);
    \draw [red,thick] (3.4,8.6) -- (3.4,8.);
    \draw [red,thick] (1.3,7.6) -- (2.,7.2);
    \draw [red,thick] (1.3,5.2) -- (2.,5.6);
    \draw [red,thick] (3.4,4.2) -- (3.4,4.8);
    \draw [red,thick] (6.2,4.8) -- (4.8,5.6);
    \draw [red,thick] (6.2,4.2) -- (6.2,4.8);
    \draw [red,thick] (7.6,5.6) -- (6.2,4.8);
    \draw [red,thick] (4.8,7.2) -- (6.2,8.);
    \draw [blue] (7.6,7.2) -- (7.6,5.6);
    \draw [blue] (9.,4.8) -- (7.6,5.6);
    \draw [blue] (9.,4.8) -- (9.,4.2);
    \draw [blue] (9.,8.) -- (7.6,7.2);
    \draw [blue] (9.,8.) -- (9.,8.6);
    \draw [blue] (9.,4.8) -- (10.4,5.6);
    \draw [blue] (9.,8.) -- (10.4,7.2);
    \draw [blue] (10.4,5.6) -- (10.4,7.2);
    \draw [blue] (10.4,5.6) -- (11.,5.2);
    \draw [blue] (11.,7.6) -- (10.4,7.2);
    \draw [blue] (6.2,8.) -- (7.6,7.2);
    \draw [blue] (6.2,8.6) -- (6.2,8.);
    \filldraw [red, opacity=0.2] (1.3,5.2) -- (1.3, 8.6) -- (3.4,8.6) -- (3.4, 8.) -- (4.8,7.2) -- (4.8,5.6) -- (6.2,4.8) -- (6.2,4.2) -- (1.3, 4.2);
    \filldraw [black, opacity=0.2] (3.4,8.6) -- (3.4, 8.) -- (4.8,7.2) -- (4.8,5.6) -- (6.2,4.8) -- (6.2,4.2) -- (9.,4.2) -- (9.,4.8) -- (7.6,5.6) -- (7.6,7.2) -- (6.2,8.) -- (6.2,8.6);
    \filldraw [blue, opacity=0.2] (9.,4.2) -- (9.,4.8) -- (7.6,5.6) -- (7.6,7.2) -- (6.2,8.) -- (6.2,8.6) --(11.,8.6) -- (11.,4.2);
    \draw [ultra thick] (3.8,6.4) -- (8.6,6.4);
    \node [centered, rotate=0.] at (2.5,6.4) {\scriptsize $(\sigma\bar\sigma,\psi)$};
    \node [centered, rotate=0.] at (9.7,6.4) {\scriptsize $(e,\psi)$};
    \node [centered, rotate=0.] at (6.2,3.5) {\scriptsize {(d)}};
\end{tikzpicture}
}

\newcommand{\FigureLoop}{
\begin{tikzpicture}[baseline={([yshift=-.8ex]current bounding box.center)},scale=.6]
    \draw [red, thick] (2.,8.) -- (2.8,8.);
    \draw [red, thick] (2.,6.4) -- (2.8,6.4);
    \draw [red, thick] (2.,4.8) -- (2.8,4.8);
    \draw [red, thick] (2.8,8.4) -- (2.8,3.6);
    \draw [red, thick] (2.8,7.2) -- (3.6,7.2);
    \draw [red, thick] (2.8,5.6) -- (3.6,5.6);
    \draw [blue] (4.4,8.4) -- (4.4,3.6);
    \draw [blue] (4.4,7.2) -- (5.2,7.2);
    \draw [blue] (4.4,5.6) -- (5.2,5.6);
    \draw [blue] (5.2,8.) -- (6.,8.);
    \draw [blue] (5.2,6.4) -- (6.,6.4);
    \draw [blue] (5.2,4.8) -- (6.,4.8);
    \draw [blue] (6.,8.4) -- (6.,3.6);
    \draw [blue] (6.,7.2) -- (6.8,7.2);
    \draw [blue] (6.,5.6) -- (6.8,5.6);
    \draw [blue] (6.8,8.4) -- (6.8,3.6);
    \draw [red, thick] (6.8,8.) -- (7.6,8.);
    \draw [red, thick] (6.8,6.4) -- (7.6,6.4);
    \draw [red, thick] (6.8,4.8) -- (7.6,4.8);
    \draw [red, thick] (7.6,8.4) -- (7.6,3.6);
    \draw [red, thick] (7.6,7.2) -- (8.4,7.2);
    \draw [red, thick] (7.6,5.6) -- (8.4,5.6);
    \draw [red, thick] (8.4,8.4) -- (8.4,3.6);
    \draw [red, thick] (8.4,8.) -- (8.8,8.);
    \draw [red, thick] (8.4,6.4) -- (8.8,6.4);
    \draw [red, thick] (8.4,4.8) -- (8.8,4.8);
    \draw [blue] (3.6,8.4) -- (3.6,3.6);
    \draw [blue] (3.6,8.) -- (4.4,8.);
    \draw [blue] (3.6,6.4) -- (4.4,6.4);
    \draw [blue] (3.6,4.8) -- (4.4,4.8);
    \draw [red, thick] (2.,8.4) -- (2.,3.6);
    \draw [red, thick] (2.,7.2) -- (1.2,7.2);
    \draw [red, thick] (2.,5.6) -- (1.2,5.6);
    \draw [red, thick] (1.2,8.4) -- (1.2,3.6);
    \draw [red, thick] (1.2,8.) -- (0.8,8.);
    \draw [red, thick] (1.2,6.4) -- (0.8,6.4);
    \draw [red, thick] (1.2,4.8) -- (0.8,4.8);
    \draw [red, thick, dashed] (8.8,8.) -- (9.2,8.);
    \draw [red, thick, dashed] (8.8,4.8) -- (9.2,4.8);
    \draw [red, thick, dashed] (0.8,8.) -- (0.4,8.);
    \draw [red, thick, dashed] (0.8,6.4) -- (0.4,6.4);
    \draw [red, thick, dashed] (0.8,4.8) -- (0.4,4.8);
    \draw [red, thick, dashed] (8.8,6.4) -- (9.2,6.4);
    \draw [red, thick, dashed] (1.2,8.4) -- (1.2,8.8);
    \draw [red, thick, dashed] (2.,8.4) -- (2.,8.8);
    \draw [red, thick, dashed] (2.8,8.4) -- (2.8,8.8);
    \draw [red, thick, dashed] (7.6,8.4) -- (7.6,8.8);
    \draw [red, thick, dashed] (8.4,8.4) -- (8.4,8.8);
    \draw [red, thick, dashed] (1.2,3.6) -- (1.2,3.2);
    \draw [red, thick, dashed] (2.,3.6) -- (2.,3.2);
    \draw [red, thick, dashed] (2.8,3.6) -- (2.8,3.2);
    \draw [red, thick, dashed] (7.6,3.6) -- (7.6,3.2);
    \draw [red, thick, dashed] (8.4,3.6) -- (8.4,3.2);
    \draw [red, thick] (1.2,4.) -- (2.,4.);
    \draw [red, thick] (7.6,4.) -- (8.4,4.);
    \draw [red, thick] (2.8,4.) -- (3.6,4.);
    \draw [blue, dashed] (3.6,8.4) -- (3.6,8.8);
    \draw [blue, dashed] (4.4,8.4) -- (4.4,8.8);
    \draw [blue, dashed] (6.,8.4) -- (6.,8.8);
    \draw [blue, dashed] (6.8,8.4) -- (6.8,8.8);
    \draw [blue, dashed] (6.8,3.6) -- (6.8,3.2);
    \draw [blue, dashed] (6.,3.6) -- (6.,3.2);
    \draw [blue, dashed] (3.6,3.6) -- (3.6,3.2);
    \draw [blue, dashed] (4.4,3.6) -- (4.4,3.2);
    \draw [blue, dashed] (5.2,8.4) -- (5.2,8.8);
    \draw [blue] (5.2,8.4) -- (5.2,3.6);
    \draw [blue, dashed] (5.2,3.6) -- (5.2,3.2);
    \draw [ultra thick] (0.4,6.8) -- (9.2,6.8);
    \draw [ultra thick] (3.2,8.8) -- (3.2,3.2);
    \filldraw [red,opacity=0.2] (0.4,3.2) -- (0.4,8.8) -- (2.8,8.8) -- (2.8,3.2);
    \filldraw [black,opacity=0.2] (3.6,3.2) -- (3.6,8.8) -- (2.8,8.8) -- (2.8,3.2);
    \filldraw [blue,opacity=0.2] (3.6,3.2) -- (3.6,8.8) -- (6.8,8.8) -- (6.8,3.2);
    \filldraw [black,opacity=0.2] (7.6,3.2) -- (7.6,8.8) -- (6.8,8.8) -- (6.8,3.2);
    \filldraw [red,opacity=0.2] (7.6,3.2) -- (7.6,8.8) -- (9.2,8.8) -- (9.2,3.2);
    \node [centered, rotate=0.] at (8.3,7.1) {\scriptsize $H$-Loop};
    \node [centered, rotate=0.] at (3.2,9.1) {\scriptsize $V$-Loop};
    \node [centered, rotate=0.] at (4.8,2.6) {\scriptsize {(b)}};
    \draw [thick,dashed,->-] (0.4,8.8) -- (9.2,8.8);
    \draw [thick,dashed,->-] (0.4,3.2) -- (9.2,3.2);
    \draw [thick,dashed] (0.4,3.2) -- (0.4,8.8);
    \draw [thick,dashed] (9.2,3.2) -- (9.2,8.8);
    \draw [thick,->] (0.4,5.4) -- (0.4,5.5);
    \draw [thick,->] (0.4,5.8) -- (0.4,5.9);
    \draw [thick,->] (9.2,5.4) -- (9.2,5.5);
    \draw [thick,->] (9.2,5.8) -- (9.2,5.9);
\end{tikzpicture}
}

\newcommand{\EqBraidingLeft}{
\hs\begin{tikzpicture}[baseline={([yshift=-.8ex]current bounding box.center)},scale=0.6]
\draw [] (.9,9.) -- (.9,4.6);
\draw [] (6.8,9.) -- (7.9,6.8);
\draw [] (7.9,6.8) -- (6.8,4.6);
\filldraw [red,opacity=0.2](1.3,8.8)--(1.3,4.8)--(3.2,4.8)--(3.2,8.8);
\filldraw [black,opacity=0.2](4.8,8.8)--(4.8,4.8)--(3.2,4.8)--(3.2,8.8);
\filldraw [blue,opacity=0.2](4.8,8.8)--(4.8,4.8)--(6.7,4.8)--(6.7,8.8);
\draw [rounded corners,ultra thick] (4.4, 7.) -- (4.4,7.6) -- (3.6,7.6) -- (3.6,5.6) -- (4.4,5.6) -- (4.4,6.6);
\draw [rounded corners,ultra thick] (3.8,6.8) -- (5.6,6.8) -- (5.6,8.) -- (2.4,8.) -- (2.4, 6.8) -- (3.4,6.8);
    \node [centered, rotate=0.] at (1.8,7.4) {\scriptsize $J_\text{DI}$};
    \node [centered, rotate=0.] at (6.2,7.4) {\scriptsize $J_\text{TC}$};
    \node [centered, rotate=0.] at (5.,6.) {\scriptsize $J_\text{DW}$};
\end{tikzpicture}\hs
}

\newcommand{\EqBraidingRight}{
\hs\begin{tikzpicture}[baseline={([yshift=-.8ex]current bounding box.center)},scale=0.6]
\draw [] (.9,9.) -- (.9,4.6);
\draw [] (6.8,9.) -- (7.9,6.8);
\draw [] (7.9,6.8) -- (6.8,4.6);
\filldraw [red,opacity=0.2](1.3,8.8)--(1.3,4.8)--(3.2,4.8)--(3.2,8.8);
\filldraw [black,opacity=0.2](4.8,8.8)--(4.8,4.8)--(3.2,4.8)--(3.2,8.8);
\filldraw [blue,opacity=0.2](4.8,8.8)--(4.8,4.8)--(6.7,4.8)--(6.7,8.8);
\end{tikzpicture}\hs
}

\newcommand{\FigureTmatrixGlobal}{
\begin{tikzpicture}[baseline={([yshift=-.8ex]current bounding box.center)},scale=1]
    \node [centered, rotate=0.] at (2.,-0.4) {\scriptsize {(a)}};
    
    \draw [red, thick] (0.8,6.8) -- (0.8,0.4);
    \draw [red, thick] (0.8,5.6) -- (1.6,5.6);
    \draw [red, thick] (0.8,4.) -- (1.6,4.);
    \draw [red, thick] (0.8,2.4) -- (1.6,2.4);
    \draw [red, thick] (0.8,0.8) -- (1.6,0.8);
    \draw [red, thick] (0.8,1.6) -- (0.4,1.6);
    \draw [red, thick] (0.8,3.2) -- (0.4,3.2);
    \draw [red, thick] (0.8,4.8) -- (0.4,4.8);
    \draw [red, thick] (0.8,6.4) -- (0.4,6.4);
    \draw [blue] (1.6,6.8) -- (1.6,0.4);
    \draw [blue] (1.6,6.4) -- (2.4,6.4);
    \draw [blue] (1.6,4.8) -- (2.4,4.8);
    \draw [blue] (1.6,3.2) -- (2.4,3.2);
    \draw [blue] (1.6,1.6) -- (2.4,1.6);
    \draw [blue] (2.4,6.8) -- (2.4,0.4);
    \draw [red, thick] (2.4,5.6) -- (3.2,5.6);
    \draw [red, thick] (2.4,4.) -- (3.2,4.);
    \draw [red, thick] (2.4,2.4) -- (3.2,2.4);
    \draw [red, thick] (2.4,0.8) -- (3.2,0.8);
    \draw [red, thick] (3.2,6.8) -- (3.2,0.4);
    \draw [red, thick] (3.2,6.4) -- (3.6,6.4);
    \draw [red, thick] (3.2,4.8) -- (3.6,4.8);
    \draw [red, thick] (3.2,3.2) -- (3.6,3.2);
    \draw [red, thick] (3.2,1.6) -- (3.6,1.6);
    \filldraw [red,opacity=0.2] (0.4,6.8) -- (0.4,0.4) -- (0.8,0.4) -- (0.8,6.8);
    \filldraw [black,opacity=0.2] (1.6,6.8) -- (1.6,0.4) -- (0.8,0.4) -- (0.8,6.8);
    \filldraw [blue,opacity=0.2] (1.6,6.8) -- (1.6,0.4) -- (2.4,0.4) -- (2.4,6.8);
    \filldraw [black,opacity=0.2] (3.2,6.8) -- (3.2,0.4) -- (2.4,0.4) -- (2.4,6.8);
    \filldraw [red,opacity=0.2] (3.2,6.8) -- (3.2,0.4) -- (3.6,0.4) -- (3.6,6.8);

    \draw [dashed, thick, ->-] (0.4,6.8) -- (3.6,6.8);
    \draw [dashed, thick, ->-] (0.4,0.4) -- (3.6,0.4);
    \draw [dashed, thick] (0.4,0.4) -- (0.4,6.8);
    \draw [dashed, thick] (3.6,0.4) -- (3.6,6.8);
    \draw [thick, ->-] (0.4,2.0) -- (0.4,2.2);
    \draw [thick, ->-] (0.4,2.6) -- (0.4,2.8);
    \draw [thick, ->-] (3.6,2.0) -- (3.6,2.2);
    \draw [thick, ->-] (3.6,2.6) -- (3.6,2.8);

    \draw [thick, black, densely dashed] (3.0,6.8) arc [start angle=-180, end angle=0, x radius=0.2, y radius=0.6];
    
    \draw [thick, black, densely dashed] (0.8,1.2) ellipse [x radius=0.2, y radius=0.6];
    \draw [thick, black, densely dashed] (0.8,2.8) ellipse [x radius=0.2, y radius=0.6];
    \draw [thick, black, densely dashed] (0.8,4.4) ellipse [x radius=0.2, y radius=0.6];
    \draw [thick, black, densely dashed] (0.8,6.0) ellipse [x radius=0.2, y radius=0.6];
    
    \draw [thick, black, densely dashed] (1.8,0.4) arc [start angle=0, end angle=180, x radius=0.2, y radius=0.6];
    \draw [thick, black, densely dashed] (1.6,2.0) ellipse [x radius=0.2, y radius=0.6];
    \draw [thick, black, densely dashed] (1.6,3.6) ellipse [x radius=0.2, y radius=0.6];
    \draw [thick, black, densely dashed] (1.6,5.2) ellipse [x radius=0.2, y radius=0.6];
    \draw [thick, black, densely dashed] (1.4,6.8) arc [start angle=-180, end angle=0, x radius=0.2, y radius=0.6];
    
    \draw [thick, black, densely dashed] (2.4,1.2) ellipse [x radius=0.2, y radius=0.6];
    \draw [thick, black, densely dashed] (2.4,2.8) ellipse [x radius=0.2, y radius=0.6];
    \draw [thick, black, densely dashed] (2.4,4.4) ellipse [x radius=0.2, y radius=0.6];
    \draw [thick, black, densely dashed] (2.4,6.0) ellipse [x radius=0.2, y radius=0.6];
    
    \draw [thick, black, densely dashed] (3.4,0.4) arc [start angle=0, end angle=180, x radius=0.2, y radius=0.6];
    \draw [thick, black, densely dashed] (3.2,2.0) ellipse [x radius=0.2, y radius=0.6];
    \draw [thick, black, densely dashed] (3.2,3.6) ellipse [x radius=0.2, y radius=0.6];
    \draw [thick, black, densely dashed] (3.2,5.2) ellipse [x radius=0.2, y radius=0.6];
    \draw [thick, black, densely dashed] (3.2,6.8) arc [start angle=-180, end angle=0, x radius=0.2, y radius=0.6];
\end{tikzpicture}
}

\newcommand{\FigureTmatrixOrigin}{
\begin{tikzpicture}[baseline={([yshift=-.8ex]current bounding box.center)},scale=1]
    \node [centered, rotate=0.] at (1.2,-2.3) {\scriptsize {(b)}};
    \draw [white] (1.1,4.8) -- (1.2,4.9);
    
    \draw [thick] (0.4,3.) -- (0.4,1.0);
    \draw [thick] (0.4,2.8) -- (1.2,2.8);
    \draw [thick] (0.4,1.2) -- (1.2,1.2);
    \draw [thick] (1.2,2.) -- (2.,2.);
    \draw [thick] (1.2,3.2) -- (1.2,0.);
    \draw [thick] (1.2,0.4) -- (2.,0.4);
    \draw [thick] (2.,2.2) -- (2.,0.2);
    \draw [thick] (0.4,2.0) -- (0.2,2.0);
    \draw [thick] (2.,1.2) -- (2.2,1.2);
    \draw [ultra thick, black!50, dashed] (0.8,2.0) -- (1.6,1.2);
    \fill [black!70] (0.8,2.0) circle (0.05);
    \fill [black!70] (1.6,1.2) circle (0.05);
    \node [centered, rotate=0.] at (0.6,2.1) {\scriptsize $P_1$};
    \node [centered, rotate=0.] at (1.8,1.0) {\scriptsize $P_2$};
    \node [centered, rotate=0.] at (0.7,1.0) {\scriptsize $E_1$};
    \node [centered, rotate=0.] at (1.3,0.8) {\scriptsize $E_2$};
    \node [centered, rotate=0.] at (1.7,0.2) {\scriptsize $E_3$};
    \draw [thick, black!50, densely dashed] (1.2,2.4) ellipse [x radius=0.2, y radius=0.6];
    \draw [thick, black!50, densely dashed] (1.2,0.8) ellipse [x radius=0.2, y radius=0.6];
\end{tikzpicture}
}

\newcommand{\FigureTmatrixShear}{
\begin{tikzpicture}[baseline={([yshift=-.8ex]current bounding box.center)},scale=1]
    \node [centered, rotate=0.] at (1.2,-2.0) {\scriptsize {(c)}};
    \draw [white] (1.1,4.5) -- (1.2,5.2);
    \draw [thick] (0.4,3.1) -- (0.4,0.9);
    \draw [thick] (0.4,2.9) -- (1.2,2.1);
    \draw [thick] (0.4,1.3) -- (1.2,0.5);
    \draw [thick] (1.2,2.9) -- (2.,2.1);
    \draw [thick] (1.2,3.3) -- (1.2,0.1);
    \draw [thick] (1.2,1.3) -- (2.,0.5);
    \draw [thick] (2.,2.5) -- (2.,0.3);
    \draw [thick] (0.4,2.1) -- (0.2,2.3);
    \draw [thick] (2.,1.3) -- (2.2,1.1);
    \draw [ultra thick, black!50, dashed] (0.8,1.7) -- (1.6,1.7);
    \fill [black!70] (0.8,1.7) circle (0.05);
    \fill [black!70] (1.6,1.7) circle (0.05);
    \node [centered, rotate=0.] at (0.6,1.8) {\scriptsize $P_1$};
    \node [centered, rotate=0.] at (1.8,1.8) {\scriptsize $P_2$};
    \node [centered, rotate=0.] at (0.6,0.8) {\scriptsize $E_1$};
    \node [centered, rotate=0.] at (1.3,0.8) {\scriptsize $E_2$};
    \node [centered, rotate=0.] at (1.7,0.6) {\scriptsize $E_3$};
    \draw [thick, black!50, densely dashed] (1.2,2.5) ellipse [x radius=0.2, y radius=0.6];
    \draw [thick, black!50, densely dashed] (1.2,0.9) ellipse [x radius=0.2, y radius=0.6];
\end{tikzpicture}
}

\newcommand{\FigureTmatrixFinal}{
\begin{tikzpicture}[baseline={([yshift=-.8ex]current bounding box.center)},scale=1]
    \node [centered, rotate=0.] at (1.2,-2.0) {\scriptsize {(d)}};
    \draw [white] (1.1,4.5) -- (1.2,5.2);
    \draw [thick] (0.4,3.) -- (0.4,1.0);
    \draw [thick] (0.4,2.8) -- (1.2,2.8);
    \draw [thick] (0.4,1.2) -- (1.2,1.2);
    \draw [thick] (1.2,2.) -- (2.,2.);
    \draw [thick] (1.2,4.0) -- (1.2,0.8);
    \draw [thick] (2.,3.8) -- (2.,1.8);
    \draw [thick] (1.2,3.6) -- (2.,3.6);
    \draw [thick] (0.4,2.) -- (0.2,2.);
    \draw [thick] (2.,2.8) -- (2.2,2.8);
    \draw [ultra thick, black!50, dashed] (0.8,2.0) -- (1.6,2.8);
    \fill [black!70] (0.8,2.0) circle (0.05);
    \fill [black!70] (1.6,2.8) circle (0.05);
    \node [centered, rotate=0.] at (0.7,1.8) {\scriptsize $P_1$};
    \node [centered, rotate=0.] at (1.7,3.0) {\scriptsize $P_2$};
    \node [centered, rotate=0.] at (0.8,1.0) {\scriptsize $E_1$};
    \node [centered, rotate=0.] at (1.3,1.6) {\scriptsize $E_2$};
    \node [centered, rotate=0.] at (1.6,2.2) {\scriptsize $E_3$};
    \draw [thick, black!50, densely dashed] (1.2,1.6) ellipse [x radius=0.2, y radius=0.6];
    \draw [thick, black!50, densely dashed] (1.2,3.2) ellipse [x radius=0.2, y radius=0.6];
\end{tikzpicture}
}

\newcommand{\FigureTmatrixLatticeOrigin}{
\begin{tikzpicture}[baseline={([yshift=-.8ex]current bounding box.center)},scale=1]
    \node [centered, rotate=0.] at (2.,-0.4) {\scriptsize {(a)}};
    
    \draw [red, thick] (0.8,6.8) -- (0.8,0.4);
    \draw [red, thick] (0.8,5.6) -- (1.6,5.6);
    \draw [red, thick] (0.8,4.) -- (1.6,4.);
    \draw [red, thick] (0.8,2.4) -- (1.6,2.4);
    \draw [red, thick] (0.8,0.8) -- (1.6,0.8);
    \draw [red, thick] (0.8,1.6) -- (0.4,1.6);
    \draw [red, thick] (0.8,3.2) -- (0.4,3.2);
    \draw [red, thick] (0.8,4.8) -- (0.4,4.8);
    \draw [red, thick] (0.8,6.4) -- (0.4,6.4);
    \draw [blue] (1.6,6.8) -- (1.6,0.4);
    \draw [blue] (1.6,6.4) -- (2.4,6.4);
    \draw [blue] (1.6,4.8) -- (2.4,4.8);
    \draw [blue] (1.6,3.2) -- (2.4,3.2);
    \draw [blue] (1.6,1.6) -- (2.4,1.6);
    \draw [blue] (2.4,6.8) -- (2.4,0.4);
    \draw [red, thick] (2.4,5.6) -- (3.2,5.6);
    \draw [red, thick] (2.4,4.) -- (3.2,4.);
    \draw [red, thick] (2.4,2.4) -- (3.2,2.4);
    \draw [red, thick] (2.4,0.8) -- (3.2,0.8);
    \draw [red, thick] (3.2,6.8) -- (3.2,0.4);
    \draw [red, thick] (3.2,6.4) -- (3.6,6.4);
    \draw [red, thick] (3.2,4.8) -- (3.6,4.8);
    \draw [red, thick] (3.2,3.2) -- (3.6,3.2);
    \draw [red, thick] (3.2,1.6) -- (3.6,1.6);
    \filldraw [red,opacity=0.2] (0.4,6.8) -- (0.4,0.4) -- (0.8,0.4) -- (0.8,6.8);
    \filldraw [black,opacity=0.2] (1.6,6.8) -- (1.6,0.4) -- (0.8,0.4) -- (0.8,6.8);
    \filldraw [blue,opacity=0.2] (1.6,6.8) -- (1.6,0.4) -- (2.4,0.4) -- (2.4,6.8);
    \filldraw [black,opacity=0.2] (3.2,6.8) -- (3.2,0.4) -- (2.4,0.4) -- (2.4,6.8);
    \filldraw [red,opacity=0.2] (3.2,6.8) -- (3.2,0.4) -- (3.6,0.4) -- (3.6,6.8);

    \draw [dashed, thick, ->-] (0.4,6.8) -- (3.6,6.8);
    \draw [dashed, thick, ->-] (0.4,0.4) -- (3.6,0.4);
    \draw [dashed, thick] (0.4,0.4) -- (0.4,6.8);
    \draw [dashed, thick] (3.6,0.4) -- (3.6,6.8);
    \draw [thick, ->-] (0.4,2.0) -- (0.4,2.2);
    \draw [thick, ->-] (0.4,2.6) -- (0.4,2.8);
    \draw [thick, ->-] (3.6,2.0) -- (3.6,2.2);
    \draw [thick, ->-] (3.6,2.6) -- (3.6,2.8);

    \filldraw [red,opacity=0.25] (0.4,6.4) -- (0.4,4.8) -- (0.8,4.8) -- (0.8,6.4);
    \filldraw [black,opacity=0.25] (0.8,5.6) -- (0.8,4.0) -- (1.6,4.0) -- (1.6,5.6);
    \filldraw [blue,opacity=0.25] (1.6,6.4) -- (1.6,4.8) -- (2.4,4.8) -- (2.4,6.4);
    \filldraw [black,opacity=0.25] (2.4,5.6) -- (2.4,4.0) -- (3.2,4.0) -- (3.2,5.6);
    \filldraw [red,opacity=0.25] (3.2,6.4) -- (3.2,4.8) -- (3.6,4.8) -- (3.6,6.4);
    
    \draw [ultra thick, dashed, black!50] (0.4,5.2) -- (3.6,5.2); 
    \fill [red] (0.4,5.2) circle (0.08);
    \fill [black] (1.2,5.2) circle (0.08);
    \fill [blue] (2.0,5.2) circle (0.08);
    \fill [black] (2.8,5.2) circle (0.08);
    \fill [red] (3.6,5.2) circle (0.08);
    \end{tikzpicture}
}

\newcommand{\FigureTmatrixLatticeFinal}{
\begin{tikzpicture}[baseline={([yshift=-.8ex]current bounding box.center)},scale=1]
    \node [centered, rotate=0.] at (2.,-0.4) {\scriptsize {(b)}};
    
    \draw [red, thick] (0.8,6.8) -- (0.8,0.4);
    \draw [red, thick] (0.8,5.6) -- (1.6,5.6);
    \draw [red, thick] (0.8,4.) -- (1.6,4.);
    \draw [red, thick] (0.8,2.4) -- (1.6,2.4);
    \draw [red, thick] (0.8,0.8) -- (1.6,0.8);
    \draw [red, thick] (0.8,1.6) -- (0.4,1.6);
    \draw [red, thick] (0.8,3.2) -- (0.4,3.2);
    \draw [red, thick] (0.8,4.8) -- (0.4,4.8);
    \draw [red, thick] (0.8,6.4) -- (0.4,6.4);
    \draw [blue] (1.6,6.8) -- (1.6,0.4);
    \draw [blue] (1.6,6.4) -- (2.4,6.4);
    \draw [blue] (1.6,4.8) -- (2.4,4.8);
    \draw [blue] (1.6,3.2) -- (2.4,3.2);
    \draw [blue] (1.6,1.6) -- (2.4,1.6);
    \draw [blue] (2.4,6.8) -- (2.4,0.4);
    \draw [red, thick] (2.4,5.6) -- (3.2,5.6);
    \draw [red, thick] (2.4,4.) -- (3.2,4.);
    \draw [red, thick] (2.4,2.4) -- (3.2,2.4);
    \draw [red, thick] (2.4,0.8) -- (3.2,0.8);
    \draw [red, thick] (3.2,6.8) -- (3.2,0.4);
    \draw [red, thick] (3.2,6.4) -- (3.6,6.4);
    \draw [red, thick] (3.2,4.8) -- (3.6,4.8);
    \draw [red, thick] (3.2,3.2) -- (3.6,3.2);
    \draw [red, thick] (3.2,1.6) -- (3.6,1.6);
    \filldraw [red,opacity=0.2] (0.4,6.8) -- (0.4,0.4) -- (0.8,0.4) -- (0.8,6.8);
    \filldraw [black,opacity=0.2] (1.6,6.8) -- (1.6,0.4) -- (0.8,0.4) -- (0.8,6.8);
    \filldraw [blue,opacity=0.2] (1.6,6.8) -- (1.6,0.4) -- (2.4,0.4) -- (2.4,6.8);
    \filldraw [black,opacity=0.2] (3.2,6.8) -- (3.2,0.4) -- (2.4,0.4) -- (2.4,6.8);
    \filldraw [red,opacity=0.2] (3.2,6.8) -- (3.2,0.4) -- (3.6,0.4) -- (3.6,6.8);

    \draw [dashed, thick, ->-] (0.4,6.8) -- (3.6,6.8);
    \draw [dashed, thick, ->-] (0.4,0.4) -- (3.6,0.4);
    \draw [dashed, thick] (0.4,0.4) -- (0.4,6.8);
    \draw [dashed, thick] (3.6,0.4) -- (3.6,6.8);
    \draw [thick, ->-] (0.4,5.2) -- (0.4,5.4);
    \draw [thick, ->-] (0.4,5.8) -- (0.4,6.0);
    \draw [thick, ->-] (3.6,5.2) -- (3.6,5.4);
    \draw [thick, ->-] (3.6,5.8) -- (3.6,6.0);
    
    \draw [ultra thick, dashed, black!50] (0.4,2.0) -- (2.8,6.8); 
    \draw [ultra thick, dashed, black!50] (2.8,0.4) -- (3.6,2.0); 
    \fill [red] (0.4,2.0) circle (0.08);
    \fill [black] (1.2,3.6) circle (0.08);
    \fill [blue] (2.0,5.2) circle (0.08);
    \fill [black] (2.8,6.8) circle (0.08);
    \fill [black] (2.8,0.4) circle (0.08);
    \fill [red] (3.6,2.0) circle (0.08);
    \filldraw [red,opacity=0.25] (0.4,3.2) -- (0.4,1.6) -- (0.8,1.6) -- (0.8,3.2);
    \filldraw [black,opacity=0.25] (0.8,2.4) -- (0.8,4.0) -- (1.6,4.0) -- (1.6,2.4);
    \filldraw [blue,opacity=0.25] (1.6,6.4) -- (1.6,4.8) -- (2.4,4.8) -- (2.4,6.4);
    \filldraw [black,opacity=0.25] (2.4,5.6) -- (2.4,6.8) -- (3.2,6.8) -- (3.2,5.6);
    \filldraw [black,opacity=0.25] (2.4,0.4) -- (3.2,0.4) -- (3.2,0.8) -- (2.4,0.8);
    \filldraw [red,opacity=0.25] (3.2,3.2) -- (3.2,1.6) -- (3.6,1.6) -- (3.6,3.2);
\end{tikzpicture}
}

\newcommand{\EqTmatrixLeft}{
\hs\begin{tikzpicture}[baseline={([yshift=-.8ex]current bounding box.center)},scale=.6]
\draw [] (1.6,9.) -- (1.6,5.4);
\draw [] (4.6,9.) -- (5.4,7.2);
\draw [] (5.4,7.2) -- (4.6,5.4);
\draw [] (3.2,8.8) -- (3.2,5.6);
\draw [] (2.,8.) -- (3.2,8.);
\draw [] (3.2,6.4) -- (4.4,6.4);
\draw [thick, black!80, densely dashed] (3.2,7.2) ellipse [x radius=0.4, y radius=1.1];
\node [centered, rotate=0.] at (3.6,8.6) {\scriptsize $a_0$};
\node [centered, rotate=0.] at (3.3,7.2) {\scriptsize $a_1$};
\node [centered, rotate=0.] at (3.6,5.8) {\scriptsize $a_2$};
\node [centered, rotate=0.] at (2.2,7.7) {\scriptsize $e_1$};
\node [centered, rotate=0.] at (4.2,6.7) {\scriptsize $e_2$};
\end{tikzpicture}\hs
}

\newcommand{\EqTmatrixRight}{
\hs\begin{tikzpicture}[baseline={([yshift=-.8ex]current bounding box.center)},scale=.6]
\draw [] (1.6,9.) -- (1.6,5.4);
\draw [] (4.6,9.) -- (5.4,7.2);
\draw [] (5.4,7.2) -- (4.6,5.4);
\draw [] (3.2,8.8) -- (3.2,5.6);
\draw [] (4.4,6.8) -- (3.2,8.);
\draw [] (3.2,6.4) -- (2.,7.6);
\draw [thick, black!80, densely dashed] (3.2,7.2) ellipse [x radius=0.4, y radius=1.1];
\node [centered, rotate=0.] at (3.6,8.6) {\scriptsize $a_0$};
\node [centered, rotate=0.] at (3.3,7.) {\scriptsize $a_1'$};
\node [centered, rotate=0.] at (3.6,5.8) {\scriptsize $a_2$};
\node [centered, rotate=0.] at (2.,7.8) {\scriptsize $e_1$};
\node [centered, rotate=0.] at (4.5,6.6) {\scriptsize $e_2$};
\end{tikzpicture}\hs
}

\newcommand{\EqTwistLeft}{
\hs\begin{tikzpicture}[baseline={([yshift=-.8ex]current bounding box.center)},scale=1]
    \draw [] (1.6,9.6) -- (1.6,5.6);
    \draw [] (5.2,9.6) -- (6.2,7.6);
    \draw [] (6.2,7.6) -- (5.2,5.6);
    \draw [red, thick] (2.4,9.2) -- (2.4,6.);
    \draw [red, thick] (2.4,8.) -- (3.2,8.);
    \draw [red, thick] (2.4,6.4) -- (3.2,6.4);
    \draw [red, thick] (3.2,9.2) -- (3.2,6.);
    \draw [red, thick] (3.2,7.2) -- (4.,7.2);
    \draw [red, thick] (3.2,8.8) -- (4.,8.8);
    \draw [blue] (4.,9.2) -- (4.,6.);
    \draw [blue] (4.,8.) -- (4.8,8.);
    \draw [blue] (4.,6.4) -- (4.8,6.4);
    \draw [blue] (4.8,9.2) -- (4.8,6.);
    \draw [blue] (4.8,8.8) -- (5.2,8.8);
    \draw [blue] (4.8,7.2) -- (5.2,7.2);
    \draw [red, thick] (2.4,8.8) -- (2.,8.8);
    \draw [red, thick] (2.4,7.2) -- (2.,7.2);
    \draw [ultra thick] (3.6,7.6) -- (3.6,8.4);
    \draw [ultra thick] (3.6,8.4) -- (2.8,8.4);
    \draw [ultra thick] (4.4,7.6) -- (5.2,7.6);
    \draw [ultra thick] (3.6,7.6) -- (2.9,7.6);
    \draw [ultra thick] (2.7,7.6) -- (2.,7.6);
    \draw [ultra thick] (2.8,8.4) -- (2.8,6.8);
    \draw [ultra thick] (2.8,6.8) -- (4.4,6.8);
    \draw [ultra thick] (4.4,6.8) -- (4.4,7.6);
    \filldraw [red,opacity=0.2] (2.,9.2) -- (2.,6.) -- (3.2, 6.) -- (3.2,9.2);
    \filldraw [black,opacity=0.2] (4.,9.2) -- (4.,6.) -- (3.2, 6.) -- (3.2,9.2);
    \filldraw [blue,opacity=0.2] (4.,9.2) -- (4.,6.) -- (5.2, 6.) -- (5.2,9.2);
    \node [centered, rotate=0.] at (2.15,7.8) {\scriptsize $J_\text{DI}$};
    \node [centered, rotate=0.] at (5.1,7.8) {\scriptsize $J_\text{TC}$};
\end{tikzpicture}\hs
}

\newcommand{\EqTwistRight}{
\hs\begin{tikzpicture}[baseline={([yshift=-.8ex]current bounding box.center)},scale=1]
    \draw [] (1.6,9.6) -- (1.6,5.6);
    \draw [] (5.2,9.6) -- (6.2,7.6);
    \draw [] (6.2,7.6) -- (5.2,5.6);
    \draw [red, thick] (2.4,9.2) -- (2.4,6.);
    \draw [red, thick] (2.4,8.) -- (3.2,8.);
    \draw [red, thick] (2.4,6.4) -- (3.2,6.4);
    \draw [red, thick] (3.2,9.2) -- (3.2,6.);
    \draw [red, thick] (3.2,7.2) -- (4.,7.2);
    \draw [red, thick] (3.2,8.8) -- (4.,8.8);
    \draw [blue] (4.,9.2) -- (4.,6.);
    \draw [blue] (4.,8.) -- (4.8,8.);
    \draw [blue] (4.,6.4) -- (4.8,6.4);
    \draw [blue] (4.8,9.2) -- (4.8,6.);
    \draw [blue] (4.8,8.8) -- (5.2,8.8);
    \draw [blue] (4.8,7.2) -- (5.2,7.2);
    \draw [red, thick] (2.4,8.8) -- (2.,8.8);
    \draw [red, thick] (2.4,7.2) -- (2.,7.2);
    \draw [ultra thick] (2.,7.6) -- (5.2,7.6);
    \filldraw [red,opacity=0.2] (2.,9.2) -- (2.,6.) -- (3.2, 6.) -- (3.2,9.2);
    \filldraw [black,opacity=0.2] (4.,9.2) -- (4.,6.) -- (3.2, 6.) -- (3.2,9.2);
    \filldraw [blue,opacity=0.2] (4.,9.2) -- (4.,6.) -- (5.2, 6.) -- (5.2,9.2);
    \node [centered, rotate=0.] at (2.2,7.8) {\scriptsize $J_\text{DI}$};
    \node [centered, rotate=0.] at (5.,7.8) {\scriptsize $J_\text{TC}$};
\end{tikzpicture}\hs
}

\newcommand{\EqGaugeLeft}{
\hs\begin{tikzpicture}[baseline={([yshift=-.8ex]current bounding box.center)},scale=.5]
    \draw [] (3.2,9.) -- (3.2,5.8);
    \draw [] (6.3,9.) -- (7.1,7.4);
    \draw [] (6.3,5.8) -- (7.1,7.4);
    \draw [] (3.4,8.8) -- (4.8,8.);
    \draw [] (6.2,8.8) -- (4.8,8.);
    \draw [] (4.8,7.) -- (4.8,8.);
    \draw [] (4.8,6.) -- (4.8,7.);
    \draw [] (4.8,7.) -- (6.,7.);
    \node [centered, rotate=0.] at (5.8,6.6) {\scriptsize $p$};
    \node [centered, rotate=0.] at (3.6,8.1) {\scriptsize $i$};
    \node [centered, rotate=0.] at (4.4,6.4) {\scriptsize $k$};
    \node [centered, rotate=0.] at (4.4,7.5) {\scriptsize $l$};
    \node [centered, rotate=0.] at (5.8,8.1) {\scriptsize $j$};
    \node [centered, rotate=0.] at (4.9,8.6) {\scriptsize $V$};
\end{tikzpicture}\hs
}
\newcommand{\EqGaugeMu}{
\hs\begin{tikzpicture}[baseline={([yshift=-.8ex]current bounding box.center)},scale=.5]
    \draw [] (3.2,9.) -- (3.2,5.8);
    \draw [] (6.3,9.) -- (7.1,7.4);
    \draw [] (6.3,5.8) -- (7.1,7.4);
    \draw [] (3.4,8.8) -- (4.8,8.);
    \draw [] (6.2,8.8) -- (4.8,8.);
    \draw [] (4.8,6.) -- (4.8,8.);
    \draw [] (5.5,8.4) -- (6.,7.);
    \node [centered, rotate=0.] at (6.,6.6) {\scriptsize $p$};
    \node [centered, rotate=0.] at (3.6,8.1) {\scriptsize $i$};
    \node [centered, rotate=0.] at (4.4,7.) {\scriptsize $k$};
    \node [centered, rotate=0.] at (5.2,7.8) {\scriptsize $m$};
    \node [centered, rotate=0.] at (6.,8.3) {\scriptsize $j$};
    \node [centered, rotate=0.] at (4.9,8.6) {\scriptsize $V$};
\end{tikzpicture}\hs
}

\newcommand{\EqGaugeNu}{
\hs\begin{tikzpicture}[baseline={([yshift=-.8ex]current bounding box.center)},scale=.5]
    \draw [] (3.2,9.) -- (3.2,5.8);
    \draw [] (6.3,9.) -- (7.1,7.4);
    \draw [] (6.3,5.8) -- (7.1,7.4);
    \draw [] (3.4,8.8) -- (4.8,8.);
    \draw [] (6.2,8.8) -- (4.8,8.);
    \draw [] (4.8,7.) -- (4.8,8.);
    \draw [] (4.8,6.) -- (4.8,7.);
    \draw [] (4.8,7.) -- (3.6,7.);
    \node [centered, rotate=0.] at (3.8,6.6) {\scriptsize $p$};
    \node [centered, rotate=0.] at (3.6,8.1) {\scriptsize $i$};
    \node [centered, rotate=0.] at (5.2,6.4) {\scriptsize $k$};
    \node [centered, rotate=0.] at (5.2,7.5) {\scriptsize $l$};
    \node [centered, rotate=0.] at (5.8,8.1) {\scriptsize $j$};
    \node [centered, rotate=0.] at (4.9,8.6) {\scriptsize $V$};
\end{tikzpicture}\hs
}

\newcommand{\FigureRibbonLowerA}{
\begin{tikzpicture}[baseline={([yshift=-.8ex]current bounding box.center)},scale=.6]
\draw [red, thick] (4.8,8.) -- (4.8,4.8);
\draw [red, thick] (4.8,8.) -- (3.6,8.8);
\draw [red, thick] (4.8,8.) -- (6.,8.8);
\draw [red, thick] (4.8,4.8) -- (3.6,4.);
\draw [red, thick] (4.8,4.8) -- (6.,4.);
\draw [red, thick] (4.8,5.4) -- (6.4,5.4);
\filldraw [red, opacity=0.2] (3.2,4.) -- (3.2,8.8) -- (6.4,8.8) -- (6.4,4.);
\node [centered, rotate=0.] at (4.8,4.4) {\scriptsize $V_1$};
\node [centered, rotate=0.] at (4.5,6.8) {\scriptsize $j_0$};
\node [centered, rotate=0.] at (4.5,5.) {\scriptsize $j_1$};
\node [centered, rotate=0.] at (6.2,5.7) {\scriptsize $r$};
\node [centered, rotate=0.] at (4.8,3.2) {\scriptsize {(a)}};
\end{tikzpicture}
}

\newcommand{\FigureRibbonLowerB}{
\begin{tikzpicture}[baseline={([yshift=-.8ex]current bounding box.center)},scale=.6]
\draw [red, thick] (4.8,8.) -- (4.8,4.8);
\draw [red, thick] (4.8,8.) -- (3.6,8.8);
\draw [red, thick] (4.8,8.) -- (6.,8.8);
\draw [red, thick] (4.8,4.8) -- (3.6,4.);
\draw [red, thick] (4.8,4.8) -- (6.,4.);
\draw [red, thick] (4.8,5.4) -- (6.4,5.4);
\draw [red, thick] (4.8,6.) -- (3.2,6.);
\draw [red, thick] (4.8,7.4) -- (6.4,7.4);
\filldraw [red, opacity=0.2] (3.2,4.) -- (3.2,8.8) -- (6.4,8.8) -- (6.4,4.);
\node [centered, rotate=0.] at (4.8,4.4) {\scriptsize $V_1$};
\node [centered, rotate=0.] at (4.5,6.8) {\scriptsize $k$};
\node [centered, rotate=0.] at (4.5,7.8) {\scriptsize $j_0$};
\node [centered, rotate=0.] at (5.1,5.8) {\scriptsize $j_0$};
\node [centered, rotate=0.] at (4.5,5.) {\scriptsize $j_1$};
\node [centered, rotate=0.] at (3.4,5.7) {\scriptsize $p$};
\node [centered, rotate=0.] at (6.2,7.8) {\scriptsize $q$};
\node [centered, rotate=0.] at (6.2,5.7) {\scriptsize $r$};
\node [centered, rotate=0.] at (4.8,3.2) {\scriptsize {(b)}};
\end{tikzpicture}
}

\newcommand{\FigureRibbonLowerC}{
\begin{tikzpicture}[baseline={([yshift=-.8ex]current bounding box.center)},scale=.6]
\draw [red, thick] (4.8,8.) -- (4.8,4.8);
\draw [red, thick] (4.8,8.) -- (3.6,8.8);
\draw [red, thick] (4.8,8.) -- (6.,8.8);
\draw [red, thick] (4.8,4.8) -- (3.6,4.);
\draw [red, thick] (4.8,4.8) -- (6.,4.);
\draw [red, thick] (4.8,6.8) -- (6.4,6.8);
\draw [red, thick] (4.8,5.4) -- (3.2,5.4);
\draw [red, thick] (4.8,7.4) -- (6.4,7.4);
\filldraw [red, opacity=0.2] (3.2,4.) -- (3.2,8.8) -- (6.4,8.8) -- (6.4,4.);
\node [centered, rotate=0.] at (4.8,4.4) {\scriptsize $V_1$};
\node [centered, rotate=0.] at (4.5,7.1) {\scriptsize $k$};
\node [centered, rotate=0.] at (4.5,7.8) {\scriptsize $j_0$};
\node [centered, rotate=0.] at (5.,6.1) {\scriptsize $l$};
\node [centered, rotate=0.] at (4.5,5.) {\scriptsize $j_1$};
\node [centered, rotate=0.] at (3.4,5.1) {\scriptsize $p$};
\node [centered, rotate=0.] at (6.2,7.8) {\scriptsize $q$};
\node [centered, rotate=0.] at (6.2,6.5) {\scriptsize $r$};
\node [centered, rotate=0.] at (4.8,3.2) {\scriptsize {(c)}};
\end{tikzpicture}
}

\newcommand{\FigureRibbonLowerD}{
\begin{tikzpicture}[baseline={([yshift=-.8ex]current bounding box.center)},scale=.6]
\draw [red, thick] (4.8,8.) -- (4.8,4.8);
\draw [red, thick] (4.8,8.) -- (3.6,8.8);
\draw [red, thick] (4.8,8.) -- (6.,8.8);
\draw [red, thick] (4.8,4.8) -- (3.6,4.);
\draw [red, thick] (4.8,4.8) -- (6.,4.);
\draw [red, thick] (4.8,5.4) -- (3.2,5.4);
\draw [red, thick] (4.8,7.4) -- (6.4,7.4);
\filldraw [red, opacity=0.2] (3.2,4.) -- (3.2,8.8) -- (6.4,8.8) -- (6.4,4.);
\node [centered, rotate=0.] at (4.8,4.4) {\scriptsize $V_1$};
\node [centered, rotate=0.] at (4.5,7.8) {\scriptsize $j_0$};
\node [centered, rotate=0.] at (5.,6.4) {\scriptsize $l$};
\node [centered, rotate=0.] at (4.5,5.) {\scriptsize $j_1$};
\node [centered, rotate=0.] at (3.4,5.8) {\scriptsize $p$};
\node [centered, rotate=0.] at (6.2,7.8) {\scriptsize $s$};
\node [centered, rotate=0.] at (4.8,3.2) {\scriptsize {(d)}};
\end{tikzpicture}
}

\newcommand{\EqRibbonUpperLeft}{
\hs\begin{tikzpicture}[baseline={([yshift=-.8ex]current bounding box.center)},scale=.4]
    \draw [] (1.,3.8) -- (1.,9.);
    \draw [] (8.4,3.8) -- (9.7,6.4);
    \draw [] (9.7,6.4) -- (8.4,9.);
    \draw [red,thick] (3.4,4.8) -- (4.8,5.6);
    \draw [red,thick] (1.3,5.2) -- (2.,5.6);
    \draw [red,thick] (3.4,4.2) -- (3.4,4.8);
    \draw [red,thick] (6.2,4.2) -- (6.2,4.8);
    \draw [red,thick] (8.3,5.2) -- (7.6,5.6);
    \draw [red,thick] (6.2,4.8) -- (4.8,5.6);
    \draw [red,thick] (7.6,5.6) -- (6.2,4.8);
    \draw [red,thick] (2.,5.6) -- (3.4,4.8);
    \draw [red,thick] (1.3,7.6) -- (2.,7.2);
    \draw [red,thick] (3.4,8.) -- (2.,7.2);
    \draw [red,thick] (3.4,8.6) -- (3.4,8.);
    \draw [red,thick] (4.8,7.2) -- (3.4,8.);
    \draw [red,thick] (8.3,7.6) -- (7.6,7.2);
    \draw [red,thick] (4.8,7.2) -- (6.2,8.);
    \draw [red,thick] (6.2,8.) -- (7.6,7.2);
    \draw [red,thick] (6.2,8.6) -- (6.2,8.);
    \draw [red,thick] (7.6,7.2) -- (7.6,5.6);
    \draw [red,thick] (4.8,5.6) -- (4.8,7.2);
    \draw [red,thick] (2.,7.2) -- (2.,5.6);
    \draw [red,thick] (4.8,6.6) -- (6.,6.6);
    \filldraw [red, opacity=0.2] (1.3,5.2) -- (2.,5.6) -- (3.4,4.8) -- (4.8,5.6) -- (6.2,4.8) -- (7.6,5.6) --(8.3,5.2) -- (8.3,4.2) -- (1.3,4.2);
    \filldraw [red, opacity=0.2] (1.3,7.6) -- (2.,7.2) -- (3.4,8.) -- (4.8,7.2) -- (6.2,8.) -- (7.6,7.2)--(8.3,7.6) --(8.3,5.2) -- (7.6,5.6) -- (6.2,4.8) -- (4.8,5.6) -- (3.4,4.8) -- (2.,5.6) -- (1.3,5.2);
    \filldraw [red, opacity=0.2] (1.3,7.6) -- (2.,7.2) -- (3.4,8.) -- (4.8,7.2) -- (6.2,8.) -- (7.6,7.2)--(8.3,7.6) -- (8.3,8.6) -- (1.3,8.6);
    \node [centered, rotate=0.] at (4.4,6.) {\scriptsize $j_0$};
    \node [centered, rotate=0.] at (4.4,7.) {\scriptsize $j_2$};
    \node [centered, rotate=0.] at (6.2,6.6) {\scriptsize $r$};
    \node [centered, rotate=0.] at (4.8,7.8) {\scriptsize $V_2$};
    \end{tikzpicture}\hs
}

\newcommand{\EqRibbonUpperRight}{
\hs\begin{tikzpicture}[baseline={([yshift=-.8ex]current bounding box.center)},scale=.4]
    \draw [] (1.,3.8) -- (1.,9.);
    \draw [] (8.4,3.8) -- (9.7,6.4);
    \draw [] (9.7,6.4) -- (8.4,9.);
    \draw [red,thick] (3.4,4.8) -- (4.8,5.6);
    \draw [red,thick] (1.3,5.2) -- (2.,5.6);
    \draw [red,thick] (3.4,4.2) -- (3.4,4.8);
    \draw [red,thick] (6.2,4.2) -- (6.2,4.8);
    \draw [red,thick] (8.3,5.2) -- (7.6,5.6);
    \draw [red,thick] (6.2,4.8) -- (4.8,5.6);
    \draw [red,thick] (7.6,5.6) -- (6.2,4.8);
    \draw [red,thick] (2.,5.6) -- (3.4,4.8);
    \draw [red,thick] (1.3,7.6) -- (2.,7.2);
    \draw [red,thick] (3.4,8.) -- (2.,7.2);
    \draw [red,thick] (3.4,8.6) -- (3.4,8.);
    \draw [red,thick] (4.8,7.2) -- (3.4,8.);
    \draw [red,thick] (8.3,7.6) -- (7.6,7.2);
    \draw [red,thick] (4.8,7.2) -- (6.2,8.);
    \draw [red,thick] (6.2,8.) -- (7.6,7.2);
    \draw [red,thick] (6.2,8.6) -- (6.2,8.);
    \draw [red,thick] (4.8,5.6) -- (4.8,7.2);
    \draw [red,thick] (7.6,7.2) -- (7.6,5.6);
    \draw [red,thick] (2.,7.2) -- (2.,5.6);
    \draw [red,thick] (4.8,6.1) -- (3.6,6.1);
    \draw [red,thick] (4.8,6.7) -- (6.,6.7);
    \draw [red,thick] (6.2,4.8) -- (4.8,5.6);
    \filldraw [red, opacity=0.2] (1.3,5.2) -- (2.,5.6) -- (3.4,4.8) -- (4.8,5.6) -- (6.2,4.8) -- (7.6,5.6) --(8.3,5.2) -- (8.3,4.2) -- (1.3,4.2);
    \filldraw [red, opacity=0.2] (1.3,7.6) -- (2.,7.2) -- (3.4,8.) -- (4.8,7.2) -- (6.2,8.) -- (7.6,7.2)--(8.3,7.6) --(8.3,5.2) -- (7.6,5.6) -- (6.2,4.8) -- (4.8,5.6) -- (3.4,4.8) -- (2.,5.6) -- (1.3,5.2);
    \filldraw [red, opacity=0.2] (1.3,7.6) -- (2.,7.2) -- (3.4,8.) -- (4.8,7.2) -- (6.2,8.) -- (7.6,7.2)--(8.3,7.6) -- (8.3,8.6) -- (1.3,8.6);
    \node [centered, rotate=0.] at (4.5,7) {\scriptsize $j_2$};
    \node [centered, rotate=0.] at (5.1,5.8) {\scriptsize $j_0$};
    \node [centered, rotate=0.] at (4.5,6.4) {\scriptsize $k$};
    \node [centered, rotate=0.] at (3.4,6.1) {\scriptsize $p$};
    \node [centered, rotate=0.] at (6.2,6.7) {\scriptsize $s$};
    \node [centered, rotate=0.] at (4.8,7.8) {\scriptsize $V_2$};
\end{tikzpicture}\hs
}

\newcommand{\EqRibbonUpperA}{
\hs\begin{tikzpicture}[baseline={([yshift=-.8ex]current bounding box.center)},scale=.6]
    \draw [] (2.4,8.8) -- (2.4,4.8);
    \draw [] (5.6,8.8) -- (6.6,6.8);
    \draw [] (6.6,6.8) -- (5.6,4.8);
    \filldraw [red, opacity=0.2] (2.6,8.6) -- (2.6,5.) -- (5.4,5.) -- (5.4,8.6);
    \draw [red, thick] (4.,8.) -- (4.,5.6);
    \draw [red, thick] (4.,8.) -- (3.1,8.6);
    \draw [red, thick] (4.,8.) -- (4.9,8.6);
    \draw [red, thick] (4.,5.6) -- (3.1,5.);
    \draw [red, thick] (4.,5.6) -- (4.9,5.);
    \draw [red, thick] (4.,7.6) -- (5.2,7.6);
    \node [centered, rotate=0.] at (3.7,6.4) {\scriptsize $j_0$};
    \node [centered, rotate=0.] at (3.7,7.8) {\scriptsize $j_2$};
    \node [centered, rotate=0.] at (5.1,7.8) {\scriptsize $r$};
    \node [centered, rotate=0.] at (4.,8.4) {\scriptsize $V_2$};
\end{tikzpicture}\hs
}

\newcommand{\EqRibbonUpperB}{
\hs\begin{tikzpicture}[baseline={([yshift=-.8ex]current bounding box.center)},scale=.6]
    \draw [] (2.4,8.8) -- (2.4,4.8);
    \draw [] (5.6,8.8) -- (6.6,6.8);
    \draw [] (6.6,6.8) -- (5.6,4.8);
    \filldraw [red, opacity=0.2] (2.6,8.6) -- (2.6,5.) -- (5.4,5.) -- (5.4,8.6);
    \draw [red, thick] (4.,8.) -- (4.,5.6);
    \draw [red, thick] (4.,8.) -- (3.1,8.6);
    \draw [red, thick] (4.,8.) -- (4.9,8.6);
    \draw [red, thick] (4.,5.6) -- (3.1,5.);
    \draw [red, thick] (4.,5.6) -- (4.9,5.);
    \draw [red, thick] (4.,7.6) -- (5.2,7.6);
    \draw [red, thick] (4.,7.2) -- (5.2,7.2);
    \draw [red, thick] (4.,6.) -- (2.8,6.);
    \node [centered, rotate=0.] at (4.4,5.8) {\scriptsize $j_0$};
    \node [centered, rotate=0.] at (3.7,7.4) {\scriptsize $j_0$};
    \node [centered, rotate=0.] at (4.4,6.6) {\scriptsize $k$};
    \node [centered, rotate=0.] at (3.7,7.8) {\scriptsize $j_2$};
    \node [centered, rotate=0.] at (3.,6.2) {\scriptsize $p$};
    \node [centered, rotate=0.] at (5.1,6.9) {\scriptsize $q$};
    \node [centered, rotate=0.] at (5.1,7.8) {\scriptsize $r$};
    \node [centered, rotate=0.] at (4.,8.4) {\scriptsize $V_2$};
\end{tikzpicture}\hs
}

\newcommand{\EqRibbonUpperC}{
\hs\begin{tikzpicture}[baseline={([yshift=-.8ex]current bounding box.center)},scale=.6]
    \draw [] (2.4,8.8) -- (2.4,4.8);
    \draw [] (5.6,8.8) -- (6.6,6.8);
    \draw [] (6.6,6.8) -- (5.6,4.8);
    \filldraw [red, opacity=0.2] (2.6,8.6) -- (2.6,5.) -- (5.4,5.) -- (5.4,8.6);
    \draw [red, thick] (4.,8.) -- (4.,5.6);
    \draw [red, thick] (4.,8.) -- (3.1,8.6);
    \draw [red, thick] (4.,8.) -- (4.9,8.6);
    \draw [red, thick] (4.,5.6) -- (3.1,5.);
    \draw [red, thick] (4.,5.6) -- (4.9,5.);
    \draw [red, thick] (4.,7.4) -- (5.2,7.4);
    \draw [red, thick] (4.,6.) -- (2.8,6.);
    \node [centered, rotate=0.] at (4.4,5.8) {\scriptsize $j_0$};
    \node [centered, rotate=0.] at (4.4,6.8) {\scriptsize $k$};
    \node [centered, rotate=0.] at (3.7,7.7) {\scriptsize $j_2$};
    \node [centered, rotate=0.] at (3.,6.2) {\scriptsize $p$};
    \node [centered, rotate=0.] at (5.1,7.6) {\scriptsize $s$};
    \node [centered, rotate=0.] at (4.,8.4) {\scriptsize $V_2$};
\end{tikzpicture}\hs
}

\newcommand{\EqRibbonLowerLeft}{
\hs\begin{tikzpicture}[baseline={([yshift=-.8ex]current bounding box.center)},scale=.4]
    \draw [] (1.,3.8) -- (1.,9.);
    \draw [] (8.4,3.8) -- (9.7,6.4);
    \draw [] (9.7,6.4) -- (8.4,9.);
    \draw [red,thick] (3.4,4.8) -- (4.8,5.6);
    \draw [red,thick] (1.3,5.2) -- (2.,5.6);
    \draw [red,thick] (3.4,4.2) -- (3.4,4.8);
    \draw [red,thick] (6.2,4.2) -- (6.2,4.8);
    \draw [red,thick] (8.3,5.2) -- (7.6,5.6);
    \draw [red,thick] (6.2,4.8) -- (4.8,5.6);
    \draw [red,thick] (7.6,5.6) -- (6.2,4.8);
    \draw [red,thick] (2.,5.6) -- (3.4,4.8);
    \draw [red,thick] (1.3,7.6) -- (2.,7.2);
    \draw [red,thick] (3.4,8.) -- (2.,7.2);
    \draw [red,thick] (3.4,8.6) -- (3.4,8.);
    \draw [red,thick] (4.8,7.2) -- (3.4,8.);
    \draw [red,thick] (8.3,7.6) -- (7.6,7.2);
    \draw [red,thick] (4.8,7.2) -- (6.2,8.);
    \draw [red,thick] (6.2,8.) -- (7.6,7.2);
    \draw [red,thick] (6.2,8.6) -- (6.2,8.);
    \draw [red,thick] (7.6,7.2) -- (7.6,5.6);
    \draw [red,thick] (4.8,5.6) -- (4.8,7.2);
    \draw [red,thick] (2.,7.2) -- (2.,5.6);
    \draw [red,thick] (4.8,6.2) -- (6.,6.2);
    \filldraw [red, opacity=0.2] (1.3,5.2) -- (2.,5.6) -- (3.4,4.8) -- (4.8,5.6) -- (6.2,4.8) -- (7.6,5.6) --(8.3,5.2) -- (8.3,4.2) -- (1.3,4.2);
    \filldraw [red, opacity=0.2] (1.3,7.6) -- (2.,7.2) -- (3.4,8.) -- (4.8,7.2) -- (6.2,8.) -- (7.6,7.2)--(8.3,7.6) --(8.3,5.2) -- (7.6,5.6) -- (6.2,4.8) -- (4.8,5.6) -- (3.4,4.8) -- (2.,5.6) -- (1.3,5.2);
    \filldraw [red, opacity=0.2] (1.3,7.6) -- (2.,7.2) -- (3.4,8.) -- (4.8,7.2) -- (6.2,8.) -- (7.6,7.2)--(8.3,7.6) -- (8.3,8.6) -- (1.3,8.6);
    \node [centered, rotate=0.] at (4.4,6.) {\scriptsize $j_1$};
    \node [centered, rotate=0.] at (4.4,6.8) {\scriptsize $j_0$};
    \node [centered, rotate=0.] at (6.2,6.2) {\scriptsize $r$};
    \node [centered, rotate=0.] at (4.8,5.) {\scriptsize $V_1$};
    \end{tikzpicture}\hs
}

\newcommand{\EqRibbonLowerRight}{
\hs\begin{tikzpicture}[baseline={([yshift=-.8ex]current bounding box.center)},scale=.4]
    \draw [] (1.,3.8) -- (1.,9.);
    \draw [] (8.4,3.8) -- (9.7,6.4);
    \draw [] (9.7,6.4) -- (8.4,9.);
    \draw [red,thick] (3.4,4.8) -- (4.8,5.6);
    \draw [red,thick] (1.3,5.2) -- (2.,5.6);
    \draw [red,thick] (3.4,4.2) -- (3.4,4.8);
    \draw [red,thick] (6.2,4.2) -- (6.2,4.8);
    \draw [red,thick] (8.3,5.2) -- (7.6,5.6);
    \draw [red,thick] (6.2,4.8) -- (4.8,5.6);
    \draw [red,thick] (7.6,5.6) -- (6.2,4.8);
    \draw [red,thick] (2.,5.6) -- (3.4,4.8);
    \draw [red,thick] (1.3,7.6) -- (2.,7.2);
    \draw [red,thick] (3.4,8.) -- (2.,7.2);
    \draw [red,thick] (3.4,8.6) -- (3.4,8.);
    \draw [red,thick] (4.8,7.2) -- (3.4,8.);
    \draw [red,thick] (8.3,7.6) -- (7.6,7.2);
    \draw [red,thick] (4.8,7.2) -- (6.2,8.);
    \draw [red,thick] (6.2,8.) -- (7.6,7.2);
    \draw [red,thick] (6.2,8.6) -- (6.2,8.);
    \draw [red,thick] (4.8,5.6) -- (4.8,7.2);
    \draw [red,thick] (7.6,7.2) -- (7.6,5.6);
    \draw [red,thick] (2.,7.2) -- (2.,5.6);
    \draw [red,thick] (4.8,6.1) -- (3.6,6.1);
    \draw [red,thick] (4.8,6.7) -- (6.,6.7);
    \draw [red,thick] (6.2,4.8) -- (4.8,5.6);
    \filldraw [red, opacity=0.2] (1.3,5.2) -- (2.,5.6) -- (3.4,4.8) -- (4.8,5.6) -- (6.2,4.8) -- (7.6,5.6) --(8.3,5.2) -- (8.3,4.2) -- (1.3,4.2);
    \filldraw [red, opacity=0.2] (1.3,7.6) -- (2.,7.2) -- (3.4,8.) -- (4.8,7.2) -- (6.2,8.) -- (7.6,7.2)--(8.3,7.6) --(8.3,5.2) -- (7.6,5.6) -- (6.2,4.8) -- (4.8,5.6) -- (3.4,4.8) -- (2.,5.6) -- (1.3,5.2);
    \filldraw [red, opacity=0.2] (1.3,7.6) -- (2.,7.2) -- (3.4,8.) -- (4.8,7.2) -- (6.2,8.) -- (7.6,7.2)--(8.3,7.6) -- (8.3,8.6) -- (1.3,8.6);
    \node [centered, rotate=0.] at (4.5,7) {\scriptsize $j_0$};
    \node [centered, rotate=0.] at (5.2,5.8) {\scriptsize $j_1$};
    \node [centered, rotate=0.] at (4.5,6.4) {\scriptsize $l$};
    \node [centered, rotate=0.] at (3.4,6.1) {\scriptsize $p$};
    \node [centered, rotate=0.] at (6.2,6.7) {\scriptsize $s$};
    \node [centered, rotate=0.] at (4.8,7.8) {\scriptsize $V_2$};
\end{tikzpicture}\hs
}

\newcommand{\FigureConExplicitA}{
\begin{tikzpicture}[baseline={([yshift=-.8ex]current bounding box.center)},scale=.3]
    \draw [red, thick] (4.8,8.) -- (2.,6.4);
    \draw [red, thick] (2.,6.4) -- (2.,3.2);
    \draw [red, thick] (4.8,8.) -- (7.6,6.4);
    \draw [red, thick] (7.6,6.4) -- (7.6,3.2);
    \draw [red, thick] (7.6,3.2) -- (4.8,1.6);
    \draw [red, thick] (4.8,1.6) -- (2.,3.2);
    \draw [red, thick] (4.8,8.) -- (4.8,9.2);
    \draw [red, thick] (2.,6.4) -- (0.8,7.);
    \draw [red, thick] (2.,3.2) -- (0.8,2.6);
    \draw [red, thick] (7.6,3.2) -- (8.8,2.6);
    \draw [red, thick] (7.6,6.4) -- (8.8,7.);
    \draw [red, thick] (4.8,1.6) -- (4.8,0.4);
    \filldraw [red, opacity=0.2] (0.4,0.4) -- (0.4,9.2) -- (9.2,9.2) -- (9.2,0.4);
    \node [centered, rotate=0.] at (1.5,5.) {\scriptsize $j_{1}$};
    \node [centered, rotate=0.] at (3.,7.7) {\scriptsize $j_{2}$};
    \node [centered, rotate=0.] at (6.6,7.7) {\scriptsize $j_{3}$};
    \node [centered, rotate=0.] at (7.,5.) {\scriptsize $j_{4}$};
    \node [centered, rotate=0.] at (1.2,7.8) {\scriptsize $j_{5}$};
    \node [centered, rotate=0.] at (5.4,8.7) {\scriptsize $j_{6}$};
    \node [centered, rotate=0.] at (8.8,7.8) {\scriptsize $j_{7}$};
\end{tikzpicture}
}

\newcommand{\FigureConExplicitB}{
\begin{tikzpicture}[baseline={([yshift=-.8ex]current bounding box.center)},scale=.3]
    \draw [red, thick] (4.8,8.) -- (2.,6.4);
    \draw [red, thick] (2.,6.4) -- (2.,3.2);
    \draw [red, thick] (4.8,8.) -- (7.6,6.4);
    \draw [red, thick] (7.6,6.4) -- (7.6,3.2);
    \draw [red, thick] (7.6,3.2) -- (4.8,1.6);
    \draw [red, thick] (4.8,1.6) -- (2.,3.2);
    \draw [red, thick] (4.8,8.) -- (4.8,9.2);
    \draw [red, thick] (2.,6.4) -- (0.8,7.);
    \draw [red, thick] (2.,3.2) -- (0.8,2.6);
    \draw [red, thick] (7.6,3.2) -- (8.8,2.6);
    \draw [red, thick] (7.6,6.4) -- (8.8,7.);
    \draw [red, thick] (4.8,1.6) -- (4.8,0.4);
    \draw [red, thick] (2.,4.1) -- (0.4,4.1);
    \draw [red, thick] (2.,5.5) -- (3.6,5.5);
    \draw [red, thick] (7.6,5.5) -- (9.2,5.5);
    \draw [red, thick] (7.6,4.1) -- (6.,4.1);
    \filldraw [red, opacity=0.2] (0.4,0.4) -- (0.4,9.2) -- (9.2,9.2) -- (9.2,0.4);
    \node [centered, rotate=0.] at (1.5,6.) {\scriptsize $j_{1}$};
    \node [centered, rotate=0.] at (2.5,3.7) {\scriptsize $j_{1}$};
    \node [centered, rotate=0.] at (1.5,5.) {\scriptsize $j_{1}'$};
    \node [centered, rotate=0.] at (3.,7.7) {\scriptsize $j_{2}$};
    \node [centered, rotate=0.] at (6.6,7.7) {\scriptsize $j_{3}$};
    \node [centered, rotate=0.] at (7.,6.1) {\scriptsize $j_{4}$};
    \node [centered, rotate=0.] at (8.1,3.6) {\scriptsize $j_{4}$};
    \node [centered, rotate=0.] at (7.,5.) {\scriptsize $j_{4}'$};
    \node [centered, rotate=0.] at (1.2,7.8) {\scriptsize $j_{5}$};
    \node [centered, rotate=0.] at (5.4,8.7) {\scriptsize $j_{6}$};
    \node [centered, rotate=0.] at (8.8,7.8) {\scriptsize $j_{7}$};
    \node [centered, rotate=0.] at (0.8,3.6) {\scriptsize $p$};
    \node [centered, rotate=0.] at (3.2,6.) {\scriptsize $q$};
    \node [centered, rotate=0.] at (6.4,3.6) {\scriptsize $q$};
    \node [centered, rotate=0.] at (8.8,6.) {\scriptsize $r$};
\end{tikzpicture}
}

\newcommand{\FigureConExplicitC}{
\begin{tikzpicture}[baseline={([yshift=-.8ex]current bounding box.center)},scale=.3]
    \draw [red, thick] (4.8,8.) -- (2.,6.4);
    \draw [red, thick] (2.,6.4) -- (2.,3.2);
    \draw [red, thick] (4.8,8.) -- (7.6,6.4);
    \draw [red, thick] (7.6,6.4) -- (7.6,3.2);
    \draw [red, thick] (7.6,3.2) -- (4.8,1.6);
    \draw [red, thick] (4.8,1.6) -- (2.,3.2);
    \draw [red, thick] (4.8,8.) -- (4.8,9.2);
    \draw [red, thick] (2.,6.4) -- (0.8,7.);
    \draw [red, thick] (2.,3.2) -- (0.8,2.6);
    \draw [red, thick] (7.6,3.2) -- (8.8,2.6);
    \draw [red, thick] (7.6,6.4) -- (8.8,7.);
    \draw [red, thick] (4.8,1.6) -- (4.8,0.4);
    \draw [red, thick] (2.,4.) -- (0.4,4.);
    \draw [red, thick] (2.,4.8) -- (3.6,4.8);
    \draw [red, thick] (2.,5.6) -- (3.6,5.6);
    \draw [red, thick] (7.6,5.2) -- (9.2,5.2);
    \filldraw [red, opacity=0.2] (0.4,0.4) -- (0.4,9.2) -- (9.2,9.2) -- (9.2,0.4);
    \node [centered, rotate=0.] at (1.5,6.) {\scriptsize $j_{1}'$};
    \node [centered, rotate=0.] at (2.5,4.2) {\scriptsize $j_{1}'$};
    \node [centered, rotate=0.] at (1.5,5.1) {\scriptsize $j_{1}$};
    \node [centered, rotate=0.] at (1.5,3.6) {\scriptsize $j_{1}$};
    \node [centered, rotate=0.] at (3.,7.7) {\scriptsize $j_{2}'$};
    \node [centered, rotate=0.] at (6.6,7.7) {\scriptsize $j_{3}'$};
    \node [centered, rotate=0.] at (6.9,5.8) {\scriptsize $j_{4}''$};
    \node [centered, rotate=0.] at (6.9,4.2) {\scriptsize $j_{4}$};
    \node [centered, rotate=0.] at (1.2,7.8) {\scriptsize $j_{5}$};
    \node [centered, rotate=0.] at (5.4,8.7) {\scriptsize $j_{6}$};
    \node [centered, rotate=0.] at (8.8,7.8) {\scriptsize $j_{7}$};
    \node [centered, rotate=0.] at (0.8,4.4) {\scriptsize $p$};
    \node [centered, rotate=0.] at (3.2,6.1) {\scriptsize $q$};
    \node [centered, rotate=0.] at (3.2,4.4) {\scriptsize $q$};
    \node [centered, rotate=0.] at (8.8,5.8) {\scriptsize $r$};
\end{tikzpicture}
}

\newcommand{\FigureConExplicitD}{
\begin{tikzpicture}[baseline={([yshift=-.8ex]current bounding box.center)},scale=.3]
    \draw [red, thick] (4.8,8.) -- (2.,6.4);
    \draw [red, thick] (2.,6.4) -- (2.,3.2);
    \draw [red, thick] (4.8,8.) -- (7.6,6.4);
    \draw [red, thick] (7.6,6.4) -- (7.6,3.2);
    \draw [red, thick] (7.6,3.2) -- (4.8,1.6);
    \draw [red, thick] (4.8,1.6) -- (2.,3.2);
    \draw [red, thick] (4.8,8.) -- (4.8,9.2);
    \draw [red, thick] (2.,6.4) -- (0.8,7.);
    \draw [red, thick] (2.,3.2) -- (0.8,2.6);
    \draw [red, thick] (7.6,3.2) -- (8.8,2.6);
    \draw [red, thick] (7.6,6.4) -- (8.8,7.);
    \draw [red, thick] (4.8,1.6) -- (4.8,0.4);
    \draw [red, thick] (2.,4.1) -- (0.4,4.1);
    \draw [red, thick] (2.,5.5) -- (3.6,5.5);
    \draw [red, thick] (7.6,5.5) -- (9.2,5.5);
    \filldraw [red, opacity=0.2] (0.4,0.4) -- (0.4,9.2) -- (9.2,9.2) -- (9.2,0.4);
    \node [centered, rotate=0.] at (1.5,6.) {\scriptsize $j_{1}''$};
    \node [centered, rotate=0.] at (2.5,3.7) {\scriptsize $j_{1}$};
    \node [centered, rotate=0.] at (1.5,4.9) {\scriptsize $j_{1}'$};
    \node [centered, rotate=0.] at (3.,7.7) {\scriptsize $j_{2}'$};
    \node [centered, rotate=0.] at (6.6,7.7) {\scriptsize $j_{3}'$};
    \node [centered, rotate=0.] at (7.,5.9) {\scriptsize $j_{4}''$};
    \node [centered, rotate=0.] at (7.,4.4) {\scriptsize $j_{4}$};
    \node [centered, rotate=0.] at (1.2,7.8) {\scriptsize $j_{5}$};
    \node [centered, rotate=0.] at (5.4,8.7) {\scriptsize $j_{6}$};
    \node [centered, rotate=0.] at (8.8,7.8) {\scriptsize $j_{7}$};
    \node [centered, rotate=0.] at (0.8,3.6) {\scriptsize $p$};
    \node [centered, rotate=0.] at (3.2,6.) {\scriptsize $s$};
    \node [centered, rotate=0.] at (8.8,6.) {\scriptsize $r$};
\end{tikzpicture}
}

\newcommand{\FigureConExplicitE}{
\begin{tikzpicture}[baseline={([yshift=-.8ex]current bounding box.center)},scale=.3]
    \draw [red, thick] (4.8,8.) -- (2.,6.4);
    \draw [red, thick] (2.,6.4) -- (2.,3.2);
    \draw [red, thick] (4.8,8.) -- (7.6,6.4);
    \draw [red, thick] (7.6,6.4) -- (7.6,3.2);
    \draw [red, thick] (7.6,3.2) -- (4.8,1.6);
    \draw [red, thick] (4.8,1.6) -- (2.,3.2);
    \draw [red, thick] (4.8,8.) -- (4.8,9.2);
    \draw [red, thick] (2.,6.4) -- (0.8,7.);
    \draw [red, thick] (2.,3.2) -- (0.8,2.6);
    \draw [red, thick] (7.6,3.2) -- (8.8,2.6);
    \draw [red, thick] (7.6,6.4) -- (8.8,7.);
    \draw [red, thick] (4.8,1.6) -- (4.8,0.4);
    \draw [red, thick] (2.,4.1) -- (0.4,4.1);
    \draw [red, thick] (7.6,5.5) -- (9.2,5.5);
    \filldraw [red, opacity=0.2] (0.4,0.4) -- (0.4,9.2) -- (9.2,9.2) -- (9.2,0.4);
    \node [centered, rotate=0.] at (2.5,3.7) {\scriptsize $j_{1}$};
    \node [centered, rotate=0.] at (2.5,5.4) {\scriptsize $j_{1}'$};
    \node [centered, rotate=0.] at (3.,7.7) {\scriptsize $j_{2}$};
    \node [centered, rotate=0.] at (6.6,7.7) {\scriptsize $j_{3}$};
    \node [centered, rotate=0.] at (7.,5.9) {\scriptsize $j_{4}''$};
    \node [centered, rotate=0.] at (7.,4.4) {\scriptsize $j_{4}$};
    \node [centered, rotate=0.] at (1.2,7.8) {\scriptsize $j_{5}$};
    \node [centered, rotate=0.] at (5.4,8.7) {\scriptsize $j_{6}$};
    \node [centered, rotate=0.] at (8.8,7.8) {\scriptsize $j_{7}$};
    \node [centered, rotate=0.] at (0.8,4.6) {\scriptsize $p$};
    \node [centered, rotate=0.] at (8.8,6.) {\scriptsize $r$};
\end{tikzpicture}
}

\newcommand{\EqConcatenationLeft}{
\hs\begin{tikzpicture}[baseline={([yshift=-.8ex]current bounding box.center)},scale=.45]
    \draw [] (1.,8.8) -- (1.,4.);
    \draw [] (11.2,8.8) -- (12.5,6.4);
    \draw [] (12.5,6.4) -- (11.2,4.);
    \draw [red,thick] (2.,5.6) -- (3.4,4.8);
    \draw [red,thick] (2.,7.2) -- (2.,5.6);
    \draw [red,thick] (2.,5.6) -- (3.4,4.8);
    \draw [red,thick] (3.4,4.8) -- (4.8,5.6);
    \draw [red,thick] (4.8,5.6) -- (4.8,7.2);
    \draw [red,thick] (4.8,7.2) -- (3.4,8.);
    \draw [red,thick] (3.4,8.) -- (2.,7.2);
    \draw [red,thick] (3.4,8.6) -- (3.4,8.);
    \draw [red,thick] (1.3,7.6) -- (2.,7.2);
    \draw [red,thick] (1.3,5.2) -- (2.,5.6);
    \draw [red,thick] (3.4,4.2) -- (3.4,4.8);
    \draw [red,thick] (6.2,4.8) -- (4.8,5.6);
    \draw [red,thick] (6.2,4.2) -- (6.2,4.8);
    \draw [red,thick] (7.6,5.6) -- (6.2,4.8);
    \draw [red,thick] (4.8,7.2) -- (6.2,8.);
    \draw [red,thick] (7.6,7.2) -- (7.6,5.6);
    \draw [red,thick] (9.,4.8) -- (7.6,5.6);
    \draw [red,thick] (9.,4.8) -- (9.,4.2);
    \draw [red,thick] (9.,8.) -- (7.6,7.2);
    \draw [red,thick] (9.,8.) -- (9.,8.6);
    \draw [red,thick] (9.,4.8) -- (10.4,5.6);
    \draw [red,thick] (9.,8.) -- (10.4,7.2);
    \draw [red,thick] (10.4,5.6) -- (10.4,7.2);
    \draw [red,thick] (10.4,5.6) -- (11.,5.2);
    \draw [red,thick] (11.,7.6) -- (10.4,7.2);
    \draw [red,thick] (6.2,8.) -- (7.6,7.2);
    \draw [red,thick] (6.2,8.6) -- (6.2,8.);
    \filldraw [red,opacity=0.2] (1.3,8.6) -- (1.3,4.2) -- (11.,4.2) -- (11.,8.6);
    \node [centered, rotate=0.] at (4.4,6.4) {\scriptsize $j_{1}$};
    \node [centered, rotate=0.] at (5.3,8.) {\scriptsize $j_{2}$};
    \node [centered, rotate=0.] at (6.7,7.3) {\scriptsize $j_{3}$};
    \node [centered, rotate=0.] at (8.,6.4) {\scriptsize $j_{4}$};
    \node [centered, rotate=0.] at (3.8,7.4) {\scriptsize $j_{5}$};
    \node [centered, rotate=0.] at (6.6,8.3) {\scriptsize $j_{6}$};
    \node [centered, rotate=0.] at (8.5,7.3) {\scriptsize $j_{7}$};
\end{tikzpicture}\hs
}

\newcommand{\EqConcatenationRight}{ 
\hs\begin{tikzpicture}[baseline={([yshift=-.8ex]current bounding box.center)},scale=.45]
    \draw [] (1.,3.8) -- (1.,9.);
    \draw [] (11.2,3.8) -- (12.5,6.4);
    \draw [] (12.5,6.4) -- (11.2,9.);
    \draw [red,thick] (2.,7.2) -- (2.,5.6);
    \draw [red,thick] (2.,5.6) -- (3.4,4.8);
    \draw [red,thick] (3.4,4.8) -- (4.8,5.6);
    \draw [red,thick] (4.8,5.6) -- (4.8,7.2);
    \draw [red,thick] (4.8,7.2) -- (3.4,8.);
    \draw [red,thick] (3.4,8.) -- (2.,7.2);
    \draw [red,thick] (3.4,8.6) -- (3.4,8.);
    \draw [red,thick] (1.3,7.6) -- (2.,7.2);
    \draw [red,thick] (1.3,5.2) -- (2.,5.6);
    \draw [red,thick] (3.4,4.2) -- (3.4,4.8);
    \draw [red,thick] (6.2,4.8) -- (4.8,5.6);
    \draw [red,thick] (6.2,4.2) -- (6.2,4.8);
    \draw [red,thick] (7.6,5.6) -- (6.2,4.8);
    \draw [red,thick] (4.8,7.2) -- (6.2,8.);
    \draw [red,thick] (7.6,7.2) -- (7.6,5.6);
    \draw [red,thick] (9.,4.8) -- (7.6,5.6);
    \draw [red,thick] (9.,4.8) -- (9.,4.2);
    \draw [red,thick] (9.,8.) -- (7.6,7.2);
    \draw [red,thick] (9.,8.) -- (9.,8.6);
    \draw [red,thick] (9.,4.8) -- (10.4,5.6);
    \draw [red,thick] (9.,8.) -- (10.4,7.2);
    \draw [red,thick] (10.4,5.6) -- (10.4,7.2);
    \draw [red,thick] (10.4,5.6) -- (11.,5.2);
    \draw [red,thick] (11.,7.6) -- (10.4,7.2);
    \draw [red,thick] (6.2,8.) -- (7.6,7.2);
    \draw [red,thick] (6.2,8.6) -- (6.2,8.);
    \filldraw [red, opacity=0.2] (1.3,5.2) -- (1.3, 8.6) -- (3.4,8.6) -- (3.4, 8.) -- (4.8,7.2) -- (4.8,5.6) -- (6.2,4.8) -- (6.2,4.2) -- (1.3, 4.2);
    \filldraw [red, opacity=0.2] (3.4,8.6) -- (3.4, 8.) -- (4.8,7.2) -- (4.8,5.6) -- (6.2,4.8) -- (6.2,4.2) -- (9.,4.2) -- (9.,4.8) -- (7.6,5.6) -- (7.6,7.2) -- (6.2,8.) -- (6.2,8.6);
    \filldraw [red, opacity=0.2] (9.,4.2) -- (9.,4.8) -- (7.6,5.6) -- (7.6,7.2) -- (6.2,8.) -- (6.2,8.6) --(11.,8.6) -- (11.,4.2);
    \draw [red,thick] (3.8,6.2) -- (4.8, 6.2);
    \draw [red,thick] (7.6,6.6) -- (8.6,6.6);
    \node [centered, rotate=0.] at (3.5,6.1) {\scriptsize $p$};
    \node [centered, rotate=0.] at (8.8,6.6) {\scriptsize $r$};
    \node [centered, rotate=0.] at (5.2,5.9) {\scriptsize $j_{1}$};
    \node [centered, rotate=0.] at (5.2,6.8) {\scriptsize $j_{1}'$};
    \node [centered, rotate=0.] at (5.2,8.) {\scriptsize $j_{2}'$};
    \node [centered, rotate=0.] at (7.4,7.9) {\scriptsize $j_{3}'$};
    \node [centered, rotate=0.] at (7.2,6.9) {\scriptsize $j_{4}''$};
    \node [centered, rotate=0.] at (7.2,6.) {\scriptsize $j_{4}$};
    \node [centered, rotate=0.] at (3.8,7.4) {\scriptsize $j_{5}$};
    \node [centered, rotate=0.] at (6.6,8.3) {\scriptsize $j_{6}$};
    \node [centered, rotate=0.] at (8.7,7.4) {\scriptsize $j_{7}$};
\end{tikzpicture}\hs
}

\newcommand{\FigureSigmaTailA}{
\begin{tikzpicture}[baseline={([yshift=-.8ex]current bounding box.center)},scale=.8]
\draw [red, thick] (1.,8.8) -- (2.4,8.);
\draw [blue] (2.4,8.) -- (3.8,8.8);
\draw [blue] (2.4,8.) -- (2.4,5.6);
\draw [blue] (2.4,5.6) -- (3.8,4.8);
\draw [blue] (2.4,5.6) -- (1.,4.8);
\draw [blue] (2.4,7.4) -- (1.2,7.4);
\filldraw [black, opacity=0.2] (1.0,8.8) -- (3.8,8.8) -- (2.4,8.) -- (2.4,5.6) -- (1.0,4.8) -- cycle;
\filldraw [blue, opacity=0.2] (1.0,4.8) -- (2.4,5.6) -- (2.4,8.) -- (3.8,8.8) -- (3.8,4.8) -- cycle;
\node [centered, rotate=0.] at (1.6,7.2) {\scriptsize $\sigma$};
\node [centered, rotate=0.] at (2.7,6.6) {\scriptsize $E_2$};
\node [centered, rotate=0.] at (1.6,8.2) {\scriptsize $E_1$};
\node [centered, rotate=0.] at (1.8,5.) {\scriptsize $E_3$};
\node [centered, rotate=0.] at (3.0,5.) {\scriptsize $E_4$};
\node [centered, rotate=0.] at (2.4,8.3) {\scriptsize $V_1$};
\node [centered, rotate=0.] at (2.4,5.3) {\scriptsize $V_2$};
\node [centered, rotate=0.] at (2.4,4.2) {\scriptsize {(a)}};
\end{tikzpicture}
}

\newcommand{\FigureSigmaTailB}{
\begin{tikzpicture}[baseline={([yshift=-.8ex]current bounding box.center)},scale=.8]
\draw [red, thick] (1.,8.8) -- (2.4,8.);
\draw [blue] (2.4,8.) -- (3.8,8.8);
\draw [blue] (2.4,8.) -- (2.4,5.6);
\draw [blue] (2.4,5.6) -- (3.8,4.8);
\draw [blue] (2.4,5.6) -- (1.,4.8);
\draw [red,thick] (2.05,8.2) -- (1.55,7.2);
\filldraw [black, opacity=0.2] (1.0,8.8) -- (3.8,8.8) -- (2.4,8.) -- (2.4,5.6) -- (1.0,4.8) -- cycle;
\filldraw [blue, opacity=0.2] (1.0,4.8) -- (2.4,5.6) -- (2.4,8.) -- (3.8,8.8) -- (3.8,4.8) -- cycle;
\node [centered, rotate=0.] at (1.8,7.4) {\scriptsize $\sigma$};
\node [centered, rotate=0.] at (2.7,6.6) {\scriptsize $E_2$};
\node [centered, rotate=0.] at (1.6,8.2) {\scriptsize $E_1$};
\node [centered, rotate=0.] at (1.8,5.) {\scriptsize $E_3$};
\node [centered, rotate=0.] at (3.0,5.) {\scriptsize $E_4$};
\node [centered, rotate=0.] at (2.4,8.3) {\scriptsize $V_1$};
\node [centered, rotate=0.] at (2.4,5.3) {\scriptsize $V_2$};
\node [centered, rotate=0.] at (2.4,4.2) {\scriptsize {(b)}};
\end{tikzpicture}
}

\newcommand{\FigureSigmaTailC}{
\begin{tikzpicture}[baseline={([yshift=-.8ex]current bounding box.center)},scale=.8]
\draw [red, thick] (1.,8.8) -- (2.4,8.);
\draw [blue] (2.4,8.) -- (3.8,8.8);
\draw [blue] (2.4,8.) -- (2.4,5.6);
\draw [blue] (2.4,5.6) -- (3.8,4.8);
\draw [blue] (2.4,5.6) -- (1.,4.8);
\draw [blue] (2.4,6.2) -- (1.2,6.2);
\filldraw [black, opacity=0.2] (1.0,8.8) -- (3.8,8.8) -- (2.4,8.) -- (2.4,5.6) -- (1.0,4.8) -- cycle;
\filldraw [blue, opacity=0.2] (1.0,4.8) -- (2.4,5.6) -- (2.4,8.) -- (3.8,8.8) -- (3.8,4.8) -- cycle;
\node [centered, rotate=0.] at (1.6,6.) {\scriptsize $\sigma$};
\node [centered, rotate=0.] at (2.7,7.) {\scriptsize $p_1$};
\node [centered, rotate=0.] at (2.7,5.9) {\scriptsize $p_2$};
\node [centered, rotate=0.] at (1.6,5.4) {\scriptsize $p_3$};
\node [centered, rotate=0.] at (3.4,5.4) {\scriptsize $p_4$};
\node [centered, rotate=0.] at (2.1,7.) {\scriptsize $E_2$};
\node [centered, rotate=0.] at (1.6,8.2) {\scriptsize $E_1$};
\node [centered, rotate=0.] at (1.8,5.) {\scriptsize $E_3$};
\node [centered, rotate=0.] at (3.0,5.) {\scriptsize $E_4$};
\node [centered, rotate=0.] at (2.4,8.3) {\scriptsize $V_1$};
\node [centered, rotate=0.] at (2.4,5.3) {\scriptsize $V_2$};
\node [centered, rotate=0.] at (2.4,4.2) {\scriptsize {(c)}};
\end{tikzpicture}
}

\newcommand{\EqLocalOperatorLeft}{
\hs\begin{tikzpicture}[baseline={([yshift=-.8ex]current bounding box.center)},scale=.5]
    \draw (1.6,9.2) -- (1.6,3.6);
    \draw (7.1,9.2) -- (8.5,6.4);
    \draw (8.5,6.4) -- (7.1,3.6);
    \draw [] (4.6,8.) -- (3.2,7.2);
    \draw [] (3.2,5.6) -- (4.6,4.8);
    \draw [] (4.6,4.8) -- (6.,5.6);
    \draw [] (6.,5.6) -- (6.,7.2);
    \draw [] (6.,7.2) -- (4.6,8.);
    \draw [] (6.6,7.6) -- (6.,7.2);
    \draw [] (6.6,5.2) -- (6.,5.6);
    \draw [] (4.6,4.4) -- (4.6,4.8);
    \draw [] (4.6,8.4) -- (4.6,8.);
    \draw [] (2.6,7.6) -- (3.2,7.2);
    \draw [] (2.6,5.2) -- (3.2,5.6);
    \draw [] (3.2,7.2) -- (3.2,6.4);
    \draw [] (3.2,6.4) -- (3.2,5.6);
    \draw [] (3.2,6.4) -- (4.4,6.4);
    \node [centered, rotate=0.] at (4.6,7.2) {\scriptsize $P$};
    \node [centered, rotate=0.] at (2.8,6.9) {\scriptsize $i_0$};
    \node [centered, rotate=0.] at (2.8,6.) {\scriptsize $i_1$};
    \node [centered, rotate=0.] at (3.6,4.9) {\scriptsize $i_2$};
    \node [centered, rotate=0.] at (5.7,4.9) {\scriptsize $i_3$};
    \node [centered, rotate=0.] at (6.5,6.4) {\scriptsize $i_4$};
    \node [centered, rotate=0.] at (5.6,7.9) {\scriptsize $i_5$};
    \node [centered, rotate=0.] at (3.6,8.) {\scriptsize $i_6$};
    \node [centered, rotate=0.] at (2.2,7.9) {\scriptsize $e_0$};
    \node [centered, rotate=0.] at (2.2,4.9) {\scriptsize $e_2$};
    \node [centered, rotate=0.] at (4.6,4.) {\scriptsize $e_3$};
    \node [centered, rotate=0.] at (7.,4.9) {\scriptsize $e_4$};
    \node [centered, rotate=0.] at (7.,7.9) {\scriptsize $e_5$};
    \node [centered, rotate=0.] at (4.6,8.7) {\scriptsize $e_6$};
    \node [centered, rotate=0.] at (4.8,6.5) {\scriptsize $p'$};
\end{tikzpicture}\hs
}

\newcommand{\EqLocalOperatorRight}{
\hs\begin{tikzpicture}[baseline={([yshift=-.8ex]current bounding box.center)},scale=.5]
    \draw (1.6,9.2) -- (1.6,3.6);
    \draw (7.1,9.2) -- (8.5,6.4);
    \draw (8.5,6.4) -- (7.1,3.6);
    \draw [] (4.6,8.) -- (3.2,7.2);
    \draw [] (3.2,5.6) -- (4.6,4.8);
    \draw [] (4.6,4.8) -- (6.,5.6);
    \draw [] (6.,5.6) -- (6.,7.2);
    \draw [] (6.,7.2) -- (4.6,8.);
    \draw [] (6.6,7.6) -- (6.,7.2);
    \draw [] (6.6,5.2) -- (6.,5.6);
    \draw [] (4.6,4.4) -- (4.6,4.8);
    \draw [] (4.6,8.4) -- (4.6,8.);
    \draw [] (2.6,7.6) -- (3.2,7.2);
    \draw [] (2.6,5.2) -- (3.2,5.6);
    \draw [] (3.2,7.2) -- (3.2,6.4);
    \draw [] (3.2,6.4) -- (3.2,5.6);
    \draw [] (3.2,6.4) -- (4.4,6.4);
    \node [centered, rotate=0.] at (2.8,6.9) {\scriptsize $j_0$};
    \node [centered, rotate=0.] at (2.8,6.) {\scriptsize $j_1$};
    \node [centered, rotate=0.] at (3.6,4.9) {\scriptsize $j_2$};
    \node [centered, rotate=0.] at (5.7,4.9) {\scriptsize $j_3$};
    \node [centered, rotate=0.] at (6.5,6.4) {\scriptsize $j_4$};
    \node [centered, rotate=0.] at (5.6,8.) {\scriptsize $j_5$};
    \node [centered, rotate=0.] at (3.6,8.) {\scriptsize $j_6$};
    \node [centered, rotate=0.] at (2.2,7.9) {\scriptsize $e_0$};
    \node [centered, rotate=0.] at (2.2,4.9) {\scriptsize $e_2$};
    \node [centered, rotate=0.] at (4.6,4.) {\scriptsize $e_3$};
    \node [centered, rotate=0.] at (7.,4.9) {\scriptsize $e_4$};
    \node [centered, rotate=0.] at (7.,7.9) {\scriptsize $e_5$};
    \node [centered, rotate=0.] at (4.6,8.7) {\scriptsize $e_6$};
    \node [centered, rotate=0.] at (4.7,6.4) {\scriptsize $q$};
\end{tikzpicture}\hs
}

\newcommand{\EqLocalActionLeft}{
\hs\begin{tikzpicture}[baseline={([yshift=-.8ex]current bounding box.center)},scale=.5]
    \draw [] (1.,3.8) -- (1.,9.);
    \draw [] (8.4,3.8) -- (9.7,6.4);
    \draw [] (9.7,6.4) -- (8.4,9.);
    \draw [red,thick] (2.,7.2) -- (2.,5.6);
    \draw [red,thick] (2.,5.6) -- (3.4,4.8);
    \draw [red,thick] (3.4,4.8) -- (4.8,5.6);
    \draw [red,thick] (4.8,5.6) -- (4.8,7.2);
    \draw [red,thick] (4.8,7.2) -- (3.4,8.);
    \draw [red,thick] (3.4,8.) -- (2.,7.2);
    \draw [red,thick] (3.4,8.6) -- (3.4,8.);
    \draw [red,thick] (1.3,7.6) -- (2.,7.2);
    \draw [red,thick] (1.3,5.2) -- (2.,5.6);
    \draw [red,thick] (3.4,4.2) -- (3.4,4.8);
    \draw [red,thick] (6.2,4.8) -- (4.8,5.6);
    \draw [red,thick] (6.2,4.2) -- (6.2,4.8);
    \draw [red,thick] (4.8,7.2) -- (6.2,8.);
    \draw [red,thick] (7.6,5.6) -- (6.2,4.8);
    \draw [ultra thick] (3.4,6.4) -- (6.2,6.4);
    \draw [red,thick] (6.2,8.) -- (7.6,7.2);
    \draw [red,thick] (7.6,7.2) -- (7.6,5.6);
    \draw [red,thick] (8.3,5.2) -- (7.6,5.6);
    \draw [red,thick] (8.3,7.6) -- (7.6,7.2);
    \draw [red,thick] (6.2,8.6) -- (6.2,8.);
    \filldraw [red, opacity=0.2] (1.3,5.2) -- (1.3, 8.6) -- (3.4,8.6) -- (3.4, 8.) -- (4.8,7.2) -- (4.8,5.6) -- (6.2,4.8) -- (6.2,4.2) -- (1.3, 4.2);
    \filldraw [red, opacity=0.2] (3.4,8.6) -- (3.4, 8.) -- (4.8,7.2) -- (4.8,5.6) -- (6.2,4.8) -- (6.2,4.2) -- (8.3,4.2) -- (8.3,5.2) -- (7.6,5.6) -- (7.6,7.2) -- (6.2,8.) -- (6.2,8.6);
    \filldraw [red, opacity=0.2] (8.3,5.2) -- (7.6,5.6) -- (7.6,7.2) -- (6.2,8.) -- (6.2,8.6) -- (8.3, 8.6);
    \node [centered, rotate=0.] at (3.3,7.2) {\scriptsize $P$};
    \node [centered, rotate=0.] at (2.2,6.4) {\scriptsize $(J_\text{DI},p')$};
    \node [centered, rotate=0.] at (7.3,6.4) {\scriptsize $(J_\text{DI},r)$};
\end{tikzpicture}\hs
}

\newcommand{\EqLocalActionRight}{
\hs\begin{tikzpicture}[baseline={([yshift=-.8ex]current bounding box.center)},scale=.5]
    \draw [] (1.,3.8) -- (1.,9.);
    \draw [] (8.4,3.8) -- (9.7,6.4);
    \draw [] (9.7,6.4) -- (8.4,9.);
    \draw [red,thick] (2.,7.2) -- (2.,5.6);
    \draw [red,thick] (2.,5.6) -- (3.4,4.8);
    \draw [red,thick] (3.4,4.8) -- (4.8,5.6);
    \draw [red,thick] (4.8,5.6) -- (4.8,7.2);
    \draw [red,thick] (4.8,7.2) -- (3.4,8.);
    \draw [red,thick] (3.4,8.) -- (2.,7.2);
    \draw [red,thick] (3.4,8.6) -- (3.4,8.);
    \draw [red,thick] (1.3,7.6) -- (2.,7.2);
    \draw [red,thick] (1.3,5.2) -- (2.,5.6);
    \draw [red,thick] (3.4,4.2) -- (3.4,4.8);
    \draw [red,thick] (6.2,4.8) -- (4.8,5.6);
    \draw [red,thick] (6.2,4.2) -- (6.2,4.8);
    \draw [red,thick] (4.8,7.2) -- (6.2,8.);
    \draw [red,thick] (7.6,5.6) -- (6.2,4.8);
    \draw [ultra thick] (3.4,6.4) -- (6.2,6.4);
    \draw [red,thick] (6.2,8.) -- (7.6,7.2);
    \draw [red,thick] (7.6,7.2) -- (7.6,5.6);
    \draw [red,thick] (8.3,5.2) -- (7.6,5.6);
    \draw [red,thick] (8.3,7.6) -- (7.6,7.2);
    \draw [red,thick] (6.2,8.6) -- (6.2,8.);
    \filldraw [red, opacity=0.2] (1.3,5.2) -- (1.3, 8.6) -- (3.4,8.6) -- (3.4, 8.) -- (4.8,7.2) -- (4.8,5.6) -- (6.2,4.8) -- (6.2,4.2) -- (1.3, 4.2);
    \filldraw [red, opacity=0.2] (3.4,8.6) -- (3.4, 8.) -- (4.8,7.2) -- (4.8,5.6) -- (6.2,4.8) -- (6.2,4.2) -- (8.3,4.2) -- (8.3,5.2) -- (7.6,5.6) -- (7.6,7.2) -- (6.2,8.) -- (6.2,8.6);
    \filldraw [red, opacity=0.2] (8.3,5.2) -- (7.6,5.6) -- (7.6,7.2) -- (6.2,8.) -- (6.2,8.6) -- (8.3, 8.6);
    \node [centered, rotate=0.] at (3.3,7.2) {\scriptsize $P$};
    \node [centered, rotate=0.] at (2.2,6.4) {\scriptsize $(J_\text{DI},q)$};
    \node [centered, rotate=0.] at (7.3,6.4) {\scriptsize $(J_\text{DI},r)$};
\end{tikzpicture}\hs
}

\newcommand{\EqMeasure}{
\hs\begin{tikzpicture}[baseline={([yshift=-.8ex]current bounding box.center)},scale=.5]
    \draw [] (1.,3.8) -- (1.,9.);
    \draw [] (8.4,3.8) -- (9.7,6.4);
    \draw [] (9.7,6.4) -- (8.4,9.);
    \draw [red,thick] (2.,7.2) -- (2.,5.6);
    \draw [red,thick] (2.,5.6) -- (3.4,4.8);
    \draw [red,thick] (3.4,4.8) -- (4.8,5.6);
    \draw [red,thick] (4.8,5.6) -- (4.8,7.2);
    \draw [red,thick] (4.8,7.2) -- (3.4,8.);
    \draw [red,thick] (3.4,8.) -- (2.,7.2);
    \draw [red,thick] (3.4,8.6) -- (3.4,8.);
    \draw [red,thick] (1.3,7.6) -- (2.,7.2);
    \draw [red,thick] (1.3,5.2) -- (2.,5.6);
    \draw [red,thick] (3.4,4.2) -- (3.4,4.8);
    \draw [red,thick] (6.2,4.8) -- (4.8,5.6);
    \draw [red,thick] (6.2,4.2) -- (6.2,4.8);
    \draw [red,thick] (4.8,7.2) -- (6.2,8.);
    \draw [red,thick] (7.6,5.6) -- (6.2,4.8);
    \draw [ultra thick] (3.4,6.4) -- (6.2,6.4);
    \draw [red,thick] (6.2,8.) -- (7.6,7.2);
    \draw [red,thick] (7.6,7.2) -- (7.6,5.6);
    \draw [red,thick] (8.3,5.2) -- (7.6,5.6);
    \draw [red,thick] (8.3,7.6) -- (7.6,7.2);
    \draw [red,thick] (6.2,8.6) -- (6.2,8.);
    \filldraw [red, opacity=0.2] (1.3,5.2) -- (1.3, 8.6) -- (3.4,8.6) -- (3.4, 8.) -- (4.8,7.2) -- (4.8,5.6) -- (6.2,4.8) -- (6.2,4.2) -- (1.3, 4.2);
    \filldraw [red, opacity=0.2] (3.4,8.6) -- (3.4, 8.) -- (4.8,7.2) -- (4.8,5.6) -- (6.2,4.8) -- (6.2,4.2) -- (8.3,4.2) -- (8.3,5.2) -- (7.6,5.6) -- (7.6,7.2) -- (6.2,8.) -- (6.2,8.6);
    \filldraw [red, opacity=0.2] (8.3,5.2) -- (7.6,5.6) -- (7.6,7.2) -- (6.2,8.) -- (6.2,8.6) -- (8.3, 8.6);
    \node [centered, rotate=0.] at (3.3,7.2) {\scriptsize $P$};
    \node [centered, rotate=0.] at (2.2,6.4) {\scriptsize $(J_\text{DI}',p')$};
    \node [centered, rotate=0.] at (7.3,6.4) {\scriptsize $(J_\text{DI},r)$};
\end{tikzpicture}\hs
}

\begin{document}

\begin{center}{\Large \textbf{
Characteristic Properties of a Composite System\\ of Topological Phases Separated by Gapped Domain Walls\\ via an Exactly Solvable Hamiltonian Model\\
}}\end{center}

% TODO: write the author list here. Use first name (+ other initials) + surname format.
% Separate subsequent authors by a comma, omit comma and use "and" for the last author.
% Mark the corresponding author with a superscript star.
\begin{center}
Yu Zhao\textsuperscript{1, 2},
Shan Huang\textsuperscript{1, 2},
Hongyu Wang\textsuperscript{1, 2},
Yuting Hu\textsuperscript{3$\star$} and
Yidun Wan\textsuperscript{1, 2$\dagger$}
\end{center}

\begin{center}
{\bf 1} State Key Laboratory of Surface Physics, Department of Physics, Center for Field Theory and Particle Physics, and Institute for Nanoelectronic devices and Quantum computing, Fudan University, Shanghai 200433, China
\\
{\bf 2} Shanghai Qi Zhi Institute, Shanghai 200030, China
\\
{\bf 3} School of Physics, Hangzhou Normal University, Hangzhou 311121, China
\\
% TODO: provide the email address of the corresponding author
${}^\star$ {\small \sf yuting.phys@gmail.com},
${}^\dagger$ {\small \sf ydwan@fudan.edu.cn}
\end{center}

\begin{center}
\today
\end{center}

% For convenience during refereeing (optional),
% you can turn on line numbers by uncommenting the next line:
%\linenumbers

\section*{Abstract}
{\bf
% TODO: write your abstract here.
In this paper, we construct an exactly solvable lattice Hamiltonian model to investigate the properties of a composite system consisting of multiple topological orders separated by gapped domain walls. There are interdomain elementary excitations labeled by a pair of anyons in different domains of this system; This system also has elementary excitations with quasiparticles in the gapped domain wall. Each set of elementary excitations corresponds to a basis of the ground states of this composite system on the torus, reflecting that the ground-state degeneracy matches the number of either set of elementary excitations. The characteristic properties of this composite system lie in the basis transformations, represented by the $S$ and $T$ matrices: The $S$ matrix encodes the mutual statistics between interdomain excitations and domain-wall quasiparticles, and the $T$ matrix encapsulates the topological spins of interdomain excitations. Our model realizes a spatial counterpart of a temporal phase transition triggered by anyon condensation, bringing the abstract theory of anyon condensation into manifestable spatial interdomain excitation states.
}

% Guideline: if your paper is longer than 6 pages, include a TOC
\vspace{10pt}
\noindent\rule{\textwidth}{1pt}
\tableofcontents\thispagestyle{fancy}
\noindent\rule{\textwidth}{1pt}
\vspace{10pt}

% TODO: write your article here.
\section{Introduction}\label{sec:intro}

Topological orders in $2+1$D dimensions, encapsulated by modular tensor categories, are characterized by their topological quantum numbers, e.g., the anyon species, the ground-state degeneracy on a torus, and the modular $S$ and $T$ matrices\cite{Kitaev2003a,Levin2004,Hu2012a}. The number of anyon species in a topological order coincides with the ground-state degeneracy on the torus, and the $S$ and $T$ matrices correlate the transformations in the ground-state space with the mutual and self statistics of the anyon species \cite{Huang2005,Nayak2008,Zhang2012,Barkeshli2013e,Teo2013a,liu2013,Cincio2013,Jiang2014a}. 

Although a categorical description provides a crucial understanding of a $2+1$D topological order, it focuses on the topological properties and omits some details of the topological system, specifically the local internal degrees of freedom of anyons that are not preserved under local perturbations. To overcome this limitation, one can represent the topological order through an exactly solvable lattice model given certain input data\cite{Bravyi1998, Kitaev2003a, Levin2004, Buerschaper2009, Hung2012, Hu2012, Hu2012a, Buerschaper2013, schulz2013, Lan2014b, Hu2017, Hu2018, cheipesh2019, Wang2020, Wang2022}. The lattice model explicitly provides the wavefunctions of elementary excitation states and uncovers the internal degrees of freedom using the input data, which are not apparent in a categorical theory.

While single topological orders have been well-explored, the composite systems consisting of multiple topological orders divided by gapped domain walls remains largely uncharted. Previous work has approached these domain walls algebraically using category theories \cite{Kitaev2012,Lan2014}. Yet, there persist essential questions demanding attention: What are the properties of such a composite system? How would the properties of a single topological phase, such as the ground-state degeneracy, and S and T matrics, be adapted in the composite system? What is the spectrum of elementary excitation states in such a system? What are quasiparticles in this system, especially those in the gapped domain wall?

These questions are not easily solvable by category theory because, as certain anyons moves across the gapped domain wall from one domain to another, the internal degrees of freedom of these anyons may turn to topological observables. Thus, to solve these questions, we require an exactly solvable lattice Hamiltonian model of two topological phases separated by a gapped domain wall. This paper explicitly presents such a model based on the extended Levin-Wen model (LW model)\cite{Hu2017} because it is the most general model of doubled $2+1$D topological orders. Our model enables us to examine the properties of the gapped domain wall and the entire system intuitively through tangible wavefunctions.

On the other hand, in addition to gapped domain walls, the relationship between topological phases can also be established via phase transitions \cite{Bais2009a,Barkeshli2010,Burnell2012,schulz2013,Barkeshli2013a,Eliens2013,Chen2013,Gu2014a,HungWan2015a,Ji2019,Hu2021}. An interesting type of phase transitions between a topological phase (the parent phase) and another (possibly trivial) topological phase (the child phase) is triggered by certain anyon condensation in the parent phase. As having been extensively studied\cite{Bais2009a,Eliens2013,Gu2014a}, interesting phenomema occur during such a phase transition: 1) Certain anyons, including the condensed ones, may split into a few sectors, so more precisely speaking, a condensed anyon may not fully condense but only one ore more sectors in its splitting condense. This is analogous to the Higgs boson condensation in breaking the electroweak symmetry, where only a one-dimensional subspace of the two-dimensional space the Higgs boson lives in condenses. 2) Since the condensed sectors become the new vacuum, two types of sectors related by fusing with a condensed anyon in the parent phase can no longer be distinguished in the child phase and thus are identified as the same type of anyons. 3) The anyons that have trivial (nontrivial) braiding statistics with the condensed anyons are unconfined (confined) in the child phase. 4) The child phase is in fact a symmetry-enriched topological phase, with a hidden global symmetry. The hidden symmetry has been made precise in Ref. \cite{Hu2021}, which also proves a generalized Goldstone theorem of anyon condensation. Therefore, the phase transition from a parent topological phase to its child phase also belongs to the Landau-Ginzburg paradigm, however in a more general sense \cite{Hung2013,Gu2014a,HungWan2015a,Hu2021}. 

Such a phase transition is temporal and is believed to correspond to a spatial composite system consisting of the parent and child topological phases separated by a gapped domain wall. In this paper, we solidify this correspondence by scrutinizing the duality between anyon condensation phenomena and the elementary excitation states within our model. Since these excitation states can be expressed through tangible wavefunctions, our model elucidates the underlying physics of anyon condensation intuitively, offering a powerful method for studying anyon condensation. 

In this paper, we develop a \emph{subsystem condensation technique} to construct a lattice model of two topological phases: Starting with the lattice model describing a single parent phase, we condense certain anyons in a subsystem (half) of the lattice, transforming the topological phase in this subsystem into a child phase, while the original parent phase remains outside this subsystem. This approach yields an exactly solvable lattice model simultaneously describing the parent phase outside the subsystem, the child phase in the subsystem, and the gapped domain wall separating them.

While this construction applies to general composite systems, for clarity and explicitness, we shall only concentrate on the model in the special case of the doubled Ising and $\mathbb{Z}_2$ toric code topological phases with a gapped domain wall in between. This specific example is simple and familiar within the research community. Moreover, the anyon condensation between these two phases encapsulates all the key phenomena --- identification, splitting, and confinement --- of the general theory of anyon condensation, making it a perfect choice for our study. Interestingly, despite the familiarity of this system, our model continues to reveal novel phenomena of this anyon condensation, such as the detailed mechanics of splitting. Nevertheless, we underscore that our construction methodology and conclusions are universally applicable and can be readily extended to any composite systems according to the specific research interests.

Analyzing the spectrum of the elementary excitation states of our model leads to the following main results.
\begin{enumerate}
    \item We find a richer spectrum of our composite system than a single topological order: There are interdomain elementary excitations labeled by a pair of anyons in different domains; There are also elementary excitations with quasiparticles in the gapped domain wall.
    \item We explicitly establish the correspondence between the transformations of anyons in anyon condensation with the manifestable interdomain elementary excitation states in our model.
    \item Both the set of interdomain elementary excitations and the set of domain-wall quasiparticles respectively correspond to a basis of the ground states of this composite system on a torus, reflecting that the ground-state degeneracy (GSD) of our model on the torus equals the number of quasiparticle species in the gapped domain wall, as well as the number of interdomain elementary excitation species.
    \item We construct the $S$ and $T$ matrices that generate the basis transformations of the ground states on the torus. Our $S$ matrix also encodes the braiding between the anyons crossing the gapped domain wall around the quasiparticles in the gapped domain wall, and the $T$ matrix records the topological spins of interdomain excitations.
\end{enumerate}

Our model has the same input data of the extended Levin-Wen model that describes the parent topological phase, without using any extra categorical data, and is thus as simple as the original extended Levin-Wen model.

% End Introduction

\section{A brief review of the extended Levin-Wen model}\label{sec:model:di}

Since our model is based on the extended LW model, we first briefly review the extended LW model. To be specific and for our purposes, we only consider the model that describes the doubled Ising phase.

\begin{figure}
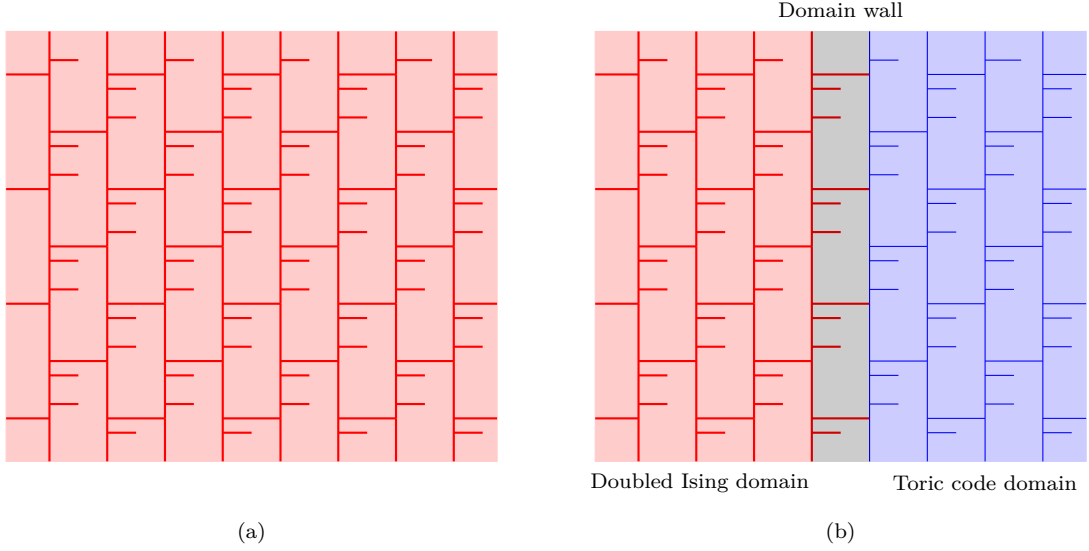
\centering
\subfloat{\FigureModelLeft\label{fig:model:di}}
\hspace{20pt}
\subfloat{\FigureModelRight\label{fig:model:ours}}
\caption{(a) The extended LW model describing the doubled Ising topological phase. (b) Our model of the doubled Ising and $\Z_2$ toric code topological phases separated by a gapped domain wall.}
\label{fig:model}
\end{figure}

The extended LW model is defined on a $2$-dimensional honeycomb lattice (see Fig. \ref{fig:model:di}). Associated with each vertex is a tail, presented as a dangling edge near the vertex. It is arbitrary to choose the edge incident at the vertex to which to attach the tail because all choices are equivalent up to gauge transformations (see Appendix \ref{appendix:gauge}).

The input data of the extended LW model describing the doubled Ising phase is a set $L_\text{DI} = \{1,\sigma,\psi\}$, equipped with three functions $N: L_\text{DI}^3 \to \mathbb{N}$, $d: L_\text{DI} \to \R$, and $G: L_\text{DI}^6 \to \mathbb{C}$. Each edge and tail of the lattice is labeled by an element in $L_\text{DI}$. 

The function $N$ sets the fusion rule and satisfies $N_{ij}^k = N_{ji}^k = N_{ik}^j$, whose nonzero independent elements are
\begin{equation}
N_{11}^1 = N_{\psi\psi}^1 = N_{\sigma\sigma}^1 = N_{\sigma\sigma}^\psi = 1.
\end{equation}
The Hilbert space $\Hil$ is spanned by all possible assignments of the labels on the edges and tails, subject to the constraint $N_{ij}^k \ne 0$ on any three incident edges (tails) labeled by $i,j,k\in L_\text{DI}$.

The function $d$ returns the quantum dimensions of the elements in $L_\text{DI}$,
\begin{equation}
d_1 = d_\psi = 1,\quad\quad d_\sigma = \sqrt{2}.
\end{equation}

The function $G$ has the symmetry $G_{kln}^{ijm} = G_{nkl}^{mij} = G_{ijn}^{klm}$. The nonzero independent elements are
\begin{equation}
G_{111}^{111} = G_{\psi\psi\psi}^{111} = G_{1\psi\psi}^{1\psi\psi} = 1,\quad G_{1\sigma\sigma}^{1\sigma\sigma} = G_{\psi\sigma\sigma}^{1\sigma\sigma} = -G_{\psi\sigma\sigma}^{\psi\sigma\sigma} = \frac{1}{\sqrt{2}},\quad G_{\sigma\sigma\sigma}^{111} = G_{\sigma\sigma\sigma}^{1\psi\psi} = \frac{1}{\sqrt[4]{2}}.
\end{equation}

The Hamiltonian of the extended LW model describing the doubled Ising phase reads
\begin{equation}
H_\text{DI} := -\sum_VQ_V - \sum_{P}B_P^\text{DI}.
\label{eq:Hamiltonian:di}
\end{equation}

The vertex operator $Q_V$ acts on the tail associated with vertex $V$ as
\begin{equation}
Q_V\EqQv = \delta_{p, 1}\EqQv,
\end{equation}
where $\delta$ denotes the Kronecker delta function that $\delta_{p,q} = 1$ if $p = q$ else $\delta_{p,q} = 0$.

The plaquette operator $B_P^\text{DI}$ acting on plaquette $P$ is a sum:
\begin{equation}
B_P^\text{DI} = \frac{d_1B_P^1 + d_\sigma B_P^\sigma + d_\psi B_P^\psi}{4},
\end{equation}
where $B_P^s$, $s\in L_\text{DI}$ is defined by
\begin{align}
B_p^s &\EqBpsLeft =\hs \delta_{p, 1}\hs\delta_{q,1}\sum_{j_0j_1j_2j_3j_4j_5j_6j_7\in L_\text{DI}}\left(\prod_{n=0}^{7}\sqrt{d_{i_n}d_{j_n}}\right)\left(G_{sj_7j_0}^{e_0i_0i_7}\right.\times\nonumber\\
&\left. G_{sj_0j_1}^{e_1i_1i_0}\hs G_{sj_1j_2}^{e_2i_2i_1}\hs G_{sj_2j_3}^{e_3i_3i_2}\hs G_{sj_3j_4}^{e_4i_4i_3}\hs G_{sj_4j_5}^{e_5i_5i_4}\hs G_{sj_5j_6}^{e_6i_6i_5}\hs G_{sj_6j_7}^{e_7i_7i_6}\right) \EqBpsRight.
\end{align}

The doubled-Ising Hamiltonian \eqref{eq:Hamiltonian:di} is exactly solvable because all the summands $Q_V$ and $B_P^\text{DI}$ therein are commuting projectors.

% End Review

\section{The lattice model with a gapped domain wall between the doubled Ising and \texorpdfstring{$\mathbb{Z}_2$}{Lg} toric code phases}\label{sec:model:ours}

We now construct our model describing a doubled Ising phase and $\Z_2$ toric code phase separated by a gapped domain wall, via partial anyon condensation explained as follows. We divide the entire lattice into two halves, left and right. See Fig. \ref{fig:model:ours}. Here, the edges and tails in the left (right) half are in red (blue); the plaquettes bounded by all red (blue) edges are in light red (blue); the plaquettes bounded by both red and blue edges are in gray. Each edge and tail of the entire lattice still take value in $L_\text{DI}$, so the Hilbert space of our model is still $\Hil$. We shall trigger anyon condensation in the right (blue) half, such that the doubled Ising phase therein will become the $\mathbb{Z}_2$ toric code phase through a phase transition.

Knowing that the $\Z_2$ toric code phase can be obtained by condensing $\psi\bar\psi$ anyons in the doubled Ising phase\cite{Bais2009, Gu2014a, Hu2021}, we are motivated to add to the doubled-Ising Hamiltonian \eqref{eq:Hamiltonian:di} the gapping term
\begin{equation}
\Delta H := -\Lambda\sum_{E\in\text{TC}}W_E^{\psi\bar\psi;1,1},\quad\quad\Lambda\gg 1,
\label{eq:Hamiltonian:delta}
\end{equation}
where $E\in\text{TC}$ represents all the blue edges, and $W_E^{\psi\bar\psi;1,1}$ is the creation operator of the $\psi\bar\psi$ anyons (to be defined in Section \ref{sec:excitation:di}). The term $\Delta H$ renders the new ground states of the system the $+1$ eigenstates of the creation operators $\left.W_E^{\psi\bar\psi;1,1}\right|_{E\in\text{TC}}$, and thus are the superpositions of the states with arbitrarily many $\psi\bar\psi$ anyons in the right half of the lattice. We say that the $\psi\bar\psi$ anyons in the right half of the lattice are condensed. The total Hamiltonian now reads
\begin{equation}
H := -\sum_VQ_V - \sum_PB_P^\text{DI} + \Delta H.
\label{eq:Hamiltonian:ours}
\end{equation}

By Eq. \eqref{eq:halfbraiding:psibarpsi}, the creation operators $W_E^{\psi\bar\psi;1,1}$ of $\psi\bar\psi$ quasiparticle pairs can be written as
\begin{equation}
W_E^{\psi\bar\psi;1,1} \EqConstraint = (-1)^{\delta_{j_E,\sigma}} \EqConstraint ,
\end{equation}
where $\delta$ is the Kronecker delta function. Hence, any blue edge has to overcome a great energy barrier ($\propto \Lambda$) to take value $\sigma$. For $\Lambda\to\infty$, the blue edges in the right half of the lattice effectively take value only in the input data $L_\text{TC} = \{1, \psi\}\subset L_\text{DI}$, equipped with the same $\delta$, $d$, and $G$ functions as that of $L_\text{DI}$ but restricted to $L_\text{TC}$:
\begin{equation}
\delta_{111} = \delta_{1\psi\psi} = 1,\quad\quad d_1 = d_\psi = 1,\quad\quad G_{111}^{111} = G_{\psi\psi\psi}^{111} = G_{1\psi\psi}^{1\psi\psi} = 1.
\end{equation}
The right half of the system therefore describes the $\Z_2$ toric code phase\cite{Kitaev2003a, Levin2004, Hu2018}, as a result of $\psi\bar\psi$ condensation.

Due to $\psi\bar\psi$ condensation, the effective Hilbert space $\Hil_\text{eff}$ of the model is the subspace of $\Hil$ in which all the blue edges can take value only in $L_\text{TC} = \{1,\psi\}$:
\begin{equation}
\Hil_\text{eff} := P_\text{eff}\Hil,
\label{eq:Hil:eff}
\end{equation}
where $P_\text{eff}$ is the projector on the blue edges
\begin{equation}
P_\text{eff} := \prod_{E\in\text{TC}}\left(1 - \delta_{j_E,\sigma}\right) = \prod_{E\in\text{TC}}\frac{I + W_E^{\psi\bar\psi;1,1}}{2}.
\label{eq:projector}
\end{equation}

Hereafter, we refer to the red (blue) edges/plaquettes the DI (TC) edges/plaquettes. The gray plaquettes, bounded by both DI and TC edges, turn out to comprise the domain wall between the doubled Ising phase and $\Z_2$ toric code phase, and are thus called the DW plaquettes. Since $P_\text{eff}B_P^\sigma P_\text{eff} = 0$ if $P$ is a TC/DW plaquette, the effective plaquette operators acting on the DW and TC plaquettes in $\Hil_\text{eff}$ become
\begin{equation}
B_P^\text{DW} := P_\text{eff}B_P^\text{DI}P_\text{eff} = \frac{B_P^1 + B_P^\psi}{4},
\end{equation}
\begin{equation}
B_P^\text{TC} := P_\text{eff}B_P^\text{DI}P_\text{eff} = \frac{B_P^1 + B_P^\psi}{4}.
\end{equation}
The effective Hamiltonian is the projection of the doubled-Ising Hamiltonian \eqref{eq:Hamiltonian:di}:
\begin{equation}
H_\text{eff} := P_\text{eff}H_\text{DI}P_\text{eff} = -\sum_VQ_V - \sum_{P\in\text{DI}}B_P^\text{DI} - \sum_{P\in\text{DW}}B_P^\text{DW} - \sum_{P\in\text{TC}}B_P^\text{TC},
\label{eq:Hamiltonian:eff}
\end{equation}
which is exactly solvable. The model describes the doubled Ising phase on the left, the $\Z_2$ toric code phase on the right, and a gapped domain wall in between. We shall refer to $H_\text{eff}$ as the Hamiltonian of our model from now on.

% End Our Model

\section{The spectrum of the elementary excitation states}\label{sec:excitation}

We now study the spectrum of our model. We assume the sphere topology, in which our model has a unique ground state $\ket\Phi$; nevertheless, the results in this section apply to other topologies. The ground state $\ket\Phi$ is defined by
\begin{equation}
Q_V\ket\Phi = B_P^\text{DI}\ket\Phi = 2B_P^\text{TC}\ket\Phi = 2B_P^\text{DW}\ket\Phi = \ket\Phi,
\label{eq:groundstate}
\end{equation}
where the factors $2$ arise from the projection. In the ground state $\ket\Phi$, all the tails have to take value $1\in L_\text{DI}$. 

An excited state $\left|\varphi\right\rangle$ is an eigenstate, in which $Q_V\left|\varphi\right\rangle = 0$ or $B_P^\mathcal{D}\left|\varphi\right\rangle = 0$ at one or more vertices $V$ or plaquettes $P$. In such a state, there are quasiparticles at vertices $V$ or plaquettes $P$. Here, the superscript $\mathcal{D}$ refers to either DW, TC, or DI. We also dub the ground state $\left|\Phi\right\rangle$ the trivial excited state, in which there are trivial quasiparticles.

For our purposes, it suffices to study the excited states with at most two quasiparticles, which we call \emph{elementary excitation states}. Since in an elementary excitation state, all tails take value $1$ except the ones where the two quasiparticles are, we can omit the tails irrelevant to these quasiparticles.

\subsection{Review of the elementary excitation states in the doubled Ising phase}
\label{sec:excitation:di}

\begin{figure}
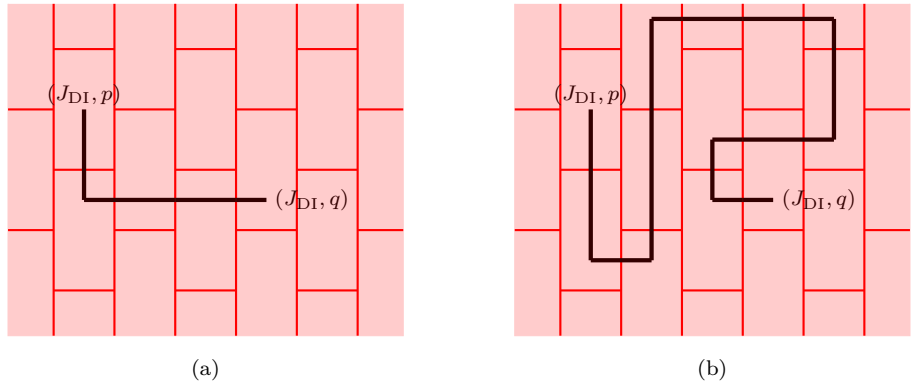
\centering
\subfloat{\FigureRibbonLeft}
\hspace{30pt}
\subfloat{\FigureRibbonRight}
\caption{Two ribbon operators in the doubled Ising phase. The two ribbon operators both create quasiparticles $(J, p)$ and $(J, q)$ at the ends of their paths, which are homotopic. Hence, they are the same operators although they take very different paths.}
\label{fig:ribbon}
\end{figure}
    
Since our model stems from the extended LW model describing the doubled Ising phase, we first focus on the elementary excitation states in the parent doubled Ising phase described by Hamiltonian $H_\text{DI}$ \eqref{eq:Hamiltonian:di}. The doubled-Ising ground state $\ket\Phi_\text{DI}\in\mathcal{H}$ satisfies
\begin{equation}
Q_V\ket\Phi_\text{DI} = B_P^\text{DI}\ket\Phi_\text{DI} = \ket\Phi_\text{DI}
\end{equation}
for all vertices $V$ and plaquettes $P$. Each doubled-Ising elementary excitation state $\ket{\varphi}_\text{DI}$ can be obtained by acting a ribbon operator $W_L$ on the ground state $\ket{\Phi}_\text{DI}$\cite{Levin2004, Hu2018}:
\begin{equation}
\ket{\varphi}_\text{DI} = W_L\ket{\Phi}_\text{DI}.
\label{eq:ribbon}
\end{equation}
The ribbon operator $W_L$ is defined along a path $L$, which crosses one or more edges in the lattice, and creates a pair of quasiparticles at the two ends of $L$ (see Fig. \ref{fig:ribbon}). The path $L$ of a ribbon operator can be homotopically deformed, with its two ends fixed.

\begin{table}\centering
\TableQuasiparticleDI
\caption{The anyon species and charges of quasiparticles in the doubled Ising phase.}
\label{table:quasiparticle:di}
\end{table}

We start with the elementary excitation states with a pair of quasiparticles in the two adjacent plaquettes with a common edge $E$. This state can be generated by ribbon operator $W_E^{J_\text{DI};p,q}$ along the shortest path that crosses only one edge $E$. This shortest ribbon operator creates in the two adjacent plaquettes a pair of quasiparticles $(J_\text{DI}, p)$ and $(J_\text{DI}, q)$, where $J_\text{DI}$ labels the anyon species of the quasiparticles, while $p$ and $q$ label the charges of the quasiparticles. Namely
\begin{equation}
\ket{J_\text{DI};p,q}_\text{DI} = \EqDoubledIsingState := W_E^{J_\text{DI};p,q}\ket\Phi_\text{DI}.
\label{eq:excitation:di}
\end{equation}
We only consider the action of ribbon operator $W_E^{J_\text{DI};p,q}$ on the ground state $\ket\Phi_\text{DI}$. The action reads
\begin{equation}
W_E^{J_\text{DI};p,q} \EqDoubledIsingRibbonLeft = \sum_{k\in L_\text{DI}}\sqrt{\frac{d_{k}}{d_{j_E}}}\cdot\overline{z_{pqj_E}^{J_\text{DI};k}} \EqDoubledIsingRibbonRight,
\end{equation}
where $j_E\in L_\text{DI}$ is the label on edge $E$. The matrix elements $z_{pqs}^{J_\text{DI};u}$ are the components of the tensor $z^{J_\text{DI}}$, and are listed in Appendix \ref{appendix:halfbraiding:di}. The tensor $z^{J_\text{DI}}$ satisfies\cite{Hu2018}
\begin{equation}
\frac{\delta_{j,t}N_{rs}^t}{d_t}z_{pqt}^{J_\text{DI};w} = \sum_{ulv\in L_\text{DI}} d_ud_v z_{lqr}^{J_\text{DI};v} z_{pls}^{J_\text{DI};u}G^{rst}_{pwu} G^{srj}_{qwv} G^{sul}_{rvw},
\label{eq:halfbraiding}
\end{equation}
where the anyon species $J_\text{DI}$ labels different minimal solutions $z^{J_\text{DI}}$ that cannot be the sum of any other nonzero tensors. There are 9 anyon species:
\begin{equation}
	1\overline{1},\quad 1\overline{\sigma},\quad 1\overline{\psi},\quad \sigma\overline{1},\quad \sigma\overline{\sigma},\quad \sigma\overline{\psi},\quad \psi\overline{1},\quad \psi\overline{\sigma},\quad \psi\bar\psi.
\end{equation} 

\begin{figure}
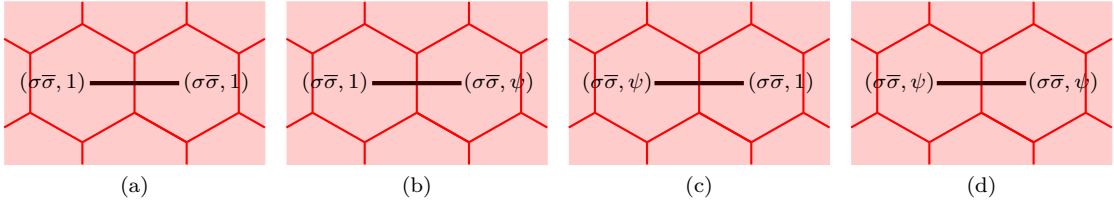
\centering
\subfloat{\FigureSigmaSigmaBarA}
\subfloat{\FigureSigmaSigmaBarB}
\subfloat{\FigureSigmaSigmaBarC}
\subfloat{\FigureSigmaSigmaBarD}
\caption{The doubled-Ising elementary excitation states with anyon species $\sigma\overline{\sigma}$. The charges of quasiparticles can take value arbitrarily in $\{1,\psi\}$.}
\label{fig:sigmasigmabar}
\end{figure}
    
Crossing any edge $E$ in the doubled Ising phase, for each $J_\text{DI}\neq \sigma\bar\sigma$, there is only one such ribbon operator $W_E^{J_\text{DI};p,q}$ with $p = q$. For $J_\text{DI}=\sigma\bar\sigma$, the corresponding charges $p$ and $q$ both can take values in $\{1,\psi\}$ (see Table \ref{table:quasiparticle:di}); hence, there are four different such ribbon operators $W^{\sigma\bar\sigma; p,q}_E$ with $p, q\in \{1,\psi\}$. All told, crossing any edge $E$, there are $12$ shortest ribbon operators.

There are four degenerate elementary excitation states $\ket{\sigma\bar\sigma;p,q}_\text{DI} = W_E^{J_\text{DI};p,q}\ket\Phi_\text{DI}$ with $p, q\in\{1,\psi\}$, as shown in Fig. \ref{fig:sigmasigmabar}. Each state has two quasiparticles, each of which can be either $(\sigma\bar\sigma, 1)$ or $(\sigma\bar\sigma,\psi)$. While $\sigma\bar\sigma$ is the topological observable of these states, the degenerate charges $1$ and $\psi$ cannot be distinguished experimentally\cite{Hu2018}, as they can be transformed into each other by local operators $B_P^{1\sigma\psi\sigma}$ and $B_P^{\psi\sigma 1\sigma}$ (defined in Eq. \eqref{eq:localoperator:di}). The elementary excitation states in the doubled Ising phase hence are characterized by their anyon species.

\begin{figure}
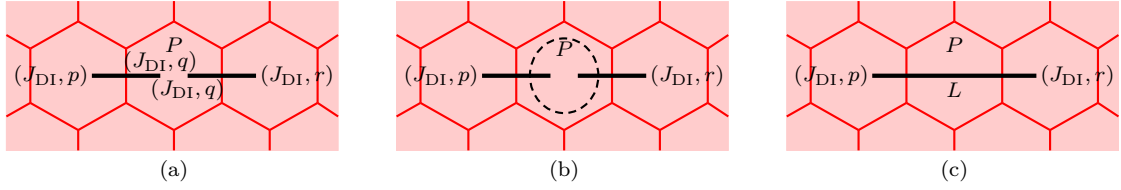
\centering
\subfloat{\FigureConcatenationA}
\hspace{8pt}
\subfloat{\FigureConcatenationB}
\hspace{8pt}
\subfloat{\FigureConcatenationC\label{fig:concatenation:ribbon}}
\caption{Concatenating two shortest ribbon operators to a longer one. (a) The state generated by two shortest ribbon operators. (b) Annihilating the two quasiparticles $(J_\text{DI},q)$ in plaquette $P$ results in (c): the state generated by the longer ribbon operator $W_L^{J_\text{DI};p,r}$.}
\label{fig:concatenation}
\end{figure}

Then we study the states with two quasiparticles in two nonadjacent plaquettes, generated by ribbon operators along longer paths. These ribbon operators result from concatenating shorter ribbon operators. For example, in Fig. \ref{fig:concatenation}, the two shortest ribbon operators create in plaquette $P$ two identical quasiparticles $(J_\text{DI}, q)$, which are then annihilated, resulting in a longer ribbon operator $W_L^{J_\text{DI}; p,r}$, which generates an elementary excitation state $W_L^{J_\text{DI}; p,r}\ket\Phi_\text{DI}$ with two quasiparticles at the end of path $L$. The matrix elements of such ribbon operators are also given by $z^{J_\text{DI}}$ tensors (see Appendix \ref{appendix:ribbon:longer}). 

% End DI Excitations

\subsection{The elementary excitation states of our model}\label{sec:excitation:ours}

Now we study the elementary excitation states of our model. These states are eigenstates of the effective Hamiltonian $H_\text{eff}$ \eqref{eq:Hamiltonian:eff} in the effective Hilbert space $\mathcal{H}_\text{eff}$ \eqref{eq:Hil:eff}. 

\begin{figure}
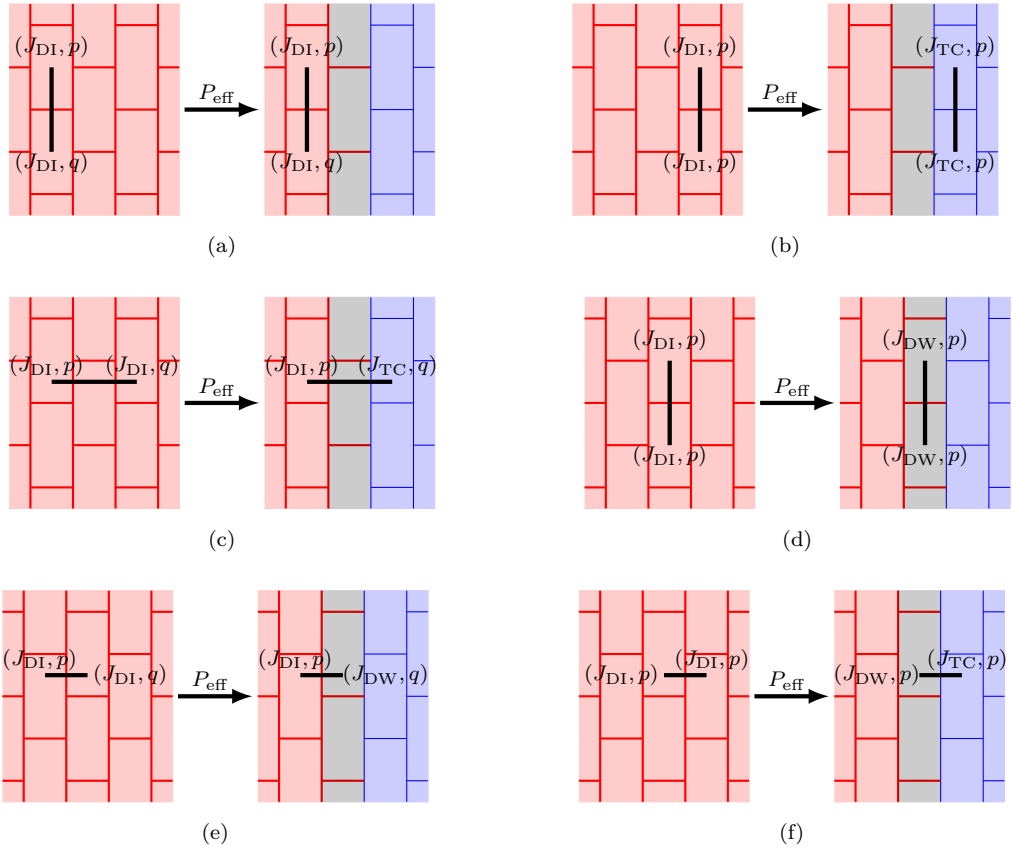
\centering
\subfloat{\FigureProjectorDoubledIsing}
\hspace{40pt}
\subfloat{\FigureProjectorToricCode\label{fig:projector:tc}}\\
\subfloat{\FigureProjectorInterDomain\label{fig:projector:inter}}
\hspace{40pt}
\subfloat{\FigureProjectorDomainWall\label{fig:projector:dw}}\\
\subfloat{\FigureProjectorDiDw\label{fig:projector:didw}}
\hspace{40pt}
\subfloat{\FigureProjectorDwTc\label{fig:projector:dwtc}}\\
\caption{Projecting the doubled-Ising elementary excitation states results in the elementary excitation states in our model. (a) The two quasiparticles in our model are all in the doubled Ising domain. (b) The two quasiparticles are all in the toric code domain. (c) The two quasiparticles are respectively in the doubled Ising domain and the toric code domain. (d) The two quasiparticles are all in the gapped domain wall. (e) The two quasiparticles are respectively in the doubled Ising domain and the gapped domain wall. (f) The two quasiparticles are respectively in the toric code domain and the gapped domain wall.}
\label{fig:projector}
\end{figure}

According to Appendix \ref{appendix:projection}, the projector \eqref{eq:projector}, $P_\text{eff}$, commutes with any doubled-Ising ribbon operator $W_L^{J_\text{DI};p,q}$ in the effective Hilbert space $\mathcal{H}_\text{eff}$:
\begin{equation}
P_\text{eff}\left[W_L^{J_\text{DI};p,q},\  P_\text{eff}\right] = 0.
\label{eq:projector:commute}
\end{equation}
Then, together with Eq. \eqref{eq:ribbon}, $P_\text{eff}$ projects the elementary excitation states $\ket\varphi_\text{DI}$ and the ribbon operators $W_L^{J_\text{DI};p,q}$ of the doubled Ising phase to those of our model. While in the doubled Ising phase, elementary excitations states do not discern the locations of the quasiparticles but only their anyon species, in our model, nevertheless, locations of the quasiparticles do matter because of the domain wall (see Fig. \ref{fig:projector}). In what follows, we shall study the elementary excitation states of our model in the cases of different quasiparticle locations.   

% End General Discussion

\subsubsection{The elementary excitation states with quasiparticle pairs in the toric code domain}\label{sec:excitation:tc}

Here, we study the elementary excitation states of our model with a pair of quasiparticles in two adjacent plaquettes completely in the toric code domain. These states result from projecting the elementary excitation states $\ket{J_\text{DI};p,q}_\text{DI}$ \eqref{eq:excitation:di} in the doubled Ising phase:
\begin{align}
&\ket{1;1,1} := P_\text{eff}\ket{1\bar 1;1,1}_\text{DI} = P_\text{eff}\ket{\psi\bar\psi;1,1}_\text{DI},\nonumber\\
&\ket{\epsilon;\psi,\psi} := P_\text{eff}\ket{\psi\bar 1;\psi,\psi}_\text{DI} = P_\text{eff}\ket{1\bar\psi;\psi,\psi}_\text{DI},\nonumber\\
&\ket{m;1,1} := P_\text{eff}\ket{\sigma\bar\sigma;1,1}_\text{DI},\nonumber\\
&\ket{e;\psi,\psi} := P_\text{eff}\ket{\sigma\bar\sigma;\psi,\psi}_\text{DI},
\label{eq:excitation:tc}
\end{align}
where we define the four nonvanishing states after the projection as $\ket{J_\text{TC};p,p}$, with $J_\text{TC} \in \{1,\epsilon,e,m\}$ the anyon species and $p\in\{1,\psi\}$ the charges of the quasiparticles. See Fig. \ref{fig:projector:tc}. But not all doubled-Ising elementary excitation states are projected to states in $\Hil_\text{eff}$:
\begin{align}
&P_\text{eff}\ket{\sigma\bar\sigma;\psi,1}_\text{DI} = P_\text{eff}\ket{\sigma\bar\sigma;1,\psi}_\text{DI} = 0,\nonumber\\
&P_\text{eff}\ket{\sigma\bar1;\sigma,\sigma}_\text{DI} = 0,\nonumber\\
&P_\text{eff}\ket{\sigma\bar\psi;\sigma,\sigma}_\text{DI} = 0,\nonumber\\ 
&P_\text{eff}\ket{1\bar\sigma;\sigma,\sigma}_\text{DI} = 0,\nonumber\\
&P_\text{eff}\ket{\psi\bar\sigma;\sigma,\sigma}_\text{DI} = 0.
\end{align}
These states are infinitely ($\Lambda\to \infty$) gapped by $\Delta H$, and should not appear in $\Hil_\text{eff}$.

\begin{table}\centering
\TableQuasiparticleTC
\caption{The anyon species and charges of quasiparticles in the toric code domain.}
\label{table:quasiparticle:tc}
\end{table}
    
The four elementary excitation states $\ket{J_\text{TC};p,p}$ are precisely the known four elementary excitation states in the $\Z_2$ toric code phase. These states are generated by the ribbon operators $W_E^{J_\text{TC};p,p}$ acting on the ground state \eqref{eq:groundstate} $\ket\Phi$ of our model 
\begin{equation}
    \ket{J_\text{TC};p,p} = \EqToricCodeState = W_E^{J_\text{TC};p,p}\ket\Phi,
\end{equation}
where the ribbon operators are the projections
\begin{align}
&W_E^{1;1,1}\hspace{5.9pt} : = P_\text{eff}W_E^{1\bar 1; 1,1}P_\text{eff} = P_\text{eff}W_E^{\psi\bar\psi; 1,1}P_\text{eff},\nonumber\\
&W_E^{\epsilon;\psi,\psi}\hspace{1pt}  : = P_\text{eff}W_E^{\psi\bar 1; \psi,\psi}P_\text{eff} = P_\text{eff}W_E^{1\bar\psi; \psi,\psi}P_\text{eff},\nonumber\\
&W_E^{m;1,1}: = P_\text{eff}W_E^{\sigma\bar\sigma; 1,1}P_\text{eff},\nonumber\\
&W_E^{e;\psi,\psi}: = P_\text{eff}W_E^{\sigma\bar\sigma; \psi,\psi}P_\text{eff}.
\end{align}
Note that the two plaquettes sharing edge $E$ must lie within the toric code domain.

These ribbon operators $W_E^{J_\text{TC};p,q}$ read
\begin{equation}
W_E^{J_\text{TC};p,q}\EqToricCodeRibbonLeft = \sum_{k\in L_\text{TC}}\sqrt{\frac{d_{k}}{d_{j_E}}}\cdot\overline{z_{ppj_E}^{J_\text{TC};k}} \EqToricCodeRibbonRight .
\end{equation}
The components $z_{ppj_E}^{J_\text{TC};k}$ are listed in Appendix \ref{appendix:halfbraiding:tc} and are precisely those comprising the ribbon operators in the extended LW model describing the $\Z_2$ toric code phase\cite{Hu2018}. 

Ribbon operators defined along longer paths crossing edges in the toric code domain can be obtained also by concatenating shorter ribbon operators. We shall not dwell on this.

% End Toric Code

\subsubsection{The elementary excitation states with interdomain quasiparticle pairs}\label{sec:excitation:inter}

Now we consider what we call \textit{interdomain elementary excitation states}, each having one quasiparticle $(J_\text{DI},p)$ in the doubled Ising domain and the other $(J_\text{TC},q)$ in the toric code domain. An interdomain state bears two different topological observables, $J_\text{DI}$ in the doubled Ising domain and $J_\text{TC}$ in the toric code domain. See Fig. \ref{fig:projector:inter}. We label interdomain elementary excitation states as $\ket{J_\text{DI}\dash J_\text{TC};p,q}$.

There are $8$ distinct interdomain elementary excitation states, as listed in Table \ref{table:quasiparticle:inter}.

\begin{table}\centering
\TableQuasiparticleInter
\caption{The interdomain elementary excitation states.}
\label{table:quasiparticle:inter}
\end{table}

These states can be generated by the ribbon operators along paths $L$ across the gapped domain wall: 
\begin{align}
\ket{J_\text{DI}\dash J_\text{TC};p,q} &= \EqInterStateLeft\nonumber\\
&= W_L^{J_\text{DI}\dash J_\text{TC};p,q}\EqInterStateRight.
\label{eq:excitation:inter}
\end{align}

The ribbon operators $W_L^{J_\text{DI}\dash J_\text{TC};p,q}$ are projected from the doubled-Ising ribbon operators along same paths $L$:
\begin{equation}
W_L^{J_\text{DI}\dash J_\text{TC};p,q} := P_\text{eff}W_L^{J_\text{DI};p,q}P_\text{eff},
\label{eq:ribbon:inter}
\end{equation}
and are explicitly written as
\begin{align}
&W_L^{J_\text{DI}\dash J_\text{TC},pq}\EqRibbonInterLeft =\hs \sum_{j_1'j_2'\in L_\text{DI}}\sum_{j_3'j_4'\in L_\text{TC}} \overline{z_{pqj_1}^{J_\text{DI};j_1'}}\hs \overline{z_{qqj_4}^{J_\text{TC};j_4'}} \hs\times \nonumber\\
&\sqrt{\frac{d_{j_1}}{d_{j_1'}}}\hs G_{qj_4j_4''}^{rj_4j_4'}\hs G_{qj_4'j_3'}^{j_7j_3j_4}\hs G_{qj_3'j_2'}^{j_6j_2j_3}\hs G_{qj_2'j_1'}^{j_5j_1j_2}\hs \EqRibbonInterRight.
\end{align}

% End Interdomain

\subsubsection{The elementary excitation states with domainwall quasiparticle pairs}\label{sec:excitation:dw}

\begin{table}\centering
\TableQuasiparticleDW
\caption{The quasiparticle species and charges of the quasiparticles in the gapped domain wall.}
\label{table:quasiparticle:dw}
\end{table}
    
Now we study the \textit{domainwall elementary excitation states}, i.e., the elementary excitation states with a pair of quasiparticles in two adjacent plaquettes within the gapped domain wall (see Fig. \ref{fig:projector:dw}). These states are projected from the doubled-Ising elementary excitation states with a pair of quasiparticles in the same plaquettes. Namely,
\begin{align}
&\ket{1;1,1}: = P_\text{eff}\ket{1\bar 1;1,1}_\text{DI} = P_\text{eff}\ket{\psi\bar\psi;1,1}_\text{DI},\nonumber \\
&\ket{\epsilon;\psi,\psi} := P_\text{eff}\ket{\psi\bar 1;\psi,\psi}_\text{DI} = P_\text{eff}\ket{1\bar\psi;\psi,\psi}_\text{DI},\nonumber\\
&\ket{\chi;\sigma,\sigma} := P_\text{eff}\ket{\sigma\bar 1;\sigma,\sigma}_\text{DI} = P_\text{eff}\ket{\sigma\bar\psi;\sigma,\sigma}_\text{DI},\nonumber\\ &\ket{\bar\chi;\sigma,\sigma} := P_\text{eff}\ket{1\bar \sigma;\sigma,\sigma}_\text{DI} = P_\text{eff}\ket{\psi\bar\sigma;\sigma,\sigma}_\text{DI},\nonumber\\
&\ket{m;1,1} := P_\text{eff}\ket{\sigma\bar\sigma;1,1}_\text{DI},\nonumber\\
&\ket{e;\psi,\psi} := P_\text{eff}\ket{\sigma\bar\sigma;\psi,\psi}_\text{DI},
\label{eq:excitation:dw}
\end{align}
where we define the six nonvanishing states after the projection as $\ket{J_\text{DW};p,p}$, with $J_\text{DW}\in\{1,\epsilon,m,e,\chi,\bar\chi\}$ the quasiparticle species and $p\in L_\text{DI}$ the charges of the quasiparticles. Graphically,
\begin{equation}
\ket{J_\text{DW};p,p} = \EqStateDomainWall.
\end{equation}

Although there are $6$ distinct domain wall elementary excitation states \eqref{eq:excitation:dw}, there are $10$ different ribbon operators across an edge $E$ in the gapped domain wall:
\begin{align}
&W_{E,1}^{1;1,1} := P_\text{eff}W_E^{1\bar 1;1,1}P_\text{eff},\quad\quad\quad W_{E,2}^{1;1,1} := P_\text{eff}W_E^{\psi\bar \psi;1,1}P_\text{eff},\nonumber\\
&W_{E,1}^{\epsilon;\psi,\psi} := P_\text{eff}W_E^{\psi\bar 1;\psi,\psi}P_\text{eff},\quad\quad\hs W_{E,2}^{\epsilon;\psi,\psi} := P_\text{eff}W_E^{1\bar \psi;\psi,\psi}P_\text{eff},\nonumber\\
&W_{E,1}^{\chi;\sigma,\sigma} := P_\text{eff}W_E^{\sigma\bar 1;\sigma,\sigma}P_\text{eff},\quad\quad\hs\hs W_{E,2}^{\chi;\sigma,\sigma} := P_\text{eff}W_E^{\sigma\bar \psi;\sigma,\sigma}P_\text{eff},\nonumber\\
&W_{E,1}^{\bar\chi;\sigma,\sigma} := P_\text{eff}W_E^{1\bar \sigma;\sigma,\sigma}P_\text{eff},\quad\quad\hs\hs W_{E,2}^{\bar\chi;\sigma,\sigma} := P_\text{eff}W_E^{\psi\bar \sigma;\sigma,\sigma}P_\text{eff},\nonumber\\
&W_{E}^{m;1,1} := P_\text{eff}W_E^{\sigma\bar \sigma;1,1}P_\text{eff},\nonumber\\
&W_{E}^{e;\psi,\psi} := P_\text{eff}W_E^{\sigma\bar \sigma;\psi,\psi}P_\text{eff}.
\label{eq:ribbon:dw}
\end{align}
Since $E$ is a DI edge taking value in $L_\text{DI} = \{1, \psi, \sigma\}$, 
\begin{equation}
W_{E,1}^{J_\text{DW};p,p} \ne W_{E,2}^{J_\text{DW};p,p}
\end{equation}
for $J_\text{DW} = 1, \epsilon, \chi$ and $\bar\chi$, but
\begin{equation}
W_{E,1}^{J_\text{DW};p,p}\ket\Phi = W_{E,2}^{J_\text{DW};p,p}\ket\Phi = \ket{J_\text{DW};p,p}.
\end{equation}

Specifically,
\begin{equation}
W_{E,i}^{J_\text{DW};p,p} \EqRibbonDomainWallLeft = \sum_{k\in L_\text{DI}}\sqrt{\frac{d_{k}}{d_{j_E}}}\cdot\overline{z_{ppj_E}^{J_\text{DW};k}} \EqRibbonDomainWallRight ,
\end{equation}
where the coefficients $z_{pqj_E}^{J_\text{DW};k}$ are listed in Appendix \ref{appendix:halfbraiding:dw}. The indices $p, q, k\in L_\text{DI}$, but the index $j_E$ is restricted to $L_\text{TC} = \{1,\psi\}$ because edge $E$ only takes value in $L_\text{TC}$ in the ground state \eqref{eq:groundstate} $\ket\Phi$.

% End Domainwall

\subsubsection{The elementary excitation states with doubled-Ising-domainwall quasiparticle pairs}\label{sec:excitation:didw}

\begin{table}\centering
\TableQuasiparticleDiDw
\caption{The elementary excitation states with one doubled-Ising quasiparticle and one domainwall quasiparticle.}
\label{table:quasiparticle:didw}
\end{table}
        
We now consider the elementary excitation states with one doubled-Ising quasiparticle $(J_\text{DI},p)$ and one domainwall quasiparticle $(J_\text{DW},q)$ in the adjacent plaquettes. See Fig. \ref{fig:projector:didw}. These elementary excitation states are defined as
\begin{equation}
\ket{J_\text{DI}\dash J_\text{DW};p,q} := W_E^{J_\text{DI};p,q}\ket\Phi,
\label{eq:excitation:didw}
\end{equation}
where $W_E^{J_\text{DI};p,q}$ is the ribbon operator across a DI edge $E$ between the doubled Ising domain and the gapped domain wall.
\begin{equation}
W_E^{J_\text{DI};p,q} \EqRibbonDiDwLeft = \sum_{k\in L_\text{DI}}\sqrt{\frac{d_{k}}{d_{j_E}}}\cdot\overline{z_{pqj_E}^{J_\text{DI};k}} \EqRibbonDiDwRight.
\label{eq:ribbon:didw}
\end{equation}
There are $12$ possible distinct elementary excitation states, as in Table \ref{table:quasiparticle:didw}. 

% End Didw

\subsubsection{The elementary excitation states with toric-code-domainwall quasiparticle pairs} \label{sec:excitation:dwtc}

We consider the case in Fig. \ref{fig:projector:dwtc}: the elementary excitation states with one toric-code quasiparticle and one domainwall quasiparticle. These elementary excitation states are defined as
\begin{equation}
\ket{J_\text{DW}\dash J_\text{TC};p,q} := W_E^{J_\text{TC}; p,p}\ket\Phi,
\label{eq:excitation:tcdw}
\end{equation}
where $W_E^{J_\text{TC};p,p}$ is the ribbon operator across TC edge $E$ between the toric code domain and the gapped domain wall. 
\begin{equation}
W_{E}^{J_\text{TC};p,p} \EqRibbonDwTcLeft = \sum_{k\in L_\text{TC}}\sqrt{\frac{d_{k}}{d_{j_E}}}\cdot\overline{z_{ppj_E}^{J_\text{TC};k}} \EqRibbonDwTcRight.
\label{eq:ribbon:tcdw}
\end{equation}

There are $4$ distinct elementary excitation states generated by $W_E^{J_\text{TC};p,q}$, as in Table \ref{table:quasiparticle:tcdw}.

Concatenating the ribbon operators in \eqref{eq:ribbon:didw} and \eqref{eq:ribbon:tcdw} results in the interdomain ribbon operators \eqref{eq:ribbon:inter}.

\begin{table}\centering
\TableQuasiparticleTcDw
\caption{The elementary excitation states with one toric-code quasiparticle and one domainwall quasiparticle.}
\label{table:quasiparticle:tcdw}
\end{table} 
    
% End Tcdw

% End Spectrum

\section{Correspondence with anyon condensation}\label{sec:correspondence}

\begin{figure}
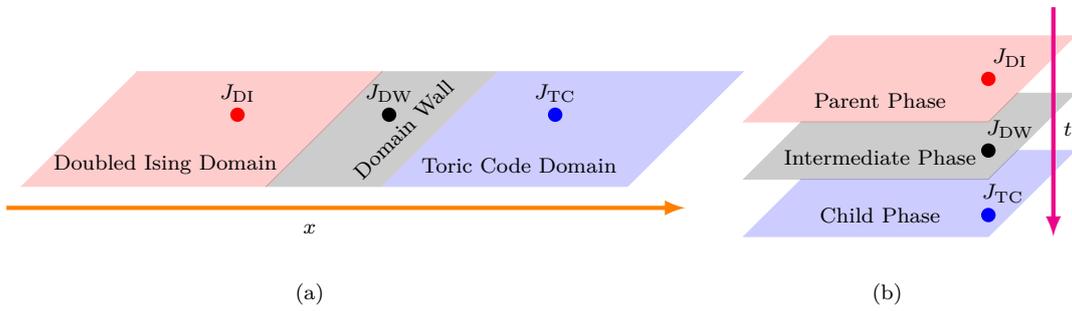
\centering
\FigureCorrespondence
\caption{The correspondence between (a): our model with gapped domain wall between the doubled Ising domain and toric code domain and (b): the anyon condensation from the parent doubled Ising phase to the child toric code phase via an auxiliary intermediate phase.}
\label{fig:correspondence}
\end{figure}

\begin{figure}
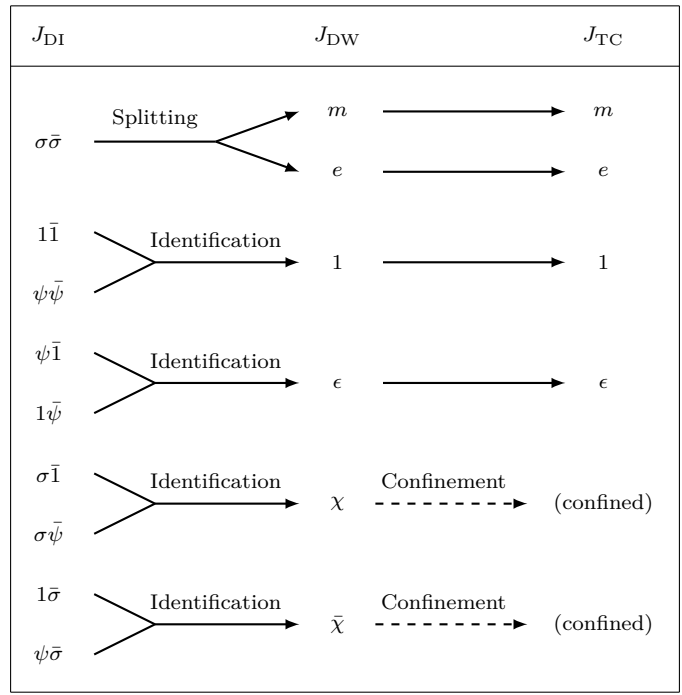
\centering
\TableAnyonCondensation
\caption{Relations between the quasiparticle species in different domains (phases).}
\label{table:anyoncondensation}
\end{figure}
            
As mentioned in the introduction, our model of two topological phases separated by a gapped domain wall can be regarded as a spatial counterpart of the phase transition (which is temporal) from one phase (the parent phase) to the other (the child phase) triggered by anyon condensation. See Fig. \ref{fig:correspondence}. 

An intermediate phase during the phase transition was introduced as merely a method to study the procedure of anyon condensation\cite{Bais2009a}. The anyon condensation in a parent phase first leads to an intermediate phase where splitting and identification have been completed, while the confinement occurs during the transition from the intermediate phase to the child phase. Interestingly, this auxiliary, virtual intermediate phase corresponds to the physical gapped domain wall between the parent and child phases. For example, Figure \ref{table:anyoncondensation} records the relations between the quasiparticles in different domains in our model, corresponding to different stages in a phase transition induced by $\psi\bar\psi$ condensation in the doubled Ising phase. Here, we shall use our model to formulate this correspondence rigorously.

The three main phenomena --- splitting, identification, and confinement --- that occur in a phase transition due to anyon condensation can find their spatial counterparts in the elementary excitation states of our model. In Eq. \eqref{eq:excitation:tc}, the doubled-Ising elementary excitation states $\ket{J_\text{DI};p,p}_\text{DI}$ in the parent phase are projected to the states $\ket{J_\text{TC};p,p}$ with quasiparticles in the toric code domain of our model. On the other hand, any interdomain elementary excitation state \eqref{eq:excitation:inter} $\ket{J_\text{DI}\dash J_\text{TC};p,q}$ bears a pair of topological observables $J_\text{DI}\dash J_\text{TC}$. The allowed pairs $J_\text{DI}\dash J_\text{TC}$ in the interdomain elementary excitation states are in one-to-one correspondence with the projections from $\ket{J_\text{DI};p,q}_\text{DI}$ to $\ket{J_\text{TC};p,q}$. Table \ref{table:corrspondence:tc} records this correspondence. We dub this correspondence the \emph{projection-state correspondence}. We now exhibit this correspondence from three aspects: splitting, identification, and confinement.

\begin{table}\centering
\TableCorrespondenceTC
\caption{The projection-state correspondence in the phenomena of splitting, identification, and confinement in the toric code domain.}
\label{table:corrspondence:tc}
\end{table}

% End General Discussion

\subsection{Splitting}\label{sec:correspondence:splitting}

Seen in Table \ref{table:corrspondence:tc}, the originally indistinguishable elementary excitation states $\ket{\sigma\bar\sigma;1,1}_\text{DI}$ and $\ket{\sigma\bar\sigma;\psi,\psi}_\text{DI}$ are projected to the topological different states $\ket{m;1,1}$ and $\ket{e;\psi,\psi}$ via $\psi\bar\psi$ condensation. It appears that the anyon species $\sigma\bar\sigma$ in the doubled Ising phase `splits' into two anyon species $e$ and $m$ in the toric code domain. This phenomenon is precisely what is known as \textit{splitting} in the language of anyon condensation. 

\begin{figure}
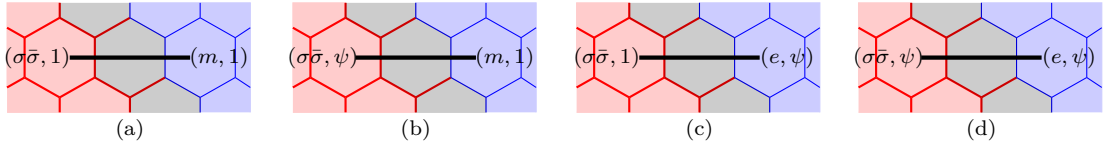
\centering
\subfloat{\FigureSplittingA}
\subfloat{\FigureSplittingB}
\subfloat{\FigureSplittingC}
\subfloat{\FigureSplittingD}
\caption{The interdomain elementary excitation states with doubled-Ising topological observable $\sigma\overline{\sigma}$.}
\label{fig:splitting}
\end{figure}

The phenomenon of splitting can also be seen spatially in the interdomain elementary excitation states under the projection-state correspondence. The projection from the doubled-Ising elementary excitation state $\ket{\sigma\bar\sigma;1,1}_\text{DI}$ to the toric-code state $\ket{m,1,1}$ corresponds to the allowed pair $\sigma\bar\sigma\dash m$ in the interdomain elementary excitation states 
\begin{equation}
\ket{\sigma\overline{\sigma}\dash m;1,1}, \quad\quad\quad\quad \ket{\sigma\overline{\sigma}\dash m;\psi, 1},
\end{equation}
while $P_\text{eff}\ket{\sigma\bar\sigma;\psi,\psi}_\text{DI} = \ket{e,\psi,\psi}$ corresponds to the allowed pair $\sigma\bar\sigma\dash e$ in
\begin{equation}
\ket{\sigma\overline{\sigma}\dash e;1,\psi}, \quad\quad\quad\quad \ket{\sigma\overline{\sigma}\dash e;\psi,\psi}.
\end{equation}
These four interdomain states all have $J_\text{DI}=\sigma\bar\sigma$, but $J_\text{TC}$ can be $m$ or $e$. See Fig. \ref{fig:splitting}. This phenomenon is the spatial counterpart of the splitting of anyons in anyon condensation. Namely, an anyon $\sigma\bar\sigma$ in the doubled Ising domain may hop into the toric code domain by crossing the gapped domain wall and become either an anyon $e$ or $m$.

With our model, splitting can also be understood dynamically as follows. The states $\ket{\sigma\bar\sigma; p, q}_\text{DI}$ with $p$, $q\in \{1, \psi\}$ in Fig. \ref{fig:sigmasigmabar} are indistinguishable in the doubled Ising phase, as they can be transformed into each other by the local operators $B^{1\sigma\psi\sigma}$ and $B^{\psi\sigma 1\sigma}$ \eqref{eq:localoperator:di}. The local operators $B_P^{p\sigma q\sigma}$ however do not commute with the condensation term $\Delta H$ \eqref{eq:Hamiltonian:delta} in Hamiltonian \eqref{eq:Hamiltonian:ours}. After $\psi\bar\psi$ condensation, the charges $1$ and $\psi$ can no longer transform into each other by local operators, and are thus associated with individual topological observables $m$ and $e$ respectively. An infinite energy barrier $\Lambda\to \infty$ prevents the toric-code states $\ket{m;1,1}$ and $\ket{e;\psi,\psi}$ from transforming into each other.

% End Splitting

\subsection{Identification}\label{sec:correspondence:identification}

Seen in Table \ref{table:corrspondence:tc}, $\ket{1\bar 1;1,1}_\text{DI}$ and $\ket{\psi\bar\psi;1,1}_\text{DI}$ in the parent phase are both projected to $\ket{1;1,1}$ in the toric code domain of our model, while $\ket{\psi\bar 1;\psi,\psi}_\text{DI}$ and $\ket{1\bar\psi;\psi,\psi}_\text{DI}$ are both projected to $\ket{\epsilon;\psi,\psi}$. This phenomenon is called \textit{identification} in anyon condensation.

The projections from elementary excitation states $\ket{1\bar 1;1,1}_\text{DI}$ and $\ket{\psi\bar\psi;1,1}_\text{DI}$ in the parent phase to $\ket{1;1,1}$ in our model individually correspond to the interdomain elementary excitation states
\begin{equation}
\ket{1\bar 1-1;1,1},\quad\quad\quad\quad \ket{\psi\bar \psi-1;1,1},
\end{equation}
which have different doubled-Ising topological obervables $1\bar 1$ and $\psi\bar\psi$ but same toric-code topological obervable $1$. The projections from $\ket{\psi\bar 1;\psi,\psi}_\text{DI}$ and $\ket{1\bar\psi;\psi,\psi}_\text{DI}$ to $\ket{\epsilon;\psi,\psi}$ respectively correspond to the interdomain elementary excitation states
\begin{equation}
\ket{\psi\bar 1-\epsilon;\psi,\psi},\quad\quad\quad\quad \ket{1\bar\psi-\epsilon;\psi,\psi}
\end{equation}
with different doubled-Ising topological observables $\psi\bar 1$ and $1\bar\psi$ but same toric-code topological observable $\epsilon$. It appears that the quasiparticles $(1\bar 1, 1)$ and $(\psi\bar\psi, 1)$ in the doubled Ising domain become the same toric-code quasiparticle $(1,1)$ when hopping into the toric code domain, while $(\psi\bar 1, 1)$ and $(1\bar\psi,\epsilon)$ are identified to be $(\epsilon,\psi)$. This phenomenon is the spatial counterpart of identification in anyon condensation. 

% End Identification

\subsection{Confinements}\label{sec:correspondence:confinement}

Seen in Table \ref{table:corrspondence:tc}, the states $\ket{\sigma\bar 1;\sigma,\sigma}_\text{DI}$, $\ket{\sigma\bar \psi;\sigma,\sigma}_\text{DI}$, $\ket{1\bar\sigma;\sigma,\sigma}_\text{DI}$ and $\ket{\psi\bar\sigma;\sigma,\sigma}_\text{DI}$ in the parent phase are all projected to $0$ in $\Hil_\text{eff}$ of our model via $\psi\bar\psi$ condensation. This phenomenon is called \textit{confinement} in anyon condensation. This is because in the toric code domain, the edges and tails cannot take value $\sigma$ in the states in $\Hil_\text{eff}$.

Correspondingly, there is no interdomain elementary excitation states $\ket{J_\text{DI}\dash J_\text{TC};p,q}$ with $J_\text{DI} = \sigma\bar 1$, $\sigma\bar\psi$, $1\bar\sigma$ or $\psi\bar\sigma$, as the quasiparticles $(\sigma\bar 1,\sigma)$, $(\sigma\bar\psi,\sigma)$, $(1\bar\sigma,\sigma)$ and $(\psi\bar\sigma,\sigma)$ in the doubled Ising domain cannot hop into the toric code domain unless overcoming infinite energy barriers $\Lambda\to\infty$. This phenomenon is the spatial counterpart of confinement in anyon condensation.

In anyon condensation, the doubled-Ising anyons $\sigma\bar 1$ and $\sigma\bar\psi$ in the doubled Ising phase become the same quasiparticle $\chi$ in the intermediate phase, and $1\bar\sigma$ and $\psi\bar\sigma$ become $\bar\chi$; however, $\chi$ and $\bar\chi$ in the intermediate phase are confined in the $\Z_2$ toric code phase because of their nontrivial braiding with the new vacuum in the intermediate phase. Now that the gapped domain wall is the spatial counterpart of the intermediate phase, we can see that domainwall quasiparticles $\chi$ and $\bar\chi$ also have nontrivial braiding with the trivial quasiparticle $1$ in the gapped domain wall (to be defined in Eq. \eqref{eq:Smatrix:braiding}). 

% End Confinement

% End Correspondence

\section{The bases of the ground states on the torus}\label{sec:basis}

\begin{figure}\centering
\includegraphics[width=4.5cm]{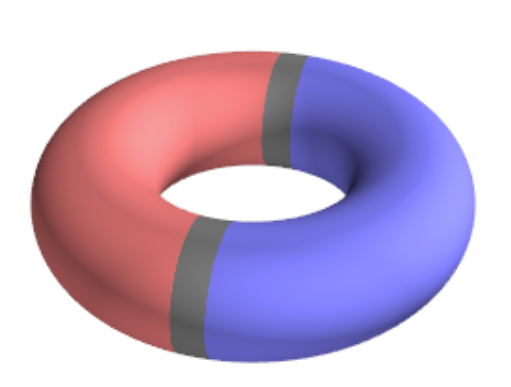}
\caption{Our model on a torus: Two gapped domain walls (gray) separate the doubled Ising domain (red) and $\Z_2$ toric code domain (blue).}
\label{fig:torus}
\end{figure}

The defining properties of a topological phase are usually obtained from the ground states of the topological phase on the torus\cite{Kitaev2003a, Levin2004, Hu2012a}. For example, on the torus, a topological phase has a ground-state degeneracy, which is a topological quantum number of the topological phase. For instances, on the torus, the doubled Ising phase has $\text{GSD}=9$, while the $\Z_2$ toric code phase has $\text{GSD}=4$. In this section, we shall find two distinct and typical ground-state bases of our model on the torus (see Fig. \ref{fig:torus}), using noncontractible loop operators to be constructed shortly. These two ground-state bases will lead us to the characteristic properties of our model, as to be shown in Sections \ref{sec:basis:gsd}, \ref{sec:stmatrix:s} and \ref{sec:stmatrix:t}.

% End Gerneral Discussion

\subsection{The domainwall basis of the ground-state subspace}\label{sec:basis:dw}

\begin{figure}\centering
\subfloat{\raisebox{-45pt}{\begin{overpic}[width=5cm]{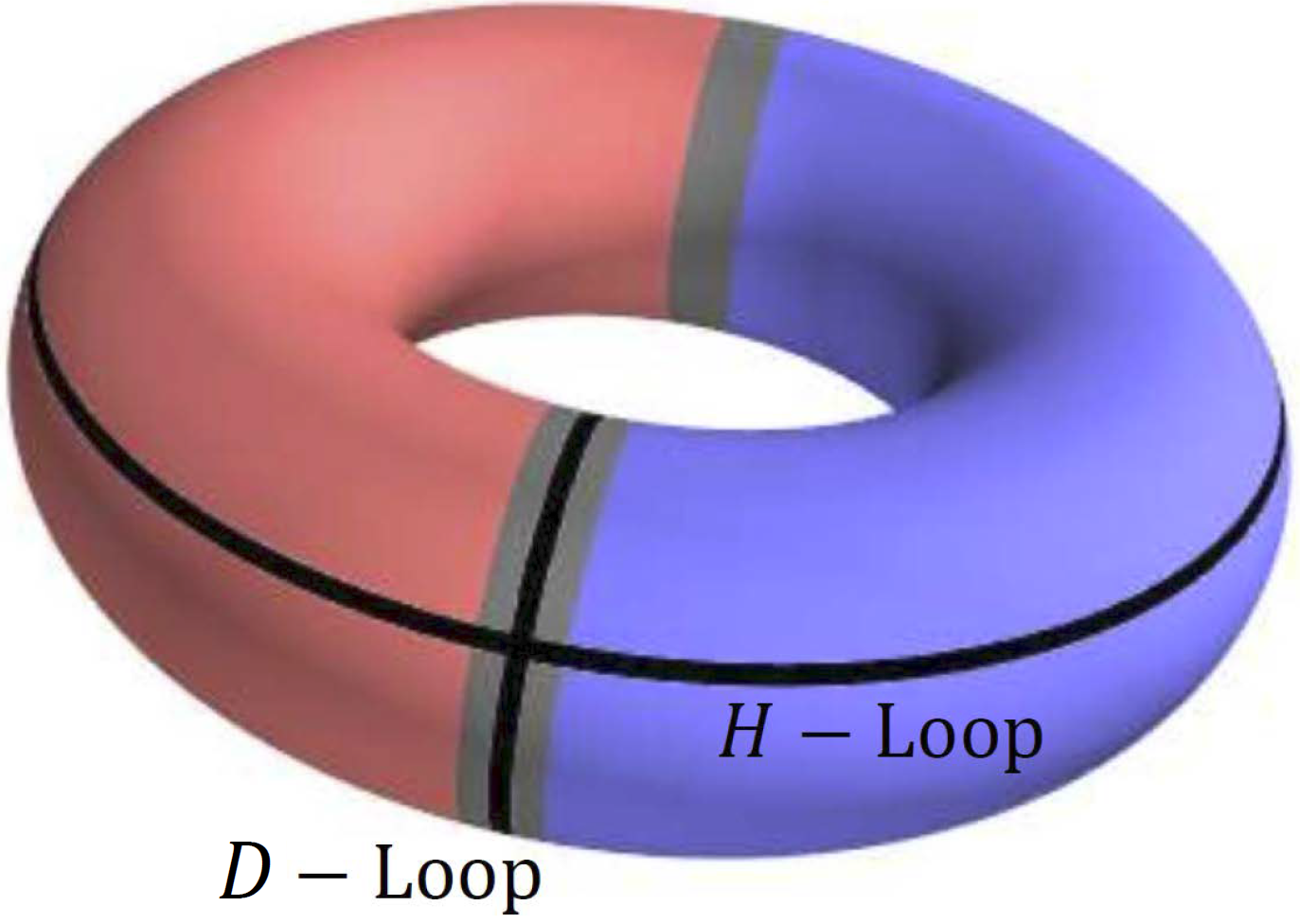}
\put(45,-6){\scriptsize {(a)}}
\end{overpic}}\label{fig:loop:ThreeD}}
\hspace{60pt}
\subfloat{\FigureLoop\label{fig:loop:TwoD}}
\caption{(a) Gapped domain walls and noncontractible loops on the torus. (b) The corresponding lattice picture of (a).}
\label{fig:loop}
\end{figure}

Sewing the two ends of a ribbon operator results in a loop operator\cite{Kitaev2003a, Levin2004}. If the loop path is noncontractible, we have a \textit{noncontractible loop operator}. Loop operators preserve the ground-state subspace because no anyons are created. On the torus, there are two homotopic classes of noncontractible loops: $V$ loop along the gapped domain wall, and $H$ loop across the gapped domain wall. Here $V$ stands for ``vertical'' and $H$ ``horizontal''. See Fig. \ref{fig:loop}.  

There are $6$ loop operators $W_H^{J_\text{DI}\dash J_\text{TC}}$ along $H$-loop, labeled by the interdomain topological observable pairs $J_\text{DI}\dash J_\text{TC}$:
\begin{equation}
W_H^{1\bar{1}\dash 1} = I,\quad\hs\hs W_H^{\psi\bar\psi\dash 1},\quad\hs\hs W_H^{\psi\bar{1}\dash \epsilon},\quad\hs\hs W_H^{1\bar{\psi}\dash \epsilon},\quad\hs\hs W_H^{\sigma\bar{\sigma}\dash m},\quad\hs\hs W_H^{\sigma\bar{\sigma}\dash e}.
\label{eq:loop:h}
\end{equation}
Similarly, there are $6$ noncontratible loop operators along the $V$-loop, labled by $J_\text{DW}$:
\begin{equation}
W_V^{1} = I,\quad\quad W_V^{{\epsilon}},\quad\quad W_V^{{m}},\quad\quad W_V^{{e}},\quad\quad W_V^{\bar{\chi}},\quad\quad W_V^{\chi}.
\label{eq:loop:v}
\end{equation}
All these operators are linearly independent. They generate an algebra denoted by $\mathcal{A}$. See Appendix \ref{appendix:loop:algebra}.

The algebra $\mathcal{A}$ generates the entire ground-state subspace $\mathcal{H}_0$ of our model given any ground state:
\begin{equation}
\Hil_0 = \mathcal{A}\ket\Phi,\quad\forall\ket\Phi\in\Hil_0\backslash\{0\}.
\label{eq:complete}
\end{equation}
We leave the full proof of Eq. \eqref{eq:complete} in Appendix \ref{appendix:loop:complete} but sketch the proof as follows. Since $P_\text{eff}$ commutes with the loop operators of the parent phase in $\Hil_\text{eff}$:
\begin{equation}
P_\text{eff}\left[W_H^{J_\text{DI}}, P_\text{eff}\right] = P_\text{eff}\left[W_V^{J_\text{DI}}, P_\text{eff}\right] = 0,
\end{equation}
it projects the doubled-Ising loop operators $W_H^{J_\text{DI}}$ and $W_V^{J_\text{DI}}$ to $W_H^{J_\text{DI}\dash J_\text{TC}}$ and $W_V^{J_\text{DW}}$ of our model, and projects the ground-state subspace $\Hil_0^\text{DI}$ of the parent phase to $\Hil_0$ of our model. Now that the doubled-Ising loop operators $W_H^{J_\text{DI}}$ and $W_V^{J_\text{DI}}$ can generate $\Hil_0^\text{DI}$ given any ground state of the parent phase, $\mathcal{A}$ can generate $\Hil_0$ of our model given any ground state of our model.

We shall construct a ground-state basis using $W^{J_\text{DW}}_V$ as follows. There exists a unique ground state $\ket{\Phi}_V\in\mathcal{H}_0$, such that
\begin{equation}
W_H^{J_\text{DI}\dash J_\text{TC}}\ket{\Phi}_V = d_{J_\text{DI}\dash J_\text{TC}}\ket{\Phi}_V
\end{equation}
for all operators $W_H^{J_\text{DI}\dash J_\text{TC}}$, which generate a largest commutative subalgebra of $\mathcal{A}$. Here $d_{J_\text{DI}\dash J_\text{TC}} = 1$ for all pairs $J_\text{DI}\dash J_\text{TC}$ are the only positive eigenvalues of $W_H^{J_\text{DI}\dash J_\text{TC}}$. This common eigenstate can be obtained up to factors by 
\begin{equation}
\ket{\Phi}_V = P_H\ket{\varphi}
\end{equation}
given arbitrary $\ket{\varphi}\in \Hil_\text{eff}$, where
\begin{align}
P_H = &\frac{I + W_H^{\psi\bar\psi\dash 1}}{2}\hs \frac{I + W_H^{\psi\bar{1}\dash \epsilon}}{2}\hs \frac{I + W_H^{1\bar{\psi}\dash \epsilon}}{2}\hs \frac{I + W_H^{\psi\bar\psi\dash 1} + 2W_H^{\sigma\bar{\sigma}\dash m}}{4}\hs \times\nonumber\\
&\frac{I + W_H^{\psi\bar\psi-1} + 2W_H^{\sigma\bar{\sigma}\dash e}}{4}\hs P_0.
\end{align}
Here, 
\begin{equation}
P_0 = \prod_{P\in\text{DI}}B_P^\text{DI}\prod_{P\in\text{DW}}B_P^\text{DW}\prod_{P\in\text{TC}}B_P^\text{TC}\hs\prod_{V}Q_V,
\end{equation}
such that $P_0\Hil_\text{eff}=\Hil_0$, where $B_P^\text{DI}$, $B_P^\text{DW}$, $B_P^\text{TC}$ and $Q_V$ are the plaquette operators and vertex operators in the Hamiltonian $H_\text{eff}$ \eqref{eq:Hamiltonian:eff} of our model.

Since $\ket{\Phi}_V$ is the common eigenstate of all $W_H^{J_\text{DI}\dash J_\text{TC}}$, according to Eq. \eqref{eq:complete},
\begin{equation}
\mathcal{H}_0 = \text{span}\left\{W_V^{J_\text{DW}}\ket{\Phi}_V\right\}.
\end{equation}
The states $W_V^{J_\text{DW}}\ket{\Phi}_V$ are orthonormal and thus form a basis of $\mathcal{H}_0$. We define
\begin{equation}
\ket{J_\text{DW}}_V := W_V^{J_\text{DW}}\ket{\Phi}_V.
\label{eq:basis:v}
\end{equation}
We call this basis the \emph{domainwall basis}, depicted in Fig. \ref{fig:Smatrix:v}. 

% End Domainwall Basis

\subsection{The interdomain basis of the ground-state subspace}\label{sec:basis:inter}

The algebra $\mathcal{A}$ has more than one largest commutative subalgebra. The $6$ $V$-loop operators $W_V^{J_\text{DW}}$ also generate a largest commutative subalgebra of $\mathcal{A}$ and determine another unique ground state $\ket{\Phi}_H$, such that
\begin{equation}
W_V^{J_\text{DW}}\ket{\Phi}_H = d_{J_\text{DW}}\ket{\Phi}_H,
\end{equation}
where $d_1 = d_\epsilon = d_m = d_e = 1$, $d_\chi = d_{\bar\chi} = \sqrt{2}$. This common eigenstate can be obtained up to factors as
\begin{equation}
\ket{\Phi}_H = P_V\ket\varphi,\quad\quad \forall\ket\varphi\in\Hil_\text{eff},
\end{equation}
where
\begin{equation}
P_V = \frac{I + W_V^{{\epsilon}}}{2}\ \frac{I + W_V^{{m}}}{2}\ \frac{I + W_V^{e}}{2}\ \frac{I + W_V^{{\epsilon}} + \sqrt{2}W_V^{\bar{\chi}}}{4}\ \frac{I + W_V^{{\epsilon}} + \sqrt{2}W_V^{\chi}}{4}\ P_0.
\end{equation}
Hence, we obtain what we call the \textit{interdomain basis} of $\Hil_0$:
\begin{equation}
\ket{J_\text{DI}-J_\text{TC}}_H := W_H^{J_\text{DI}-J_\text{TC}}\ket{\Phi}_H,
\label{eq:basis:h}
\end{equation}
as depicted in Fig. \ref{fig:Smatrix:h}.

% End Interdomain Basis

\subsection{Ground-state degeneracy on the torus}\label{sec:basis:gsd}

According to the domainwall basis $\ket{J_\text{DW}}_V $ \eqref{eq:basis:v} or interdomain basis $\ket{J_\text{DI}\dash J_\text{TC}}_H$ \eqref{eq:basis:h} of the ground-state subspace on the torus, our model of the doubled Ising and $\Z_2$ toric code phases separated by two gapped domain walls on the torus has
\begin{equation}
\text{GSD}_\text{torus} = 6.
\label{eq:gsd}
\end{equation}
This GSD agrees with the number of the domainwall quasiparticle species $J_\text{DW}$, as well as the number of interdomain topological observable pairs $J_\text{DI}\dash J_\text{TC}$. This is a generalization of the correspondence between the GSD of a topological phase on the torus and the number of anyon species of this topological phase. We can simply replace the input data of our model with that of any other parent and child phases: for any two domain-wall-separated topological phases related by anyon condensation, the following correspondence holds.
\begin{equation}
\begin{tikzpicture}[baseline={([yshift=-.8ex]current bounding box.center)},scale=1]
    \draw [] (-1.2, 0.8) -- (-2.4, -0.8);
    \draw [] (-1.15, 0.75) -- (-2.35, -0.85);
    \draw [] (1.2, 0.8) -- (2.4, -0.8);
    \draw [] (1.15, 0.75) -- (2.35, -0.85);
    \draw [] (-0.6, -1.22) -- (0.6, -1.22);
    \draw [] (-0.6, -1.15) -- (0.6, -1.15);
    \node [centered, rotate=0.] at (0,1.1) {\scriptsize {GSD on the torus}};
    \node [centered, rotate=0.] at (-2.4,-1) {\scriptsize {The number of domain}};
    \node [centered, rotate=0.] at (-2.4,-1.3) {\scriptsize {wall excitation states}};
    \node [centered, rotate=0.] at (2.6,-1) {\scriptsize {The number of inter-domain }};
    \node [centered, rotate=0.] at (2.6,-1.3) {\scriptsize {topological observable pairs}};
\end{tikzpicture}
\end{equation}

Note that the GSD in Eq. \eqref{eq:gsd} has also been obtained before by algebraic methods\cite{Lan2014}. 

% End GSD

% End Ground States

\section{The \texorpdfstring{$S$}{Lg} and \texorpdfstring{$T$}{Lg} matrices}\label{sec:stmatrix}

Besides the ground-state degeneracy, another fingerprint of the topological phase consists of the $S$ and $T$ matrices, which generate the basis transformations in the ground-state Hilbert space $\Hil_0$ on the torus. We shall construct the $S$ and $T$ matrices of our model and show their physical significance. 

% End General Discussion

\subsection{The \texorpdfstring{$S$}{Lg} matrix on the torus}\label{sec:stmatrix:s}

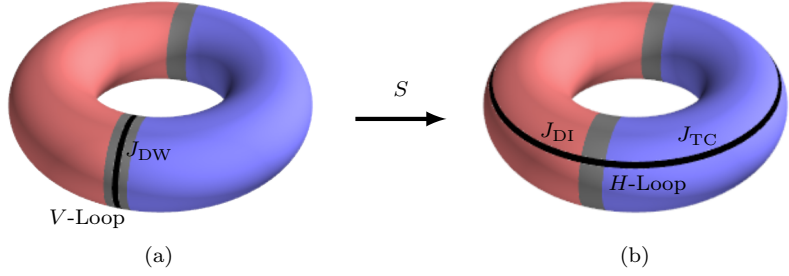
\begin{figure}\centering
\subfloat{\raisebox{10pt}{\begin{overpic}[width=5cm]{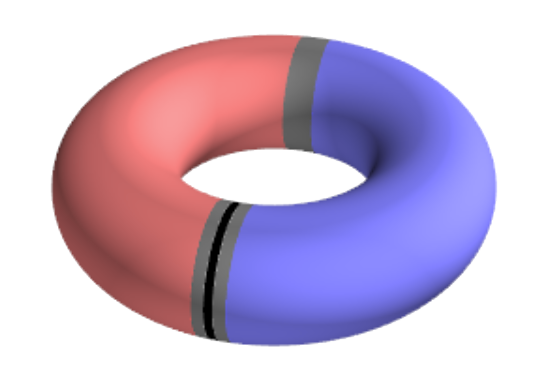}
    \put(20,4){\scriptsize {$V$-Loop}}
    \put(40,22){\scriptsize {$J_\text{DW}$}}
    \put(45,-6){\scriptsize {(a)}}
\end{overpic}}\label{fig:Smatrix:v}}
\subfloat{\raisebox{60pt}{\begin{tikzpicture}[baseline={([yshift=-.8ex]current bounding box.center)},scale=1]
    \draw [ultra thick, ->] (4.2,9.6) -- (5.4,9.6);
    \node [centered, rotate=0.] at (4.8,10.) {\scriptsize $S$};
\end{tikzpicture}}\label{fig:Smatrix:h}}
\subfloat{\raisebox{10pt}{\begin{overpic}[width=5cm]{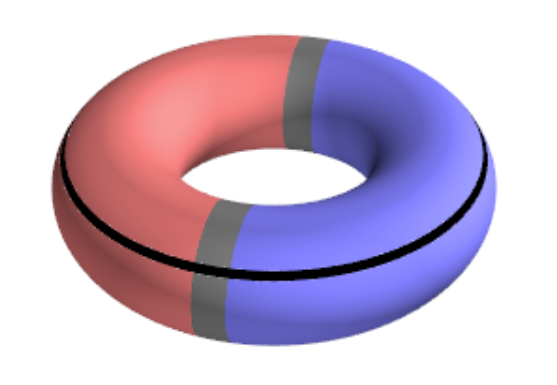}
    \put(42,13){\scriptsize {$H$-Loop}}
    \put(60,26){\scriptsize {$J_\text{TC}$}}
    \put(24,27){\scriptsize {$J_\text{DI}$}}
    \put(45,-6){\scriptsize {(b)}}
\end{overpic}}}
\caption{The $S$ matrix: the basis transformation from (a) the domainwall basis $\{\left|J_\text{DW}\right\rangle_V\}$ to (b) the interdomain basis $\{\left|J_\text{DI}-J_\text{TC}\right\rangle_H\}$.}
\label{fig:Smatrix}
\end{figure}
    
We define the $S$ matrix as the basis transformation (see Fig. \ref{fig:Smatrix}):
\begin{equation}
S_{J_\text{DW},J_\text{DI}-J_\text{TC}} :=\, _V\hspace{-3pt}\left\langle J_\text{DW}\middle|J_\text{DI}-J_\text{TC}\right\rangle_H,
\label{eq:Smatrix:basis}
\end{equation}
which up to phase factors reads

\begin{align}
S = \frac{\sqrt{2}}{4}\quad
\begin{pNiceMatrix}[first-row,first-col]
& 1\bar{1}-1 & \psi\bar\psi-1 & \psi\bar{1}-\epsilon & 1\bar{\psi}-\epsilon & \sigma\bar{\sigma}-m & \sigma\bar{\sigma}-e\\
1 & 1 & 1 & 1 & 1 & 1 & 1\\
{\epsilon} & 1 & 1 & 1 & 1 & -1 & -1\\
{m} & 1 & 1 & -1 & -1 & 1 & -1\\
{e} & 1 & 1 & -1 & -1 & -1 & 1\\
\bar{\chi} & \sqrt{2} & -\sqrt{2} & \sqrt{2} & -\sqrt{2} & 0 & 0\\
\chi & \sqrt{2} & -\sqrt{2} & -\sqrt{2} & \sqrt{2} & 0 & 0
\end{pNiceMatrix}.
\label{eq:Smatrix:value}
\end{align}

The Levin-Wen model of a single topological phase on the torus is invariant under rotations generated by a $\tfrac{\pi}{2}$ rotation of the lattice\cite{Zhang2012}; the $S$ matrix of the model represents the $\frac{\pi}{2}$ rotation and is thus symmetric and unitary. Nevertheless, our model on the torus does not have this rotation invariance due to the gapped domain walls, so the $S$ matrix \eqref{eq:Smatrix:basis} has nothing to do with rotations. Since the domainwall basis and interdomain basis are labeled by different sets of quasiparticle species, our $S$ matrix is neither symmetric nor unitary.

The $S$ matrix of a single topological phase not only transforms the ground-state bases on the torus but also characterizes the mutual statistics of the anyons in the topological phase. This feature of the $S$ matrix is generalized in our model. That is, the matrix elements of our $S$ matrix \eqref{eq:Smatrix:value} can be understood in the following sense of braiding
\begin{equation}
    \EqBraidingLeft = \hs\frac{S_{J_\text{DW}, J_\text{DI}-J_\text{TC}}}{S_{1, 1\bar{1}-1}} \EqBraidingRight .
\end{equation}

Note that in Eq. \eqref{eq:Smatrix:value}, domainwall quasiparticles $\chi$ and $\bar\chi$ have nontrivial mutual statistics with the trivial domainwall quasiparticle $1$:
\begin{equation}
    \frac{S_{\chi, \psi\bar\psi-1}}{S_{1,1\bar 1-1}} = \frac{S_{\bar\chi, \psi\bar\psi-1}}{S_{1,1\bar 1-1}} = -\sqrt{2}.
\label{eq:Smatrix:braiding}
\end{equation}
See Section \ref{sec:correspondence:confinement}.

% End S Matrix

\subsection{The \texorpdfstring{$T$}{Lg} matrix on the torus}\label{sec:stmatrix:t}

Although our model on the torus is not invariant under the $\tfrac{\pi}{2}$ rotation of the lattice, it is still invariant under the shear, i.e., the $\mathcal{T}$ transformation, of the lattice along the vertical direction. 

\begin{figure}\centering
\subfloat{\FigureTmatrixGlobal\label{fig:Tmatrix:global}}\hspace{30pt}
\subfloat{\FigureTmatrixOrigin\label{fig:Tmatrix:origin}}\hspace{3pt}
\subfloat{\raisebox{20pt}{\begin{tikzpicture}[baseline={([yshift=-.8ex]current bounding box.center)},scale=1]
    \draw [ultra thick, ->] (0,0) -- (1,0);
    \node [centered, rotate=0.] at (0.4,0.3) {\scriptsize $\mathcal{T}$};
\end{tikzpicture}}\label{fig:Tmatrix:shear}}\hspace{3pt}
\subfloat{\FigureTmatrixShear\label{fig:Tmatrix:final}}\hspace{3pt}
\subfloat{\raisebox{12pt}{\begin{tikzpicture}[baseline={([yshift=-.8ex]current bounding box.center)},scale=1]
    \draw [ultra thick] (0,0) -- (1,0);
    \draw [ultra thick] (0,0.2) -- (1,0.2);
    \draw [ultra thick] (0,0.4) -- (1,0.4);
\end{tikzpicture}}}\hspace{3pt}
\subfloat{\FigureTmatrixFinal}
\caption{The $\mathcal{T}$ transformation and change of perspectives. (a) The original lattice on the torus. Each oval encircles the vertices to be exchanged under the $\mathcal{T}$ transformation. (b) Two horizontally adjacent plaquettes $P_1$ and $P_2$. (c) The plaquettes $P_1$ and $P_2$ after the $\mathcal{T}$ transformation. (d) is (c) in a more convenient perspective.}
\label{fig:Tmatrix}
\end{figure}
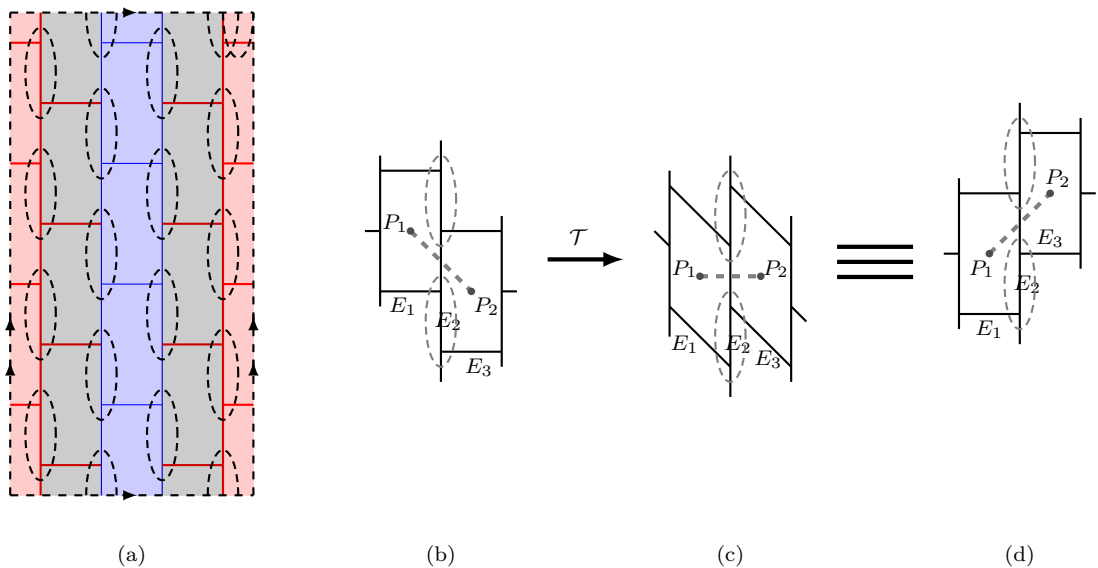

The $\mathcal{T}$ transformation exchanges the positions of two vertically neighboring vertices in the ovals in Fig. \ref{fig:Tmatrix}. After the $\mathcal{T}$ transformation, for any two horizontally adjacent plaquettes in the original lattice, the one on the right is shifted by one plaquette upward relative to the one on the left. See for example the plaquettes $P_1$ and $P_2$ in Fig. \ref{fig:Tmatrix:origin}. 

\begin{figure}\centering
\subfloat{\FigureTmatrixLatticeOrigin\label{fig:Tmatrix:lattice:origin}}\hspace{20pt}
\subfloat{\raisebox{30pt}{\begin{tikzpicture}[baseline={([yshift=-.8ex]current bounding box.center)},scale=1]\draw [ultra thick, ->] (0,0) -- (2,0);\node [centered, rotate=0.] at (1.,0.6) {\scriptsize $\mathcal{T}$};\end{tikzpicture}}\label{fig:Tmatrix:lattice:final}}\hspace{20pt}
\subfloat{\FigureTmatrixLatticeFinal}
\caption{The $\mathcal{T}$ transformation on the entire torus. The number of columns and the number of plaquettes in each column are both $4$. (a) The original lattice. (b) The lattice after the $\mathcal{T}$ transformation and the change of perspective. The plaquettes in deeper colors illustrate how the plaquettes shift under the $\mathcal{T}$ transformation.}
\label{fig:Tmatrix:lattice}
\end{figure}
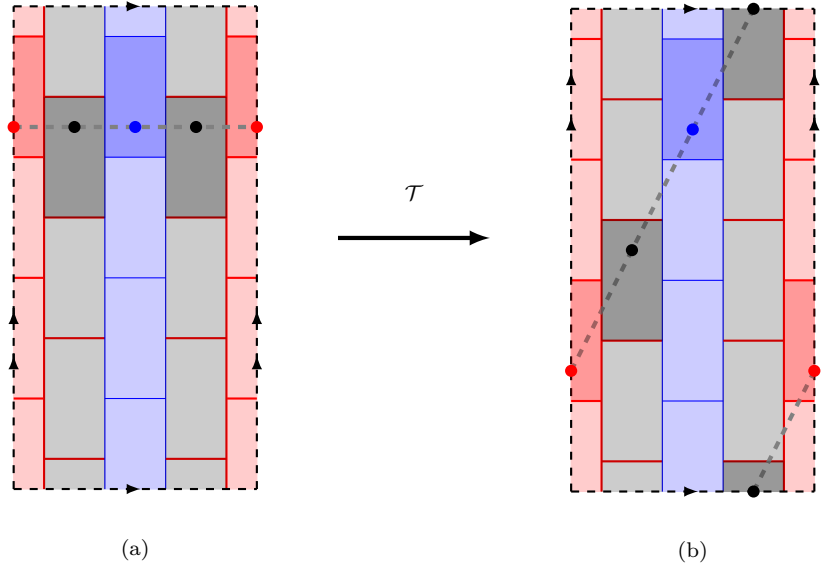

Figure \ref{fig:Tmatrix:lattice} shows how the $\mathcal{T}$ transformation acts on the entire lattice. In the lattice, the number of columns is equal to the number of plaquettes in each column (e.g., the number is $4$ for the lattice in Fig. \ref{fig:Tmatrix:lattice}), so the configuration of the lattice on the torus is invariant and thus the Hilbert space of our model is unchanged under the $\mathcal{T}$ transformation. The $\mathcal{T}$ transformation can be represented in this invariant Hilbert space. Note that any loop remains a loop under the $\mathcal{T}$ transformation; hence, the $\mathcal{T}$ transformation preserves the ground-state Hilbert space $\Hil_0$.

To see how the $\mathcal{T}$ transformation acts on a basis state of the Hilbert space, we can zoom in to see how $\mathcal{T}$ acts in the vicinity of a dashed oval:
\begin{equation}
    \EqTmatrixLeft\quad\Longrightarrow\quad \sqrt{d_{a_1}d_{a_1'}}\hs G^{a_0e_1a_1}_{a_2e_2a_1'}\EqTmatrixRight.
\end{equation}
We define the $T$ matrix as a representation of the $\mathcal{T}$ transformation over the interdomain basis:
\begin{equation}
T_{J_\text{DW}-J_\text{TC}, K_\text{DW}-K_\text{TC}} :=\hs _H\hspace{-3pt}\left\langle J_\text{DI}-J_\text{TC}\middle|\hs\mathcal{T}\hs\middle|K_\text{DW}-K_\text{TC}\right\rangle_H,
\label{eq:Tmatrix:basis}
\end{equation}

The matrix $T$ is diagonal and reads
\begin{equation}
\begin{array}{c|ccccccccccc}
    J_{\text{DI}}-J_{\text{TC}} & 1\bar{1}-1 & \psi\bar\psi-1 & \psi\bar{1}-\epsilon & 1\bar{\psi}-\epsilon & \sigma\bar{\sigma}-m & \sigma\bar{\sigma}-e\\ \hline
    T_{J_\text{DI}-J_\text{TC},J_\text{DI}-J_\text{TC}} & 1 & 1 & -1 & -1 & 1 & 1
\end{array}
\end{equation}
The diagonal elements $T_{J_\text{DI}-J_\text{TC}, J_\text{DI}-J_\text{TC}}$ of the $T$ matrix on the torus are also the topological spins $\theta_{J_\text{DI}} = \theta_{J_\text{TC}} = T_{J_\text{DI}-J_\text{TC}, J_\text{DI}-J_\text{TC}}$ of the anyons $J_\text{DI}$ and $J_\text{TC}$:
\begin{equation}
    \EqTwistLeft =\hs T_{J_\text{DI}-J_\text{TC}, J_\text{DI}-J_\text{TC}} \EqTwistRight .
\end{equation}

The $S$ matrix \eqref{eq:Smatrix:basis} and $T$ matrix \eqref{eq:Tmatrix:basis} generate all possible basis transformations of the ground states of our model on the torus.

% End T Matrix

% End ST Matrix

\section{Conclusion}\label{sec:conclusion}

In this paper, we construct an exactly solvable lattice Hamiltonian model to investigate the properties of a composite system consisting of multiple topological orders separated by gapped domain walls. We develop a \emph{subsystem condensation technique} to construct a lattice model of two topological phases: Starting with the lattice model describing a single parent phase, we condense certain anyons in a subsystem (half) of the lattice, transforming the topological phase in this subsystem into a child phase, while the original parent phase remains outside this subsystem. This approach yields an exactly solvable lattice model simultaneously describing the parent phase outside the subsystem, the child phase in the subsystem, and the gapped domain wall separating them.

Analyzing the spectrum of the elementary excitation states of our model leads to the following main results.
\begin{enumerate}
\item We find a richer spectrum of our composite system than a single topological order: There are interdomain elementary excitations labeled by a pair of anyons in different domains; There are also elementary excitations with quasiparticles in the gapped domain wall.
\item We explicitly establish the correspondence between the transformations of anyons in anyon condensation with the manifestable interdomain elementary excitation states in our model.
\item Both the set of interdomain elementary excitations and the set of domain-wall quasiparticles respectively correspond to a basis of the ground states of this composite system on a torus, reflecting that the ground-state degeneracy (GSD) of our model on the torus equals the number of quasiparticle species in the gapped domain wall, as well as the number of interdomain elementary excitation species.
\item We construct the $S$ and $T$ matrices that generate the basis transformations of the ground states on the torus. Our $S$ matrix also encodes the braiding between the anyons crossing the gapped domain wall around the quasiparticles in the gapped domain wall, and the $T$ matrix records the topological spins of interdomain excitations.
\end{enumerate}

Our construction methods and conclusions can be directly utilized for any composite system including a parent topological phase and a child phase separated by topological interfaces, inspiring further novel results about composite topological systems consisting of multiple topological orders separated by gapped domain walls.

% End Conclusion

\section*{Acknowledgements}
YW is supported by the General Program of Science and Technology of Shanghai No. 21ZR1406700, and Shanghai Municipal Science and Technology Major Project (Grant No.2019SHZDZX01). YW is grateful for the Hospitality of the Perimeter Insitute during his visit, where the main part of this work is done. YH is supported by Zhejiang Provincial Natural Science Foundation of China (No. LY23A050001).

% End Mainbody

\begin{appendix}

\section{Gauge transformations of the positions of tails\label{appendix:gauge}}

In the lattice of the extended LW model, each tail associated with vertex $V$ is chosen to attach to any one of the three edges incident at $V$. Different choices lead to different lattice configurations and hence different Hilbert spaces of the extended LW model. Nevertheless, since tails are internal degrees of freedom that cannot be probed, the different Hilbert spaces underline the same topological phase. Specifically, these Hilbert spaces are equivalent up to the gauge transformation $\mu$:
\begin{equation}
    \mu_p\EqGaugeLeft = \sum_{m\in L_\text{DI}}\sqrt{d_ld_m}\hs G_{jim}^{kpl}\EqGaugeMu.
\end{equation}

Besides the gauge transformation of the positions of tails, the directions of tails are also defined up to gauge transformations\cite{Hu2018}. For example, the following two states are equivalent up to a gauge transformation:
\begin{equation}
\EqGaugeLeft \quad \begin{tikzpicture}[baseline={([yshift=-.8ex]current bounding box.center)},scale=1]
    \draw [thick] (0.0,0.0) -- (1.0,0.0);
    \draw [thick] (0.0,0.15) -- (1.0,0.15);
    \draw [thick] (0.0,0.3) -- (1.0,0.3);
\end{tikzpicture}\quad \EqGaugeNu.
\end{equation}

% End Gauge

\section{The matrix elements of the doubled-Ising ribbon operators}\label{appendix:ribbon}

In Section \ref{sec:excitation:di}, we have defined the action of shortest ribbon operators $W_E^{J_\text{DI};p,q}$ on the states where all tails take value $1\in L_\text{DI}$:
\begin{equation}
W_E^{J_\text{DI};p,q}\EqDoubledIsingRibbonLeft = \sum_{k\in L_\text{DI}}\sqrt{\frac{d_k}{d_{j_E}}}z_{pqj_E}^{J_\text{DI};k}\EqDoubledIsingRibbonRight,
\end{equation}
where $j_E\in L_\text{DI}$ is the label on edge $E$, and $z_{pqj_E}^{J_\text{DI};k}$ are listed in Appendix \ref{appendix:halfbraiding:di}. 

In this appendix, we define the actions of any ribbon operators on any states in $\Hil$ of the doubled Ising phase.

% End General Discussion

\subsection{Matrix elements of shortest ribbon operators}\label{appendix:ribbon:shortest}

Here, we define the action of the shortest ribbon operator $W_E^{J_\text{DI};p,q}$ on the states, in which there are nontrivial tails attached on edge $E$.

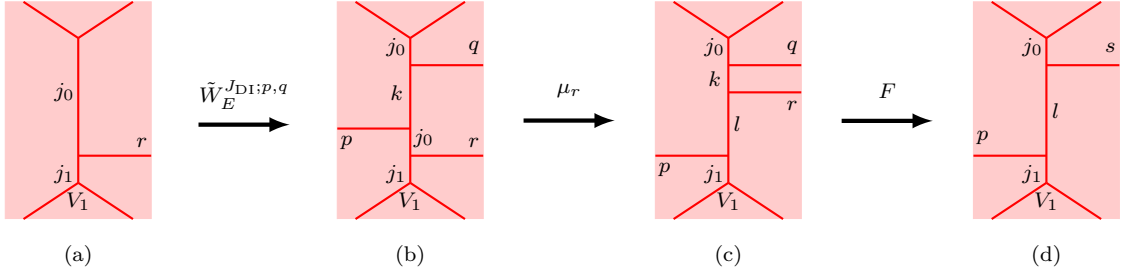
\begin{figure}\centering
\subfloat{\FigureRibbonLowerA}\hspace{5pt}
\subfloat{\raisebox{15pt}{\begin{tikzpicture}[baseline={([yshift=-.8ex]current bounding box.center)},scale=1]
    \draw [ultra thick, ->] (4.2,10.4) -- (5.4,10.4);
    \node [centered, rotate=0.] at (4.8,10.8) {\scriptsize $\tilde{W}_E^{J_\text{DI};p,q}$};
\end{tikzpicture}}}\hspace{5pt}
\subfloat{\FigureRibbonLowerB}\hspace{5pt}
\subfloat{\raisebox{15pt}{\begin{tikzpicture}[baseline={([yshift=-.8ex]current bounding box.center)},scale=1]
    \draw [ultra thick, ->] (4.2,10.) -- (5.4,10.);
    \node [centered, rotate=0.] at (4.8,10.4) {\scriptsize $\mu_r$};
\end{tikzpicture}}}\hspace{5pt}
\subfloat{\FigureRibbonLowerC}\hspace{5pt}
\subfloat{\raisebox{15pt}{\begin{tikzpicture}[baseline={([yshift=-.8ex]current bounding box.center)},scale=1]
    \draw [ultra thick, ->] (4.2,10.) -- (5.4,10.);
    \node [centered, rotate=0.] at (4.8,10.4) {\scriptsize $F$};
\end{tikzpicture}}}\hspace{5pt}
\subfloat{\FigureRibbonLowerD}
\caption{The action of $W_E^{J_\text{DI};p,q}$ on the state where a tail $r$ is associated with $V_1$.}
\label{fig:ribbon:lower}
\end{figure}
    
We start with the simplest case where the tail is associated with the upper vertex $V_2$ of edge $E$ and carries a charge $r$ and points to the right:
\begin{equation}
W_E^{J_\text{DI};p,q}\EqRibbonUpperLeft = \sum_{ks\in L_\text{DI}}\sqrt{d_kd_s}\hs z_{pqj_0}^{J_\text{DI},k}\hs G_{kqs}^{rj_2j_0}\EqRibbonUpperRight.
\label{eq:ribbon:upper}
\end{equation}
This action formally is the composition of two operators $\tilde{W}_E^{J_\text{DI};p,q}$ and $F$: 
\begin{equation}
W_E^{J_\text{DI};p,q}\EqRibbonUpperLeft = F\tilde{W}_E^{J_\text{DI};p,q}\EqRibbonUpperLeft.
\end{equation}
First, the operator $\tilde{W}_{E}^{J_\text{DI};p,q}$ acts as
\begin{equation}
\tilde{W}_{E}^{J_\text{DI};p,q}\EqRibbonUpperA = \sum_{k\in L_\text{DI}}\sqrt{\frac{d_{k}}{d_{j_0}}}\hs z_{pqj_0}^{J_\text{DI};k}\EqRibbonUpperB .
\end{equation}
Now there are two tails ($q$ and $r$) associated with vertex $V_2$ on edge $E$, which can then be fused by operator $F$:
\begin{equation}
F\EqRibbonUpperB := \sum_{s\in L_\text{DI}}\sqrt{d_{j_0}d_{s}}\hs G^{rj_2j_0}_{kqs} \EqRibbonUpperC .
\end{equation}
The result is Eq. \eqref{eq:ribbon:upper}.

Similarly, when edge $E$ has one tail ($r$) associated with the lower vertex $V_1$ and pointing right, $W_E^{J_\text{DI};p,q}$ acts on the state as
\begin{equation}
W_E^{J_\text{DI};p,q}\EqRibbonLowerLeft = \sum_{kls\in L_\text{DI}}d_{k}\sqrt{d_ld_s}\hs z_{pqj_0}^{J_\text{DI},k}G^{kpj_0}_{j_1rl}G^{qj_0k}_{lrs}\EqRibbonLowerRight,
\end{equation}
which is also a composition of the actions of two operators:
\begin{equation}
W_E^{J_\text{DI};p,q}\EqRibbonLowerLeft = F\mu_r\tilde{W}_E^{J_\text{DI};p,q}\EqRibbonLowerLeft .
\end{equation}
See Fig. \ref{fig:ribbon:lower}.

All other matrix elements of ribbon operators $W_E^{J_\text{DI};p,q}$ can be obtained likewise.

% End Matrix Elements

\subsection{Concatenating shorter ribbon operators to a longer ribbon operator}\label{appendix:ribbon:longer}

\begin{figure}\centering
\subfloat{\FigureConExplicitA}
\subfloat{\begin{tikzpicture}[baseline={([yshift=-.8ex]current bounding box.center)},scale=1]
    \draw [ultra thick, ->] (3.6,9.6) -- (6.,9.6);
    \node [centered, rotate=0.] at (4.8,10.) {\scriptsize $W_{E_4}^{J_\text{DI};q,r}W_{E_1}^{J_\text{DI};p,q}$};
\end{tikzpicture}}
\subfloat{\FigureConExplicitB}
\subfloat{\begin{tikzpicture}[baseline={([yshift=-.8ex]current bounding box.center)},scale=1]
    \draw [ultra thick, ->] (3.6,9.6) -- (6.,9.6);
    \node [centered, rotate=0.] at (4.8,10.) {\scriptsize $\left(\mu_q^*\right)^4$};
\end{tikzpicture}}
\subfloat{\FigureConExplicitC}\\
\subfloat{\begin{tikzpicture}[baseline={([yshift=-.8ex]current bounding box.center)},scale=1]
    \draw [ultra thick, ->] (3.6,9.6) -- (6.,9.6);
    \node [centered, rotate=0.] at (4.8,10.) {\scriptsize $F$};
\end{tikzpicture}}
\subfloat{\FigureConExplicitD}
\subfloat{\begin{tikzpicture}[baseline={([yshift=-.8ex]current bounding box.center)},scale=1]
    \draw [ultra thick, ->] (3.6,9.6) -- (6.,9.6);
    \node [centered, rotate=0.] at (4.8,10.) {\scriptsize $B_P^\text{DI}$};
\end{tikzpicture}}
\subfloat{\FigureConExplicitE}
\caption{Concatenating shorter ribbon operators to a longer one.}
\label{fig:con:explicit}
\end{figure}
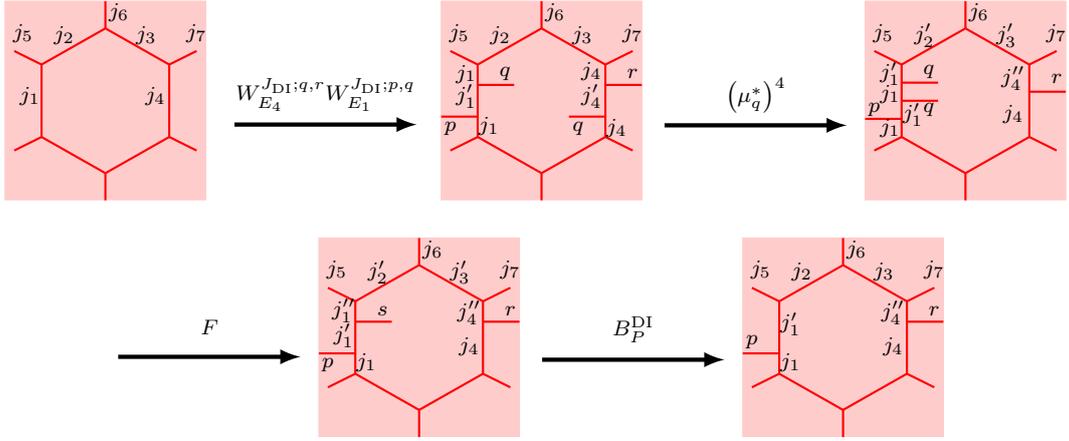

Now we define the ribbon operators along longer paths. Consider ribbon operator $W_L^{J_\text{DI};p,q}$ in Fig. \ref{fig:concatenation:ribbon}, whose path $L$ crosses two edges labeled by $j_1$ and $j_4$ respectively:
\begin{align}
&W_L^{J_\text{DI};p,r}\EqConcatenationLeft = \hs \sum_{qj_1'j_2'j_3'j_4'j_4''\in L_\text{DI}} \overline{z_{pqj_1}^{J_\text{DI};j_1'}}\hs \overline{z_{qrj_4}^{J_\text{DI};j_4'}} \hs\times\nonumber \\
&\sqrt{\frac{d_{j_1}}{d_{j_1'}}}\hs G_{qj_4j_4''}^{rj_4j_4'}\hs G_{qj_4'j_3'}^{j_7j_3j_4}\hs G_{qj_3'j_2'}^{j_6j_2j_3}\hs G_{qj_2'j_1'}^{j_5j_1j_2}\hs \EqConcatenationRight .
\end{align}
The operator $W_L^{J_\text{DI};p,r}$ can be formally written as
\begin{equation}
    W_L^{J_\text{DI};p,r} = B_P^\text{DI}\sum_{q\in L_\text{DI}}F(\mu_q^*)^4W_{E_4}^{J_\text{DI};q,r}W_{E_1}^{J_\text{DI};p,q},
\end{equation}
see Fig. \ref{fig:con:explicit}.

All matrix elements of the doubled-Ising ribbon operators taking any paths can be obtained likewise.

% End Ribbon Operators

\section{Proof of the commutation in \texorpdfstring{$\Hil_\text{eff}$}{Lg} of \texorpdfstring{$P_\text{eff}$}{Lg} and the doubled-Ising ribbon operators}\label{appendix:projection}

\begin{figure}\centering
\subfloat{\FigureSigmaTailA\label{fig:sigmatail:a}}
\hspace{10pt}\subfloat{\raisebox{10pt}{\begin{tikzpicture}[baseline={([yshift=-.8ex]current bounding box.center)},scale=1]
\draw [ultra thick] (0.0,0.0) -- (1.0,0.0);
\draw [ultra thick] (0.0,0.2) -- (1.0,0.2);
\draw [ultra thick] (0.0,0.4) -- (1.0,0.4);
\end{tikzpicture}\label{fig:sigmatail:b}}}\hspace{10pt}
\subfloat{\FigureSigmaTailB\label{fig:sigmatail:c}}
\hspace{100pt}
\subfloat{\FigureSigmaTailC}
\caption{Two different cases where a tail with charge $\sigma$ is attached to a TC edge $E_2$. (a) The tail $\sigma$ is associated with the vertex $V_1$ with an incident DI edge $E_1$. (b) is equivalent to (a) up to a $\mu_\sigma^*$ gauge transformation. (c) The tail $\sigma$ is associated with the vertex $V_2$ with two incident TC edges $E_3$ and $E_4$.}
\label{fig:sigmatail}
\end{figure}
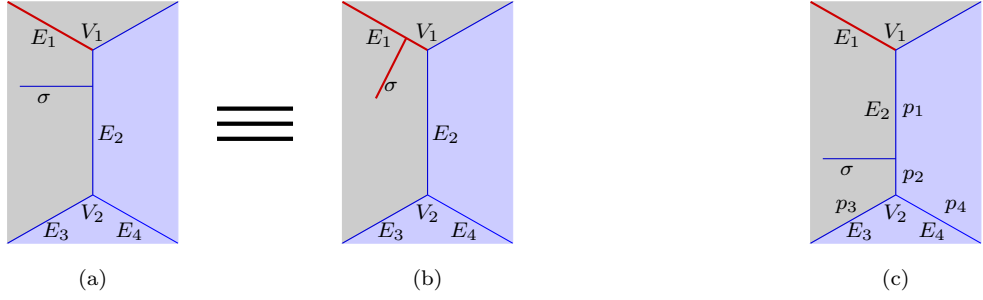

In this section we prove Eq. \eqref{eq:projector:commute}:
\begin{equation}
    P_\text{eff}\left[W_L^{J_\text{DI};p,q},P_\text{eff}\right] = 0.
\end{equation}

Obviously, Eq. \eqref{eq:projector:commute} holds when $J_\text{DI} = 1\bar 1$, $\psi\bar\psi$, $\psi\bar 1$, $1\bar\psi$ and $\sigma\bar\sigma$ because these anyon species all have charges in $\{1,\psi\}$ that are preserved under the projection.

We then consider the ribbon operators $W_L^{J_\text{DI};p,q}$ with $J_\text{DI} = \sigma\bar 1$, $\sigma\bar\psi$, $1\bar\sigma$ and $\psi\bar\sigma$. These operators create quasiparticles with charge $\sigma$ at the ends of path $L$. Note that Eq. \eqref{eq:projector:commute} holds when path $L$ crosses DI edges because $P_\text{eff}$ only acts on the TC edges, we only need to consider the projections of states $\ket\varphi = W_L^{J_\text{DI};p,q}\ket\Phi$ with nontrivial tails $\sigma$ on TC edges. There are two cases of these states, depicted in Fig. \ref{fig:sigmatail:a} and Fig. \ref{fig:sigmatail:c}.

In the first case, the tail $\sigma$ is associated with vertex $V_1$ with an incident DI edge $E_1$. Up to the gauge transformations introduced in Appendix \ref{appendix:gauge}, this state is equivalent to the state with tail $\sigma$ on DI edge $E_1$ (see Fig. \ref{fig:sigmatail:b}) and therefore satisfies Eq. \eqref{eq:projector:commute}.

In the second case, the tail $\sigma$ is associated with vertex $V_2$ with two other incident TC edges $E_3$ and $E_4$. Note that
\begin{equation}
    N_{\sigma\sigma}^1 = N_{\sigma\sigma}^\psi = 1,\quad\quad N_{11}^\sigma = N_{1\psi}^\sigma = N_{\psi 1}^\sigma = N_{\psi\psi}^\sigma = 0,
\end{equation}
one of the labels $p_1$ and $p_2$ on edge $E_2$ must be $\sigma$. If $p_1 = \sigma$, $W_{E_2}^{\psi\bar\psi;1,1}\ket\varphi = -\ket\varphi$; otherwise, if $p_2 = \sigma$, one of the labels $p_3$ and $p_4$ must be $\sigma$. Since associated with vertex $V$ there is at most one nontrivial tail that has been on edge $E_2$, we have $W_{E_3}^{\psi\bar\psi;1,1} = -\ket\varphi$ or $W_{E_4}^{\psi\bar\psi;1,1}\ket\varphi = -\ket\varphi$. Therefore,
\begin{equation}
    P_\text{eff}\ket\varphi = 0,
\end{equation}
which leads to
\begin{equation}
    P_\text{eff}W_L^{J_\text{DI};p,q}P_\text{eff} = P_\text{eff}W_L^{J_\text{DI};p,q} = 0.
\end{equation}

% End Proof

\section{The components of \texorpdfstring{$z$}{Lg} tensors}\label{appendix:halfbraiding}

\subsection{Nonzero components of \texorpdfstring{$z^{J_\text{DI}}$}{Lg} tensors in the doubled Ising domain}\label{appendix:halfbraiding:di}

Equation \eqref{eq:halfbraiding}
\begin{equation}
\frac{\delta_{j,t}N_{rs}^t}{d_t}z_{pqt}^{J_\text{DI};w} = \sum_{ulv\in L_\text{DI}} d_ud_v z_{lqr}^{J_\text{DI};v} z_{pls}^{J_\text{DI};u} G^{rst}_{pwu} G^{srj}_{qwv} G^{sul}_{rvw}
\end{equation}
has $9$ minimal solutions $z^{J_\text{DI}}$, labeled by the $9$ double-Ising anyon species. The nonzero components of these tensors are
\begin{equation}
z _{111}^{1\bar{1},1} = z _{11\psi}^{1\bar{1},\psi} = z _{11\sigma}^{1\bar{1},\sigma} = 1,
\end{equation}
\begin{equation}
z _{111}^{\psi\bar\psi, 1} = z _{11\psi}^{\psi\bar\psi, \psi} = 1,\quad\quad z _{11\sigma}^{\psi\bar\psi, \sigma} = -1,
\label{eq:halfbraiding:psibarpsi}
\end{equation}
\begin{equation}
z _{\psi\psi 1}^{\psi\bar{1},\psi} =  1,\quad\quad z _{\psi\psi \psi}^{\psi\bar{1},1} = -1,\quad\quad z _{\psi\psi \sigma}^{\psi\bar{1}, \sigma} = i,
\end{equation}
\begin{equation}
z _{\psi\psi 1}^{1\bar{\psi},\psi} = 1,\quad\quad z _{\psi\psi \psi}^{1\bar{\psi},1} = -1,\quad\quad z _{\psi\psi \sigma}^{1\bar{\psi},\sigma} = -i,
\end{equation}
\begin{equation}
z _{111}^{\sigma\bar{\sigma},1} = z _{\psi\psi 1}^{\sigma\bar{\sigma},\psi} =  z _{\psi\psi \psi}^{\sigma\bar{\sigma},1} = 1,\quad\quad z _{11\psi}^{\sigma\bar{\sigma},\psi} = -1,\quad\quad z _{1\psi\sigma}^{\sigma\bar{\sigma},\sigma} = z _{\psi 1\sigma}^{\sigma\bar{\sigma},\sigma} = 1,
\label{eq:halfbraiding:sigmasigmabar}
\end{equation}
\begin{equation}
z _{\sigma\sigma 1}^{\sigma\bar{1},\sigma} = 1,\quad\quad z _{\sigma\sigma\psi}^{\sigma\bar{1},\sigma} = i,\quad\quad z _{\sigma\sigma\sigma}^{\sigma\bar{1},1} = e^{\frac{i\pi}{8}},\quad\quad z _{\sigma\sigma\sigma}^{\sigma\bar{1},\psi} = e^{-\frac{3i\pi}{8}},
\end{equation}
\begin{equation}
z _{\sigma\sigma 1}^{\sigma\bar{\psi},\sigma} = 1,\quad\quad\ z _{\sigma\sigma\psi}^{\sigma\bar{\psi},\sigma} = i,\quad\quad z _{\sigma\sigma\sigma}^{\sigma\bar{\psi},1} = e^{-\frac{7i\pi}{8}},\quad\quad z _{\sigma\sigma\sigma}^{\sigma\bar{\psi},\psi} = e^{\frac{5i\pi}{8}},
\end{equation}
\begin{equation}
z _{\sigma\sigma 1}^{1\bar{\sigma},\sigma} = 1,\quad\quad\ z _{\sigma\sigma\psi}^{1\bar{\sigma},\sigma} = -i,\quad\quad z _{\sigma\sigma\sigma}^{1\bar{\sigma},1} = e^{-\frac{i\pi}{8}},\quad\quad z _{\sigma\sigma\sigma}^{1\bar{\sigma},\psi} = e^{\frac{3i\pi}{8}},
\end{equation}
\begin{equation}
z _{\sigma\sigma 1}^{\psi\bar{\sigma},\sigma} = 1,\quad\quad z _{\sigma\sigma\psi}^{\psi\bar{\sigma},\sigma} = -i,\quad\quad z _{\sigma\sigma\sigma}^{\psi\bar{\sigma},1} = e^{\frac{7i\pi}{8}},\quad\quad z _{\sigma\sigma\sigma}^{\psi\bar{\sigma},\psi} = e^{-\frac{5i\pi}{8}}.
\end{equation}

% End DI Z

\subsection{Nonzero components of \texorpdfstring{$z^{J_\text{TC}}$}{Lg} tensors in the toric code domain}\label{appendix:halfbraiding:tc}

The tensors $z^{J_\text{TC}}$, $J_\text{TC}\in\{1,e,m,\epsilon\}$, are
\begin{align}
&z_{pqs}^{1,u} = z_{pqs}^{1\bar 1,u} = z_{pqs}^{\psi\bar \psi,u},\quad\quad z_{pqs}^{\epsilon,u} = z_{pqs}^{\psi\bar 1,u} = z_{pqs}^{1\bar\psi,u},\quad\quad z_{pqs}^{m,u} = \delta_{p,1}z_{pqs}^{\sigma\bar\sigma,u},\nonumber\\ 
& z_{pqs}^{e,u} = \delta_{p,\psi}z_{pqs}^{\sigma\bar\sigma,u},\quad\quad\quad\quad\quad p,q,s,u\in L_\text{TC}.
\end{align}
The nonzero components of $z^{J_\text{TC}}$ tensors are
\begin{equation}
z _{111}^{1,1} = z _{11\psi}^{1,\psi} = 1,
\end{equation}
\begin{equation}
z _{\psi\psi 1}^{e,\psi} = z _{\psi\psi\psi}^{e,1} = 1,
\end{equation}
\begin{equation}
z _{111}^{m,1} = - z _{11\psi}^{m,\psi} = 1,
\end{equation}
\begin{equation}
z _{\psi\psi 1}^{\epsilon,\psi} = - z _{\psi\psi\psi}^{\epsilon,1} = 1.
\end{equation}

These $z^{J_\text{TC}}$ tensors are the four minimal solutions to the equation
\begin{equation}
\frac{\delta_{j,t}N_{rs}^t}{d_t}z_{pqt}^{J_\text{TC};w} = \sum_{ulv\in L_\text{TC}}d_ud_v z_{lqr}^{J_\text{TC};v} z_{pls}^{J_\text{TC};u} G^{rst}_{pwu} G^{srj}_{qwv} G^{sul}_{rvw},
\label{eq:halfbraiding:tc}
\end{equation}
with all indices in $L_\text{TC} = \{1,\psi\}$. 

Note that although the doubled-Ising tensor $z^{\sigma\bar\sigma}$ \eqref{eq:halfbraiding:sigmasigmabar} also solves Eq. \eqref{eq:halfbraiding:tc}, it is not a minimal solution but the sum of two minimal solutions $z^e$ and $z^m$:
\begin{equation}
z_{pqs}^{\sigma\bar\sigma,u} = z_{pqs}^{m,u} + z_{pqs}^{e,u},\quad\quad p,q,r,s\in L_\text{TC}.
\end{equation}

% End TC Z

\subsection{Nonzero components of \texorpdfstring{$z^{J_\text{DW}}$}{Lg} tensors in the gapped domain wall}\label{appendix:halfbraiding:dw}

The tensors $z^{J_\text{DW}}$, $J_\text{DW}\in\{1,e,m,\epsilon,\chi,\bar\chi\}$, are
\begin{align}
&z_{pqs}^{1,u} = z_{pqs}^{1\bar 1,u} = z_{pqs}^{\psi\bar\psi,u},\quad\hs\hs z_{pqs}^{{\epsilon},u} = z_{pqs}^{\psi\bar 1,u} = z_{pqs}^{1\bar\psi,u},\quad\hs\hs z_{pqs}^{{m},u} = \delta_{p,1}z_{pqs}^{\sigma\bar\sigma,u},\quad\hs\hs z_{pqs}^{{e},u} = \delta_{p,\psi}z_{pqs}^{\sigma\bar\sigma,u},\nonumber\\
&z_{pqs}^{\chi,u} = z_{pqs}^{\sigma\bar 11,u} = z_{pqs}^{\sigma\bar\psi,u},\quad\hs\hs z_{pqs}^{\bar\chi,u} = z_{pqs}^{1\bar\sigma,u} = z_{pqs}^{\psi\bar\sigma,u},\quad\quad p,q,u\in L_\text{DI},\quad\quad s\in L_\text{TC},
\end{align}
where the nonzero components are
\begin{equation}
z _{111}^{1,1} = z _{11\psi}^{1,\psi} = 1,
\end{equation}
\begin{equation}
z _{\psi\psi 1}^{{e},\psi} = z _{\psi\psi\psi}^{{e},1} = 1,
\end{equation}
\begin{equation}
z _{111}^{{m},1} = - z _{11\psi}^{{m},\psi} = 1,
\end{equation}
\begin{equation}
z _{\psi\psi 1}^{{\epsilon},\psi} = - z _{\psi\psi\psi}^{{\epsilon},1} = 1,
\end{equation}
\begin{equation}
z _{\sigma\sigma 1}^{\chi,\sigma} = 1,\quad\quad z _{\sigma\sigma\psi}^{\chi,\sigma} = i,
\end{equation}
\begin{equation}
z _{\sigma\sigma 1}^{\bar{\chi},\sigma} = 1,\quad\quad z _{\sigma\sigma\psi}^{\bar{\chi},\sigma} = -i.
\end{equation}

The tensors $z^{J_\text{DW}}$ are the $6$ minimal solutions to the equation
\begin{equation}
    \frac{\delta_{j,t}N_{rs}^t}{d_t}z_{pqt}^{J_\text{DW};w} = \sum_{ulv\in L_\text{DI}} d_ud_v z_{lqr}^{J_\text{DW};v} z_{pls}^{J_\text{DW};u} G^{rst}_{pwu} G^{srj}_{qwv} G^{sul}_{rvw},
\end{equation}
with all indices in $L_\text{DI}$ except that $r, s, t\in L_\text{TC} = \{1,\psi\}$. 

% End DW Z

% End Z Tensor

\section{Measuring elementary excitation states by local operators}\label{appendix:measure}

\subsection{Local operators in the doubled Ising phase}\label{appendix:measure:local}

Since our model stems from the extended LW model describing the doubled Ising phase, we first focus on the local operators in the doubled Ising phase. In the doubled Ising phase, the local operators $B_P^{psqu}$ are defined by
\begin{align}
B_P^{psqu}&\EqLocalOperatorLeft = \delta_{p,p'}\sum_{j_0j_1j_2j_3j_4j_5j_6\in L_\text{DI}}\left(\prod_{n=0}^6\sqrt{d_{i_n}d_{j_n}}\right)\left(G_{i_0i_1j_1}^{sup}\ G_{j_1j_0i_0}^{suq}\right)\times\nonumber\\
&\left(G_{sj_1j_2}^{e_2i_2i_1}\ G_{sj_2j_3}^{e_3i_3i_2}\ G_{sj_3j_4}^{e_4i_4i_3}\ G_{sj_4j_5}^{e_5i_5i_4}\ G_{sj_5j_6}^{e_6i_6i_5}\ G_{sj_6j_0}^{e_0i_0i_6}\right)\hs\EqLocalOperatorRight .
\label{eq:localoperator:di}
\end{align}

The local operators $B_P^\text{psqu}$ preserve the anyon species $J_\text{DI}$ of the elementary excitation states $\ket{J_\text{DI};p,r}_\text{DI}$ but change the charges of the doubled-Ising quasiparticles $(J_\text{DI}, p)$ in plaquettes $P$.
\begin{equation}
B_P^{psqu}\EqLocalActionLeft \hs\propto\hs \delta_{p,p'}\EqLocalActionRight.
\end{equation}

There are in total $12$ local operators $B_P^{psqu}$ acting on plaquette $P$:
\begin{align}
&B_P^{1111},\quad\quad B_P^{1\psi 1\psi},\quad\quad B_P^{1\sigma 1\sigma},\quad\quad B_P^{1\sigma \psi\sigma},\quad\quad B_P^{\psi 1\psi \psi},\quad\quad B_P^{\psi\psi\psi 1},\nonumber\\
&B_P^{\psi\sigma \psi\sigma},\quad\quad B_P^{\psi\sigma 1\sigma},\quad\quad B_P^{\sigma 1\sigma\sigma},\quad\quad B_P^{\sigma \psi\sigma\sigma},\quad\quad B_P^{\sigma\sigma\sigma 1},\quad\quad B_P^{\sigma\sigma\sigma\psi}.
\end{align}

% End DI Local

\subsection{Measurement operators in the doubled Ising phase}\label{appendix:measure:di}

Now we define the measurement operators of the elementary excitation states in the doubled Ising phase via the local operators defined above. Since the doubled-Ising elementary excitation states $\ket{J_\text{DI};p,q}_\text{DI}$ are determined by quasiparticles $(J_\text{DI}, p)$ and $(J_\text{DI}, q)$ therein, to measure the elementary excitation states, we only need to detect the quasiparticles in the plaquettes.

The measurement operators $\Pi_P^{J_\text{DI}, p}$ of quasiparticles $(J_\text{DI},p)$ in plaquette $P$ is a linear composition of local operators
\begin{equation}
\Pi_P^{J_\text{DI},p} := \sum_{su}\pi_{psu}^{J_\text{DI}}B_P^{pspu}.
\label{eq:measure:di}
\end{equation}
Here the coefficients $\pi_{psu}^{J_\text{DI}}$ satisfy
\begin{equation}
\frac{\pi_{psu}^{J_\text{DI}}}{\pi_{p1p}^{J_\text{DI}}} = \frac{d_sd_u}{d_p}z_{pps}^{J_\text{DI};u}
\end{equation}
where $\pi_{p1p}^J$ is a normalization factor, such that
\begin{equation}
\Pi_P^{J_\text{DI},p}\EqMeasure = \delta_{p,p'}\delta_{J_\text{DI},J_\text{DI}'}\EqMeasure.
\end{equation}

The measurement operators $\Pi_P^{J_\text{DI}}$ of anyon species $J_\text{DI}$ are thus
\begin{equation}
\Pi_P^{J_\text{DI}} = \sum_{p\in J_\text{DI}}\Pi_P^{J_\text{DI}, p},
\end{equation}
where $p\in J_\text{DI}$ are the charges of $J_\text{DI}$ anyons, i.e., there exist $q, s, u\in L_\text{DI}$, such that $z_{pqs}^{J_\text{DI}, u}\ne 0$.

% End Measure DI

\subsection{Local operators and measurement operators in the toric code domain and the gapped domain wall}\label{appendix:measure:ours}

The local operators in our model are projected from the local operators \eqref{eq:localoperator:di} in the doubled Ising phase, while the measurement operators of quasiparticles in our model are projected from the doubled-Ising measurement operators \eqref{eq:measure:di}. 

There are four local operators acting on the TC plaquette $P$:
\begin{equation}
P_\text{eff}B^{1111}_PP_\text{eff},\quad\quad P_\text{eff}B^{1\psi 1\psi}_PP_\text{eff},\quad\quad P_\text{eff}B^{\psi 1\psi\psi}_PP_\text{eff},\quad\quad P_\text{eff}B^{\psi\psi\psi 1}_PP_\text{eff},
\end{equation}
which comprise the measurement operators of the four quasiparticles $(J_\text{TC},p)$ in the toric code domain
\begin{align}
\Pi^{1,1}_P &= \frac{1}{2}P_\text{eff}\left(B^{1111}_P + B^{1\psi 1\psi}_P \right)P_\text{eff},\nonumber\\ 
\Pi^{m,1}_P &= \frac{1}{2}P_\text{eff}\left(B^{1111}_P - B^{1\psi 1\psi}_P \right)P_\text{eff},\nonumber\\ 
\Pi^{e,\psi}_P &= \frac{1}{2}P_\text{eff}\left(B^{\psi 1\psi\psi}_P + B^{\psi\psi\psi 1}_P\right)P_\text{eff},\nonumber\\
\Pi^{\epsilon,\psi}_P &= \frac{1}{2}P_\text{eff}\left(B^{\psi 1\psi\psi}_P - B^{\psi\psi\psi 1}\right)_PP_\text{eff}.
\label{eq:measure:tc}
\end{align}

In the gapped domain wall, since $s$ is restricted to $L_\text{TC}$, there are $6$ local operators:
\begin{align}
&P_\text{eff}B^{1111}_PP_\text{eff},\quad\quad\quad\quad P_\text{eff}B^{1\psi 1\psi}_PP_\text{eff},\quad\quad\quad\quad P_\text{eff}B^{\psi 1\psi\psi}_PP_\text{eff},\nonumber\\ &P_\text{eff}B^{\psi\psi\psi 1}_PP_\text{eff},\quad\quad\quad\quad P_\text{eff}B^{\sigma 1\sigma \sigma}_PP_\text{eff},\quad\quad\quad\quad P_\text{eff}B^{\sigma\psi\sigma \sigma}_PP_\text{eff}.
\end{align}
They comprise the measurement operators of the domainwall quasiparticles $(J_\text{DW},p)$.
\begin{align}
\Pi^{1,1}_P &= \frac{1}{2}P_\text{eff}\left(B^{1111}_P + B^{1\psi 1\psi}_P \right)P_\text{eff},\nonumber\\ \Pi^{m,1}_P &= \frac{1}{2}P_\text{eff}\left(B^{1111}_P - B^{1\psi 1\psi}_P \right)P_\text{eff},\nonumber\\ 
\Pi^{e,\psi}_P &= \frac{1}{2}P_\text{eff}\left(B^{\psi 1\psi\psi}_P + B^{\psi\psi\psi 1}_P\right)P_\text{eff},\nonumber\\ \Pi^{\epsilon,\psi}_P &= \frac{1}{2}P_\text{eff}\left(B^{\psi 1\psi\psi}_P - B^{\psi\psi\psi 1}_P\right)P_\text{eff},\nonumber\\
\Pi^{\chi,\sigma}_P &= \frac{\sqrt{2}}{2}P_\text{eff}\left(B^{\sigma 1\sigma\sigma}_P + i B^{\sigma\psi\sigma\sigma}_P\right)P_\text{eff},\nonumber\\
\Pi^{\bar\chi,\sigma}_P &= \frac{\sqrt{2}}{2}P_\text{eff}\left(B^{\sigma 1\sigma\sigma}_P - i B^{\sigma\psi\sigma \sigma}_P\right)P_\text{eff}.
\label{eq:measure:dw}
\end{align}

Using the measurement operators \eqref{eq:measure:di}, \eqref{eq:measure:tc} and \eqref{eq:measure:dw} in different areas of our model, we can measure the quasiparticle species of our model. See Table \ref{table:quasiparticle:di}, \ref{table:quasiparticle:tc}, \ref{table:quasiparticle:dw}, \ref{table:quasiparticle:didw} and \ref{table:quasiparticle:tcdw}.

% End DW/TC Measurements

% End Measurement

\section{The algebra of the noncontractible loop operators}\label{appendix:loop}

\subsection{The multiplications of noncontractible loop operators}\label{appendix:loop:algebra}

The loop operators $W_H^{J_\text{DI}-J_\text{TC}}$ \eqref{eq:loop:h} and $W_V^{J_\text{DW}}$ \eqref{eq:loop:v} generate a $36$-dimensional algebra $\mathcal{A}$. Here, we list the multiplications of these loop operators, which completely determine this algebra.

The six $H$-loop operators $W_H^{J_\text{DI}-J_\text{TC}}$ are commutative:
\begin{align}
    &\left(W_H^{1\bar{\psi}-\epsilon}\right)^2 = \left(W_H^{\psi\bar{1}-\epsilon}\right)^2 = W_H^{1\bar{1}-1} = I,\quad\quad\quad\quad W_H^{\psi\bar{1}-\epsilon}W_H^{1\bar{\psi}-\epsilon} = W_H^{\psi\bar\psi-1},\nonumber\\ 
    & W_H^{\psi\bar{1}-\epsilon} W_H^{\sigma\bar{\sigma}-e} = W_H^{1\bar{\psi}-\epsilon} W_H^{\sigma\bar{\sigma}-e} = W_H^{\sigma\bar{\sigma}-m},\quad\quad \left(W_H^{\sigma\bar{\sigma}-e}\right)^2 = \frac{W_H^{1\bar{1}-1} + W_H^{\psi\bar\psi-1}}{2}.
\end{align}

The six $V$-loop operators $W_V^{J_\text{DW}}$ along the gapped domain wall are also commutative:
\begin{align}
    &\left(W_V^{{e}}\right)^2 = \left(W_V^{{m}}\right)^2 = W_V^{1} = I,\quad\quad W_V^{{e}}W_V^{{m}} = W_V^{{\epsilon}},\nonumber\\
    & W_V^{{e}}W_V^{\chi} = W_V^{{m}}W_V^{\chi} = W_V^{\bar{\chi}},\quad\quad \left(W_V^{\chi}\right)^2 = W_V^{1} + W_D^{{\epsilon}}.
\end{align}

Multiplying $W_V^{J_\text{DW}}$ and $W_H^{J_\text{DI}-J_\text{TC}}$ generate the additional $25$ linearly independent symmetry operators.

\begin{equation*}
    W_H^{\sigma\bar{\sigma}\dash m}W_V^{{m}} = W_V^{{m}}W_H^{\sigma\bar{\sigma}\dash m},
\end{equation*}
\begin{equation*}
    W_H^{\sigma\bar{\sigma}\dash e}W_V^{{e}} = W_V^{{e}}W_H^{\sigma\bar{\sigma}\dash e},
\end{equation*}
\begin{equation*}
    W_H^{\sigma\bar{\sigma}\dash e}W_V^{{m}} = -W_V^{{m}}W_H^{\sigma\bar{\sigma}\dash e},
\end{equation*}
\begin{equation*}
    W_H^{\sigma\bar{\sigma}\dash m}W_V^{{e}} = -W_V^{{e}}W_H^{\sigma\bar{\sigma}\dash m},
\end{equation*}
\begin{equation*}
    W_H^{\sigma\bar{\sigma}\dash e}W_V^{{\epsilon}} = -W_V^{{\epsilon}}W_H^{\sigma\bar{\sigma}\dash e},
\end{equation*}
\begin{equation*}
    W_H^{\sigma\bar{\sigma}\dash m}W_V^{{\epsilon}} = -W_V^{{\epsilon}}W_H^{\sigma\bar{\sigma}\dash m},
\end{equation*}
\begin{equation*}
    W_H^{\psi\bar\psi\dash 1}W_V^{\chi} = -W_V^{\chi}W_H^{\psi\bar\psi\dash 1},
\end{equation*}
\begin{equation*}
    W_H^{\psi\bar\psi\dash 1}W_V^{\bar{\chi}} = -W_V^{\bar{\chi}}W_H^{\psi\bar\psi\dash 1},
\end{equation*}
\begin{equation*}
    W_H^{\psi\bar\psi\dash 1}W_V^{{m}} = W_V^{{m}}W_H^{\psi\bar\psi\dash 1},
\end{equation*}
\begin{equation*}
    W_H^{\psi\bar{1}\dash\epsilon}W_V^{{\epsilon}} = W_V^{{\epsilon}}W_H^{\psi\bar{1}\dash\epsilon},
\end{equation*}
\begin{equation*}
    W_H^{\psi\bar{1}\dash\epsilon}W_V^{\chi} = -W_V^{\chi}W_H^{\psi\bar{1}\dash\epsilon},
\end{equation*}
\begin{equation*}
    W_H^{\psi\bar{1}\dash\epsilon}W_V^{\bar{\chi}} = -W_V^{\bar{\chi}}W_H^{\psi\bar{1}\dash\epsilon},
\end{equation*}
\begin{equation*}
    W_H^{1\bar{\psi}\dash\epsilon}W_V^{\chi} = -W_V^{\chi}W_H^{1\bar{\psi}\dash\epsilon},
\end{equation*}
\begin{equation*}
    W_H^{1\bar{\psi}\dash\epsilon}W_V^{\bar{\chi}} = -W_V^{\bar{\chi}}W_H^{1\bar{\psi}\dash\epsilon},
\end{equation*}
\begin{equation*}
    W_H^{\psi\bar{1}\dash\epsilon}W_V^{{m}} = -W_V^{{m}}W_H^{\psi\bar{1}\dash\epsilon},
\end{equation*}
\begin{equation*}
    W_H^{1\bar{\psi}\dash\epsilon}W_V^{{m}} = -W_V^{{m}}W_H^{1\bar{\psi}\dash\epsilon},
\end{equation*}
\begin{equation*}
    W_H^{\psi\bar{1}\dash\epsilon}W_V^{{e}} = -W_V^{{e}}W_H^{\psi\bar{1}\dash\epsilon},
\end{equation*}
\begin{equation*}
    W_H^{\sigma\bar{\sigma}\dash m}W_V^{\chi},
\end{equation*}
\begin{equation*}
    W_V^{\chi}W_H^{\sigma\bar{\sigma}\dash m},
\end{equation*}
\begin{equation*}
    W_H^{\sigma\bar{\sigma}\dash e}W_V^{\chi},
\end{equation*}
\begin{equation*}
    W_V^{\chi}W_H^{\sigma\bar{\sigma}\dash e},
\end{equation*}
\begin{equation*}
    W_H^{\sigma\bar{\sigma}\dash m}W_V^{\bar{\chi}},
\end{equation*}
\begin{equation*}
    W_V^{\bar{\chi}}W_H^{\sigma\bar{\sigma}\dash m},
\end{equation*}
\begin{equation*}
    W_H^{\sigma\bar{\sigma}\dash e}W_V^{\bar{\chi}},
\end{equation*}
\begin{equation}
    W_D^{\bar{\chi}}W_H^{\sigma\bar{\sigma}\dash e}.
\label{eq:algebra}
\end{equation}

All other multiplications of operators are not linearly
independent:
\begin{equation*}
    W_H^{\psi\bar\psi-1}W_V^{{e}} = W_V^{{e}}W_H^{\psi\bar\psi-1} = W_H^{\psi\bar\psi-1}W_V^{{m}} + W_V^{{e}} - W_V^{{m}},
\end{equation*}
\begin{equation*}
    W_H^{1\bar{\psi}-\epsilon}W_V^{{\epsilon}} = W_V^{{\epsilon}}W_H^{1\bar{\psi}-\epsilon} = W_H^{\psi\bar{1}-\epsilon}W_V^{{\epsilon}} + W_H^{1\bar{\psi}-\epsilon} - W_H^{\psi\bar{1}-\epsilon},
\end{equation*}
\begin{equation*}
    W_H^{1\bar{\psi}-\epsilon}W_V^{{e}} = - W_V^{{e}}W_H^{1\bar{\psi}-\epsilon} = W_H^{\psi\bar{1}-\epsilon}W_V^{{e}} + W_H^{1\bar{\psi}-\epsilon}W_V^{{e}} - W_H^{\psi\bar{1}-\epsilon}W_V^{{m}},
\end{equation*}
\begin{equation}
    W_V^{\chi}W_H^{\sigma\bar{\sigma}-e}W_V^{\chi} = W_V^{\chi}W_H^{\sigma\bar{\sigma}-m}W_V^{\chi} = 0.
\end{equation}

% End Algebra

\subsection{Generating the entire ground-state subspace}\label{appendix:loop:complete}

Finally, we prove Eq. \eqref{eq:complete}: For any given two ground states $\ket\Phi\in\Hil_\text{eff}$ and $\ket{\Phi'}\in\Hil_\text{eff}$ of our model on the torus, there exists an operator $W\in\mathcal{A}$, such that
\begin{equation}
    \ket{\Phi'} = W\ket\Phi.
\end{equation}

Since the ribbon operators in our model are projected from the doubled-Ising ribbon operators in the doubled Ising phase, the doubled-Ising loop operators in our model on the torus --- the special cases of ribbon operators --- are also projections of the doubled-Ising loop operators along the same paths:
\begin{align}
    &W_V^{1} = P_\text{eff}{W}_V^{1\bar{1}}P_\text{eff} = P_\text{eff}{W}_V^{\psi\bar\psi}P_\text{eff},\nonumber\\
    &W_V^{{\epsilon}} = P_\text{eff}{W}_V^{\psi\bar{1}} P_\text{eff}= P_\text{eff}{W}_V^{1\bar{\psi}}P_\text{eff},\nonumber\\
    &W_V^{m} + W_V^{e} = P_\text{eff}{W}_V^{\sigma\bar{\sigma}}P_\text{eff},\nonumber\\
    &W_V^{{\chi}} = P_\text{eff}{W}_V^{\sigma\bar{1}}P_\text{eff} = P_\text{eff}{W}_V^{\sigma\bar{\psi}}P_\text{eff},\nonumber\\
    &W_V^{{\bar\chi}} = P_\text{eff}{W}_V^{1\bar\sigma}P_\text{eff} = P_\text{eff}{W}_V^{\psi\bar\sigma}P_\text{eff},\nonumber\\
    &W_H^{1\bar 1-1} = P_\text{eff}{W}_H^{1\bar{1}}P_\text{eff},\nonumber\\
    &W_H^{{\psi\bar\psi}} = P_\text{eff}{W}_H^{\psi\bar{\psi}}P_\text{eff},\nonumber\\
    &W_H^{\sigma\bar\sigma-m} + W_H^{\sigma\bar\sigma-e} = P_\text{eff}{W}_H^{\sigma\bar{\sigma}}P_\text{eff},\nonumber\\
    &W_H^{{\psi\bar 1-\epsilon}} = P_\text{eff}{W}_H^{\psi\bar{1}}P_\text{eff},\nonumber\\
    &W_H^{{1\bar\psi-\epsilon}} = P_\text{eff}{W}_H^{1\bar\psi}P_\text{eff}.
\end{align}
Therefore, the algebra $\mathcal{A}$ in our model satisfies
\begin{equation}
    \mathcal{A} = P_\text{eff}\mathcal{A}^\text{DI}P_\text{eff},
\end{equation}
where the algebra $\mathcal{A}^\text{DI}$ is generated by all noncontractible loop operators ${W}_D^{J_\text{DI}}, {W}_H^{J_\text{DI}}$ in the doubled Ising phase on the torus along $H$-loop and $V$-loop.

On the other hand, in the doubled Ising phase, the projector
\begin{equation}
    P_0^\text{DI} = \prod_PB_P^\text{DI}\hs\prod_VQ_V
\end{equation}
projects the total Hilbert space $\Hil$ to the doubled-Ising ground-state subspace $\Hil_0^\text{DI}$:
\begin{equation}
    P_0^\text{DI}\mathcal{H} = \Hil_0^\text{DI}.
\end{equation}

In our model, the projector
\begin{equation}
    P_0 = \prod_{P\in\text{DI}}B_P^\text{DI}\hs\prod_{P\in\text{DW}}B_P^\text{DW}\hs\prod_{P\in\text{TC}}B_P^\text{TC}\hs\prod_VQ_V
\end{equation}
projectes the effecitve Hilbert space $\Hil_\text{eff}$ to the ground-state subspace $\Hil_0$. Note that $P_0 = P_\text{eff}P_0^\text{DI}P_\text{eff}$, the projector $P_\text{eff}$ projects $\Hil_0^\text{DI}$ to $\Hil_0$:
\begin{equation}
    P_\text{eff}\Hil_0^\text{DI} = P_\text{eff}\Hil_0^\text{DI}.
\end{equation}

Therefore, for any two ground states $\ket\Phi$ and $\ket{\Phi'}$ of our model, there exist doubled-Ising ground states $\ket\Phi_\text{DI}$ and $\ket{\Phi'}_\text{DI}$, such that
\begin{equation}
    \ket\Phi = \ket\Phi_\text{DI},\quad\quad \ket{\Phi'} = \ket{\Phi'}_\text{DI}.
\end{equation}
Since $\Hil_0^\text{DI}$ is generated by the algebra $\mathcal{A}^\text{DI}$ given any doubled-Ising ground state $\ket\Phi_\text{DI}$, there exists a doubled-Ising operator $W^\text{DI}\in\mathcal{A}^\text{DI}$, such that
\begin{equation}
    \ket{\Phi'}_\text{DI} = W^\text{DI}\ket{\Phi}_\text{DI}.
\end{equation}
As their projections, 
\begin{equation}
    \ket{\Phi'} = \left(P_\text{eff}W^\text{DI}P_\text{eff}\right)\ket{\Phi}_\text{DI},
\end{equation}
where $P_\text{eff}W^\text{DI}P_\text{eff}\in\mathcal{A}$. Therefore, the algebra $\mathcal{A}$ generates the entire ground-state subspace $\Hil_0$ of our model given any ground state $\ket\Phi$.

\end{appendix}

% TODO:
% Provide your bibliography here. You have two options:

% FIRST OPTION - write your entries here directly, following the example below, including Author(s), Title, Journal Ref. with year in parentheses at the end, followed by the DOI number.
%\begin{thebibliography}{99}
%\bibitem{1931_Bethe_ZP_71} H. A. Bethe, {\it Zur Theorie der Metalle. i. Eigenwerte und Eigenfunktionen der linearen Atomkette}, Zeit. f{\"u}r Phys. {\bf 71}, 205 (1931), \doi{10.1007\%2FBF01341708}.
%\bibitem{arXiv:1108.2700} P. Ginsparg, {\it It was twenty years ago today... }, \url{http://arxiv.org/abs/1108.2700}.
%\end{thebibliography}

% \bibliographystyle{SciPost_bibstyle} % Include this style file here only if you are not using our template
\bibliography{StringNet.bib}

\begin{thebibliography}{10}
\providecommand{\url}[1]{\texttt{#1}}
\providecommand{\urlprefix}{URL }
\expandafter\ifx\csname urlstyle\endcsname\relax
  \providecommand{\doi}[1]{doi:\discretionary{}{}{}#1}\else
  \providecommand{\doi}{doi:\discretionary{}{}{}\begingroup
  \urlstyle{rm}\Url}\fi
\providecommand{\eprint}[2][]{\url{#2}}

\bibitem{Kitaev2003a}
A.~Kitaev,
\newblock \emph{{Fault-tolerant quantum computation by anyons}},
\newblock Annals of Physics \textbf{303}(1), 2 (2003),
\newblock \doi{10.1016/S0003-4916(02)00018-0}.

\bibitem{Levin2004}
M.~Levin and X.-g. Wen,
\newblock \emph{{String-net condensation: A physical mechanism for topological
  phases}},
\newblock Physical Review B \textbf{71}(4), 21 (2005),
\newblock \doi{10.1103/PhysRevB.71.045110},
\newblock \eprint{0404617}.

\bibitem{Hu2012a}
Y.~Hu, Y.~Wan and Y.-S. Wu,
\newblock \emph{{Twisted quantum double model of topological phases in two
  dimensions}},
\newblock Physical Review B \textbf{87}(12), 125114 (2013),
\newblock \doi{10.1103/PhysRevB.87.125114},
\newblock \eprint{1211.3695}.

\bibitem{Huang2005}
Y.-Z. Huang,
\newblock \emph{{Vertex operator algebras, fusion rules and modular
  transformations}},
\newblock Contemporary Mathematics  (2005),
\newblock \eprint{0502558}.

\bibitem{Nayak2008}
C.~Nayak, A.~Stern, M.~Freedman and S.~{Das Sarma},
\newblock \emph{{Non-Abelian anyons and topological quantum computation}},
\newblock Reviews of Modern Physics \textbf{80}(3), 1083 (2008),
\newblock \doi{10.1103/RevModPhys.80.1083}.

\bibitem{Zhang2012}
Y.~Zhang, T.~Grover, A.~Turner, M.~Oshikawa and A.~Vishwanath,
\newblock \emph{{Quasiparticle statistics and braiding from ground-state
  entanglement}},
\newblock Physical Review B \textbf{85}(23), 235151 (2012),
\newblock \doi{10.1103/PhysRevB.85.235151}.

\bibitem{Barkeshli2013e}
M.~Barkeshli, C.-M. Jian and X.-L. Qi,
\newblock \emph{{Twist defects and projective non-Abelian braiding
  statistics}},
\newblock Physical Review B \textbf{87}(4), 045130 (2013),
\newblock \doi{10.1103/PhysRevB.87.045130}.

\bibitem{Teo2013a}
J.~C.~Y. Teo, A.~Roy and X.~Chen,
\newblock \emph{{Braiding statistics and congruent invariance of twist defects
  in bosonic bilayer fractional quantum Hall states}},
\newblock Physical Review B \textbf{90}(15), 155111 (2014),
\newblock \doi{10.1103/PhysRevB.90.155111},
\newblock \eprint{1308.5984}.

\bibitem{liu2013}
F.~Liu, Z.~Wang, Y.-Z. You and X.-G. Wen,
\newblock \emph{Modular transformations and topological orders in two
  dimensions},
\newblock arXiv preprint arXiv:1303.0829  (2013),
\newblock \doi{10.48550/arXiv.1303.0829}.

\bibitem{Cincio2013}
L.~Cincio and G.~Vidal,
\newblock \emph{{Characterizing topological order by studying the ground states
  on an infinite cylinder}},
\newblock Physical Review Letters \textbf{110}(6), 1 (2013),
\newblock \doi{10.1103/PhysRevLett.110.067208},
\newblock \eprint{1208.2623}.

\bibitem{Jiang2014a}
S.~Jiang, A.~Mesaros and Y.~Ran,
\newblock \emph{{Generalized modular transformations in 3+1D topologically
  ordered phases and triple linking invariant of loop braiding}},
\newblock Physical Review X  (2014),
\newblock \doi{10.1103/PhysRevX.4.031048},
\newblock \eprint{1404.1062}.

\bibitem{Bravyi1998}
S.~B. Bravyi and A.~Y. Kitaev,
\newblock \emph{{Quantum codes on a lattice with boundary}},
\newblock arXiv preprint quant-ph/9811052  (1998),
\newblock \eprint{9811052}.

\bibitem{Buerschaper2009}
O.~Buerschaper and M.~Aguado,
\newblock \emph{{Mapping Kitaev's quantum double lattice models to Levin and
  Wen's string-net models}},
\newblock Physical Review B \textbf{80}(15), 155136 (2009),
\newblock \doi{10.1103/PhysRevB.80.155136},
\newblock \eprint{arXiv:0907.2670v2}.

\bibitem{Hung2012}
L.-Y. Hung and Y.~Wan,
\newblock \emph{{String-net models with Z{\_}{\{}N{\}} fusion algebra}},
\newblock Physical Review B \textbf{86}(23), 235132 (2012),
\newblock \doi{10.1103/PhysRevB.86.235132},
\newblock \eprint{1207.6169}.

\bibitem{Hu2012}
Y.~Hu, S.~D. Stirling and Y.-s. Wu,
\newblock \emph{{Ground State Degeneracy in the Levin-Wen Model for Topological
  Phases}},
\newblock Physical Review B \textbf{85}(7), 075107 (2011),
\newblock \doi{10.1103/PhysRevB.85.075107},
\newblock \eprint{1105.5771}.

\bibitem{Buerschaper2013}
O.~Buerschaper, M.~Christandl, L.~Kong and M.~Aguado,
\newblock \emph{{Electric–magnetic duality of lattice systems with
  topological order}},
\newblock Nuclear Physics B \textbf{876}(2), 619 (2013),
\newblock \doi{10.1016/j.nuclphysb.2013.08.014},
\newblock \eprint{1006.5823}.

\bibitem{schulz2013}
M.~D. Schulz, S.~Dusuel, K.~P. Schmidt and J.~Vidal,
\newblock \emph{Topological phase transitions in the golden string-net model},
\newblock Physical review letters \textbf{110}(14), 147203 (2013),
\newblock \doi{10.1103/PhysRevLett.110.147203}.

\bibitem{Lan2014b}
T.~Lan and X.~G. Wen,
\newblock \emph{{Topological quasiparticles and the holographic bulk-edge
  relation in (2+1) -dimensional string-net models}},
\newblock Physical Review B - Condensed Matter and Materials Physics
  \textbf{90}(11), 1 (2014),
\newblock \doi{10.1103/PhysRevB.90.115119},
\newblock \eprint{1311.1784}.

\bibitem{Hu2017}
Y.~Hu, Y.~Wan and Y.-s. Wu,
\newblock \emph{{Boundary Hamiltonian theory for gapped topological orders}},
\newblock Chinese Physics Letters \textbf{34}(7), 077103 (2017),
\newblock \doi{10.1088/0256-307X/34/7/077103},
\newblock \eprint{1706.00650}.

\bibitem{Hu2018}
Y.~Hu, N.~Geer and Y.-S. Wu,
\newblock \emph{{Full dyon excitation spectrum in extended Levin-Wen models}},
\newblock Physical Review B \textbf{97}(19), 195154 (2018),
\newblock \doi{10.1103/PhysRevB.97.195154}.

\bibitem{cheipesh2019}
Y.~Cheipesh, L.~Cevolani and S.~Kehrein,
\newblock \emph{Exact description of the boundary theory of the kitaev toric
  code with open boundary conditions},
\newblock Physical Review B \textbf{99}(2), 024422 (2019).

\bibitem{Wang2020}
H.~Wang, Y.~Li, Y.~Hu and Y.~Wan,
\newblock \emph{{Electric-magnetic duality in the quantum double models of
  topological orders with gapped boundaries}},
\newblock Journal of High Energy Physics \textbf{2020}(2), 30 (2020),
\newblock \doi{10.1007/JHEP02(2020)030}.

\bibitem{Wang2022}
H.~Wang, Y.~Hu and Y.~Wan,
\newblock \emph{{Extend The Levin-Wen Model To Two-dimensional Topological
  Orders With Gapped Boundary Junctions}},
\newblock arXiv preprint arXiv:2201.04072  (2022),
\newblock \eprint{2201.04072}.

\bibitem{Kitaev2012}
A.~Kitaev and L.~Kong,
\newblock \emph{{Models for Gapped Boundaries and Domain Walls}},
\newblock Communications in Mathematical Physics \textbf{313}(2), 351 (2012),
\newblock \doi{10.1007/s00220-012-1500-5}.

\bibitem{Lan2014}
T.~Lan, J.~C. Wang and X.-G. Wen,
\newblock \emph{{Gapped Domain Walls, Gapped Boundaries, and Topological
  Degeneracy}},
\newblock Physical Review Letters \textbf{114}, 9 (2015),
\newblock \doi{10.1103/PhysRevLett.114.076402},
\newblock \eprint{1408.6514}.

\bibitem{Bais2009a}
F.~A. Bais and J.~Slingerland,
\newblock \emph{{Condensate-induced transitions between topologically ordered
  phases}},
\newblock Physical Review B \textbf{79}(4), 045316 (2009),
\newblock \doi{10.1103/PhysRevB.79.045316}.

\bibitem{Barkeshli2010}
M.~Barkeshli and X.-g. Wen,
\newblock \emph{{Anyon Condensation and Continuous Topological Phase
  Transitions in Non-Abelian Fractional Quantum Hall States}},
\newblock Physical Review Letters \textbf{105}(21), 216804 (2010),
\newblock \doi{10.1103/PhysRevLett.105.216804}.

\bibitem{Burnell2012}
F.~J. Burnell, S.~H. Simon and J.~K. Slingerland,
\newblock \emph{{Phase transitions in topological lattice models via
  topological symmetry breaking}},
\newblock New Journal of Physics \textbf{14}(1), 015004 (2012),
\newblock \doi{10.1088/1367-2630/14/1/015004}.

\bibitem{Barkeshli2013a}
M.~Barkeshli,
\newblock \emph{{Transitions Between Chiral Spin Liquids and Z2 Spin Liquids}},
\newblock arXiv preprint arXiv:1307.8194  (2013),
\newblock \eprint{1307.8194}.

\bibitem{Eliens2013}
I.~S. Eli{\"{e}}ns, J.~C. Romers and F.~A. Bais,
\newblock \emph{{Diagrammatics for Bose condensation in anyon theories}},
\newblock Physical Review B \textbf{90}(19), 195130 (2014),
\newblock \doi{10.1103/PhysRevB.90.195130},
\newblock \eprint{1310.6001}.

\bibitem{Chen2013}
X.~Chen, F.~Wang, Y.-m. Lu and D.-h. Lee,
\newblock \emph{{Critical theories of phase transition between symmetry
  protected topological states and their relation to the gapless boundary
  theories}},
\newblock Nuclear Physics B  (2013),
\newblock \doi{10.1016/j.nuclphysb.2013.04.015},
\newblock \eprint{1302.3121}.

\bibitem{Gu2014a}
Y.~Gu, L.-Y. Hung and Y.~Wan,
\newblock \emph{{Unified Framework of Topological Phases with Symmetry}},
\newblock Phys. Rev. B \textbf{90}, 245125 (2014),
\newblock \doi{10.1103/PhysRevB.90.245125},
\newblock \eprint{1402.3356}.

\bibitem{HungWan2015a}
L.-Y. Hung and Y.~Wan,
\newblock \emph{{Generalized ADE Classification of Gapped Domain Walls}},
\newblock Journal of High Energy Physics \textbf{1507}, 120 (2015),
\newblock \doi{10.1007/JHEP07(2015)120}.

\bibitem{Ji2019}
W.~Ji and X.-G. Wen,
\newblock \emph{{Categorical symmetry and non-invertible anomaly in
  symmetry-breaking and topological phase transitions}},
\newblock Physical Review Research  (2019),
\newblock \doi{10.1103/PhysRevResearch.2.033417},
\newblock \eprint{1912.13492}.

\bibitem{Hu2021}
Y.~Hu, Z.~Huang, L.-y. Hung and Y.~Wan,
\newblock \emph{{Anyon condensation: coherent states, symmetry enriched
  topological phases, Goldstone theorem, and dynamical rearrangement of
  symmetry}},
\newblock Journal of High Energy Physics \textbf{2022}(3), 26 (2022),
\newblock \doi{10.1007/JHEP03(2022)026},
\newblock \eprint{2109.06145v1}.

\bibitem{Hung2013}
L.-Y. Hung and Y.~Wan,
\newblock \emph{{Symmetry-enriched phases obtained via pseudo anyon
  condensation}},
\newblock International Journal of Modern Physics B \textbf{28}(24), 1450172
  (2014),
\newblock \doi{10.1142/S0217979214501720}.

\bibitem{Bais2009}
F.~A. Bais, J.~Slingerland and S.~Haaker,
\newblock \emph{{Theory of Topological Edges and Domain Walls}},
\newblock Physical Review Letters \textbf{102}(22), 220403 (2009),
\newblock \doi{10.1103/PhysRevLett.102.220403},
\newblock \eprint{arXiv:0812.4596v1}.

\end{thebibliography}
\nolinenumbers

\end{document}